\documentclass[11pt,a4paper,twoside,openright]{book}
\usepackage{mythesis}

\usepackage{amsmath, amsthm, amsfonts, amssymb, latexsym}
\usepackage{graphicx}
\usepackage{color}
\usepackage{appendix}

\usepackage{caption}
\usepackage[perpage,symbol*]{footmisc}

\usepackage{paralist}

\usepackage{epigraph}

\usepackage{url}

\usepackage[pagebackref,unicode]{hyperref}
\hypersetup{
    unicode=true,          
    pdftitle={Radiation from a $D$-dimensional collision of gravitational shock waves},
    pdfauthor={Fl\'avio de Sousa Coelho},
    pdfsubject={PhD Thesis},
    pdfcreator={LaTeX},          
    pdfproducer={LaTeX},         
    pdfkeywords={General Relativity, Gravitational Radiation, Black Holes, Higher Dimensions, Shock Waves},
    colorlinks=false,            
    bookmarks=true,              
    linktoc=page,                
  }
\renewcommand*{\backref}[1]{}
\renewcommand*{\backrefalt}[4]{%
  \ifcase #1 { } \or (Cited on page~#2.) \else %
  (Cited on pages~#2.) \fi %
}

\let\origdoublepage\cleardoublepage
\newcommand{\clearemptydoublepage}{%
  \clearpage
  {\pagestyle{empty}\origdoublepage}%
}

\let\cleardoublepage\clearemptydoublepage

\title{
Radiation from a $D$-dimensional collision of gravitational shock waves}
\author{Fl\'avio de Sousa Coelho}
\date{Novembro de 2014}

\submitionplace{
Disserta\c{c}\~ao apresentada \`a Universidade de Aveiro no \^ambito do Programa Doutoral MAP-fis para
  cumprimento dos requisitos necess\'arios \`a obten\c{c}\~ao do grau de Doutor em F\'isica,
  realizada sob a orienta\c{c}\~ao cient\'ifica do Doutor Carlos Herdeiro, Professor Auxiliar com Agrega\c{c}\~ao do
  Departamento de F\'isica da Universidade de Aveiro.}

\begin{document}

\pagestyle{plain}

\coverp
\titlep
\dedication
\jury


\chapter*{Agradecimentos}

\epigraph{Live as if you were to die tomorrow.\\
  Learn as if you were to live forever.}{Mahatma Gandhi}

A minha aventura com a F\'isica come\c{cou} em 2005 com a minha participa\c{c}\~ao numa s\'erie de eventos realizados no \^ambito do Ano Internacional da F\'isica, em especial nas Olimp\'iadas de F\'isica. Tal n\~ao teria sido poss\'ivel sem o incentivo e entusiasmo do professor Carlos Azevedo, bem como o apoio dos meus colegas de turma.

No ano seguinte tive o privil\'egio de participar naquilo que se viria a chamar projecto Quark!, no Departamento de F\'isica da Universidade de Coimbra, e de representar Portugal na XXVII Olimp\'iada Internacional de F\'isica (IPhO) 2006 em Singapura. N\~ao posso, pois, deixar de agradecer a toda a equipa, em especial aos Profs. Jos\'e Ant\'onio Paix\~ao e Fernando Nogueira, bem como aos meus colegas ol\'impicos, por essa experi\^encia inesquec\'ivel.

De 2006 a 2009 frequentei o curso de licenciatura em F\'isica na Faculdade de Ci\^encias da Universidade do Porto. Agrade\c{c}o a todos os meus colegas e amigos pela dimens\~ao humana que trouxeram a esse per\'iodo. Um especial obrigado \`aqueles que me acompanharam na direc\c{c}\~ao da Physis, bem como ao Departamento de F\'isica e Astronomia pela ced\^encia de espa\c{c}os e apoio na realiza\c{c}\~ao de eventos.

Ap\'os um ano de estudo intensivo na Universidade da Cambridge, tive o privil\'egio de ensinar na Universidade de Catemandu no Nepal. Agrade\c{c}o ao Dr Dipak Subedi pela oportunidade e hospitalidade, e aos meus alunos pelo que me ensinaram. 

Oportunidades de viagem e enriquecimento, ali\'as, n\~ao faltaram durante o meu doutoramento. Agrade\c{c}o ao Shinji Hirano e ao Yuki Sato pela sua hospitalidade nas Universidades de Nagoya (Jap\~ao) e Witwatersrand (Joanesburgo, \'Africa do Sul), e pela colabora\c{c}\~ao e amizade que fomos construindo. Tamb\'em ao Luis Crispino por ter sempre abertas as portas da Amaz\'onia em Bel\'em do Par\'a (Brasil).

Porque tudo isto custa dinheiro, foi indispens\'avel o apoio financeiro de diversas institui\c{c}\~oes: a Funda\c{c}\~ao para a Ci\^encia e a Tecnologia, atrav\'es da Bolsa de Doutoramento SFRH/ BD/60272/2009; a Funda\c{c}\~ao Calouste Gulbenkian, atrav\'es do Pr\'emio de Est\'imulo \`a Investiga\c{c}\~ao 2012; a Funda\c{c}\~ao Luso-Americana pela bolsa \emph{`Papers'}; e o Cambridge European Trust durante o meu mestrado.

Duas pessoas contribu\'iram de forma essencial para o meu doutoramento, e para o projecto em que esta disserta\c{c}\~ao se insere. Em primeiro lugar, o meu orientador, Carlos Herdeiro, por me ter cativado para o estudo da f\'isica gravitacional, por me ter dado este projecto e por ter criado em Aveiro um grupo de refer\^encia nesta \'area do conhecimento. A independ\^encia e autonomia com que me deixou trabalhar foram muito importantes para mim. Agrade\c{c}o tamb\'em pela amizade e compreens\~ao face a outros projectos pessoais, e pelas caipirinhas que bebemos nas praias do Brasil e do M\'exico.

Em segundo lugar, o meu colega Marco Sampaio, pela dedica\c{c}\~ao e perseveran\c{c}a no desenvolvimento deste projecto, mesmo em momentos de frustra\c{c}\~ao e des\^animo (sei que ele diria o mesmo de mim). Obrigado pelas longas discuss\~oes, por verificares as minhas contas horr\'iveis, pelos momentos `Eureka!', e pelo esfor{c}o que fizeste nesta recta final, mesmo com sacrif\'icio da vida pessoal.

A ambos devo tamb\'em um agradecimento pelos gr\'aficos e ilustra\c{c}\~oes que inclu\'i nesta tese: ao Carlos, pela sua extraordin\'aria capacidade de visualizar e ilustrar estruturas multidimensionais, e ao Marco pela apresenta\c{c}\~ao colorida dos resultados num\'ericos (e pelo c\'odigo que os produziu). 

Por \'ultimo, agrade\c{c}o a todos os meus familiares e amigos, por me aturarem. Um abra\c{c}o especial para o Clube Taekwondo Little Dragon e para todos os que l\'a suam comigo.

\cleardoublepage

\chapter*{Abstract}
Classically, if two highly boosted particles collide head-on, a black hole is expected to form whose mass may be inferred from the gravitational radiation emitted during the collision. If this occurs at trans-Planckian energies, it should be well described by general relativity. Furthermore, if there exist hidden extra dimensions, the fundamental Planck mass may well be of the order of the TeV and thus achievable with current or future particle accelerators.
By modeling the colliding particles as Aichelburg-Sexl shock waves on a flat, $D$-dimensional background, we devise a perturbative framework to compute the space-time metric in the future of the collision. Then, a generalisation of Bondi's formalism is employed to extract the gravitational radiation and compute the \emph{inelasticity} of the collision: the percentage of the initial centre-of-mass energy that is radiated away. Using the axial symmetry of the problem, we show that this information is encoded in a single function of the transverse metric components - the \emph{news function}.
We then unveil a hidden conformal symmetry which exists at each order in perturbation theory and thus makes the problem effectively two-dimensional. Moreover, it allows for the factorisation of the angular dependence of the news function, i.e. the radiation pattern at null infinity, and clarifies the correspondence between the perturbative series of the metric and an angular expansion off the collision axis.
The first-order estimate, or isotropic term, is computed analytically and yields a remarkable simple formula for the inelasticity for any $D$. Higher-order terms, however, require the use of a computer for numerical integration. We study the integration domain and compute, numerically, the Green's functions and the sources, thus paving the way for the computation of the inelasticity in a future work.

{\bf Keywords:} black holes, shock waves, gravitational radiation, extra dimensions, trans-Planckian scattering.

\cleardoublepage

\chapter*{Resumo}
Classicamente, se duas part\'iculas altamente energ\'eticas colidirem frontalmente, espera-se a forma\c c\~ao de um buraco negro cuja massa pode ser inferida a partir da radia\c c\~ao gravitacional emitida durante a colis\~ao. Se isto ocorrer a energias trans-Planckianas, dever\'a ser bem descrito pela relatividade geral. Al\'em disso, se existirem dimens\~oes extra escondidas, a massa de Planck fundamental pode bem ser da ordem do TeV e portanto alcan\c c\'avel em actuais ou futuros aceleradores de part\'iculas.
Modelando essas part\'iculas como ondas de choque de Aichelburg-Sexl num fundo plano $D$-dimensional, estabelecemos um m\'etodo perturbativo para calcular a m\'etrica do espa\c co-tempo no futuro da colis\~ao. Uma generaliza\c c\~ao do formalismo de Bondi \'e ent\~ao empregue para extrair a radia\c c\~
ao gravitacional e calcular a \emph{inelasticidade} da colis\~ao: a percentagem da energia inicial do centro-de-massa que \'e radiada. Usando a simetria axial do problema, mostramos que essa informa\c c\~ao est\'a codificada numa s\'o fun\c c\~ao das componentes transversas da m\'etrica -  a `\emph{news function}'.
De seguida desvendamos uma simetria conforme que existe escondida em cada ordem da teoria de perturba\c c\~oes e assim torna o problema efectivamente bidimensional. Adicionalmente, permite a factoriza\c c\~ao da depend\^encia angular da `news function' e clarifica a correspond\^encia entre a s\'erie perturbativa da m\'etrica e uma expans\~ao angular a partir do eixo.
A estimativa de primeira ordem, ou o termo isotr\'opico, \'e calculada analiticamente e produz uma f\'ormula simples para a inelasticidade em qualquer $D$. Termos de ordem superior, no entanto, requerem o uso de um computador para integra\c c\~ao num\'erica. Estudamos o dom\'inio de integra\c{c}\~ao e calculamos, numericamente, as fun\c{c}\~oes de Green e as fontes, pavimentando assim o caminho para o c\'alculo da inelasticidade num trabalho futuro.

{\bf Palavras-chave:} buracos negros, ondas de choque, radia\c{c}\~ao gravitacional, dimens\~oes extra, difus\~ao trans-Planckiana.



\cleardoublepage
\tableofcontents
\cleardoublepage
\listoftables
\cleardoublepage
\listoffigures
\cleardoublepage

\pagestyle{headings}
\numberwithin{equation}{section}

\chapter{Introduction}
\label{ch:intro}
\renewcommand{\textflush}{flushepinormal}


\epigraph{I am going on an adventure!}{Bilbo Baggins\\\emph{The Hobbit}}

\section{Why $D$ instead of four?}
\label{sc:whyD}
The theory of general relativity, presented by Einstein almost a century ago in 1916 \cite{Einstein:1916vd}, revolutionised physics by proposing that the gravitational interaction could be described in purely geometric terms. It did not take too long for others to realise that this geometrisation of apparently non-geometric degrees of freedom could equally describe other fundamental interactions if a higher number of spatial dimensions was invoked: if the world is $D$-dimensional, with $D>4$, motion in the `extra' apparently unseen dimensions is not perceived as motion, but rather as some other degree of freedom. Since the proposals of Kaluza \cite{Kaluza:1921tu} and Klein \cite{Klein:1926tv} (preceded by Nordstr\"om \cite{Nordstrom:1914fn}), the converse idea that observed non-motion degrees of freedom may be transformed into motion if a higher-dimensional space is considered has been a powerful attractor in the quest for a unified description of fundamental interactions.

In the last few decades, the idea of extra dimensions regained popularity due to its naturalness in string theory, which is most naturally formulated in $D=26$ (bosonic), $D=10$ (perturbative superstring) or $D=11$ (non-perturbative superstring or M-theory). Since this framework promised to be a fundamental description of all interactions, and in particular a quantum theory of gravity, the study of $D$-dimensional gravity was motivated by its appearance as an effective low-energy limit of string theory.

The AdS/CFT correspondence, introduced in 1997 by Maldacena \cite{Maldacena:1997zz}, established a duality between conformal (quantum) field theories in $D$ dimensions and classical gravity in a $(D+1)$-dimensional anti de Sitter space. This gauge/gravity duality is a realisation of the more general \emph{holographic principle} originally due to t'Hooft \cite{'tHooft:1993gx} and later developed by Susskind \cite{Susskind:1994vu}. In this context, higher-dimensional gravity is motivated by what it can teach us about otherwise untreatable problems in gauge theories (both qualitatively and as a computing tool).

However, the study of general relativity regarding the number of space-time dimensions $D$ as a parameter can, by itself, be qualitatively and quantitatively informative for the understanding of our (apparently) four-dimensional world. Already fifty years ago, Tangherlini \cite{Tangherlini:1963bw} showed that no stable bound states exist in the $D>4$ Schwarzschild solution. A similar argument for solutions of the Schr\"odinger equation provides a reason why atoms or planetary systems could not exist in more than four \emph{infinite} dimensions. This is an interesting lesson, although compact dimensions (or brane constructions) can provide a way around this argument. More recently, Emparan and Reall \cite{Emparan:2001wn} demonstrated that the uniqueness theorems for black holes do not hold in higher dimensions or, at best, need more assumptions to define uniqueness (rod structure, stability, etc). This exemplifies that general relativity has special properties in $D=4$ which can only be appreciated if $D>4$ is considered.

\section{The classical trans-Planckian problem}
\label{sc:transP}
Relativistic particle collisions are in the realm of quantum field theory and if energies are high enough such that gravity becomes relevant they should enter the domain of quantum gravity. Moreover, if a black hole forms, as expected in a trans-Planckian head-on collision, non-perturbative processes should become relevant and therefore we find ourselves with a hopeless problem in non-linear, non-perturbative, strongly time-dependent quantum gravity. However, as first argued by 't Hooft \cite{'tHooft:1987rb}, \emph{well above} the fundamental Planck scale this process should be well described by classical gravity (general relativity), the reason being that the Schwarzschild radius for the collision energy becomes much larger than the de Broglie wavelength (for the same energy) or any other interaction scale. 

Thorne's \emph{hoop conjecture} \cite{Thorne:1972ji} further tells us that if an amount of energy $E$ is trapped in a region of space such that a circular hoop with radius $R$ encloses this matter in all directions, a black hole (i.e. an event horizon) is formed if its Schwarzschild radius $R_S>R$ (the classical version of this conjecture has been supported by numerical relativity simulations, see for example the work by Choptuik and Pretorius on boson stars \cite{Choptuik:2009ww}).

All complex field theoretical interactions will then be cloaked by an event horizon and therefore causally disconnected from the exterior. This horizon, in turn, will be sufficiently classical if large enough, in the sense that quantum corrections will be small on and outside of it. If gravity is the dominant interaction, the total energy of the colliding particles is the dominant parameter of the collision. All other constituent details such as gauge charges should have a sub-dominant role, i.e. \emph{matter does not matter} \cite{Giddings:2004xy}. This idea has been verified within numerical relativity in several setups, namely the collision of highly boosted black holes, boson stars and self-gravitating fluid spheres \cite{Sperhake:2008ga,East:2012mb,Sperhake:2012me}.

\subsection{TeV gravity}
\label{ssc:TeVgravity}
The enormity of the Planck mass, $m_P\sim 1/\sqrt{G_N}\sim 10^{19}$ GeV, seems to render utopical any experimental realisation of this scenario. However, if one admits the possibility of existence of extra \emph{hidden} dimensions, with standard model interactions confined to a four-dimensional brane, the fundamental Planck mass may be well below its effective four-dimensional value. Such models were proposed to address what came to be known as the \emph{hierarchy problem}: the relative weakness of gravity by about forty orders of magnitude when compared to the other fundamental interactions. Pictorially, it is as if gravity is diluted in the extra dimensions, which may be large up to a sub-millimetre scale (models exist with both compact and infinite extra spatial dimensions \cite{ArkaniHamed:1998rs,Randall:1999ee,Randall:1999vf,Antoniadis:1998ig}).

If the fundamental Planck mass is of the order of the TeV, then high-energy particle colliders such as CERN's Large Hadron Collider (LHC) \cite{Ahn:2002mj,Chamblin:2004zg}, or collisions of ultra-high-energy cosmic rays with the Earth's atmosphere \cite{Feng:2001ib,Ahn:2003qn}, or even astrophysical black hole environments \cite{Banados:2009pr,Berti:2009bk,Jacobson:2009zg}, may realise the above scenario with formation and evaporation of microscopic black holes. 

Since 2011, both the CMS \cite{Khachatryan:2010wx,Savina:2013lga,Savina:2013eja} and ATLAS \cite{ATLAS:2011fga,ATLAS:2011cga,ATLAS:2014} collaborations have set bounds on such physics beyond the standard model based on the analysis of the 7 TeV LHC data. However, these bounds are extremely dependent on regions of parameter space where black holes would be produced with masses, at best, close to the unknown Planck scale \cite{Park:2011je}. If we require the produced objects to be in the semi-classical (and thus computable) regime, the cross-sections become negligible at 7 TeV, and only after the upgrade to 13-14 TeV, planned to take place in 2015, will the scenario be properly tested. Any improvement in the phenomenology of these models is therefore quite timely.

The two most important quantities for the modeling of black hole production in high-energy particle collisions are the critical impact parameter for black hole formation in parton-parton scattering and the energy lost in this process, emitted as gravitational radiation. If the latter is large and dominates the missing energy, and if it can be calculated with enough precision, it could be an important signature for discovery or exclusion. The event generators used at the LHC to look for signatures of black hole production and evaporation, {\sc charybdis2} \cite{Frost:2009cf} and {\sc blackmax} \cite{Dai:2007ki}, are very sensitive to these two parameters. 

\section{Current estimates of the inelasticity}
\label{sc:estimates}
In the highly trans-Planckian limit, the colliding particles are greatly boosted, traveling very close to the speed of light. This has motivated the study of gravitational shock wave collisions as a model for the gravitational fields of highly boosted particles.

\subsection{Apparent horizon bounds}
\label{ssc:AH bounds}
The gravitational field of an ultra-relativistic particle of energy $E$ is obtainable from a boost of the Schwarzschild metric. As the boost increases, the gravitational field becomes increasingly Lorentz contracted and in the limit in which the velocity goes to $c$ (but keeping the energy fixed) the gravitational field (i.e. tidal forces, described by the Riemann tensor) becomes planar and has support only on a null surface. This \emph{shock wave} is described by the Aichelburg-Sexl metric \cite{Aichelburg:1970dh}. Due to their flatness outside this null surface, it is possible to superimpose two oppositely moving waves and the geometry, as an exact solution of general relativity, is completely known everywhere except in the future light cone of the collision.

Strikingly, such knowledge is enough to actually show the existence of an apparent horizon for this geometry and thus provide strong evidence that a black hole (i.e. an event horizon) forms in the collision. Thus, if cosmic censorship holds, and since the sections of the event horizon must lie outside the apparent horizon, the size of the latter yields a lower bound on the size of the black hole. An energy balance argument then provides an upper bound on the \emph{inelasticity} $\epsilon$, i.e. the percentage of initial centre-of-mass energy that is radiated away as gravitational radiation.

Penrose pioneered this computation in $D=4$, and by finding an apparent horizon on the past light cone of the collision concluded that no more than $1-1/\sqrt{2}$, or about $29.3\%$, of the energy was lost into gravitational waves. His method was later generalised to arbitrary $D$ dimensions by Eardley and Giddings \cite{Eardley:2002re}, who obtained
\begin{equation}
\epsilon_{\text{AH}}\leq 1-\frac{1}{2}\left(\frac{D-2}{2}\frac{\Omega_{D-2}}{\Omega_{D-3}}\right)^{\frac{1}{D-2}}\,,\label{AHbound}
\end{equation}
where $\Omega_n$ is the volume of the unit $n$-sphere. Note that this bound increases monotonically approaching $50\%$ as $D\rightarrow\infty$. Later, Yoshino and Rychkov found an apparent horizon in the future light cone \cite{Yoshino:2005hi}. Their analysis coincided with the bound in Eq.~\eqref{AHbound} for head-on collisions, but the critical impact parameters they obtained were larger than previous estimates \cite{Yoshino:2002tx}.

\subsection{A perturbative approach}
\label{ssc:pert_app}
Instead of computing a bound one may decide to compute the precise inelasticity by solving Einstein's equations in the future of the collision. This is a \emph{tour de force}. A method, first developed by D'Eath and Chapman and later by D'Eath and Payne \cite{D'Eath:1992hb,D'Eath:1992hd,D'Eath:1992qu}, is to set up a perturbative approach: considering the collision in a highly boosted frame, one shock becomes much stronger than the other and the latter can be considered as a perturbation of the former. This is how they justified, conceptually, the perturbative expansion. In four dimensions, they obtained a value of $25\%$ for the inelasticity at first order in perturbation theory, in agreement with Smarr's Zero Frequency Limit \cite{Smarr:1977fy}. This was originally thought to be the exact value, but Payne \cite{Payne} showed that Smarr's formula is in fact a linearised approximation valid only when the gravitational radiation is weak, and cannot predict the strong-field radiation generated by fully non-linear gravitational interactions. Second-order perturbation theory lowered the inelasticity to $16.4\%$. This is within the range $14\pm3\%$ obtained by considering the collision of highly boosted black holes in numerical relativity \cite{Sperhake:2008ga}, which confers further validity to the approach. Moreover, these values are smaller than the apparent horizon bound. If less energy is lost into gravitational radiation, the final black hole is more massive and hence more consistent with the semi-classical analysis used for the potentially observable Hawking radiation.

\section{Collision of gravitational shock waves in $D$ dimensions}
\label{sc:shocksD}
Phenomenologically interesting TeV gravity models occur in dimensions greater than six. Therefore, it would be helpful to extend the above-mentioned methods to higher dimensions. Numerical relativity for higher-dimensional space-times has seen an increasing amount of development in recent years \cite{Zilhao:2013gu,Cardoso:2014uka}, but so far black hole collisions have only been successfully computed at low energies \cite{Witek:2010az,Witek:2010xi}.

Extending the method of D'Eath and Payne to higher generic $D$ is a demanding task involving analytical and numerical methods. This was the goal set out by the Gr@v group at the University of Aveiro in 2011 just prior to the author joining as a PhD student. The remainder of this thesis will describe our findings and conclusions so far. Its structure is as follows.

We start in Chapter~\ref{ch:kinematics} by studying the geometrical properties of Aichelburg-Sexl shock waves and setting up a space-time where two such waves collide head-on.

Then, in Chapter~\ref{ch:dynamics}, we devise a perturbative strategy to compute the metric in the future of the collision; using the axial symmetry of the problem, we reduce it to three dimensions and write the general solution for the metric at each order in perturbation theory.

Chapter~\ref{ch:radiation} is about gravitational radiation and how to extract it in our problem, using a generalisation of the Bondi formalism.

Then, in Chapter~\ref{ch:analytics}, we unveil a hidden symmetry of the problem which allows for the factorisation of the angular dependence of the radiation pattern at null infinity.

With this, we compute analytically the first-order estimate for the inelasticity in Chapter~\ref{ch:surface}.

Next, in Chapter~\ref{ch:volume}, we use this hidden symmetry to reduce the problem to two-dimensions and prepare it for numerical evaluation in a computer; we find characteristic coordinates and produce a conformal diagram of the effective two-dimensional space-time.

In Chapter~\ref{ch:second_order} we pave the way for the second-order calculation by computing numerically the Green's functions and the sources, and discussing the asymptotic behaviour of these functions at the boundaries of the integration domain.

We conclude in Chapter~\ref{ch:conclusion} with some final remarks and an outlook on the second-order calculation, which we hope to finish in a future work.



Many technical details are given in Appendices.

\chapter{Kinematics of shock wave collisions}
\label{ch:kinematics}
\renewcommand{\textflush}{flushepinormal}

\epigraph{If you can't explain it simply, you don't understand it well enough.}{Albert Einstein}

\section{The Aichelburg-Sexl shock wave}
\label{se:ASwave}
The Aichelburg-Sexl (AS) metric is an exact solution of general relativity representing the gravitational field of a point-like particle moving at the speed of light. It was first obtained in $D=4$ by Aichelburg and Sexl \cite{Aichelburg:1970dh} by boosting the Schwarzschild solution and taking the speed of light limit while keeping the energy finite. 

For higher dimensions, the starting point is the $D$-dimensional Schwarzschild-Tangherlini solution, i.e. a spherically-symmetric, static black hole with mass $M$ in $D$ dimensions,
\begin{equation}
ds^2=-\left(1-\frac{16\pi G_D M}{(D-2)\Omega_{D-2}}\frac{1}{r^{D-3}}\right)dt^2+\left(1-\frac{16\pi G_D M}{(D-2)\Omega_{D-2}}\frac{1}{r^{D-3}}\right)^{-1}dr^2+r^2d\Omega_{D-2}^2\,,\label{tangherlini}
\end{equation}
where $d\Omega^2_{D-2}$ and $\Omega_{D-2}$ are the line element and volume of the unit $(D-2)$-sphere and $G_D$ is the $D$-dimensional Newton constant. Before taking the boost, it is convenient to rewrite this in isotropic coordinates by changing to a new radial coordinate $R$,
\begin{equation}
r=R(1+Z)^{\frac{2}{D-3}}\,,\qquad Z\equiv\frac{1}{4}\left(\frac{r_s}{R}\right)^{D-3}\,,\qquad r_s^{D-3}\equiv\frac{16\pi G_D M}{(D-2)\Omega_{D-2}}\,.
\end{equation}
Here $r_s$ is the Schwarzschild radius of the black hole and $Z$ is simply a short-hand. Then Eq. \eqref{tangherlini} becomes $\left(R=\sqrt{\sum_{i=1}^{D-1}x_i^2}\right)$
\begin{equation}
ds^2=-\left(\frac{1-Z}{1+Z}\right)^2dt^2+(1+Z)^{\frac{4}{D-3}}\sum_{i=1}^{D-1}dx_i^2\,.\label{isotropic}
\end{equation}
Next, we choose a spatial direction (say along $x^{D-1}\equiv z$) and perform a boost with velocity $v$, i.e.
\begin{equation}
t\rightarrow\gamma(t-v z)\,,\qquad z\rightarrow\gamma(z-v t)\,,
\end{equation}
where $\gamma=(1-v^2)^{-\frac{1}{2}}$ is the Lorentz factor. 

Meanwhile the transverse coordinates $x_T\equiv(x_1,\ldots,x_{D-2})$ are unaffected by the boost. The quantity $Z$ now reads
\begin{equation}
Z=\frac{4\pi G_D E}{(D-2)\Omega_{D-2}}\frac{1}{\gamma\left(\gamma^2(z-vt)^2+\rho^2\right)^{\frac{D-3}{2}}}\,,
\end{equation}
where $E\equiv\gamma M$ is the energy and $\rho\equiv\sqrt{\sum_{i=1}^{D-2}x_i^2}$ is the radius in the transverse plane. 

In the limit $v\rightarrow1$, $Z\rightarrow0$ and the metric can be expanded as
\begin{equation}
ds^2=-dt^2+dz^2+dx_T^2+4\frac{D-2}{D-3}\gamma^2 Z (dt-v dz)^2+\ldots\,,\label{after_limit}
\end{equation} 
where $\dots$ denotes terms that vanish in the limit. 

\begin{figure}
\includegraphics[width=\linewidth]{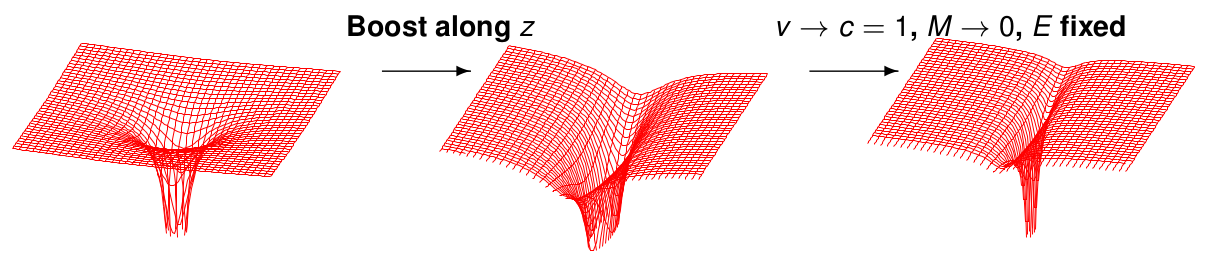}
\caption[Schematic representation of the qualitative behaviour of the gravitational field of a black hole subject to a boost at some fixed time.]{\label{boost}Schematic representation of the qualitative behaviour of the gravitational field (Kretschmann scalar $-|R^{\mu\nu\alpha\beta}R_{\mu\nu\alpha\beta}|$) of a black hole subject to a boost (horizontal directions represent the isotropic spatial coordinates $x_i$) at some fixed time. As the boost increases, the field becomes increasingly concentrated on the transverse plane containing the particle. \emph{From \cite{Sampaio:2013faa}.} }
\end{figure}

Fig.~\ref{boost} provides an illustration of this procedure: in a frame where the black hole is moving with velocity $v$, the gravitational field is Lorentz contracted along $z$ and the curvature becomes increasingly concentrated on a plane perpendicular to the motion; transverse directions are not affected. Indeed, in the limit $v\rightarrow1$ the term proportional to $Z$ in Eq.~\eqref{after_limit} goes to zero off the plane $z=t$ (which is moving at the speed of light) and diverges on it. 

In Appendix~\ref{app:boost} we show that the end result is simply flat Minkowski space-time plus a Dirac delta function with a radial profile on the transverse moving plane,
\begin{equation}
ds^2=-dudv+d\rho^2+\rho^2d\Omega_{D-3}^2+\kappa\Phi(\rho)\delta(u)du^2\,,\label{ASmetric}
\end{equation}
where $\kappa\equiv 8\pi G_D E/\Omega_{D-3}$ and $u=t-z$, $v=t+z$ are null coordinates. The profile function $\Phi$ depends only on $\rho$ and takes the form\footnote{In $D=4$ the profile contains an arbitrary length scale, $\Phi(\rho)=-2\ln(\rho/\ell)$, but since all physical quantities will be independent from it, we set $\ell=1$.}
\begin{equation}
\Phi(\rho)=\left\{
\begin{array}{ll}
 -2\ln(\rho)\ , &  D=4\  \vspace{2mm}\\
\displaystyle{ \frac{2}{(D-4)\rho^{D-4}}}\ , & D>4\ \label{Phi}
\end{array} \right. \ .
\end{equation} 
Clearly, a shock wave moving in the opposite direction with the same energy is obtained by replacing $z\leftrightarrow-z$ or, equivalently, exchanging $u$ and $v$.

\section{Properties of the Aichelburg-Sexl solution}
\label{properties_AS}
The Aichelburg-Sexl metric, Eq.~\eqref{ASmetric}, has the following properties:
\begin{itemize}
\item axial symmetry: it is invariant under rotations ($d\Omega_{D-3}$) on the transverse plane;
\item $v$-translational symmetry: the metric components do not depend on $v$;
\item transformation under boosts: a further boost of the solution moving along $+z$ with velocity $v=\tanh\beta$ amounts to a rescaling of the energy parameter $E\rightarrow e^\beta E$, as expected from the transformation law of the $D$-momentum of a null particle.
\end{itemize}
It is also instructive to look at the Riemann tensor in order to better understand the gravitational field. One can either compute it directly from Eq.~\eqref{ASmetric} or from Eq.~\eqref{tangherlini} and then applying the boost and taking the limit. The only independent non-vanishing component is
\begin{equation}
R_{uiuj}=-\frac{\kappa}{2}\delta(u)\left[\frac{\Phi'}{\rho}\delta_{ij}+\left(\nabla^2\Phi-(D-2)\frac{\Phi'}{\rho}\right)\frac{x_ix_j}{\rho^2}\right]\,,\label{riemann}
\end{equation}
where $\nabla^2$ is the Laplacian on the transverse plane. In Eq.~\eqref{riemann} it is manifest that the gravitational field only has support on the plane $u=0$, decaying away from its centre as an inverse power of $\rho$.
\begin{figure}
\includegraphics[width=\linewidth]{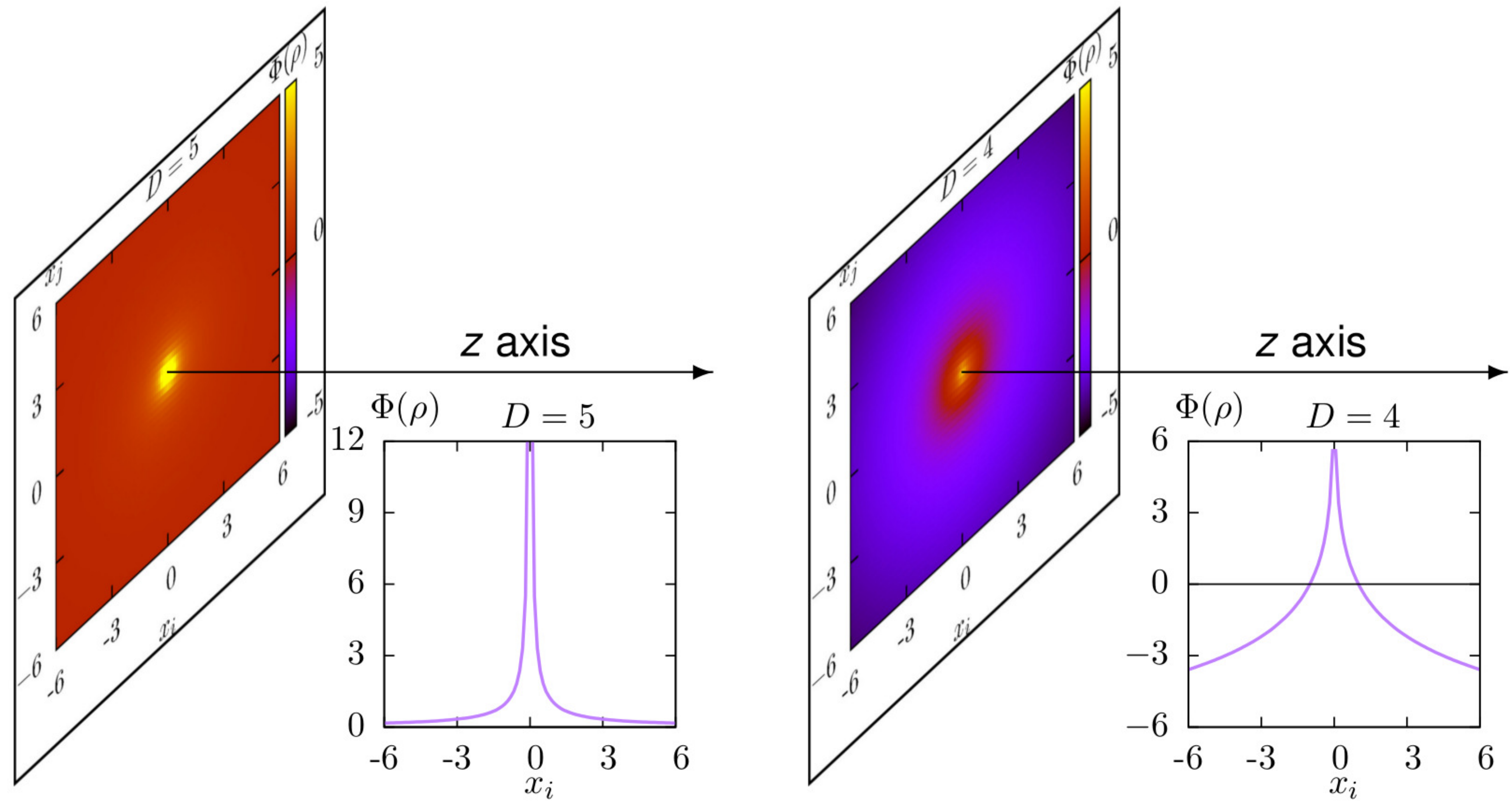}
\caption[Profile function $\Phi(\rho)$ of the gravitational field on the transverse plane.]{\label{FigProfilesColor}Profile function $\Phi(\rho)$ of the gravitational field on the transverse plane, for $D=5$ (left) and $D=4$ (right). The plane is shown in perspective and in the inset smaller figures we plot $\Phi$ as a function of a transverse direction $x_i$. \emph{From \cite{Sampaio:2013faa}.} }
\end{figure}
This can be better visualised in Fig.~\ref{FigProfilesColor} showing plots of the profile function on the transverse plane. For $D=5$ ($D>4$ in general), $\Phi(\rho)$ decays for large $\rho$, whereas in $D=4$ it goes to $-\infty$. This special case does not contradict the above statements on the qualitative behaviour of the gravitational field, since $\Phi$ is not a gauge-invariant quantity. Any gauge-invariant object contains at least one derivative like $\Phi'$ in Eq.~\eqref{riemann}, so the gravitational field does indeed decay for large $\rho$. 

From the Riemann tensor one constructs the Einstein tensor or, equivalently, the energy-momentum tensor, whose only non-vanishing component reads
\begin{equation}
T_{uu}=-\frac{E}{2\Omega_{D-3}}\delta(u)\nabla^{2}\Phi=E\delta(u)\delta^{(D-2)}(x_i)\,.
\end{equation}
It is now clear that this solution describes the gravitational field of a point-like particle moving at the speed of light with energy $E$. Furthermore, the impulsive nature of this shock wave, encoded in the $\delta(u)$ and $\delta^{(D-2)}(x_i)$ distributions, can be seen as an idealisation of a more realistic distribution replacing the delta function by a smeared out form, as would be the case if the matter distribution had some spatial extension. As long as this source is sufficiently localised, we expect the gravitational field away from the centre to be insensitive to these details.

\section{Geometric optics}
\label{geometric_optics}
We will now look at the null geodesics of this space-time in order to understand the gravitational deflection and redshift suffered by null test particles. Later, when considering the collision of two AS shock waves, such null rays will also be useful to define the causal structure of the post-collision space-time. 

The parametric equations of null geodesics can be given as
\begin{eqnarray}
u&=&u\,,\nonumber\\
v&=&v_0+\kappa\Phi(\rho_0)\theta(u)+u\sum_{i}\left[{x'}_0^i+\frac{\kappa}{2}\nabla_i\Phi(\rho_0)\theta(u)\right]^2\,,\label{null_geo}\\
x^i&=&x_0^i+{x'}_0^i u+\frac{\kappa}{2}\nabla_i\Phi(\rho_0)u\theta(u)\,,\nonumber
\end{eqnarray}
where the coordinate $u$ is itself an affine parameter, $v_0$, $x_0^i$ and ${x'}_0^i$ are integration constants and $\theta(u)$ is the Heaviside step function. In these coordinates (known as Brinkmann coordinates \cite{Brinkmann:1925fr}) geodesics and their tangent vectors are discontinuous across the shock wave. This problem is solved by identifying the parameters of null geodesics which are incident perpendicularly to the shock plane (${x'}_0^i=0$) with new coordinates $(\bar{u},\bar{v},\bar{x}^i)$,
\begin{eqnarray}
u&=&\bar{u}\,,\nonumber\\
v&=&\bar{v}+\kappa\theta(\bar{u})\left(\Phi+\frac{\kappa\bar{u}{\Phi'}^2}{4}\right)\,,\label{Rosen_coords}\\
x^i&=& \bar{x}^i + \frac{\kappa}{2} \bar{\nabla}_i \Phi(\bar{x})\bar{u}\theta(\bar{u}) \Rightarrow \left\{\begin{array}{rcl} \rho&=& \bar{\rho}\Big(1+ \frac{\kappa \bar{u} \, \theta(\bar{u})}{2 \bar \rho}\Phi'\Big)  \nonumber\\
\phi_{a}&=&\bar \phi_a  \end{array}\right.\nonumber
\ ,
\end{eqnarray}
where $\Phi$ and $\Phi'$ are evaluated at $\bar{\rho}$ and $\phi_a$, $a=1\ldots D-3$, are the angles on the $(D-3)$-sphere. These are known as Rosen coordinates \cite{Eardley:2002re}\cite{Einstein:1937qu}. 

\begin{figure}
\includegraphics[width=0.335\linewidth]{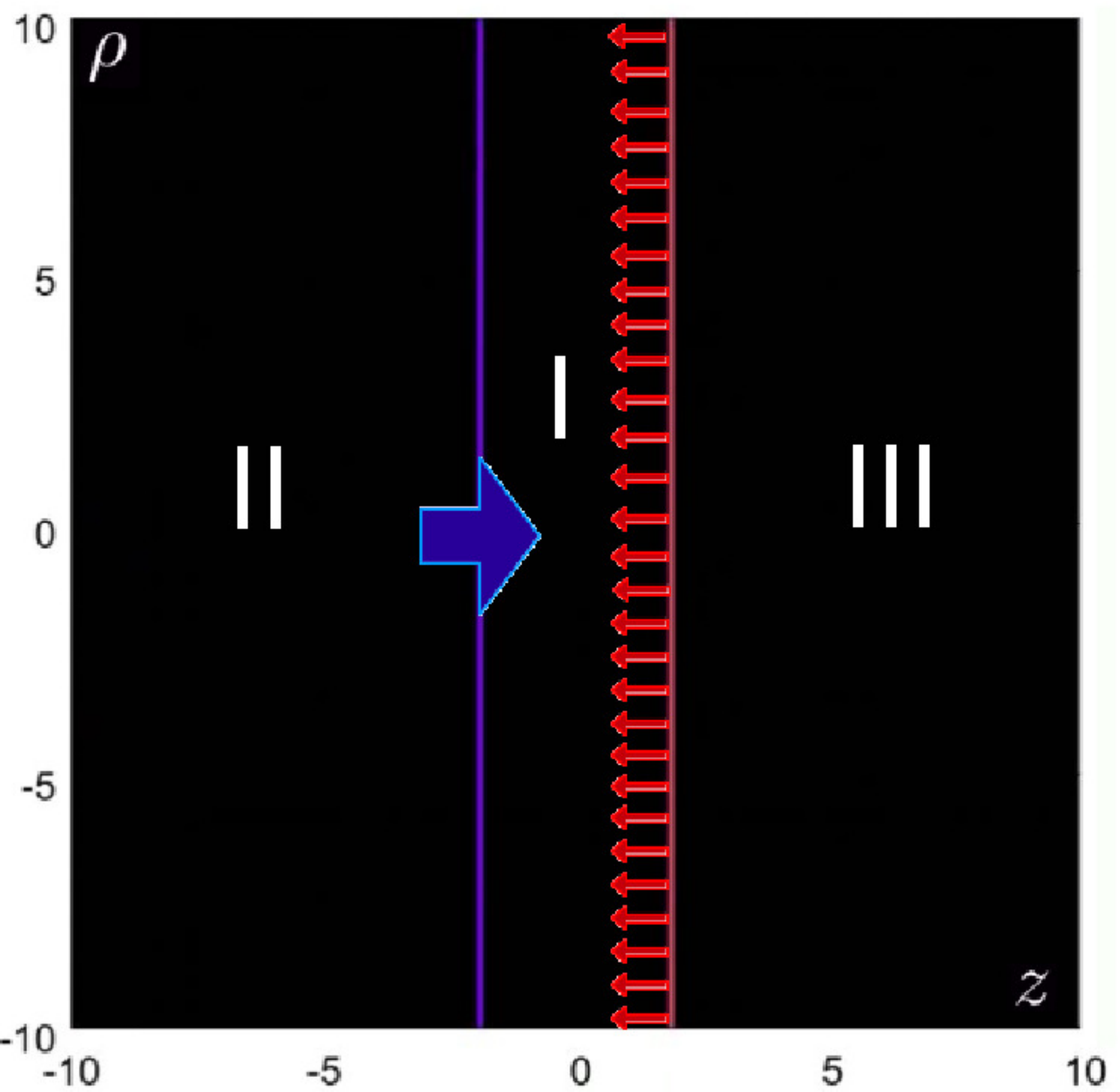}\includegraphics[width=0.336\linewidth]{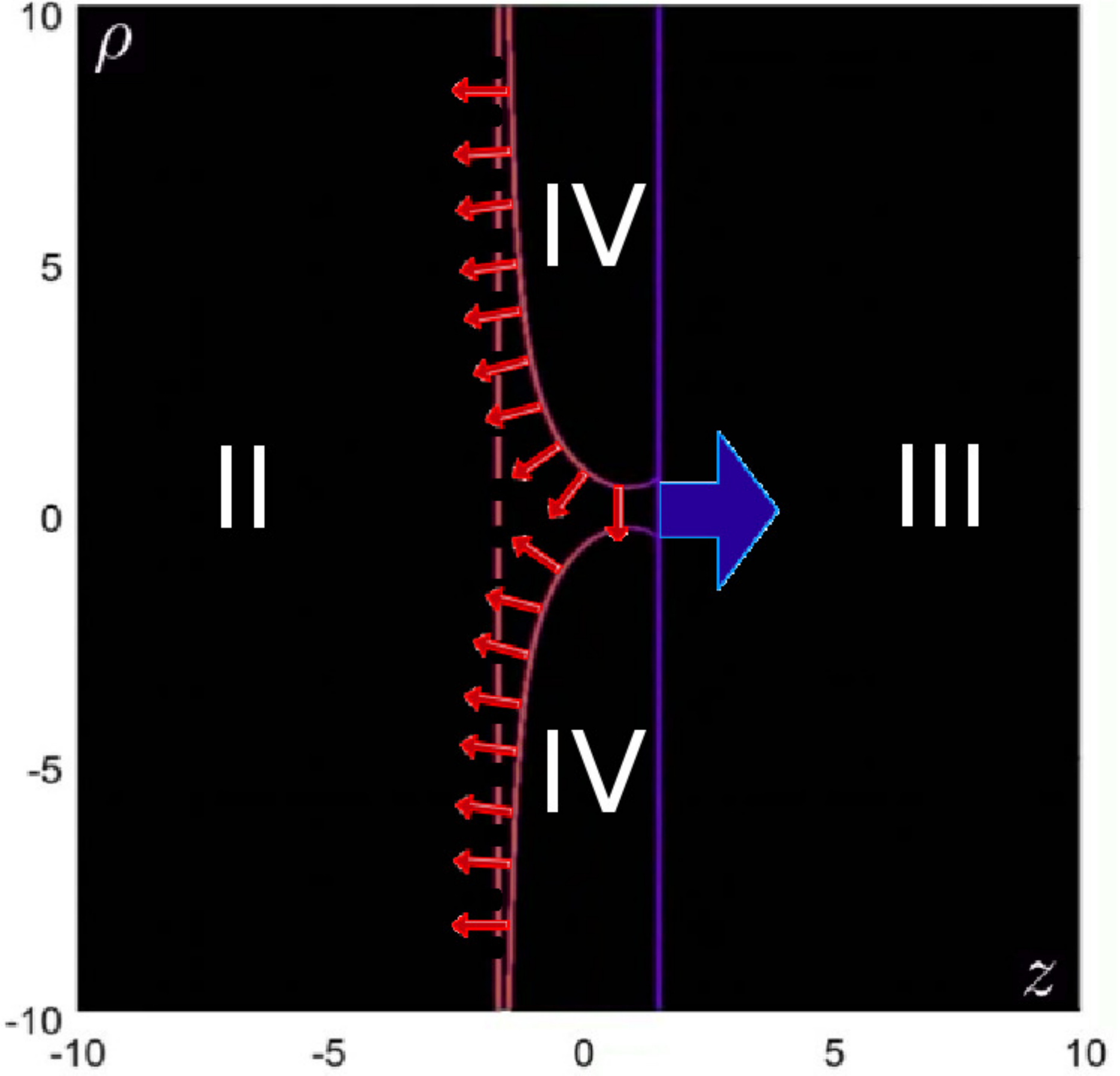}\includegraphics[width=0.335\linewidth]{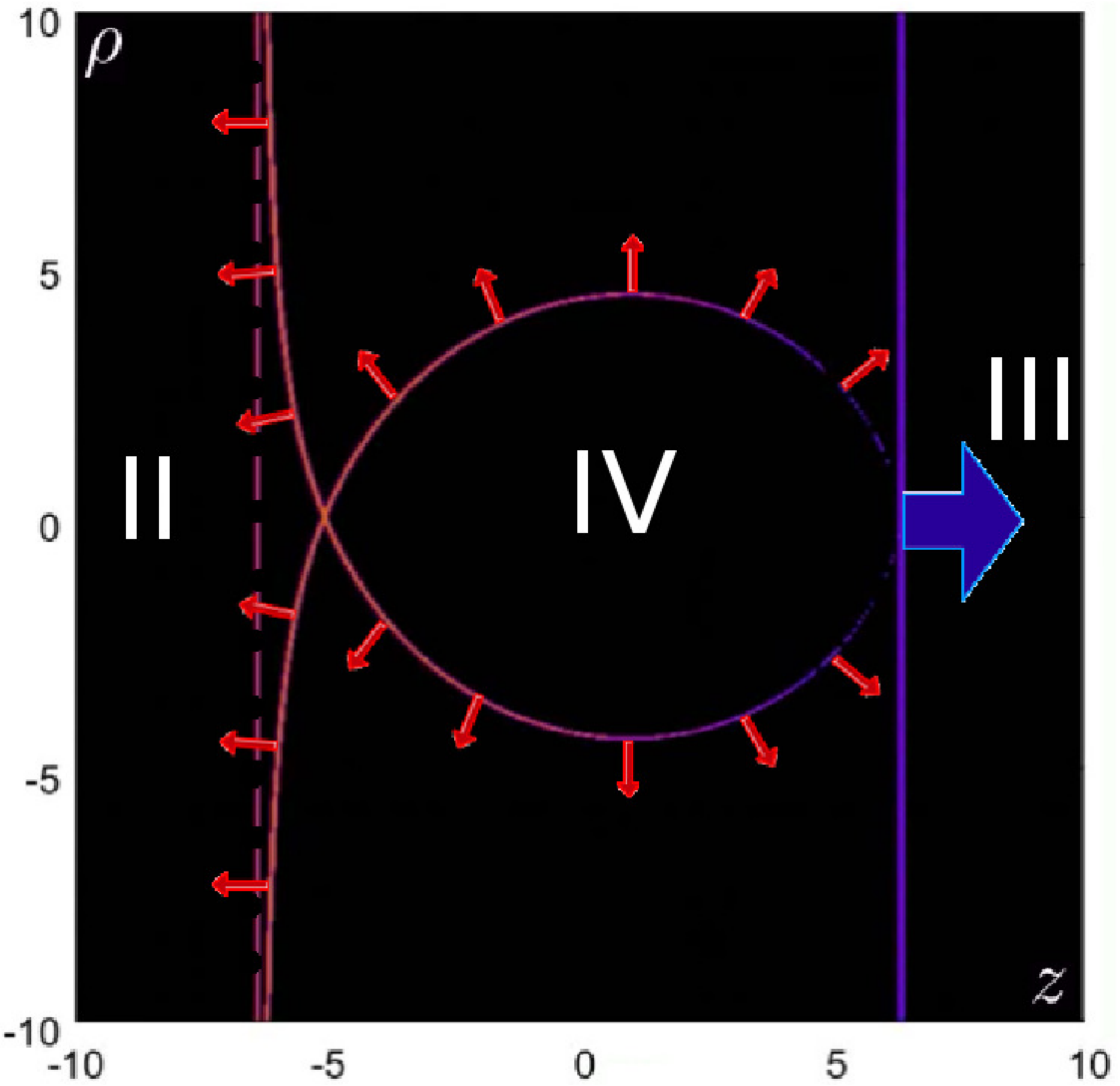}
\caption[Representation of a test wavefront of null rays scattering through the transverse plane of the AS shock wave travelling along the positive $z$ direction.]{\label{FigRaysMovie1_5D}Representation of a test wavefront of null rays (small red arrows) scattering through the transverse plane of the AS shock wave travelling along the positive $z$ direction (big blue arrow), for $D=5$: i) before scattering (left), ii) a small time after scattering (middle) and iii) a later time after scattering (right). The dashed red line (right) represents the plane wavefront the rays would have followed if no scattering had occurred. \emph{From \cite{Sampaio:2013faa}.} }
\end{figure}

From inspecting Eqs.~\eqref{Rosen_coords} we can draw the following conclusions:
\begin{enumerate}
\item \emph{Redshift:} null rays incident perpendicularly to the plane of the shock wave ($u=\bar{u}=0$) suffer a discontinuous jump in $v$, which is always positive for $D>4$ but becomes negative for large $\rho$ in $D=4$.
\item \emph{Focusing:} they are further deflected by an angle obeying $\tan\alpha=\kappa\rho^{-(D-3)}$ and focus, generating a caustic, at
\begin{equation}
\rho=0\,,\qquad v=\bar{v}+\kappa\left\{
\begin{array}{ll}
 1-\ln(\kappa u)\ , &  D=4\  \vspace{2mm}\\
\displaystyle{ \frac{D-2}{D-4}(\kappa u)^{-\frac{D-4}{D-2}}}\ , & D>4
\end{array} \right. \ .
\end{equation}
This focusing increases (decreases) with $D$ for short (long) distances, as expected from the behaviour of the gravitational force.
\end{enumerate}
These properties are illustrated in Fig.~\ref{FigRaysMovie1_5D}, where the small red arrows represent the scattered trajectories of a plane of null rays and the big blue arrow represents an AS shock wave moving to the right. The rays propagate freely (i.e. in a straight line) before crossing the shock (left panel). They are scattered instantaneously at $t=0$ when the two planes meet and thereafter continue on another straight line. 

The two effects described above are clearly seen in the middle and right panels. In the middle one, the tangent vectors are bent inwards towards the axis, as expected from the gravitational attraction of the source at the centre - this is the focusing effect. Moreover, the farther the rays are from the axis, the lesser they are bent. Later (right panel), we see there is an increasingly circular envelope of rays around the scattering centre, and another outermost one which asymptotes the initial wavefront (dashed line). 

The redshift effect can be inferred from the middle panel: we see that there are rays which have not yet emerged from the central region of the shock wave - they get stuck for some time until emerging later in their bent trajectory. Again, this effect is stronger the closer the rays are to the centre of the shock wave. 

\section{Superposition and the apparent horizon}
\label{superposition}
In the previous two sections we showed that an AS shock wave with support on $u=0$ represents a massless particle moving along $z=t$ at the speed of light. Moreover, we saw that its gravitational field is restricted to the transverse plane. In particular, an incident ray is not affected before it crosses the shock. This is simply due to causality: since no signal can travel faster than the speed of light, the incident ray has no means of ``knowing'' what is coming before it hits the shock and thus experiences no gravitational field. 

Therefore, if we superpose another shock wave travelling in the opposite direction, i.e. along $z=-t$, the space-time metric is simply the superposition of the two metrics everywhere but in the future of the collision. 

In Rosen coordinates, the line element \eqref{ASmetric} becomes
\begin{equation}
ds^2 = -d\bar{u} d\bar{v} + \left(1+\dfrac{\kappa \bar{u} \theta(\bar{u})}{2}\Phi''\right)^2d\bar{\rho}^{2} + \bar{\rho}^2\left(1+ \frac{\kappa \bar{u} \, \theta(\bar{u})}{2 \bar \rho}\Phi'\right)^2 d\bar{\Omega}^2_{D-3} \ .\label{Rosen_metric}
\end{equation}

The geometry for an identical shock wave traveling in the opposite direction is obtained by exchanging $\bar{u}\leftrightarrow \bar{v}$. If the shocks have different energy parameters $\nu$ and $\lambda$, the superposed metric reads
\begin{eqnarray}
ds^2& =& -d\bar{u} d\bar{v} + \left[\left(1+\dfrac{\nu \bar{u} \theta(\bar{u})}{2}\Phi''\right)^2+\left(1+\dfrac{\lambda \bar{v} \theta(\bar{v})}{2}\Phi''\right)^2-1\right]d\bar{\rho}^{2}\nonumber\\&& + \bar{\rho}^2\left[\left(1+ \frac{\nu \bar{u} \, \theta(\bar{u})}{2 \bar \rho}\Phi'\right)^2+\left(1+ \frac{\lambda \bar{v} \, \theta(\bar{v})}{2 \bar \rho}\Phi'\right)^2-1\right] d\bar{\Omega}^2_{D-3} \ ,\label{Rosen_metric_2}
\end{eqnarray}
and is valid everywhere except in the future light cone of $\bar{u}=\bar{v}=0$. 

\begin{figure}
\begin{center}
\mbox{\includegraphics[width=0.49\linewidth]{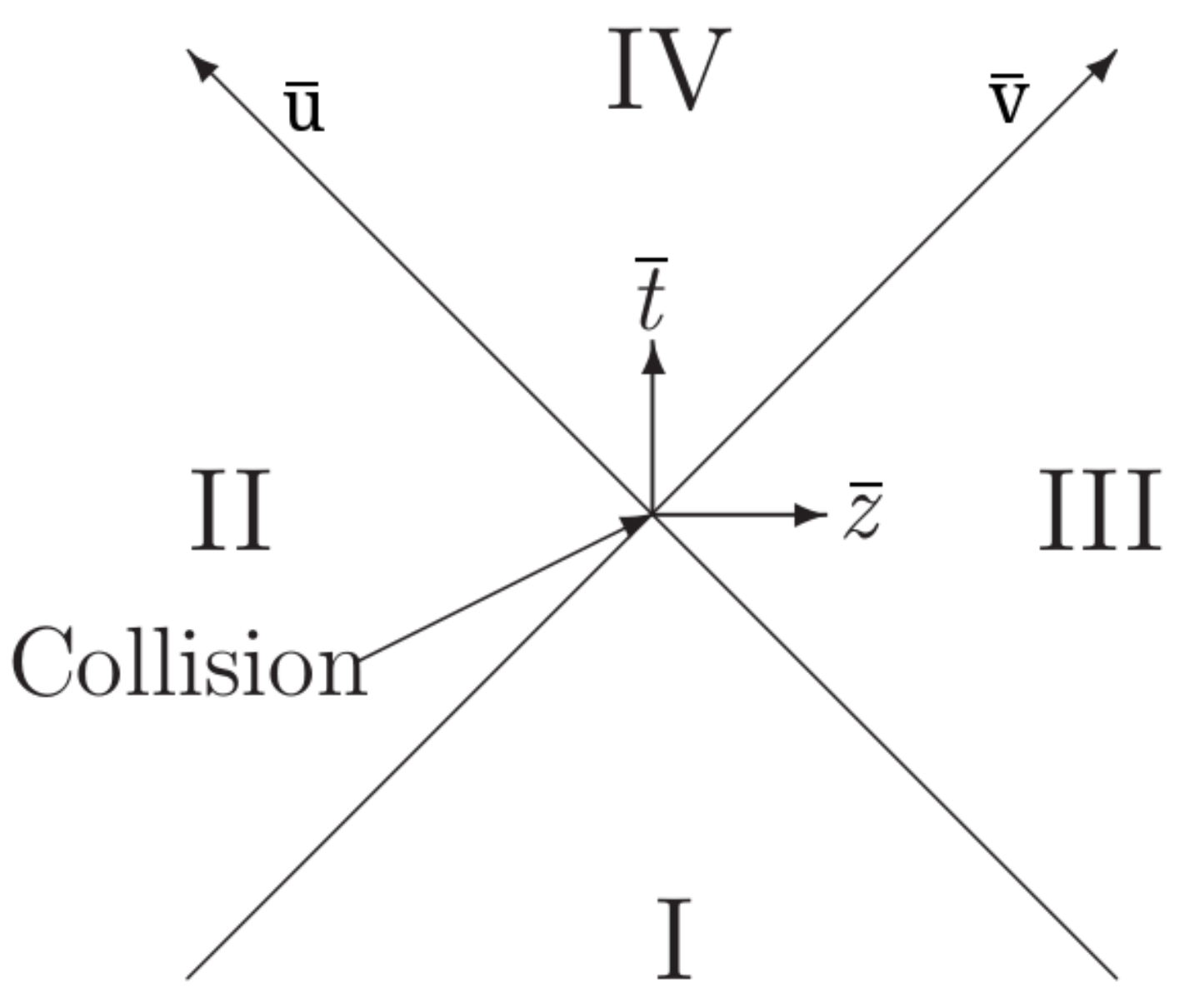}\hspace{0.02\linewidth}\includegraphics[width=0.49\linewidth]{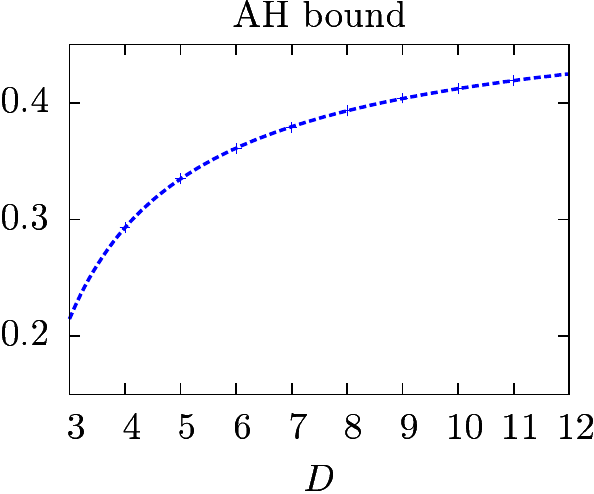}}
\end{center}
\caption[Space-time diagram of the collision and apparent horizon upper bound on the inelasticity.]{\label{barredDiagAH} Space-time diagram of the collision (left) and apparent horizon upper bound on the inelasticity (right) as a function of $D$. \emph{From \cite{Sampaio:2013faa}.}}
\end{figure}

Fig.~\ref{barredDiagAH} is a space-time diagram, in Rosen coordinates, where the transverse dimensions have been suppressed. The two shock waves are represented by the $\bar{u},\bar{v}$ axes where they have support. Regions I, II and III are flat. Region I is the region in-between the shocks, before the collision, whereas II and III are, respectively, the regions behind the $\bar{u}$ and $\bar{v}$ shocks. 

The apparent horizon mentioned in Chapter~\ref{ch:intro} was found, first by Penrose in $D=4$ and later by Eardley and Giddings \cite{Eardley:2002re} in higher $D$, on the union of the two null surfaces $\bar{u}=0\wedge\bar{v}\leq0$ and $\bar{v}=0\wedge\bar{u}\leq0$. Their upper bound on the inelasticity, Eq.~\eqref{AHbound}, is plotted in the right panel of Fig.~\ref{barredDiagAH}, where we see that it increases monotonically with $D$, approaching $50\%$ in the limit $D\rightarrow\infty$.

The existence of an apparent horizon, which traps the colliding particles, further supports the independence of the results on the details of the localised sources. In particular, their point-like nature (as opposed to a small but finite volume) is not expected to affect the formation of a black hole in the trans-Planckian limit. 

In the next chapter we shall devise a perturbative framework to compute the metric in the future of the collision, i.e. in region IV.

\chapter{Dynamics: a perturbative approach}
\label{ch:dynamics}
\renewcommand{\textflush}{flushepinormal}

\epigraph{Physicists like to think that all you have to do is say, `These are the conditions, now what happens next?'}{Richard Feynman \\ \emph{The Character of Physical Law}}

\section{The collision in a boosted frame}
\label{boosted_frame}
The original argument of D'Eath and Payne consisted in looking at the collision in a boosted frame, in which one of the shocks appears much stronger than the other. This allowed them to consider the latter as a perturbation of the former, and set up an iterative process of solving Einstein's equations order by order in perturbation theory. 

Starting from the Rosen form of the metric of one shock wave, Eq.~\eqref{Rosen_metric}, a boost with velocity $v$ in the $\pm z$ direction amounts to a scaling of the energy parameter,
\begin{equation}
\kappa\rightarrow e^{\pm\beta}\kappa,\qquad \beta=\tanh v\,.
\end{equation}
If two identical but oppositely travelling shocks are superposed, this change of reference frame makes them appear to have different energy parameters. The metric, in this case, would be given (except in the future of the collision) by Eq.~\eqref{Rosen_metric_2}, with
\begin{equation}
\nu\equiv e^\beta\kappa\,,\qquad \lambda\equiv e^{-\beta}\kappa\,.
\end{equation}
In this way, the $\nu$-shock (which we call the \emph{strong shock}) carries much more energy than the $\lambda$-shock (which we call the \emph{weak shock}), since
\begin{equation}
\frac{\nu}{\lambda}=e^{2\beta}=\frac{1+v}{1-v}\xrightarrow{v\rightarrow1}\infty\,.
\end{equation}
Thus one may face the shock wave travelling in the $-z$ direction as a small perturbation of the shock wave travelling in the $+z$ direction. Since the geometry of the latter is flat for $\bar{u}>0$ (regions II and IV), we can make a perturbative expansion of Einstein's equations around flat space, in order to compute the metric in the future of the collision ($\bar{v},\bar{u}>0$, i.e. region IV). 

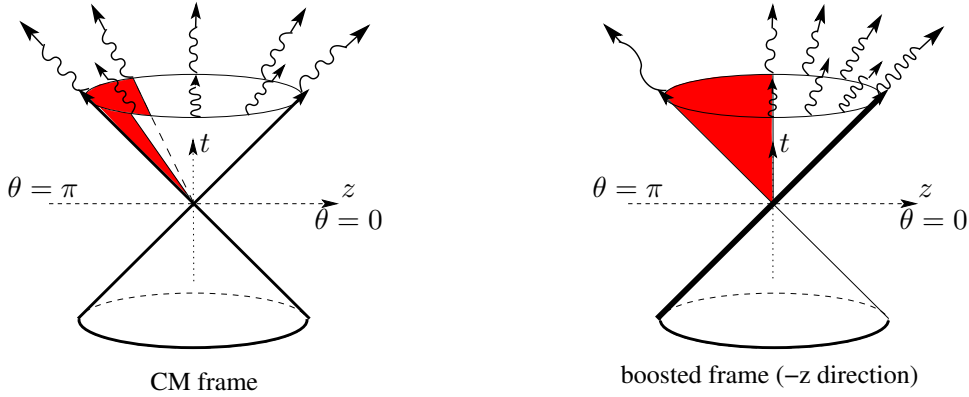
\begin{figure}[t]
\begin{center}
\begin{picture}(0,0)
\put(72,90){$t$}
\put(125,73){$z$}
\put(115,61){$\theta=0$}
\put(0,73){$\theta=\pi$}
\put(289,90){$t$}
\put(341,73){$z$}
\put(334,61){$\theta=0$}
\put(219,73){$\theta=\pi$}
\end{picture}
\includegraphics[scale=0.6]{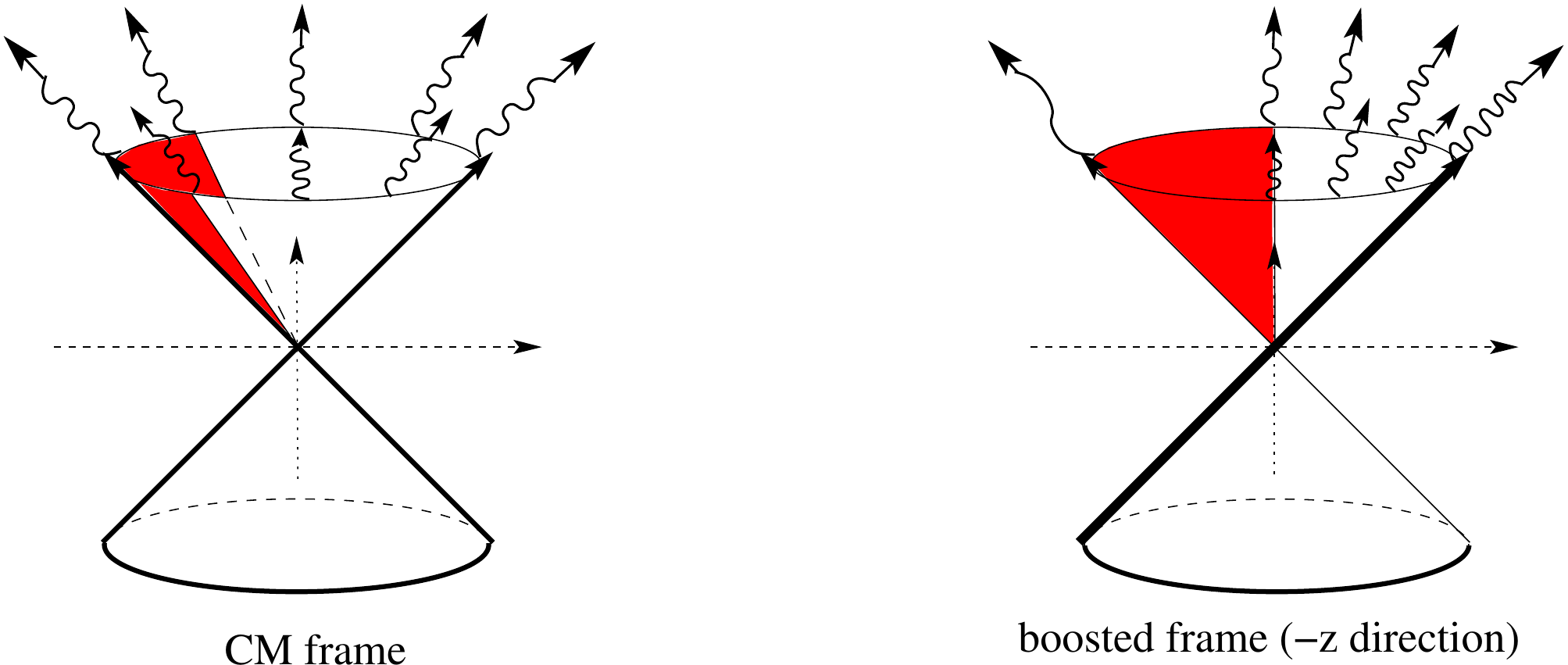}
\end{center}
\caption[Space-time diagram of the collision in the centre-of-mass and boosted frames.]{Space-time diagram of the collision in the centre-of-mass (left panel) and boosted (right panel) frames. The transverse $x^i$ directions are orthogonal to the $t$ and $z$ axes.  In the centre-of-mass frame, the radiation is symmetric under $z\rightarrow -z$. However, only the radiation that propagates along a small solid angle around $\theta=\pi$ in that frame (red area, delimited by the polar angle $\tan \theta=\sqrt{v^{-2}-1}$) is still propagating in the $-z$ direction in the boosted frame. All the radiation in this solid angle is redshifted in that frame. \emph{From \cite{Herdeiro:2011ck}.}}
\label{spacetime}
\end{figure}

Fig.~\ref{spacetime} illustrates the effect of the boost in the pattern of gravitational radiation. In the boosted frame, the whole $z<0$ hemisphere maps to a region around $\theta=\pi$ in the centre-of-mass frame. Thus that is the region where perturbation theory is expected to be valid, with higher orders needed to extrapolate off the axis. This statement will be made clear in Section~\ref{meaning}, where we shall derive an exact correspondence between a perturbative expansion of the metric (in the sense here described), and an angular expansion of the radiation pattern at null infinity. 

\section{The collision in the centre-of-mass frame}
\label{CM_frame}
The independence of physics from the reference frame or the coordinates used is one of the pillars of general relativity. Therefore, one must be able to construct the perturbative approach of the previous section in the centre-of-mass frame in which the two colliding shock waves have the same energy parameter. 

We start by rewriting the superposed metric, Eq.~\eqref{Rosen_metric}, for identical shocks ($\nu=\lambda=\kappa$), in a form which resembles a perturbation of flat space,
\begin{multline}ds^2 = -d\bar{u} d\bar{v} +\delta_{ij}d\bar{x}^id\bar{x}^j + \left[2\left(\dfrac{\kappa\Phi'}{2\bar{\rho}}\right)\left(\bar{u}\theta(\bar{u})+\bar{v}\theta(\bar{v})\right)\bar{\Delta}_{ij}+\phantom{\left(\dfrac{\kappa}{\bar{\rho}^{D-2}}\right)^2}\right. \\\left.+\left(\dfrac{\kappa\Phi'}{2\bar{\rho}}\right)^2\left(\bar{u}^2\theta(\bar{u})+\bar{v}^2\theta(\bar{v})\right)\left((D-3)\delta_{ij}-(D-4)\bar{\Delta}_{ij}\right)\right]d\bar{x}^id\bar{x}^j 
 \ ,\label{collision}
\end{multline}
where $\bar{\Delta}_{ij}\equiv \delta_{ij}-(D-2)\bar{\Gamma}_i\bar{\Gamma}_j$ is a traceless tensor on the transverse plane and $\bar{\Gamma}_i=\bar{x}_i/\bar{\rho}$ are angular factors. The first two terms in Eq.~\eqref{collision} are simply the Minkowski line element, and those inside the square brackets constitute a small perturbation in regions where
\begin{equation}
\bar\rho>>(\kappa\bar{u})^{\frac{1}{D-2}}\qquad \wedge \qquad \bar\rho>>(\kappa\bar{v})^{\frac{1}{D-2}}\,.
\end{equation}
These two conditions highlight the main shortcoming of Rosen coordinates: in regions II and III the space-time is \emph{exactly} flat, but the metric is not in standard Minkowski form, and actually depends on the energy $\kappa$ of the shock waves. This can be partially fixed by returning to Brinkmann coordinates through Eq.~\eqref{Rosen_coords}. Since this transformation is adapted to the shock wave with support on $\bar{u}=u=0$, the $z\leftrightarrow-z$ symmetry of the problem is no longer apparent. 

By splitting the $u$-shock from the $v$-shock, Eq.~\eqref{collision} can be written as
\begin{multline}ds^2 = -du dv +\delta_{ij}dx^idx^j+\kappa \Phi(\rho) \delta(u) du^2+\\ \left\{-2\bar{h}(u,v,\rho)\bar{\Delta}_{ij}+\bar{h}(u,v,\rho)^2\left((D-3)\delta_{ij}-(D-4)\bar{\Delta}_{ij}\right)\right\}d\bar{x}^id\bar{x}^j 
 \ ,\label{collisionBrink}
\end{multline}
where the barred coordinates in the second line are to be understood as functions of the unbarred coordinates through Eq.~\eqref{Rosen_coords}, and
\begin{equation}
\bar{h}(u,v,\rho)=-\frac{\kappa}{2}\frac{\Phi'}{\bar{\rho}}\bar{v}\theta(\bar{v})\,.
\end{equation}
As long as $\bar{h}(u,v,\rho)<<1$, the $v$-shock now looks like a small perturbation on a $u$-shock background (which is itself flat everywhere except on $u=0$). This is analogous to the picture of Section \ref{boosted_frame} where we saw that, in a boosted frame, one shock looks much stronger than the other.
\begin{figure}
\begin{center}
\includegraphics[width=\linewidth]{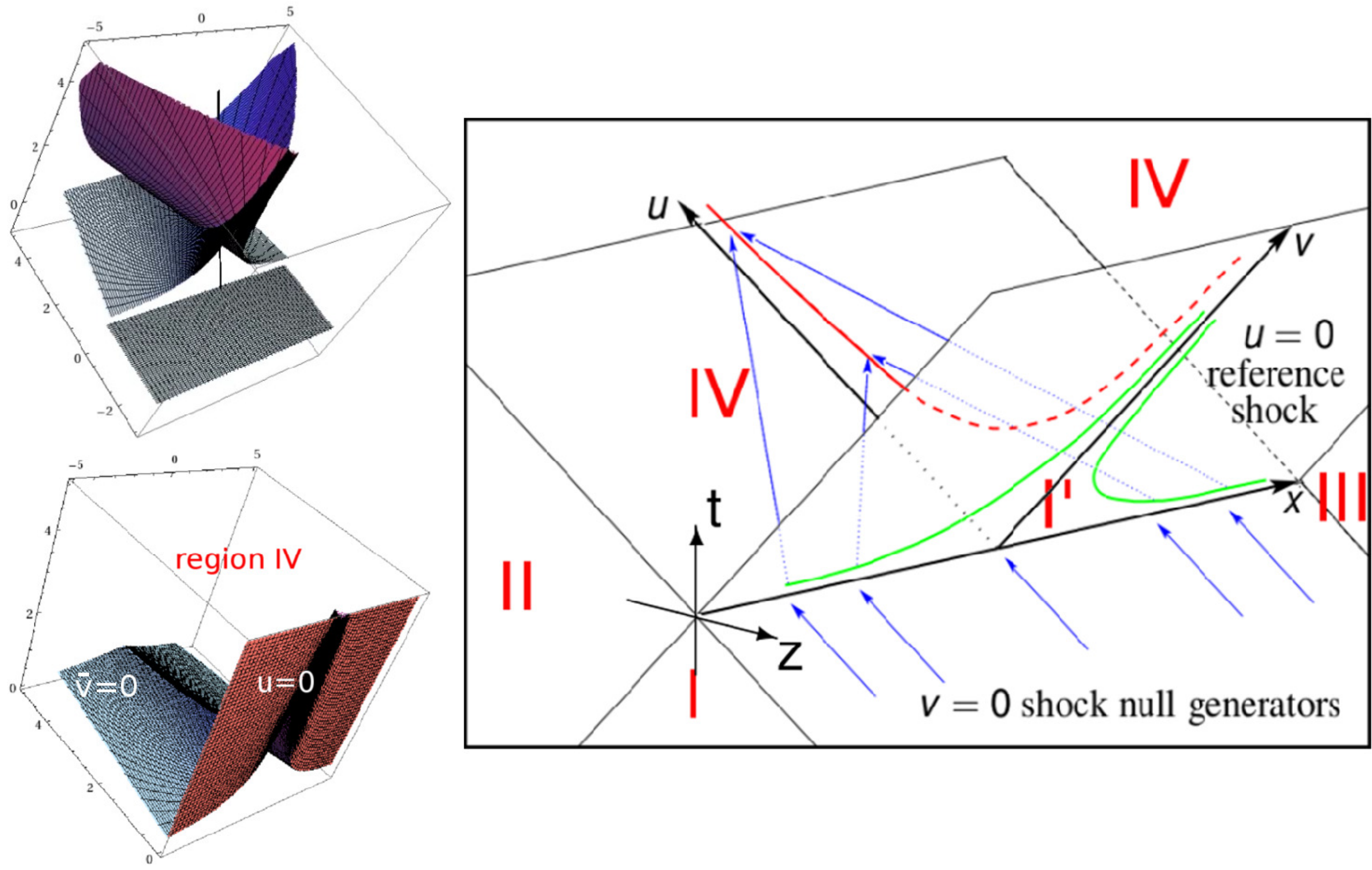}
\end{center}
\caption[3D space-time diagram showing the $(t,z,\rho\equiv x)$ axes, the numbered regions, the generators of the $v$-shock and the collision surface.]{\label{5DspacetimeDiag}\emph{Right:} 3D space-time diagram showing the $(t,z,\rho\equiv x)$ axes, the numbered regions (in red), the generators of the $v$-shock (in blue) and the collision surface (in green). \emph{From \cite{Herdeiro:2011ck}.} \\ \emph{Left:} Top diagram shows the surface defined by the null generators of the $v$-shock as they scatter through the $u$-shock. Bottom one shows the causal surfaces ($u=0\wedge v>0$ and $\bar{v}=0\wedge u>0$) defining the future light cone of the collision (region IV). \emph{From \cite{Sampaio:2013faa}.}}
\end{figure}

At this point, it is instructive to revisit Section \ref{geometric_optics} and repeat the analysis of the null rays in the new adapted Brinkmann coordinates. First note that, from the $u$-shock point of view, the generators of the $v$-shock (i.e. the null rays travelling on $v=0$) will scatter through the $u$-shock exactly as the incident `test' rays of Fig.~\ref{FigRaysMovie1_5D}. Thus the outermost envelope of rays, better seen in the right panel of Fig.~\ref{FigRaysMovie1_5D}, defines the causal boundary between the curved region (IV) and the flat region (II). 

The right panel of Fig.~\ref{5DspacetimeDiag} represents a space-time diagram of the collision in $u$-adapted Brinkmann coordinates. The generators of the $u$-shock travel along $u=0$ without any discontinuity. The generators of the $v$-shock, however, travel along $v=0$ while $u<0$ (blue arrows) but suffer a discontinuous jump to the green line upon hitting the `strong' shock at $u=0$. The green line is the collision surface defined by $\bar{u}=\bar{v}=0$, or
\begin{equation}
u=0\qquad\wedge\qquad v=\Phi(\rho)\,.
\end{equation}
After that they focus towards the axis and cross, forming a caustic (red line). Region I' represents a patch of flat space below region IV in $u>0$ where the null rays have not yet crossed and thus space-time is still flat. 

The 3D diagrams on the left complement the visualisation. The top one shows the surface formed by the generators of the $v$-shock, together with the world-line of a time-like observer for comparison. The bottom one represents the boundary of the curved region (IV).

\section{An initial value problem}
\label{initial_value}
In this section, we shall use the results of the previous one to set up an initial value problem for the dynamical equations that determine the metric in region IV. From Eq.~\eqref{collision} we see that on $\bar{v}=0\wedge u>0$ (blue surface on the bottom left diagram of Fig.~\ref{5DspacetimeDiag}), the metric has a standard Minkowski form,
\begin{equation}
g_{\mu\nu}(u>0,\bar{v}=0,x^i)=\eta_{\mu\nu}\,.\label{v_zero}
\end{equation}
On the other hand, on $u=0^+$ (red surface),
\begin{equation}
g_{\mu\nu}(u=0^+,v,x^i)=\eta_{\mu\nu}+h_{\mu\nu}^{(1)}+h_{\mu\nu}^{(2)}\,,\label{u_zero}
\end{equation}
where the $k$ in $h_{\mu\nu}^{(k)}$ is defined by the power of $\bar{h}(u=0^+,v,x^i)$ it contains. In natural units where $\kappa=1$\footnote{$\kappa$ has dimensions of [Lenght]$^{D-3}$.},
\begin{eqnarray}
h_{uu}^{(1)}&=&\frac{D-3}{2}\Phi'^2 h(v,\rho)\,,\qquad h_{uu}^{(2)}=\frac{(D-3)^2}{4}\Phi'^2 h(v,\rho)^2\,,\nonumber\\
h_{ui}^{(1)}&=&-(D-3)\Phi' h(v,\rho)\Gamma_i\,,\qquad h_{ui}^{(2)}=-\frac{(D-3)^2}{2}\Phi' h(v,\rho)^2\Gamma_i\,,\label{initial_data}\\
h_{ij}^{(1)}&=&-2 h(v,\rho)\Delta_{ij}\,,\qquad h_{ij}^{(2)}=h(v,\rho)^2\left((D-3)\delta_{ij}-(D-4)\Delta_{ij}\right)\,,\nonumber
\end{eqnarray}
where
\begin{equation}
h(v,\rho)\equiv\bar{h}(0,v,\rho)=-\frac{\Phi'}{2\rho}(v-\Phi(\rho))\theta(v-\Phi(\rho))\,.
\end{equation}
These initial conditions, Eqs.~\eqref{v_zero} and \eqref{u_zero}, have the form of a perturbation of a flat Minkowski background which is \emph{exact} at second order (on $u=0^+$). 

By adopting the reference frame of the $u$-shock, we have successfully encoded the information on the scattering of the $v$-shock in Eq.~\eqref{u_zero}. It is important to stress that no approximation has yet been made - all we did was chose a convenient frame for the superposition. 

To find the metric in the future of the collision (region IV) we must solve Einstein's equations in vacuum, $R_{\mu\nu}=0$, subject to initial conditions on $u=0^+$. The form of Eq.~\eqref{u_zero} suggests the use of perturbation theory, in which one obtains each new term of an infinite series in an iterative process. However, $h_{\mu\nu}^{(1)}$ and $h_{\mu\nu}^{(2)}$ are only small in regions where
\begin{equation}
|h(v,\rho)|\ll 1 \Rightarrow g(\rho)\equiv \left|\Phi(\rho)+\dfrac{2\rho}{\Phi'(\rho)}\right|\gg v\,.
\end{equation}

\begin{figure}
\includegraphics[width=0.34\linewidth,height=0.35\linewidth]{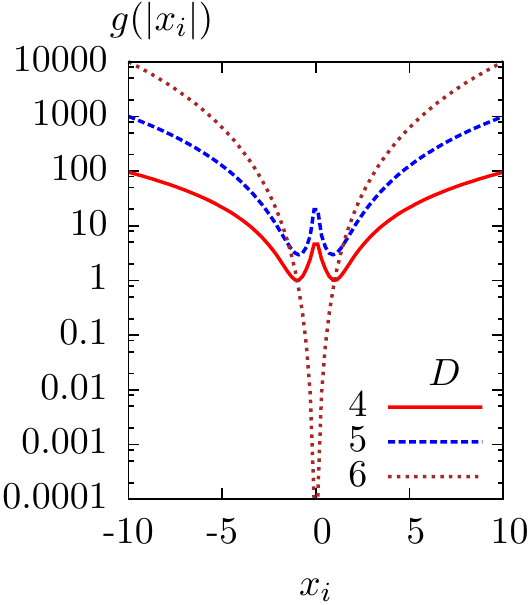}\includegraphics[width=0.35\linewidth]{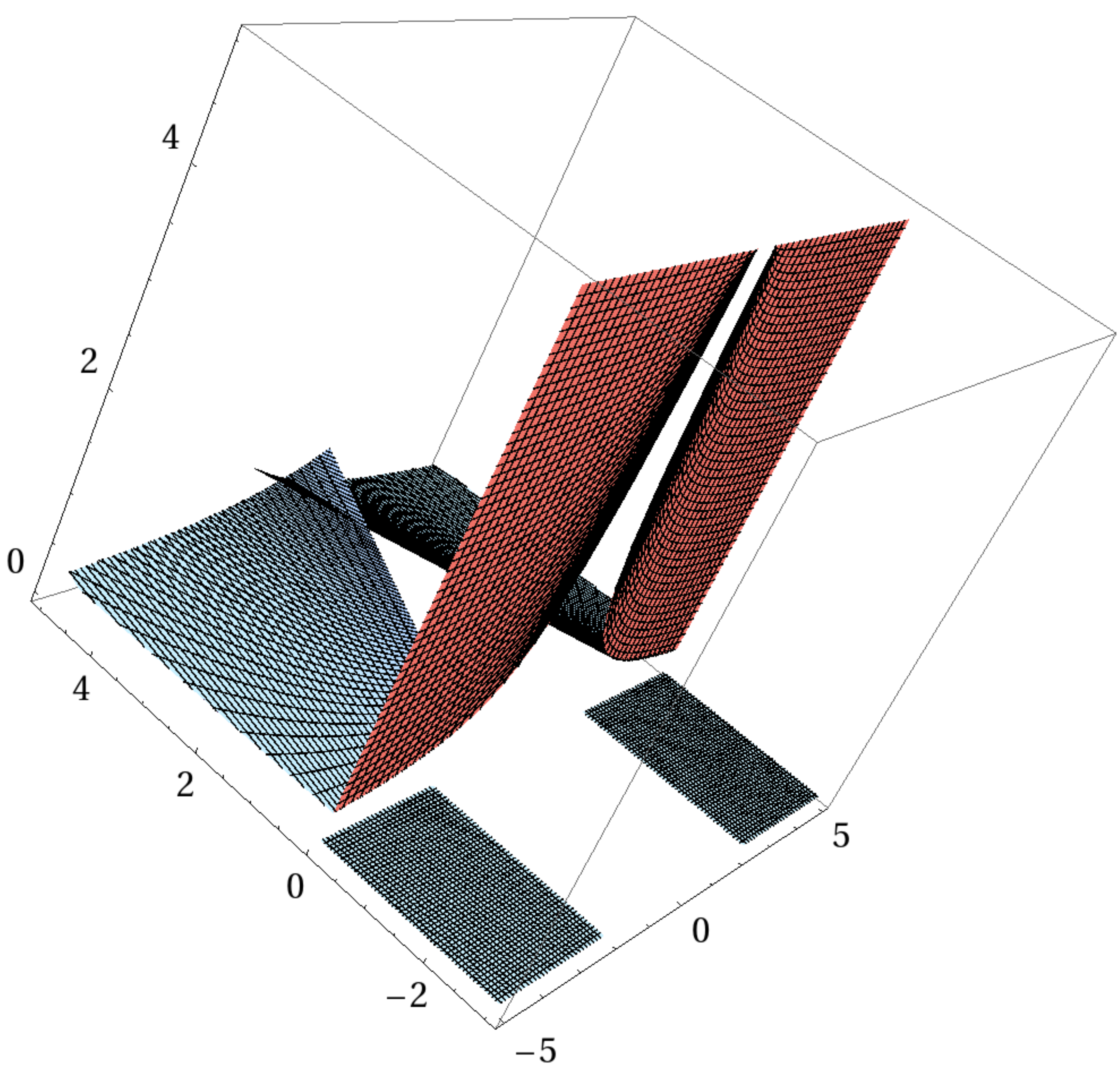}\includegraphics[width=0.33\linewidth]{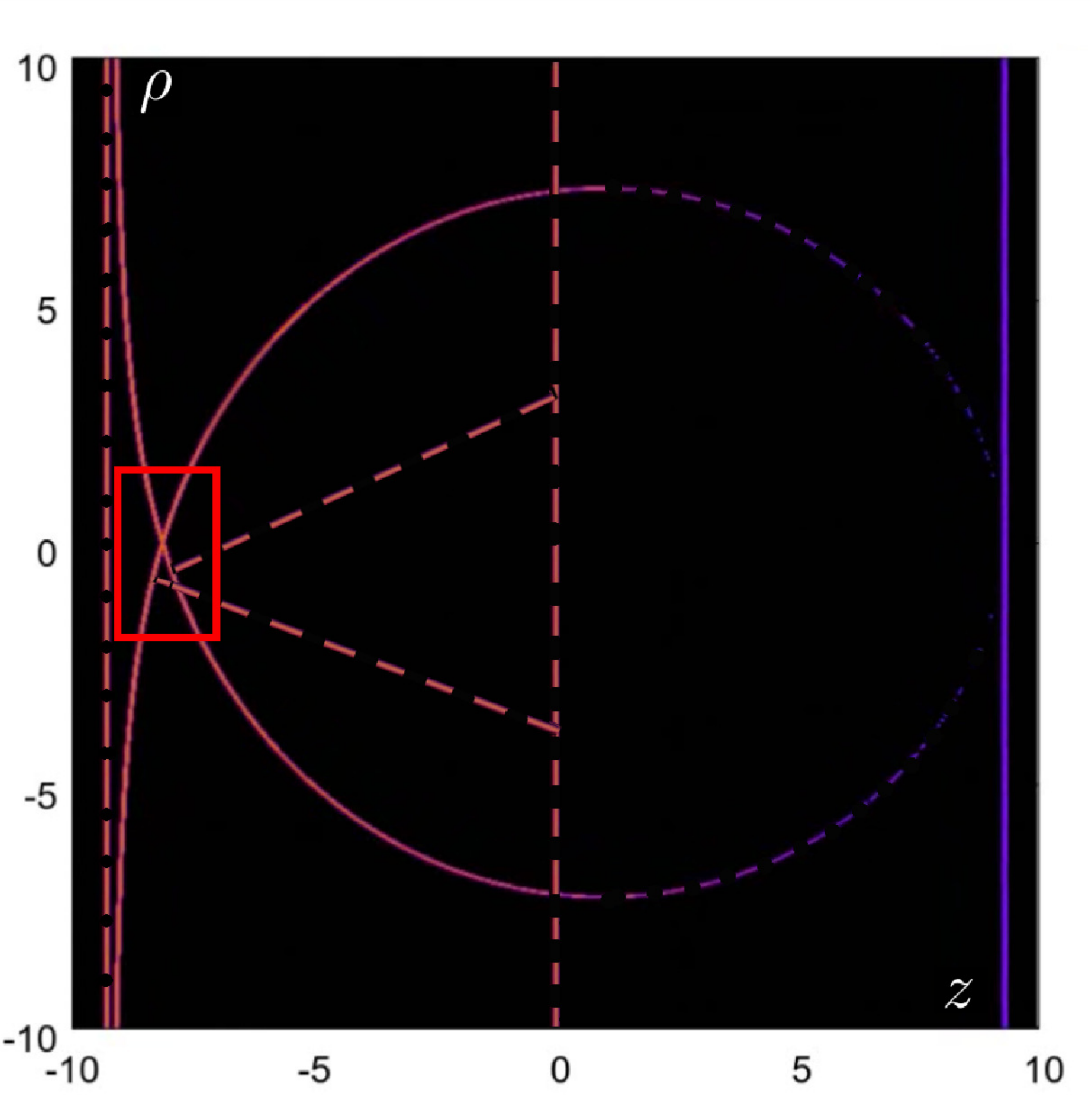}
\caption[Illustration of the region of validity of perturbation theory.]{\emph{Left:} Function $g(\rho)$ which determines the validity of the perturbative approximation. \emph{Middle:} Diagrammatic representation of the $v$-shock null generators for large $\rho$ in $D=5$. \emph{Right:} Spatial slice (in $D=5$) showing the observation region behind the collision event reached by such rays with large $\rho$, where perturbation theory is expected to yield a good result. \emph{From \cite{Sampaio:2013faa}.}}
\label{Validity}
\end{figure}

This function $g(\rho)$ is plotted in Fig.~\ref{Validity}, where we see it grows very fast for large $\rho$. Combined with the fact that rays incident close to the axis are trapped inside the apparent horizon (which is at $\bar\rho=\rho=1$), this suggests that the outside region of the resulting black hole might be well described by perturbation theory.

\begin{figure}[t]
\begin{center}
\includegraphics[scale=0.7]{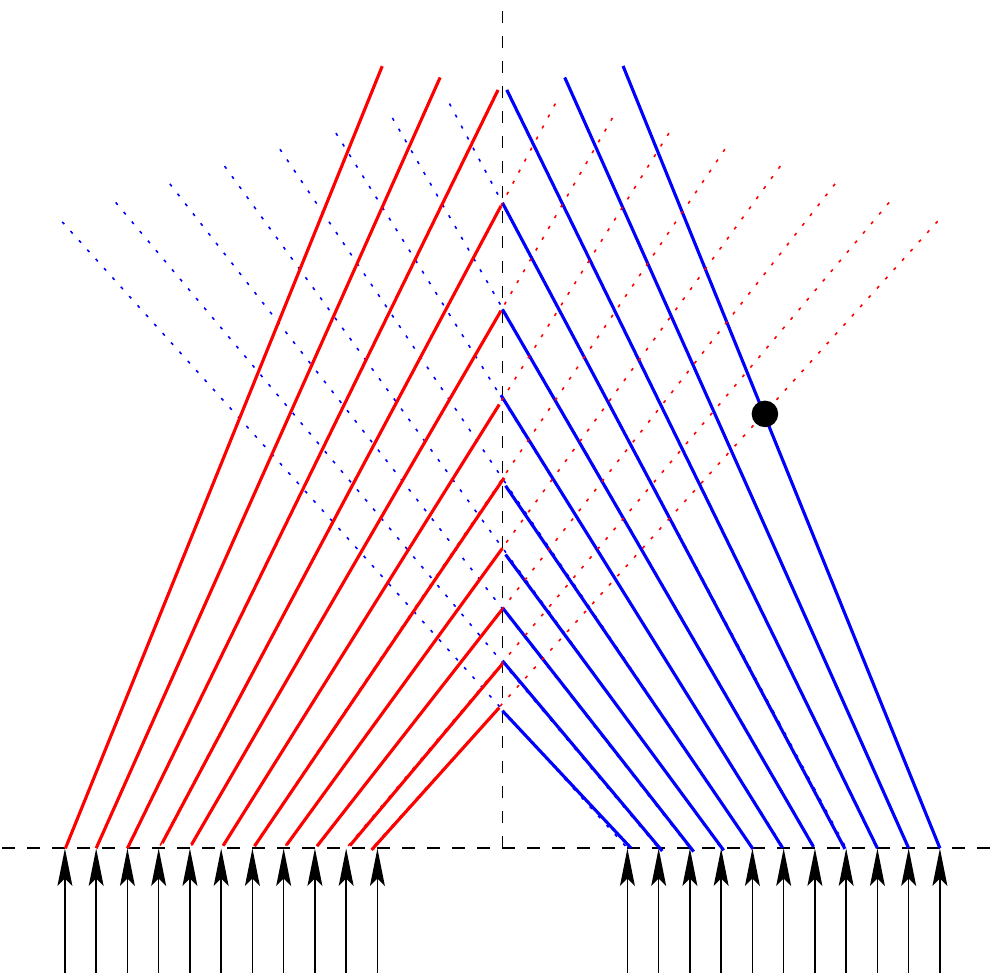}
\begin{picture}(0,0)
\put(-40,90){${\rm ray \ 1}$}
\put(-30,129){${\rm ray \ 2}$}
\put(-116,199){${\rho=0}$}
\put(-46,110){${\mathcal{P}}$}
\put(-155,4){${\rm weak \ shock \ null \ generators}$}
\put(-5,28){$u=0$}
\put(-5,18){${\rm strong \ shock}$}
\end{picture}
\end{center}
\caption[Illustration of the \textit{spatial} trajectories of the weak shock null generators, exhibiting their focusing, and interpretation of the rays seen by a far-away observer.]{Illustration of the \textit{spatial} trajectories of the weak shock null generators, exhibiting their focusing after $u=0$. A point on the axis ($\rho=0$) will be hit, simultaneously, by an infinite number of null generators (two of each are in the represented section). This occurs at the caustic and marks the beginning of the curved region of space-time. Therefore, the continuation of the rays beyond the axis, represented by a dotted line, is merely illustrative. For points outside the axis, such as point $\mathcal{P}$, this diagram suggests an initial radiation signal, associated with ray 1, followed by a later burst, associated with ray 2. The closer the observation point $\mathcal{P}$ is to the axis, the farther those rays come from. \emph{From \cite{Herdeiro:2011ck}.}}
\label{rays}
\end{figure}

Furthermore, we saw in Section \ref{boosted_frame} that perturbation theory should be valid in the region close to $\theta=\pi$, i.e. close to the $-z$ axis. Indeed, due to focusing, the rays reaching such an observer do come from the far field region, as shown in the right panel of Fig.~\ref{Validity} and in Fig.~\ref{rays}.
This is the true meaning of perturbation theory in this problem: it gives an angular expansion of the metric in the future of the collision, with higher orders needed to extrapolate further off the axis or $\theta=\pi$. However, a quantitative restatement of this idea will have to wait until Chapter~\ref{ch:analytics}.

\subsection{The dynamical equations}
\label{field_eqs}
The standard way of applying perturbation theory in general relativity starts with an ansatz for the metric,
\begin{equation}
g_{\mu\nu}=\eta_{\mu\nu}+\sum_{k=1}^\infty h_{\mu\nu}^{(k)}\,,\label{metric_series}
\end{equation}
where each term $h_{\mu\nu}^{(k)}$ is assumed to be sufficiently smaller than the previous ones such that the series converges or a consistent truncation can be made. Then Einstein's equations produce a tower of linear tensorial equations, one for each order, where the source for each $h_{\mu\nu}^{(k)}$ is given by an effective energy-momentum tensor built out of lower order terms. 

Each of these equations generally couples different components, but it is well known that coordinates can be chosen such that they decouple. Defining the trace-reversed metric perturbation
\begin{equation}
\bar{h}_{\mu\nu}^{(k)}\equiv h_{\mu\nu}^{(k)}-\frac{1}{2}\eta_{\mu\nu}\eta^{\alpha\beta}h_{\alpha\beta}^{(k)}\,,
\end{equation}
the de Donder or Lorentz gauge is obtained by imposing $D$ conditions
\begin{equation}
\bar{h}^{(k)\mu\nu}_{\phantom{(k)\mu\nu},\nu}=0\,.\label{de_Donder}
\end{equation}
Then the field equations become a set of wave equations for each component,
\begin{equation}
\Box h_{\mu\nu}^{(k)}=T_{\mu\nu}^{(k-1)}\left[h_{\alpha\beta}^{(j<k)}\right]\,,\label{wave_eq}
\end{equation}
where $\Box\equiv\eta^{\mu\nu}\partial_\mu\partial_\nu$ is the $D$-dimensional d'Alembertian operator and the source on the right-hand side can be computed explicitly order by order.

Here and throughout this thesis, the superscript index $(k)$ denotes the order of the space-time perturbation theory used (i.e. needed) to compute the given object. In particular observe that, for objects which are not linear in the metric, this does not correspond to the perturbation theory order of the object itself. For example, in Eq.~\ref{wave_eq} at $O(k)$, the right-hand side is denoted $T_{\mu\nu}^{(k-1)}$ because, being quadratic in the metric, it is built out of lower order metric perturbations.

\section{Formal solution}
\label{formal_solution}
The formal solution to Eq.~\eqref{wave_eq} can be written using the Green's function for the $\Box$ operator. In Appendix~\ref{app:Green} we define it as the solution to the equation
\begin{equation}
\Box G(x^\mu)=\delta(x^\mu)\,,\label{Green_def}
\end{equation}
and show that, in our coordinates $x^\mu=\{u,v,x^i\}$, it is given by\footnote{$\delta^{(n)}$ denotes the $n$-th derivative of the delta function. See Appendix~\ref{app:Green} for the meaning of non-integer $n$.}
\begin{equation}
G(x)=-\frac{1}{4\pi^{\frac{D-2}{2}}}\delta^{\left(\frac{D-4}{2}\right)}(\chi)\,,\qquad \chi\equiv-\eta_{\mu\nu}x^\mu x^\nu\,.\label{Greens}
\end{equation}
Then, using Theorem 6.3.1 of \cite{Friedlander:112411}, the metric perturbations in de Donder gauge are given by
\begin{equation}
h_{\mu\nu}^{(k)}(x)=F.P. \int d^Dx'\, G(x-x')\left[4\delta(u')\partial_{v'}h^{(k)}_{\mu\nu}(x')+T_{\mu\nu}^{(k-1)}(x')\right]\,,\label{formal_sol}
\end{equation}
where $F.P.$ denotes the finite part of the integral and the source only has support in $u>0$. Using Eq.~\eqref{Green_def}, it is straightforward to check that Eq.~\eqref{formal_sol} obeys Eqs.~\eqref{u_zero} and \eqref{wave_eq} (see Appendix~\eqref{app:char_data}).

The first term inside the square brackets of Eq.~\eqref{formal_sol}, which we shall call the \emph{surface term}, simply propagates the initial data on $u=0^+$ and makes the solution compliant with Eq.~\eqref{u_zero}. The second one, which we shall call the \emph{volume term}, encodes the non-linearities resulting from the interference with the background radiation sources generated by lower order terms.

Note that at first order there is no source: the solution is simply the propagation of the initial conditions on a flat background. Moreover, since the initial data are exact at second order, for $k\geq3$ there are only volume terms. This might be the reason why the second order result of D'Eath and Payne in $D=4$ is already in good agreement with the predictions of numerical relativity (see Chapter \ref{ch:intro}).

\begin{figure}[t]
\begin{center}
\begin{picture}(0,0)
\put(100,120){$u$}
\put(217,57){$(0,v',\vec{x}')$}
\put(260,44){$\vec{x}$}
\put(266,120){$v$}
\put(194,157){$\mathcal{P}=(u,v,\vec{x})$}
\put(323,104){$u=0$}
\end{picture}
\includegraphics[scale=0.4]{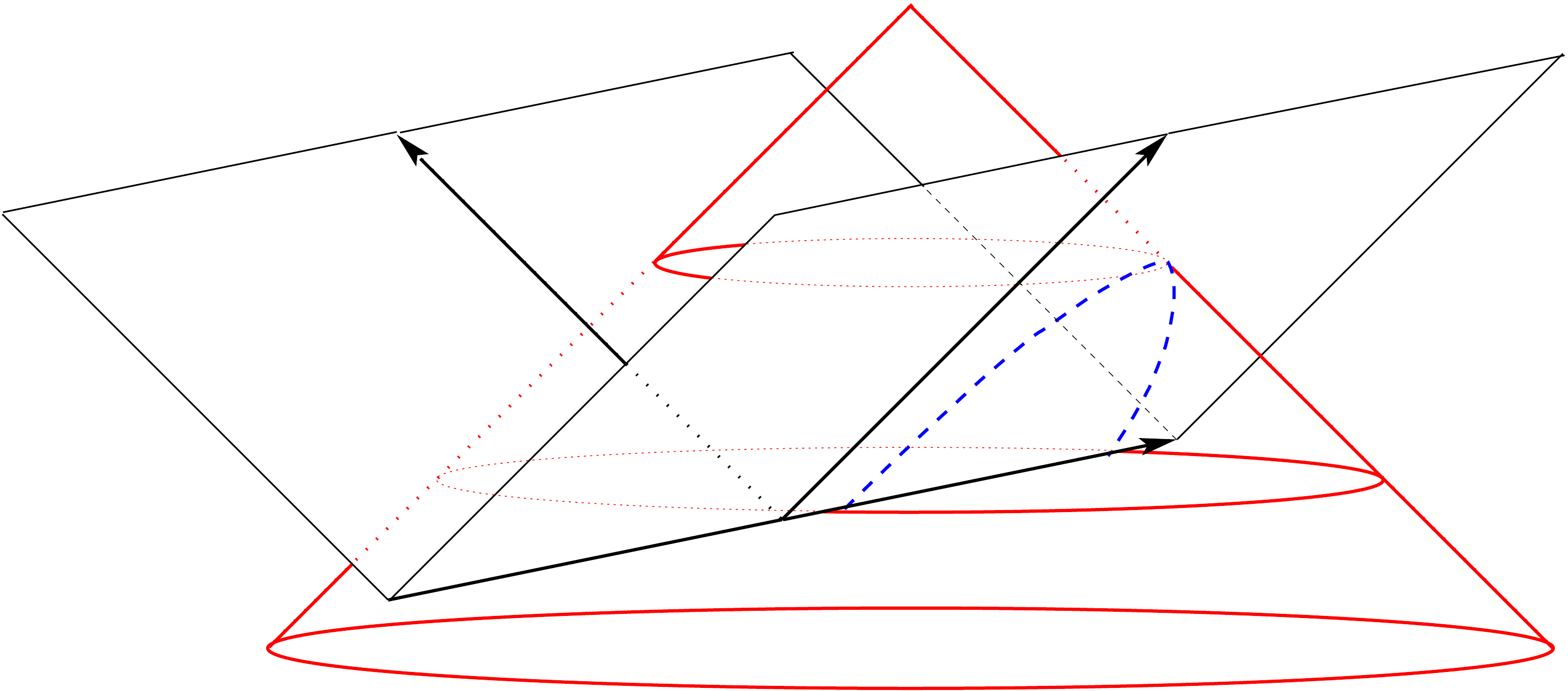}
\end{center}
\caption[Intersection of the past light cone of a space-time event with the null hypersurface where the initial data have support.]{The dashed blue parabola is the intersection of the past light cone of a space-time event $\mathcal{P}(u,v,\vec{x})$, to the future of $u=0=v$, with the null hypersurface $u=0$. \emph{From~\cite{Herdeiro:2011ck}.}}
\label{pastlightcone}
\end{figure}
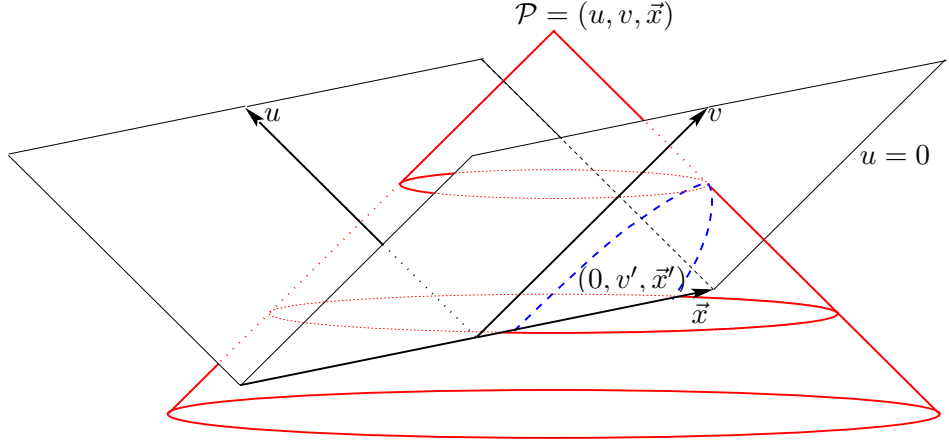

Fig.~\ref{pastlightcone} is a space-time diagram showing the past light cone of an observer $\mathcal{P}$ after the collision. This is defined by $\chi=0$ and its intersection with the $u=0$ plane is a parabola,
\begin{equation}
v'=v-\frac{|x^i-{x'}^i|^2}{u}\,.
\end{equation}
The initial data on $u=0$ have support above the green line of Fig.~\ref{5DspacetimeDiag} (where the collision takes place).

\subsection{Gauge fixing}
\label{formal_gauge}
The initial data, Eqs.~\eqref{initial_data}, are not in de Donder gauge. Therefore, before inserting them in the formal solution, Eq.~\eqref{formal_sol}, we need to make a coordinate transformation
\begin{equation}
x^\mu\rightarrow x^\mu+\sum_{k=1}^\infty \xi^{(k)^\mu}(x^\alpha)\,,\label{gauge_x}
\end{equation}
where $\xi^\mu$ will be determined order by order through the de Donder gauge condition, Eq.~\eqref{de_Donder}. In what follows we derive a condition to be imposed on the initial data, such that the de Donder gauge is preserved by the evolution equations. From Eq.~\eqref{wave_eq},
\begin{equation}
\Box\bar{h}_{\mu\nu}^{(k),\nu}=\partial^\nu\bar{T}_{\mu\nu}^{(k-1)}=0\,,
\end{equation}
where, as before, the bar denotes trace-reversed quantities. The conservation of the effective energy-momentum tensor was checked explicitly up to third order, but it must also hold at higher orders since it is a consequence of the symmetries of the (flat vacuum) background. The formal solution to the de Donder gauge condition is then
\begin{equation}
\bar{h}_{\mu\nu}^{(k),\nu}(x)=4\int d^Dx'\, G(x-x')\delta(u')\partial_{v'}\bar{h}^{(k),\nu}_{\mu\nu}(x')\,,\label{formal_sol_gauge}
\end{equation}
where the finite part prescription was left implicit. Thus we conclude that
\begin{equation}
\left.\partial_{v}\bar{h}^{(k),\nu}_{\mu\nu} \right|_{u=0^+}=0 \,, \label{deDonderV2}
\end{equation}
is a sufficient condition for the de Donder gauge to be maintained for all $u$.

\section{Reduction to three dimensions}
\label{reduction_3D}
The general solution of Eq.~\eqref{formal_sol} does not take into account the axial symmetry of the problem (rotations around the $z$ axis). Indeed, if we work with spherical coordinates on the $(D-2)$-dimensional transverse space, we should be able to integrate out the $D-3$ angles $\phi^a$, and be left with an integral in $\rho=\sqrt{x^ix_i}$ only.

In the coordinates we have been using so far, $x^\mu=\{u,v,x^i\}$, the metric has the generic form
\begin{equation}\label{eq:metricDeDonder}
ds^2=ds^2_{Flat}+h_{uu}du^2+2h_{ui}dx^{i}du+h_{vv}dv^2+2h_{uv}dudv+2h_{vi}dx^idv+h_{ij} dx^i dx^j \,.
\end{equation}

Defining a basis of tensors in that plane,
\begin{equation}\label{eq:basis_tensors}
\delta_{ij}\,,\qquad \Gamma_i\equiv \frac{x_i}{\rho}\,,\qquad \Delta_{ij}\equiv \delta_{ij}-(D-2)\Gamma_i\Gamma_j \,, 
\end{equation} 
where we have chosen the last one to be traceless, we can decompose the metric perturbations $h_{ij}$ into seven functions of $(u,v,\rho)$, here denoted $A,B,C,E,F,G,H$, in the following way:
\begin{eqnarray}
&h_{uu}\equiv A=A^{(1)}+A^{(2)}+\ldots \qquad &h_{ui}\equiv B \,\Gamma_i =(B^{(1)}+B^{(2)}+\ldots)\Gamma_i  \nonumber\\
&h_{uv}\equiv C=C^{(1)}+C^{(2)}+\ldots \qquad &h_{vi}\equiv F \,\Gamma_i =(F^{(1)}+F^{(2)}+\ldots)\Gamma_i  \label{gen_perts}\\
&h_{vv}\equiv G=G^{(1)}+G^{(2)}+\ldots \qquad &h_{ij}\equiv E \,\Delta_{ij}+H\, \delta_{ij} \nonumber\\&&\phantom{h_{ij}}= (E^{(1)}+\ldots) \Delta_{ij}+(H^{(1)}+\ldots) \delta_{ij} \ .\nonumber 
\end{eqnarray}

A similar decomposition can be applied to the sources $T^{(k-1)}_{\mu\nu}$. For example, for the transverse components,
\begin{eqnarray}
T^{(k-1)}_{ij}&=&T_H^{(k-1)}(u,v,\rho)\delta_{ij}+T^{(k-1)}_E(u,v,\rho)\Delta_{ij}\,.\label{TE_TH}
\end{eqnarray}

Now let $F$ denote generically any of those functions and let $S$ be its associated source, including both volume and surface terms,
\begin{equation}
S^{(k)}(u,v,\rho)\equiv T^{(k-1)}(u,v,\rho)+4\delta(u)\partial_{v}F^{(k)}(0,v,\rho) \ .\label{def_S}
\end{equation}

In Section~\ref{app:Green3D} of Appendix~\ref{app:Green} we show that Eq.~\eqref{formal_sol} implies
\begin{equation}
F(u,v,\rho)=\int du'\int dv'\int d\rho'\,G_m(u-u',v-v';\rho,\rho')S(u',v',\rho')\,,\label{sol_3D}
\end{equation}
where the new Green's function $G_m$ depends on the rank $m=\{0,1,2\}$ of the invariant tensor multiplying $F$ in the respective component of $h_{\mu\nu}$. For example, $m=0$ for $A$ and $H$, $m=1$ for $B$ and $m=2$ for $E$ as can be observed in Eqs.~\eqref{gen_perts}.

It is given by
\begin{equation}
G_m(u-u',v-v';\rho,\rho')=-\frac{1}{2\rho}\left(\frac{\rho'}{\rho}\right)^{\frac{D-4}{2}}I_m^{D,0}(x_\star)\,
\end{equation}
where
\begin{equation}
x_\star=\frac{\rho^2+\rho'^2-(u-u')(v-v')}{2\rho\rho'}\,,\label{def_x_star}
\end{equation} 
and the one-dimensional functions $I^{D,n}_m$ are defined as
\begin{equation}
I_m^{D,n}(x)=\frac{\Omega_{D-4}}{(2\pi)^{\frac{D-2}{2}}}\lambda_m^{-1}\int_{-1}^1dz\,\partial^{m}(1-z^2)^{\frac{D-5}{2}+m}\delta^{\left(\frac{D-4}{2}-n\right)}(z-x)\,,
\end{equation}
with $\lambda_m=\{1,-(D-3),(D-1)(D-3)\}$.

Thus, the problem of finding the post-collision metric is now formally solved. The extraction of physically relevant information from that solution will be the subject of the next chapter.

\chapter{Extraction of gravitational radiation}
\label{ch:radiation}
\renewcommand{\textflush}{flushepinormal}

\epigraph{{\bf energy} $\bullet$ the property of matter and radiation which is manifest as a capacity to perform work; from the greek \emph{energeia}.}{Oxford English Dictionary}

In Chapter \ref{ch:kinematics} we analysed the geometric properties of shock waves and in Chapter \ref{ch:dynamics} we developed a perturbative framework which allowed us to obtain a formal expression for the metric in the future of the collision. It is now time for us to see how this can be used to extract physically meaningful information, namely the energy that is radiated away by gravitational waves. If the end product of the collision is a black hole at rest, its mass must be the difference between the initial energy of the colliding particles and the total energy radiated. We shall see that the power or flux of gravitational energy that reaches a sphere at `infinity' is characterised by a single function of a retarded time $\hat{\tau}$ and angular coordinate $\hat{\theta}$, called the \emph{news function}, which is then used to derive a formula for the inelasticity.

We start in Section~\ref{mass_AF} by discussing the concept of energy in asymptotically flat space-times and introducing the Bondi mass, which is then detailed in Section~\ref{bondi_mass} for arbitrary $D$ dimensions. Section~\ref{Donder_Bondi} establishes the relationship between de Donder and Bondi coordinates. Due to its technical character, a summary is provided in the end, Section~\ref{news_Donder}, to which a reader may safely jump without loss of comprehensibility.

\section{Mass and energy in asymptotically flat space-times}
\label{mass_AF}
The notion of mass, or energy, in general relativity has always been a delicate topic due to the difficulty in defining it precisely in generic terms. In particular, there is no local definition of the energy density of the gravitational field (as there is, for instance, for the electromagnetic field). However, satisfactory definitions of total energy and radiated energy have been found for isolated systems, for which the behaviour of the gravitational field at large distances away from the source can provide a notion of `total gravitational mass'. 

Ideally isolated systems live in what are known as \emph{asymptotically flat} space-times. Several definitions exist, with varying degrees of mathematical sophistication (see for example \cite{Wald:1984rg}), but they all rely on some notion of `infinity' and on appropriate fall-off behaviour of the metric components (in suitable coordinates).

For stationary, asymptotically flat space-times (which are vacuum near infinity), the time translation symmetry implies there is an associated conserved quantity, called the Komar mass \cite{Komar}, which is given by an integral over a distant sphere of the corresponding Killing vector field. In the non-stationary case, alternative definitions must be found. There are two main options.

The Arnowitt-Deser-Misner (ADM) energy-momentum was motivated by the Hamiltonian formulation of general relativity \cite{ADM}, and gives a well-defined notion of energy/mass and momentum at \emph{spatial infinity}. It is usually interpreted as the total energy available in the space-time. In particular, it cannot differentiate between rest mass and energy stored in gravitational waves, which makes it inappropriate for our purposes.

If the limiting procedure is taken along a null direction instead, we get a notion of mass at \emph{null infinity} known as the Bondi or Bondi-Sachs mass \cite{Bondi:1962zz,Sachs:1962wk}. This mass is time-dependent but never increasing, which means that it takes into account the energy that is carried away by gravitational waves (which is itself always positive). Furthermore, the positivity of the Bondi mass itself, and of the ADM mass, have been proved for physically reasonable space-times (\cite{Wald:1984rg} and references therein).

\begin{figure}
$\phantom{.\qquad.\qquad.\qquad.}$\includegraphics[width=0.8\linewidth]{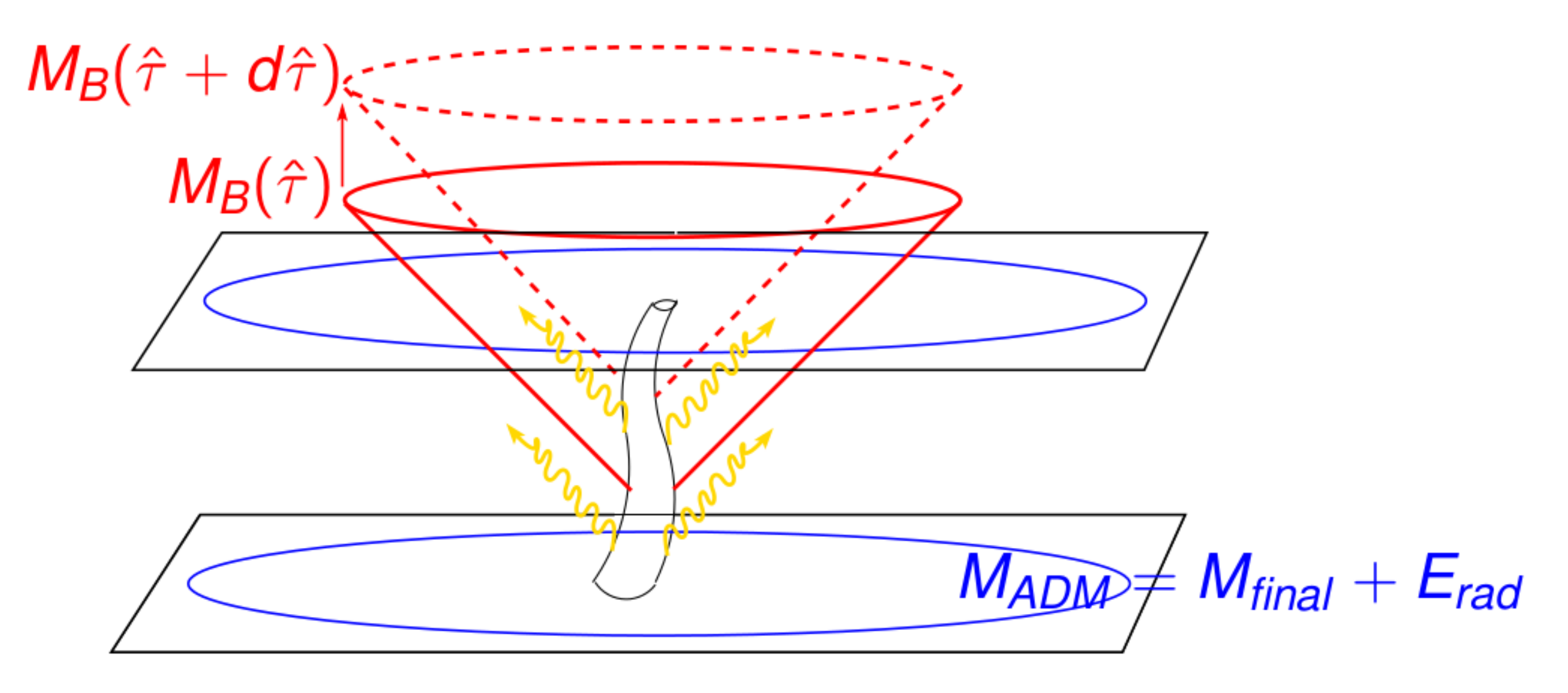}
\caption[Schematic illustration of the different slices where the ADM and Bondi masses are computed.]{\label{BondiADM}Schematic illustration of the different slices where the ADM and Bondi masses are computed: whereas the ADM mass is computed on a spatial slice (in blue), the Bondi mass is defined on a null slice (in red) and thus is sensitive to gravitational radiation (in yellow). \emph{From \cite{Sampaio:2013faa}.}}
\end{figure}

Fig.~\ref{BondiADM} illustrates the difference between the Bondi and ADM definitions. While the ADM mass is computed on the spatial (blue) slices, and thus measures the total (constant) energy available in the space-time, the Bondi mass is computed on null slices (red cones) which previously emitted radiation never reaches. Therefore, the Bondi energy is the energy remaining in the space-time at some retarded time, after the emission of gravitational radiation, and differs from the ADM mass precisely by the integral of such energy flux. All these three definitions of mass agree wherever their applicability regimes overlap (for instance, for a Schwarzschild black hole).

Finally, another notion of energy-momentum exists in linearised theory. The Landau-Lifshitz pseudo-tensor \cite{Landau} is essentially $T^{(1)}_{\mu\nu}$ in Eq.~\eqref{wave_eq}, i.e. it is an effective energy-momentum tensor generated by first-order metric perturbations. Despite not being unique nor gauge invariant, a well-defined gauge-invariant energy can be defined by a suitable integral at infinity. Its generalisation to higher dimensions \cite{Tanabe:2011es} was actually the method employed in \cite{Herdeiro:2011ck} for the first-order calculation. However, in what follows we shall focus exclusively on the Bondi formalism which is valid non-perturbatively.

\section{Bondi mass in $D$ dimensions}
\label{bondi_mass}
The Bondi formalism was only recently extended to higher dimensions \cite{Tanabe:2011es,Tanabe:2012fg}. The starting point is the choice of Bondi coordinates $(\hat{\tau},\hat{r},\hat{x}^{\hat{I}})$, where $\hat{\tau}$ is a retarded time, $\hat{r}$ is a radial (null) coordinate and $\hat{x}^{\hat{I}}$ are $D-2$ angles, such that the metric can be written in Bondi form,
\begin{equation}
ds^2=g_{\hat \tau \hat \tau}d\hat\tau^2+2g_{\hat \tau \hat r}d\hat \tau d\hat r+2g_{\hat I \hat \tau}dx^{\hat I}d\hat\tau+g_{\hat I \hat J}dx^{\hat I}dx^{\hat J} \ . \label{Bondi_metric}
\end{equation}
The radial coordinate $\hat{r}$ is chosen to be an areal radius, i.e. such that
\begin{equation}\label{eq:arealR}
\sqrt{|g_{\hat I\hat J}|}=\hat r^{D-2}\omega_{D-2}\ ,
\end{equation}
where $\omega_{D-2}$ is the volume element of the unit $(D-2)$-sphere (not be be confused with the total volume $\Omega_{D-2}$). In these coordinates, null infinity is located at $\hat{r}=\infty$.

Asymptotic flatness is then defined by imposing an appropriate fall-off behaviour with outgoing boundary conditions at null infinity,
\begin{equation}
\dfrac{g_{\hat I \hat J}}{\hat r^2}=\omega_{\hat I \hat J}+\mathfrak{h}_{\hat I\hat J} =\omega_{\hat I \hat J}+\sum_{k\geq 0}\dfrac{\mathfrak{h}_{\hat I \hat J}^{[k+1]}}{\hat r^{D/2+k-1}} \ ,\label{eq:BondiDecay}
\end{equation}
where $\omega_{\hat I \hat J}$ is the metric on the unit $(D-2)$-sphere and $k$ runs over all (non-negative) integers for $D$ even and semi-integers for $D$ odd.

This, together with Einstein's equations (in vacuum), is enough to fix the asymptotic behaviour of all metric components. In particular,
\begin{equation}
g_{\hat \tau \hat \tau}=-1-\sum_{k=0}^{k<\frac{D-2}{2}}\frac{a_{k+1}}{\hat{r}^{\frac{D-2}{2}+k}}+\frac{m(\hat{\tau},\hat{x}^{\hat{I}})}{\hat{r}^{D-3}}+O\left(\hat{r}^{-\left(D-\frac{5}{2}\right)}\right)\,.
\end{equation}
The coefficients $a_{k+1}$ are unimportant but $m(\hat{\tau},\hat{x}^{\hat{I}})$ defines the Bondi mass,
\begin{equation}
M_B(\hat{\tau})\equiv\frac{D-2}{16\pi G_D}\int_{S^{D-2}}m(\hat{\tau},\hat{x}^{\hat{I}})\,d\Omega_{D-2}\,.
\end{equation}
Einstein's equations guarantee its finiteness and yield an evolution equation,
\begin{equation}\label{eq:BondiMassLoss}
\dfrac{dM_B}{d\hat\tau}=-\dfrac{1}{32\pi G_D}\int_{S^{D-2}}\dot{\mathfrak{h}}_{\hat I\hat J}^{[1]}\dot{\mathfrak{h}}^{[1]\hat I\hat J}d\Omega_{D-2}\leq0 \ ,
\end{equation}
where the latin indices $\hat{I}$ are raised and lowered with $g_{\hat{I}\hat{J}}$, $\mathfrak{h}_{\hat I \hat J}^{[1]}$ was defined in Eq.~\eqref{eq:BondiDecay} and the dot denotes a derivative with respect to retarded time $\hat{\tau}$. 

In the remainder of this chapter we shall apply this mass-loss formula to our problem of shock wave collisions and derive an expression for the inelasticity, valid non-perturbatively. Before diving into the details, however, we will specialise Eq.~\eqref{eq:BondiMassLoss} to axisymmetric space-times.

\subsection{Bondi mass-loss formula in axisymmetric space-times}
\label{bondi_axisym}
In axisymmetric space-times one can choose one of the $D-2$ angles to be the angle with the axis of symmetry, such that the remaining $D-3$ angles lie on a transverse plane. So we split
\begin{equation}
\hat{x}^{\hat{I}}=(\hat{\theta},\hat{\phi}^a)\,,
\end{equation}
and the metric components only depend on $\hat{\theta}$. Moreover,
\begin{equation}
\mathfrak{h}_{\hat I \hat J}dx^{\hat I} dx^{\hat J}=\mathfrak{h}_{\hat \theta \hat \theta}\,d{\hat\theta}^2+\mathfrak{h}_{\hat \phi \hat \phi}\sin^2\hat\theta\,d\Omega^2_{D-3} \ ,
\end{equation}
and the areal radius condition, Eq.~\eqref{eq:arealR}, becomes
\begin{equation}
\Big(1+\mathfrak{h}_{\hat \theta \hat \theta}\Big)\left(1+\mathfrak{h}_{\hat \phi \hat \phi}\right)^{D-3}=1\,.
\end{equation}
Thus $\mathfrak{h}_{\hat \theta \hat \theta}$ can be eliminated in favour of $\mathfrak{h}_{\hat \phi \hat \phi}$. In particular, asymptotically we have
\begin{equation}
\mathfrak{h}_{\hat \theta \hat \theta}\rightarrow-(D-3)\mathfrak{h}_{\hat \phi \hat \phi}\,.
\end{equation}
Then the integration over the transverse angles $\hat{\phi}^a$ in Eq.~\eqref{eq:BondiMassLoss} can be made, and we get the following angular power flux
\begin{equation}\label{eq:BondiMassSimpler}
\dfrac{dM_B}{d\hat\tau d\cos\hat\theta}=-\dfrac{(D-2)(D-3)\Omega_{D-3}}{32\pi G_D}\lim_{\hat r\rightarrow +\infty}\left[\hat r\hat \rho^{\frac{D-4}{2}} \dot{\mathfrak{h}}_{\hat \phi \hat \phi}\right]^2 \; ,
\end{equation}
where $\hat{\rho}=\hat{r}\sin\hat{\theta}$ is the radius on the transverse plane orthogonal to the axis of symmetry. We now have all we need to define the inelasticity of the collision in Bondi coordinates.

\subsection{Formula for the inelasticity of the collision}
\label{inelasticity}
The \emph{inelasticity} of the collision, $\epsilon$, is the ratio between the energy radiated and the total initial energy. Since each of the colliding shock waves has energy $E$, and the radiated energy is the difference between the initial and final Bondi masses,
\begin{equation}
\epsilon\equiv1- \lim_{\hat{\tau}\rightarrow\infty}\frac{M_B(\hat{\tau})}{2E}\,.
\end{equation}
Recall that the energy $E$ relates to the energy parameter $\kappa$ through
\begin{equation}
\kappa=\frac{8\pi G_D E}{\Omega_{D-3}}\,,
\end{equation}
and in Section \ref{initial_value} we have chosen units in which $\kappa=1$. From Eq.~\eqref{eq:BondiMassSimpler} we then conclude that the inelasticity is given by the following time and angular integral
\begin{equation}
\epsilon=\int_{-1}^1\frac{d\cos\hat\theta}{2}\epsilon(\hat\theta)\equiv\int_{-1}^1\frac{d\cos\hat\theta}{2}\int d\hat\tau\,\dot{\mathcal{E}}(\hat\tau,\hat\theta)^2\,,\label{e1}
\end{equation}
where
\begin{equation}
\dot{\mathcal{E}}(\hat\tau,\hat\theta)=\sqrt{\frac{(D-2)(D-3)}{4}}\lim_{\hat r\rightarrow +\infty}\left[\hat r\hat \rho^{\frac{D-4}{2}} \dot{\mathfrak{h}}_{\hat \phi \hat \phi}\right]\,,\label{news1}
\end{equation}
is essentially the generalisation of Bondi's \emph{news function} to higher dimensions.

In the next section we shall study the relationship between de Donder and Bondi coordinates, and rewrite Eqs.~\eqref{e1} and \eqref{news1} in terms of quantities we know. We will see that Bondi coordinates actually asymptote to (a combination of) de Donder coordinates, and that the news function is determined by the transverse metric perturbations $h_{ij}$ only.

\section{Relationship between de Donder and Bondi coordinates}
\label{Donder_Bondi}
In the de Donder coordinates of Chapter~\ref{ch:dynamics}, the metric has the generic form
\begin{equation}\label{eq:metricDeDonder}
ds^2=ds^2_{Flat}+h_{uu}du^2+2h_{ui}dx^{i}du+h_{vv}dv^2+2h_{uv}dudv+2h_{vi}dx^idv+h_{ij} dx^i dx^j \,.
\end{equation}

First we change to Bondi-like coordinates $(\tau,r,\theta)$, where $\tau=t-r$ is a retarded time, $r=\sqrt{\rho^2+z^2}$ is the radius of a $(D-1)$-sphere and $\theta$ is the angle with the $z$ axis. The other angles $\phi^a$ are ignorable so we omit them in this discussion. Eq.~\eqref{eq:metricDeDonder} now reads
\begin{equation}\label{eq:metricDeDonderAngular}
ds^2=ds^2_{Flat}+h_{\tau\tau}d\tau^2+2h_{\tau r}d\tau dr+h_{rr}dr^2+2h_{\tau \theta}d\tau d\theta+h_{\theta\theta}d\theta^2+2h_{r\theta}drd\theta+r^2\sin^2\theta h_{\phi\phi}d\Omega_{D-3}^2 \,,
\end{equation}
where, from Eqs.~\eqref{gen_perts},
\begin{eqnarray}
h_{\tau\tau}&=&A+G+2C \ ,\nonumber\\
h_{\tau r} &=&A(1-\cos\theta)+G(1+\cos\theta)+2C+(B+F)\sin\theta \ , \nonumber\\
h_{rr}&=&A(1-\cos\theta)^2+G(1+\cos\theta)^2+(2C+H-(D-3)E)\sin^2\theta+\nonumber \nonumber\\
&&+2\left(B(1-\cos\theta)+F(1+\cos\theta)\right)\sin\theta \ ,\nonumber\\
h_{\tau \theta}&=&r\left[(A-G)\sin\theta+(B+F)\cos\theta\right] \ ,\label{eq:MetricTauR}\\
h_{\theta \theta}&=&r^2\left[(A+G-2C )\sin^2\theta+2(B-F)\sin\theta\cos\theta+(H-(D-3)E)\cos^2\theta\right] \ ,\nonumber\\
h_{r\theta}&=&r\left[\left(A(1-\cos\theta)-G(1+\cos\theta)\right)\sin\theta+(2C+H-(D-3)E)\sin\theta\cos\theta+\right. \nonumber\\
&&+\left.(B+F)\cos\theta+(B-F)\left(\sin^2\theta-\cos^2\theta\right)\right]\ ,\nonumber\\
h_{\phi\phi}&=&E+H \ .\nonumber
\end{eqnarray}

Next we do the actual transformation to Bondi coordinates,
\begin{equation}
x^\mu=\hat x^{\mu}+\xi^{\mu}(\hat x^{\alpha})\,.
\end{equation}
To make contact with Section~\ref{bondi_mass}, we denote the metric perturbations in these hatted coordinates by $\mathfrak{h}_{\mu\nu}$, where the indices now run over all coordinates. The Bondi gauge is then defined by imposing $\mathfrak{h}_{\hat r \hat r}=\mathfrak{h}_{\hat r\hat\theta}=0$ and the areal radius condition, Eq.~\eqref{eq:arealR}.

Finding the vector $\xi^{\mu}$ that satisfies these three conditions is a demanding task. However, since we are only interested in the asymptotic behaviour at null infinity, we shall assume that $\xi^{\mu}$ decays sufficiently fast with some power of $1/\hat r\sim 1/r$, such that both coordinate systems match asymptotically\footnote{We shall see later in Chapter~\ref{ch:second_order} that this is not true in $D=4$. As discussed extensively by Payne~\cite{Payne}, $\xi^{\mu}$ contains $\ln \hat r$ terms in $D=4$. Here, however, we focus on $D>4$, following \cite{{Coelho:2012sy}}.}. Then the Bondi metric perturbations are given by
\begin{equation}
\mathfrak{h}_{\mu\nu}=h_{\mu\nu}+\eta_{\mu\nu,\alpha}\xi^{\alpha}+2\eta_{\alpha \left(\nu\right.}\xi^{\alpha}_{,\left.\mu\right)} +O(\xi^2)\,,
\end{equation}
and the solution (with $\xi^{\hat{\phi}^a}=0$) is
\begin{eqnarray}
\xi^{\hat \tau}&=&\int \dfrac{h_{r r}}{2}+\gamma(\hat \tau,\hat \theta) \ , \\
\xi^{\hat \theta}&=&\int \dfrac{1}{\hat r^2}\int \dfrac{h_{r r,\hat\theta}}{2}-\int \dfrac{h_{ r \theta}}{\hat r^2}-\dfrac{\gamma(\hat \tau,\hat\theta)_{,\hat \theta}}{\hat r}+\beta(\hat \tau, \hat \theta)\ ,\\
\xi^{\hat r}&=&-\dfrac{\hat r}{2(D-2)}\left[\dfrac{h_{\theta \theta}}{\hat r^2}+(D-3)h_{ \phi \phi}-2\left((D-3)\cot \hat \theta\, \xi^{\hat\theta}+\xi^{\hat\theta}_{,\hat \theta}\right)\right] \ ,
\end{eqnarray}
where $\int$ denotes primitivation with respect to $\hat{r}$. 

$\gamma(\hat\tau,\hat\theta)$ and $\beta(\hat\tau,\hat\theta)$ are two arbitrary integration functions, constrained only by requiring the Bondi metric to be asymptotically flat, Eq.~\eqref{eq:BondiDecay}. In particular,
\begin{equation}\label{eq:PhiPhi1st}
(D-2)\mathfrak{h}_{\hat \phi \hat \phi}=h_{\phi \phi}-\dfrac{h_{\theta \theta}}{\hat r^2}+2(\cot \hat \theta-\partial_{\hat \theta})\left(\int \dfrac{1}{\hat r^2}\int \dfrac{h_{ r r,\hat\theta}}{2}-\int \dfrac{h_{ r \theta}}{\hat r^2}-\dfrac{\gamma_{,\hat \theta}}{\hat r}+\beta\right) \,.
\end{equation}

To have the required asymptotic decay, the contribution from $\beta$ must vanish. Equating the differential operator acting on $\beta$ to zero gives $\beta=a(\hat\tau)\sin\hat\theta$.

The same applies to the $\gamma$ contribution in $D>4$. In $D=4$, however, $\gamma$ remains arbitrary. This is the well known \emph{supertranslation} freedom \cite{D'Eath:1992hd,Bondi:1962zz,Tanabe:2011es}, but we shall ignore it nonetheless (amounting to a choice of a particular \emph{supertranslation state}).

All that is left is to write Eq.~\eqref{eq:PhiPhi1st} in terms of the scalar functions in Eqs.~\eqref{eq:MetricTauR} (evaluated in Bondi coordiantes) and take a $\hat\tau$ derivative. The result, which depends on many of them, can be simplified by making use of the de Donder gauge conditions, Eq.~\eqref{de_Donder}, which near null infinity read
\begin{eqnarray}
\lim_{\hat r\rightarrow+\infty} \hat r^{\frac{D-2}{2}}\left[(1-\cos\hat\theta)\dot A+\sin\hat\theta\dot B+\tfrac{D-2}{4}(1+\cos\theta)\dot H\right]&=&0 \ ,\nonumber\\
\lim_{\hat r\rightarrow+\infty} \hat r^{\frac{D-2}{2}}\left[(1+\cos\hat\theta)\dot G+\sin\hat\theta\dot F+\tfrac{D-2}{4}(1-\cos\theta)\dot H\right]&=&0 \ ,\nonumber\label{eq:GaugeAsympt}\\
\lim_{\hat r\rightarrow+\infty} \hat r^{\frac{D-2}{2}}\left[(1-\cos\hat\theta)\dot B+(1+\cos\hat\theta)\dot F+\sin\hat\theta\left(2\dot C-(D-3)\dot E-\tfrac{D-4}{2}\dot H\right)\right]&=&0\nonumber \,.\\
\end{eqnarray}

These, together with Eqs.~\eqref{eq:MetricTauR}, imply that
\begin{eqnarray}\label{eq:AsymptoticDeDonder}
\lim_{\hat r\rightarrow+\infty} \hat r^{\frac{D-2}{2}} \dot h_{r r}&=&\lim_{\hat r\rightarrow+\infty} \hat r^{\frac{D-2}{2}} \dot h_{ r \theta}=0 \\
\lim_{\hat r\rightarrow+\infty} \hat r^{\frac{D-2}{2}} \dfrac{\dot h_{ \theta  \theta}}{\hat r^2}&=&-(D-3)\left(\dot E+\dot H\right) \,,
\end{eqnarray}
which, inserted in Eq.~\eqref{eq:PhiPhi1st}, yields
\begin{equation}
\dot{\mathfrak{h}}_{\hat\phi\hat\phi}=\dot{E}+\dot{H}\,.
\end{equation}
Therefore, the inelasticity is determined solely by the transverse metric perturbations.

In $D>4$, we may drop the hats since $(\hat\tau,\hat\theta)\rightarrow(\tau,\theta)$. However, in $D=4$, the two coordinate systems do not match asymptotically. Nevertheless, we shall use the same expression, i.e. work with de Donder coordinates asymptotically. Later in Chapter~\ref{ch:second_order} we will see that this gives rise to logarithmically divergent terms in the second-order news function when $r\rightarrow\infty$. Then, following Payne~\cite{Payne}, we shall simply remove them.

\subsection{News function in de Donder coordinates}
\label{news_Donder}
In summary, if the transverse metric perturbation, in de Donder gauge, is decomposed into its trace and trace-free parts,
\begin{equation}
h_{ij}=H\delta_{ij}+E\Delta_{ij}\,,
\end{equation}
the news function defined in Eq.~\eqref{news1} reads
\begin{equation}
\dot{\mathcal{E}}(\tau,\theta)=\sqrt{\frac{(D-2)(D-3)}{4}}\lim_{r\rightarrow +\infty}\left[ r \rho^{\frac{D-4}{2}} \left(\dot{E}+\dot{H}\right)\right]\,,\label{news2}
\end{equation}
where the dot denotes a derivative with respect to $\tau$ and the relationship between $(\tau,r,\theta)$ and $(u,v,\rho)$ is
\begin{equation}
u=\tau+r(1-\cos\theta)\,,\qquad v=\tau+r(1+\cos\theta)\,,\qquad \rho=r\sin\theta\,.
\end{equation}
Finally, the inelasticity, Eq.~\eqref{e1}, becomes
\begin{equation}
\epsilon=\int_{-1}^1\frac{d\cos\theta}{2}\epsilon(\theta)\equiv\int_{-1}^1\frac{d\cos\theta}{2}\int d\tau\,\dot{\mathcal{E}}(\tau,\theta)^2\,.\label{e2}
\end{equation}

\section{Integration domain}
\label{integration_domain}
The domain of integration in Eq.~\eqref{e2} is, in principle, $\tau\in\left]-\infty,+\infty\right[$ and $\theta\in\left[0,\pi\right]$. However, since no gravitational radiation is expected prior to the collision, one may ask if the news function in Eq.~\eqref{news2} is actually non-vanishing in that entire domain. This is particularly important if one wants to compute, and integrate, the news function numerically. Furthermore, since the transformation to Brinkmann coordinates adapted to the $u$-shock broke the $\mathbb{Z}_2$ symmetry of the problem, one should check what effect, if any, does that have on the $\theta$ domain.

\subsection{Time domain}
\label{time_domain}
Regarding the time domain, let us go back to Fig.~\ref{rays}. The radiation signal at point $\mathcal{P}$ starts when it is hit by ray 1 and is expected to peak around the time it is hit by ray 2, which comes from the other side of the axis and has already crossed the caustic at $\rho=0$.

The trajectories of such rays are given by Eqs.~\eqref{Rosen_coords} with $\bar{v}=0$ (and $\kappa=1$),
\begin{equation}
v=\Phi(\bar{\rho})+u\frac{\Phi'(\bar\rho)^2}{4}\,,\qquad \rho=\pm\,\bar{\rho}\left(1+u\frac{\Phi'(\bar\rho)}{2\bar{\rho}}\right)\,,\label{eq_rays12}
\end{equation}
where the $+$ sign gives ray 1 and the $-$ sign gives ray 2.

Thus if the observation point $\mathcal{P}$ has coordinates $(r,\theta)$, the times $\tau_1$ and $\tau_2$ at which it is hit by those rays are the solutions of
\begin{eqnarray}
\tau(\bar{\rho})&=&\frac{\left(2\,\Phi-\bar\rho\,\Phi'\right)\mp2\,\bar\rho\cot\frac{\theta}{2}}{2\pm\cot\frac{\theta}{2}\,\Phi'}\simeq\mp\cot\frac{\theta}{2}\,\bar{\rho}+O(\Phi)\,,\\
r(\bar{\rho})&=&\frac{\left(\Phi'^2-4\right)\bar\rho-2\Phi\Phi'}{\pm(\Phi'^2-4)\sin\theta-\,4\,\Phi'\cos\theta}\simeq\csc\frac{\theta}{2}\,\bar\rho+O(\Phi)\,.
\end{eqnarray}

The asymptotic limit $r\rightarrow\infty$ is achieved by $\bar\rho\rightarrow\infty$ (we had already seen in Chapter~\ref{ch:dynamics} that a far-away observer gets rays coming from the far-field region where the gravitational field is weak). Therefore it is clear that $\tau_1\rightarrow-\infty$ and $\tau_2\rightarrow+\infty$ for all $D$ unless $\theta=\pi$\,: an observer at the axis gets both rays at $\tau=-\infty$ in $D=4$ and at $\tau=0$ in $D>4$.
\footnote{The discussion of this qualitative difference in \cite{Herdeiro:2011ck} only applies to an observer exactly at the axis. In practice, we extract the radiation off the axis (even if only slightly), thus the integration limits to be considered are $-\infty$ to $+\infty$, both in $D=4$ and $D>4$.}

The past light cone of an observer sitting on the axis at a finite distance $r$ is represented in Fig.~\ref{pastlightcone3}, where it is manifest that the radiation signal will start as soon as it becomes tangent to the collision line $v=\Phi(\rho)$. Fig.~\ref{tangent} details these intersections for various points along the caustic.

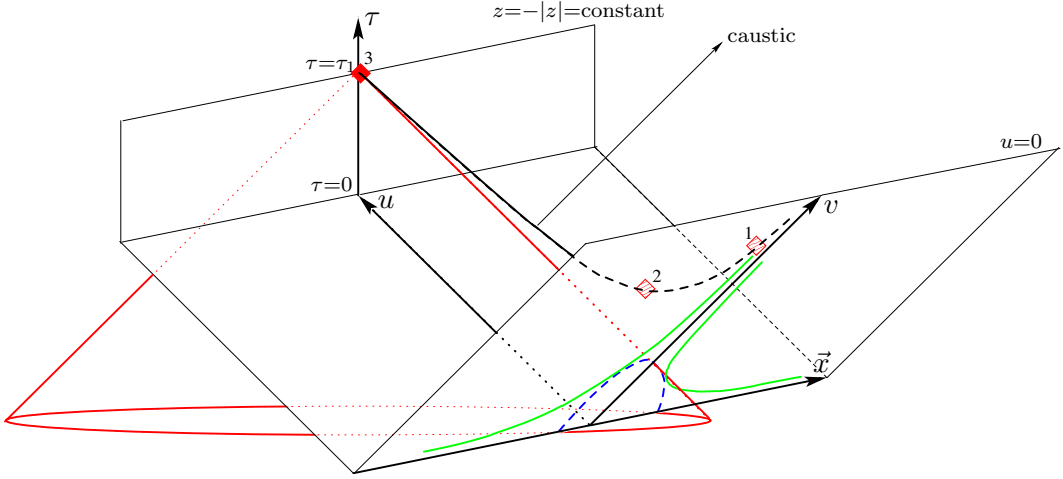
\begin{figure}[t]
\begin{center}
\begin{picture}(0,0)
\put(144,101){$u$}
\put(311,99){$v$}
\put(377,125){$_{u=0}$}
\put(119,109){$_{\tau=0}$}
\put(307,39){$\vec{x}$}
\put(117,155){$_{\tau=\tau_1}$}
\put(139,170){$\tau$}
\put(187,175){$_{z=-|z|= {\rm constant}}$}
\put(275,165){$_{\rm caustic}$}
\end{picture}
\includegraphics[scale=0.4]{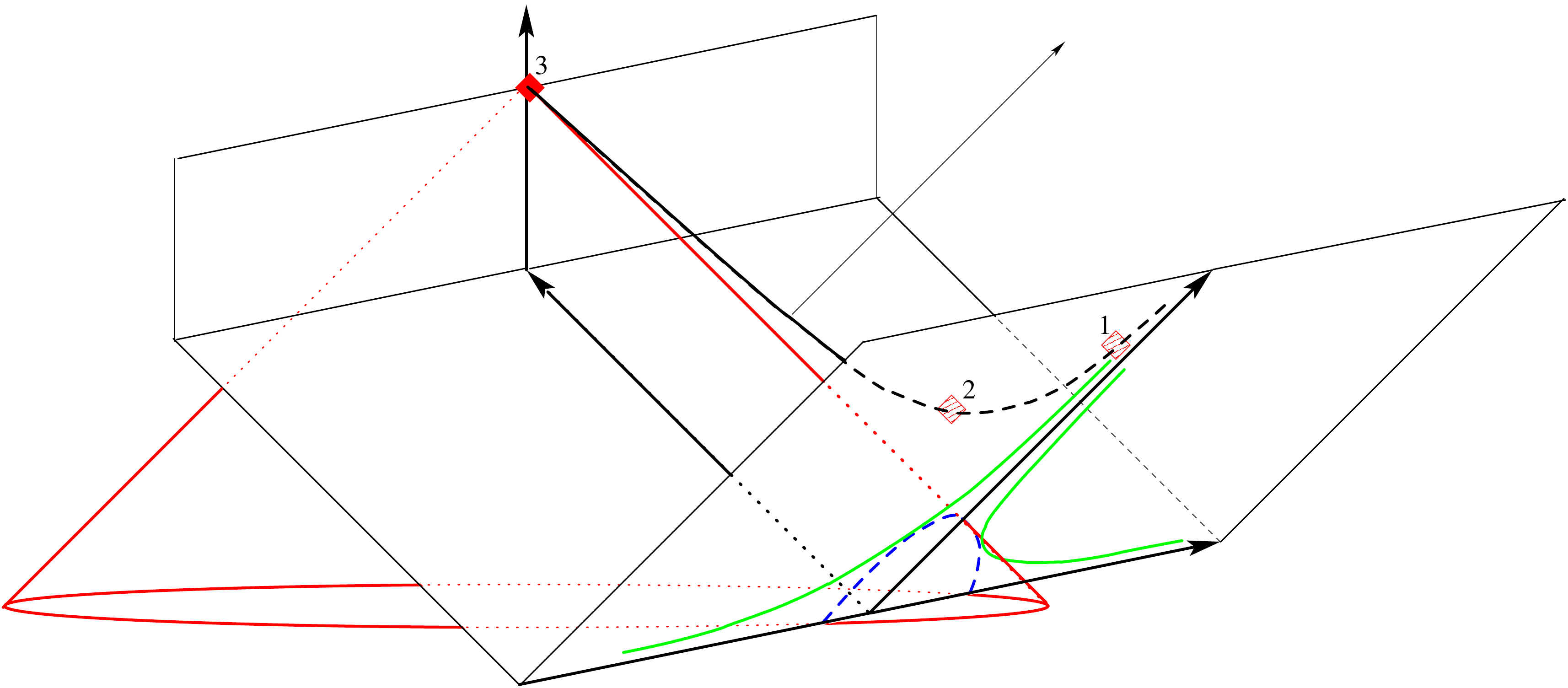}
\end{center}
\caption[3D space-time diagram showing the caustic and three selected events on it.]{3D space-time diagram (for the case $D>4$) showing the caustic (in black) and three selected events, $1$-$3$, on it. The past light cone of event $3$ is drawn in red. Its intersection with the surface $u=0$ (blue dashed parabola) is tangent to the collision surface $u=0 \wedge v=\Phi$ (in green). The $v$-shock generators travelling along $u$ that emerge from the intersection points will focus and converge at $3$. The $\tau$ axis is the world-line of an observer at fixed $z$, which will see no gravitational radiation before $\tau=\tau_1$. The collision lines and the intersections with $u=0$ of the past light cones of other points along the caustic are represented in the right panel of Fig.~\ref{tangent}. \emph{From \cite{Herdeiro:2011ck}.}}
\label{pastlightcone3}
\end{figure}
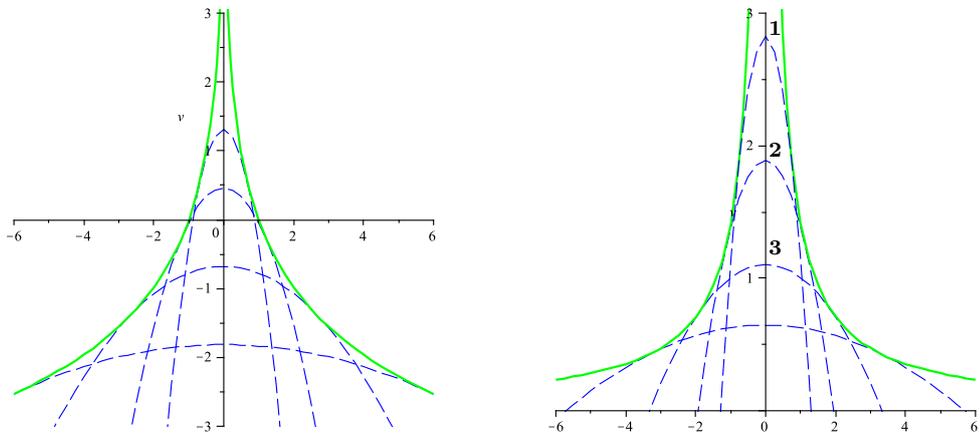
\begin{figure}[t]
\begin{center}
\includegraphics[scale=0.3]{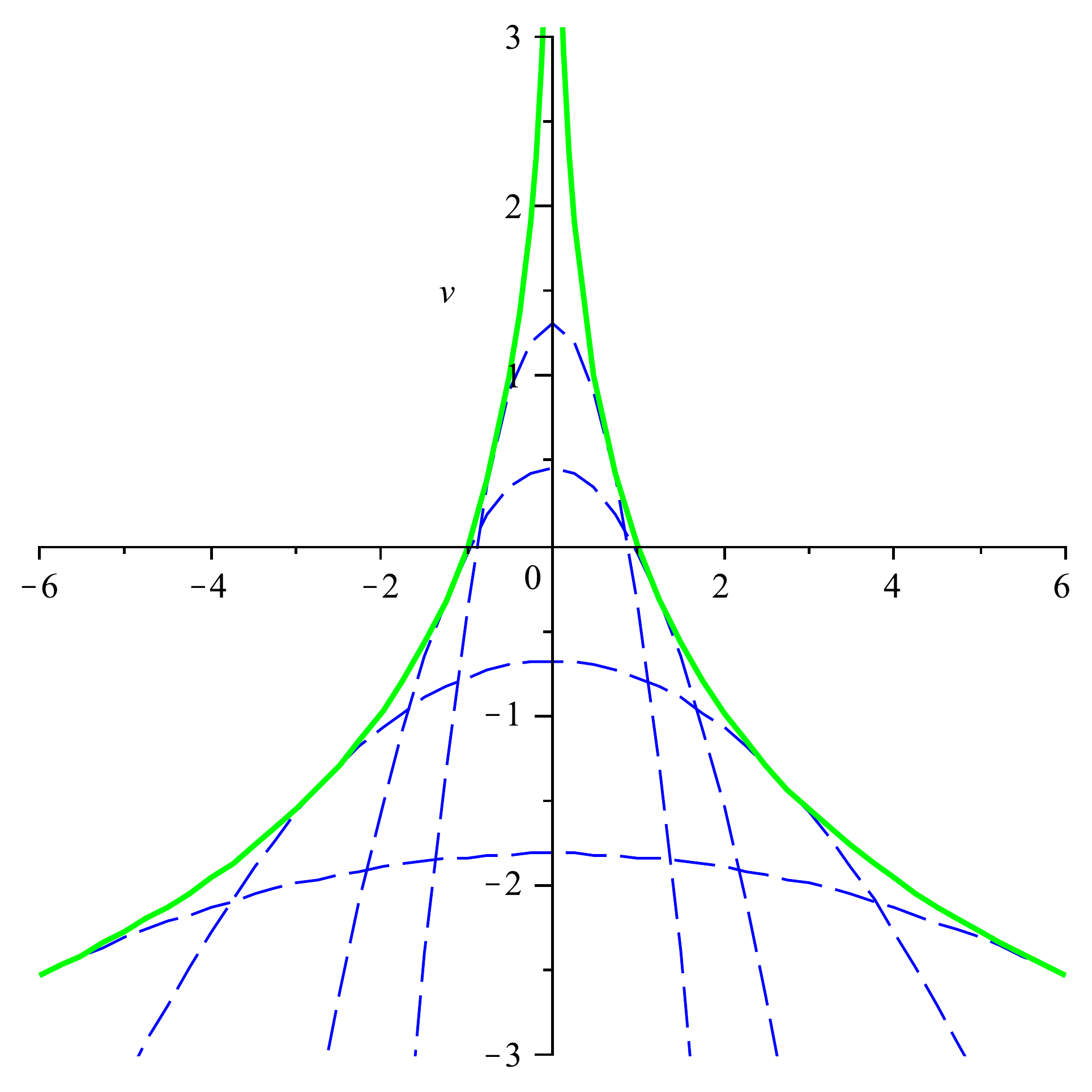} \hspace{1cm}
\includegraphics[scale=0.3]{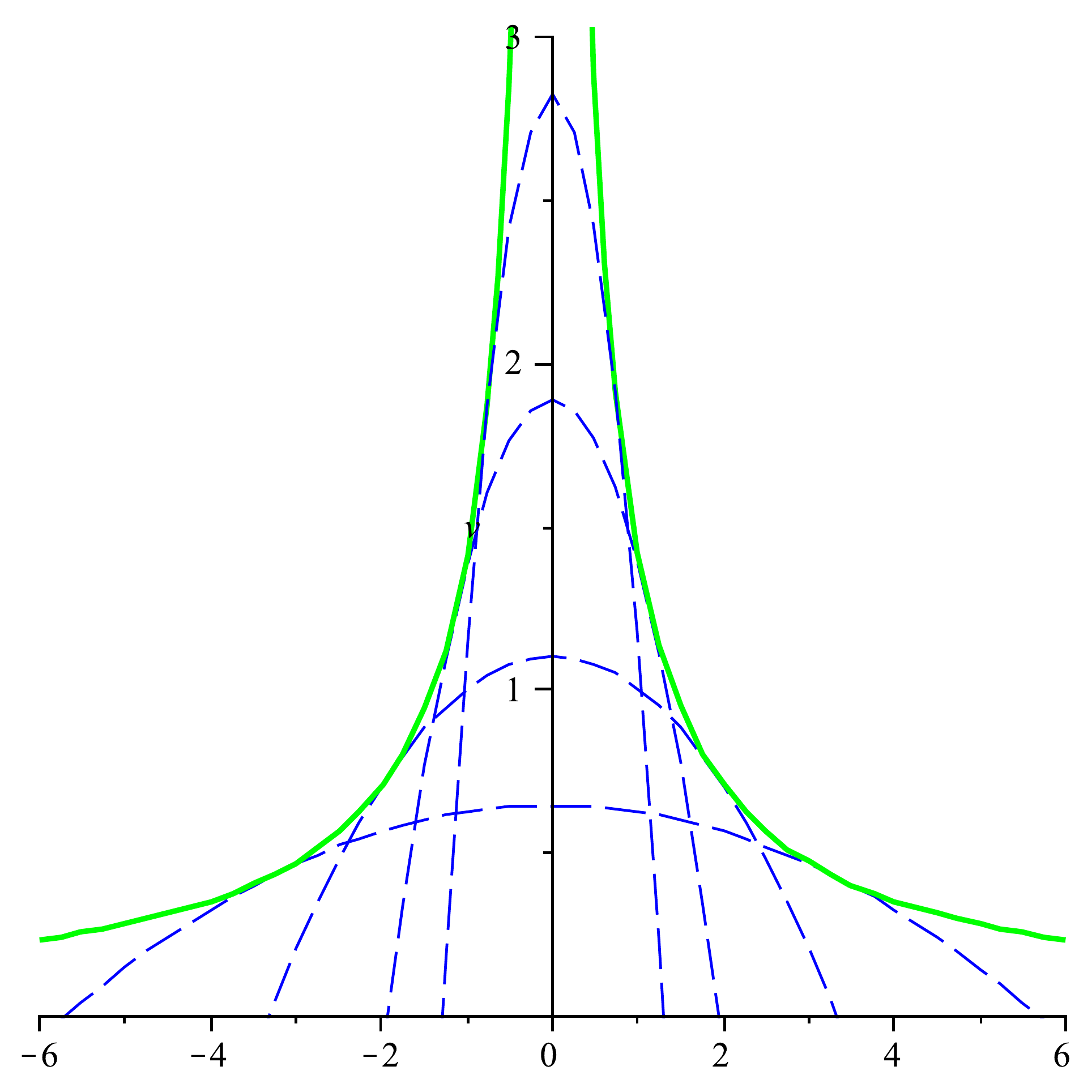}
\begin{picture}(0,0)
\put(-85,155){$_{\bf 1}$}
\put(-85,109){$_{\bf 2}$}
\put(-85,72){$_{\bf 3}$}
\end{picture}
\end{center}
\caption[Collision surface and the intersection of past light cones of various space-time points along the caustic with $u=0$.]{Collision surface, $u=0\wedge v=\Phi$, (green lines) and the intersection of past light cones (dashed blue lines) of various space-time points along the caustic with $u=0$ for $D=4$ (left panel) and $D=5$ (right panel). The past light cones are tangent to the collision surface. The latter case is qualitatively similar to any $D>5$. The numbering $1$-$3$ in the right panel corresponds to the space-time points with similar numbering in Fig. \ref{pastlightcone3}.  \emph{From \cite{Herdeiro:2011ck}.}}
\label{tangent}
\end{figure}

\subsection{Angular domain}
With respect to the $\theta$ domain, we note that the determinant of the transformation matrix given by Eqs.~\eqref{Rosen_coords} is
\begin{equation}
\det\left[\frac{\partial(u,v,\rho)}{\partial(\bar u,\bar v,\bar\rho)}\right]=1+\frac{\kappa}{2}\bar u\theta(\bar u)\Phi''(\bar\rho)\,.
\end{equation}
Thus the transformation is well-defined for all $\bar u\leq0$, and for all $\bar u>0$ except at the axis $\bar\rho=0$. Actually, in the asymptotic limit $\bar r\rightarrow\infty$ both coordinate systems coincide since $\bar\rho\rightarrow\infty$ off the axis and hence $\Phi\rightarrow0$.

Furthermore, the radiation signal is expected to vanish for an observer at the axis since he sees no variation of the quadrupole moment. We shall see later that this is indeed true.

Therefore we conclude that the domain of angular integration is $\theta\in\left]0,\pi\right[$. Given the $\mathbb{Z}_2$ symmetry, this should yield twice the integral over half the set (one of the hemispheres). Since our perturbative method consists of an expansion off $\theta=\pi$, we anticipate that the limit $\theta\rightarrow0$ might be ill-defined at each order, in which case it would be appropriate to choose the $-z$ hemisphere.

\chapter{Angular dependence of the news function}
\label{ch:analytics}

\epigraph{\emph{Sym\'etrie.} En ce qu'on voit d'une vue; fond\'ee sur ce qu'il n'y a pas de raison de faire autrement.}{Blaise Pascal\\ \emph{Pens\'ees}}

In the previous chapter we showed that the radiation signal seen by an infinitely far away observer, which when squared and integrated gives the energy radiated, is encoded in a sole function of retarded time $\tau$ and angular coordinate $\theta$. Furthermore, this so called news function depends exclusively on the transverse metric components $h_{ij}$, whose formal solution was given in integral form in Chapter~\ref{ch:dynamics}.

In this chapter we shall simplify that solution and study the angular dependence of the news function. We start in Section~\ref{CL} by showing that the metric possesses a conformal symmetry at each order in perturbation theory which, at null infinity, implies a factorisation of the angular dependence of the news function at each order. By actually computing that angular factor, we shall write the inelasticity's angular distribution as a power series in $\sin^2\theta$. In particular, we show that a consistent truncation of this angular series at $O(n)$ requires knowledge of the metric perturbations up to $O(n+1)$. This clarifies the meaning of perturbation theory in this problem: it allows for an angular expansion of the news function off the collision axis. In Section~\ref{asymp_waveforms} we simplify the integral solution and verify, explicitly, that it becomes effectively one-dimensional.

\section{A hidden symmetry}
\label{CL}
D'Eath and Payne observed, in $D=4$, that if one starts with a shock wave and performs a boost followed by an appropriate scaling of all coordinates, the end result is just an overall conformal factor \cite{D'Eath:1992hd}. This is easily understood in physical terms: since the only scale of the problem is the energy of the shock wave, the effect of the boost can be undone by rescaling the coordinates. In what follows we shall generalise this to $D\geq4$.

Recall the metric of one Aichelburg-Sexl shock wave in Rosen coordinates,
\begin{equation}
ds^2 = -d\bar{u} d\bar{v} + \left(1+\dfrac{\kappa \bar{u} \theta(\bar{u})}{2}\Phi''\right)^2d\bar{\rho}^{2} + \bar{\rho}^2\left(1+ \frac{\kappa \bar{u} \, \theta(\bar{u})}{2 \bar \rho}\Phi'\right)^2 d\bar{\Omega}^2_{D-3} \ .
\end{equation}
Now let $L$ be the Lorentz transformation
\begin{equation}
(\bar{u},\bar{v},\bar{x}^i)\xrightarrow{L}(e^{-\beta}\bar{u},e^\beta\bar{v},\bar{x}^i)\,,
\end{equation}
and $C$ the conformal scaling
\begin{equation}
(\bar{u},\bar{v},\bar{x}^i)\xrightarrow{C}e^{-\frac{1}{D-3}\beta}(\bar{u},\bar{v},\bar{x}^i)\,.
\end{equation}
Then, under the combined action of $CL$,
\begin{eqnarray}
(\bar{u},\bar{v},\bar{x}^i)&\xrightarrow{CL}&(e^{-\frac{D-2}{D-3}\beta}\bar{u},e^{\frac{D-4}{D-3}\beta}\bar{v},e^{-\frac{1}{D-3}\beta}\bar{x}^i)\,,\\
g_{\mu\nu}(\bar X)&\xrightarrow{CL}&e^{\frac{2}{D-3}\beta}g_{\mu\nu}(\bar X')\; ,
\end{eqnarray}
where $\bar X$ and $\bar X'$ denote the coordinates before and after the transformation respectively, i.e. $\bar X\xrightarrow{CL}\bar X'$. The metric simply scales by an overall factor.

Since this is a one parameter symmetry, one can find coordinates on $(D-1)$-dimensional sheets which are invariant under the transformation, and a normal coordinate parameterising inequivalent sheets. A suitable set of invariant coordinates on such sheets is $\left\{p,q,\phi^a\right\}$, with
\begin{equation}
p\equiv \bar{v}\bar{\rho}^{D-4}\,,\qquad q\equiv \bar{u}\bar{\rho}^{-(D-2)}\,, 
\end{equation}
and $\phi^a$ the angles on the transverse plane. The normal coordinate along the orbits of the symmetry is simply $\bar{\rho}$ which transforms as
\begin{equation}
\bar{\rho}\rightarrow\bar{\rho}'=e^{-\frac{1}{D-3}\beta}\bar{\rho}\,.
\end{equation} 

Now consider the superposition of two shocks. In the boosted frame with Rosen coordinates, the action of $CL$ is
\begin{equation}
g_{\mu\nu}(\nu,\lambda;\bar X)\xrightarrow{CL}e^{\frac{2}{D-3}\beta}g_{\mu\nu}(\nu,e^{-2\beta}\lambda;\bar X')\,.
\end{equation}
Thus the perturbative expansion remains the same except for $\lambda\rightarrow e^{-2\beta}\lambda$. In the centre-of-mass frame with Brinkmann coordinates, Eq.~\eqref{metric_series} becomes
\begin{eqnarray}\label{eq:CLonMetric}
g_{\mu\nu}(X) \xrightarrow{CL}g_{\mu\nu}(X')=e^{\frac{2}{D-3}\beta}\left[\eta_{ \mu \nu}+\sum_{k=1}^\infty e^{-2k\beta}  h_{\mu\nu}^{(k)}(X')\right]\,.
\end{eqnarray}
Quite remarkably, in the future of the collision the metric possesses a conformal symmetry at each order in perturbation theory.

Note that the metric functions $h_{\mu\nu}^{(k)}$ in Eq.~\eqref{eq:CLonMetric} are the same before and after the transformation, the only change being the coordinates they are evaluated on. On the other hand, for a generic coordinate transformation, the metric transforms as a rank-2 tensor,
\begin{equation}
g_{\mu \nu}(X)=\frac{\partial x^{\mu'}}{\partial x^\mu}\frac{\partial x^{\nu'}}{\partial x^\nu}g_{\mu' \nu'}(X')\,.
\end{equation}
Together with Eq.~\eqref{eq:CLonMetric} this implies that
\begin{equation}\label{CL_transf}
h_{\mu\nu}^{(k)}(X')=e^{(2k+N_u-N_v)\beta}h_{\mu\nu}^{(k)}(X)\,,
\end{equation}
where $N_u$ and $N_v$ are the number of $u$-indices and $v$-indices. Thus a given metric function evaluated on the transformed coordinates is the same function evaluated on the initial coordinates multiplied by an appropriate factor.

In these coordinates, the invariants $p,q$ read
\begin{equation}
p=(v-\Phi(\rho))\rho^{D-4}\,,\qquad q=u\rho^{-(D-2)}\,,\label{def_pq}
\end{equation}
while $\rho$ transforms as
\begin{equation}
\rho\rightarrow \rho'=e^{-\frac{1}{D-3}\beta}\rho\,.
\end{equation}

\subsection{Reduction to two dimensions}

Since $\rho$ is the only coordinate that transforms under the action of $CL$, Eq.~\eqref{CL_transf} implies a separation of variables in the form
\begin{equation}\label{eq:separation_rho}
h_{\mu\nu}^{(k)}(p,q,\rho,\phi^a)=\rho^{-(D-3)(2k+N_u-N_v)}f^{(k)}_{\mu\nu}(p,q,\phi^a)\,.
\end{equation}
Since the angles $\phi^a$ are ignorable, the problem becomes two dimensional at each order in perturbation theory. Let us see how.

In Section~\ref{reduction_3D} we defined scalar functions of $(u,v,\rho)$ by factoring out the trivial dependence of each metric component on $\phi^a$, Eqs.~\eqref{gen_perts}, and similarly for the sources $T^{(k-1)}_{\mu\nu}$. Then Eq.~\eqref{eq:separation_rho} implies that, for each of those functions generically denoted by $F(u,v,\rho)$ as before,
\begin{equation}
F^{(k)}(u,v,\rho)= \dfrac{f^{(k)}(p,q)}{\rho^{(D-3)(2k+N_u-N_v)}}\,.\label{f_pq}
\end{equation}
For its respective source as defined in Eq.~\eqref{def_S}, since the d'Alembertian operator scales as $\rho^{-2}$ under $CL$,
\begin{equation}
S^{(k)}(u,v,\rho)=\dfrac{s^{(k)}(p,q)}{\rho^{(D-3)(2k+N_u-N_v)+2}}\,.\label{def_s}
\end{equation}

Thus our problem is effectively two-dimensional in $(p,q)$ coordinates: at each order, $f^{(k)}(p,q)$ is the solution of a differential equation (inherited from the wave equation) with a source $s^{(k)}(p,q)$ and subject to initial conditions on $q=0$. This is an enormous computational advantage. In Section~\ref{app:Green2D} of Appendix~\ref{app:Green} we obtain the differential equation obeyed by $f^{(k)}(p,q)$ and the reduced Green's function $G^k_m(p,q;p',q')$. However, there is another consequence of this symmetry which only becomes manifest when one goes to null infinity.

\subsection{The $CL$ symmetry at null infinity}
In terms of $(r,\tau,\theta)$ the new coordinates $p$ and $q$ read
\begin{eqnarray}
p&=&(\tau+r(1+\cos\theta)-\Phi(r\sin\theta))(r\sin\theta)^{D-4}\,,\\
q&=&(\tau+r(1-\cos\theta))(r\sin\theta)^{-(D-2)}\,.
\end{eqnarray}
One can see they are not well-defined in the limit $r\rightarrow\infty$ since
\begin{equation}
p\rightarrow\infty\,,\qquad q\rightarrow0\,.
\end{equation}

However, it is reasonable to expect that the finite, non-trivial dependence of the news function in $(\tau,\theta)$ should be given in terms of combinations of $p,q$ that remain finite and non-trivial at null infinity. An appropriate choice is
\begin{equation}
\hat{p}\equiv
\left\{
\begin{array}{ll}
 \frac{pq-1}{q}+2\log q\ , &  D=4\  \vspace{2mm}\\
\displaystyle{ \frac{pq-1}{q^{\frac{1}{D-3}}}}\ , & D>4\ \label{tau_transform}\,
\end{array} \right. \ ,
\qquad \hat{q}\equiv q^{\frac{1}{D-3}}\,.
\end{equation}

This coordinate transformation, $(p,q)\rightarrow(\hat{p},\hat{q})$, has a constant non-vanishing determinant,
\begin{equation}
\det \left[\frac{\partial (p,q)}{\partial(\hat{p},\hat{q})}\right]=D-3\,,
\end{equation}
hence is well-defined everywhere.

When $r\rightarrow\infty$,
\begin{eqnarray}
\hat{p}&\rightarrow&2\bar{\tau}(\tau,\theta)+O\left(r^{-1}\right)\,,\\
\hat{q}&\rightarrow&(1-\cos\theta)^{\frac{1}{D-3}}(\sin\theta)^{-\frac{D-2}{D-3}}\times r^{-1}+O\left(r^{-2}\right)\,,
\end{eqnarray}
where we have defined a new time coordinate
\begin{eqnarray}
\bar{\tau}(\tau,\theta)=\left\{
\begin{array}{ll}
 \tau\times(1-\cos\theta)^{-1}+\log\left(\frac{1-\cos\theta}{\sin\theta}\right)\ , &  D=4\  \vspace{2mm}\\
\displaystyle{ \tau\times(1-\cos\theta)^{-\frac{1}{D-3}}(\sin\theta)^{-\frac{D-4}{D-3}}}\ , & D>4\ \label{tau_transform}\,
\end{array} \right. \ .
\end{eqnarray}

Recall that we had already factored out the $\rho$ dependence in Eq.~\eqref{f_pq}, leaving only a function of $(p,q)$ to compute. But since $\bar\tau$ is the only surviving quantity in the asymptotic limit, we expect the news function to become effectively one-dimensional at null infinity: besides the trivial dependence on $\theta$ coming from the known powers of $r$ and $\rho$ (remember that $\rho=r\sin\theta$), it must be a function of $\bar\tau(\tau,\theta)$ only.

From Eqs.~\eqref{news2} and \eqref{f_pq},
\begin{equation}
\dot{\mathcal{E}}^{(k)}(\tau,\theta)=(\sin\theta)^{\frac{D-4}{2}-2k(D-3)}\times\lim_{r\rightarrow\infty}\frac{d}{d\tau}\left[r^{\frac{D-2}{2}-2k(D-3)}\hat{f}^{(k)}(\hat{p},\hat{q})\right]\,,
\end{equation}
where $\hat{f}^{(k)}$ contains the relevant contribution from $E^{(k)}$ and $H^{(k)}$, as a function of $(\hat{p},\hat{q})$. 

If the quantity inside brackets is to remain finite at null infinity, then necessarily
\begin{equation}
\hat{f}^{(k)}(\hat{p},\hat{q})\simeq\alpha^{(k)}(\hat{p})\hat{q}^{\frac{D-2}{2}-2k(D-3)}+\dots\,,
\end{equation}
where $\dots$ denotes higher powers of $\hat{q}$ and $\alpha^{(k)}(\hat{p})$ some unknown function of $\hat{p}$. 

Thus, after taking the limit $r\rightarrow\infty$,
\begin{equation}
\dot{\mathcal{E}}^{(k)}(\tau,\theta)=\frac{1}{1-\cos\theta}\left(\frac{1+\cos\theta}{1-\cos\theta}\right)^{k-1+\frac{1}{4}\frac{D-4}{D-3}}\frac{d\bar\tau}{d\tau}(\theta)\times\frac{d}{d\bar\tau}\alpha^{(k)}(2\bar\tau)\,.
\end{equation}

Using Eq.~\eqref{tau_transform}, and evaluating this expression at $\theta=\tfrac{\pi}{2}$ noting that $\bar\tau\left(\tau,\tfrac{\pi}{2}\right)=\tau$, we conclude that
\begin{equation}
\dot{\mathcal{E}}^{(k)}(\tau,\theta)=\left(\frac{1}{1-\cos\theta}\right)^2\left(\frac{1+\cos\theta}{1-\cos\theta}\right)^{k-1-\frac{1}{4}\frac{D-4}{D-3}}\dot{\mathcal{E}}^{(k)}\left(\bar\tau(\tau,\theta),\frac{\pi}{2}\right)\,,\label{Eq:angular_factorization}
\end{equation}
which proves our proposition.

\subsection{The meaning of perturbation theory}
\label{meaning}

The perturbative expansion of the metric, Eq.~\eqref{metric_series}, implies an analogous series for the inelasticity's angular distribution $\epsilon(\theta)$ defined in Eq.~\eqref{e1},
\begin{equation}
\epsilon(\theta)=\sum_{N=1}^\infty\epsilon^{(N)}(\theta)\,,
\end{equation}
where
\begin{equation}
\epsilon^{(N)}(\theta)=\sum_{k=1}^{N}\int d\tau\, \dot{\mathcal{E}}^{(k)}(\tau,\theta)\dot{\mathcal{E}}^{(N+1-k)}(\tau,\theta)\,.\label{N_N-k}
\end{equation}

Inserting the result of Eq.~\eqref{Eq:angular_factorization} and changing the integration variable from $\tau$ to $\bar{\tau}$, we conclude that
\begin{equation}
\epsilon^{(N)}(\theta)=\left(\frac{1}{1-\cos\theta}\right)^3 \left(\frac{1+\cos\theta}{1-\cos\theta}\right)^{N-1}\epsilon^{(N)}\left(\frac{\pi}{2}\right)\, . \label{seriesi2}
\end{equation}
Thus it suffices to compute the news function on the symmetry plane.

The whole series reads
\begin{eqnarray}
\epsilon(\theta)&=&\sum_{N=1}^\infty \epsilon^{(N)}(\theta) =\left(\frac{1}{1-\cos\theta}\right)^3\sum_{N=1}^\infty\left(\frac{1+\cos\theta}{1-\cos\theta}\right)^{N-1}\epsilon^{(N)}\left(\frac{\pi}{2}\right)\,,\\
&\equiv&\sum_{n=0}^\infty\alpha_n(\theta) \epsilon^{(n+1)}\left(\frac{\pi}{2}\right)\,.
\label{series2}
\end{eqnarray}

Observe that each individual $\alpha_n(\theta)$ is regular at $\theta=\pi$, where we saw that perturbation theory should be valid, but does not obey $\alpha_n(\theta)=\alpha_n(\pi-\theta)$ since the transformation to Brinkmann coordinates broke the $\mathbb{Z}_2$ symmetry. However, $\epsilon(\theta)$ should respect that symmetry. In particular, if it has a regular limit at the axis, it can be written as a power series in $\sin^2\theta$,
\begin{equation}
\epsilon(\theta)=\sum_{n=0}^\infty \epsilon_n(\sin\theta)^{2n}\,.\label{angular_series}
\end{equation}

Indeed, writing $\cos\theta=-\sqrt{1-\sin^2\theta}$, we see that near $\theta=\pi$,
\begin{equation}
\alpha_n(\theta)\sim (\sin\theta)^{2n}\,.
\end{equation}

For example, the first three terms are given by
\begin{eqnarray}
\epsilon_0&=&\frac{1}{8}\epsilon^{(1)}\left(\frac{\pi}{2}\right)\,,\label{e0}\\
\epsilon_1&=&\frac{1}{32}\left[3\epsilon^{(1)}\left(\frac{\pi}{2}\right)+\epsilon^{(2)}\left(\frac{\pi}{2}\right)\right]\,,\\
\epsilon_2&=&\frac{1}{128}\left[9\epsilon^{(1)}\left(\frac{\pi}{2}\right)+5\epsilon^{(2)}\left(\frac{\pi}{2}\right)+\epsilon^{(3)}\left(\frac{\pi}{2}\right)\right] \ .
\end{eqnarray}

Thus we conclude that a consistent truncation of the series in Eq.~\eqref{angular_series}, i.e. the extraction of the coefficient $\epsilon_n$, requires knowledge of $\epsilon^{(n+1)}(\tfrac{\pi}{2})$ and hence, from Eq.~\eqref{N_N-k}, of the metric up to $O(n+1)$. This clarifies the meaning of perturbation theory in this problem: it allows for the extraction of successive coefficients $\epsilon_n$, thus amounting to an angular expansion off the collision axis.

It should be stressed that this result is a kinematical consequence of the $CL$ symmetry only: the dynamical (wave) equations have not yet been used (except for the background solution of course).

Finally, the inelasticity is then given by
\begin{eqnarray}
\epsilon&=&\sum_{n=0}^\infty \epsilon_n \int_{-1}^1\frac{d\cos\theta}{2}(\sin\theta)^{2n}\,,\\
&=&\sum_{n=0}^\infty \frac{2^n n!}{(2n+1)!!}\epsilon_n \,,\\
&=&\epsilon_0+\frac{2}{3}\epsilon_1+\ldots\,.\label{eq_second_order_e}
\end{eqnarray}

In the next section we shall obtain simplified expressions for the integrals giving the asymptotic metric functions that contribute to the news function. By working in Fourier space with respect to the retarded time $\tau$, we shall also confirm, explicitly, that the coordinate transformation of Eq.~\eqref{tau_transform} factorises the angular dependence out of the integrals.

\section{Asymptotic integral solution for the metric functions}
\label{asymp_waveforms}
Back in Chapter~\ref{ch:dynamics} we wrote the formal solution for each metric function $F(u,v,\rho)$ in integral form, Eq.~\eqref{sol_3D}. Since we are ultimately interested in the news function $\dot{\mathcal{E}}(\tau,\theta)$, we define the asymptotic waveform $\dot{F}(\tau,\theta)$, obtained from $F(u,v,\rho)$, according to Eq.~\eqref{news2},
\begin{equation}
\dot{F}(\tau,\theta)\equiv\lim_{r\rightarrow\infty} r\rho^{\frac{D-4}{2}}\frac{d}{d\tau}F(u,v,\rho)\,.
\end{equation}

Actually, the Dirac delta in the Green's function is better dealt with in Fourier space, so we define
\begin{equation}
\hat{\dot{F}}(\omega,\theta)\equiv\int d\tau\,\dot{F}(\tau,\theta)e^{-i\omega\tau}\,.
\end{equation}
This is equally useful to compute the inelasticity since, by the Parseval-Plancherel theorem,
\begin{equation}
\int d\tau\, \dot{F}(\tau,\theta)^2=\frac{1}{2\pi}\int d\omega\, |\hat{\dot{F}}(\omega,\theta)|^2\,.
\end{equation}

The next steps are detailed in Appendix~\ref{app:asympWF} due to their tedious and technical nature. 

In short, one needs to take the asymptotic limit and integrate in $\tau$ to obtain the Fourier transform. Dropping primes on integration variables for ease of notation, we get
\begin{equation}
\hat{\dot{F}}(\omega,\theta)=-i^{\frac{D-2}{2}+m}\omega\int_0^\infty d\rho\,\rho^{\frac{D-2}{2}}J_{\frac{D-4}{2}+m}(\omega\rho\sin\theta)\hat{S}\left(\omega\frac{1+\cos\theta}{2},\omega\frac{1-\cos\theta}{2};\rho\right)\,,\label{initial_volume}
\end{equation}
where $J_\nu(z)$ is the $\nu$-th order Bessel function of the first kind and
\begin{equation}
\hat{S}(x,y;\rho)=\frac{1}{2}\int du\int dv\, e^{-iux}e^{-ivy}\, S(u,v,\rho)\,.\label{S_hat}
\end{equation}

Next, one finds it convenient to transform to a new frequency $\Omega$,
\begin{equation}
\omega\rightarrow\Omega\equiv\omega^{D-3}\left(\frac{\sin\theta}{2}\right)^{D-4}\frac{1-\cos\theta}{2}\,,\label{omega_transformation}
\end{equation}
together with a new waveform
\begin{equation}
\hat{\mathcal{F}}(\Omega,\theta)\equiv\sqrt{\omega'(\Omega)}\,\hat{\dot{F}}(\omega(\Omega),\theta)\,,
\end{equation}
such that
\begin{equation}
\int d\omega\, |\hat{\dot{F}}(\omega,\theta)|^2=\int d\Omega\, |\hat{\mathcal{F}}(\Omega,\theta)|^2\,.\label{oO_int}
\end{equation}

Then, using the $CL$ symmetry, namely Eq.~\eqref{def_s}, one shows that, at each order $k$ in perturbation theory,
\begin{equation}
\hat{\mathcal{F}}^{(k)}(\Omega,\theta)=\left(\frac{1}{1-\cos\theta}\right)^{\frac{3}{2}}\left(\frac{1+\cos\theta}{1-\cos\theta}\right)^{k-1}\hat{\mathcal{F}}^{(k)}\left(\Omega,\frac{\pi}{2}\right)\,, \label{angular_factorization}
\end{equation}
where the angular dependence is now completely factored out of the integral, and the (new) Fourier space waveform evaluated at $\theta=\tfrac{\pi}{2}$, which we shall abbreviate to
\begin{equation}
\hat{\mathcal{F}}(\Omega)\equiv\hat{\mathcal{F}}\left(\Omega,\frac{\pi}{2}\right)\,,
\end{equation}
is given by
\begin{equation}
\hat{\mathcal{F}}(\Omega)\equiv -\sqrt{\frac{8}{D-3}}i^{\frac{D-2}{2}+m}\,\Omega^{2k-1}\int_0^\infty dR\,R^{\frac{D-2}{2}}J_{\frac{D-4}{2}+m}(2R)\hat{S}(\Omega^{-1},\Omega;R) \; . \label{Omega_form}
\end{equation}

This formulation in terms of the new frequency $\Omega$ will prove to be useful later on in Chapter~\ref{ch:surface} for the evaluation of surface terms. However, for the time being, we can invert the transformation in Eq.~\eqref{omega_transformation} at $\theta=\tfrac{\pi}{2}$, i.e.
\begin{equation}
\Omega\rightarrow\bar{\omega}\equiv2\,\Omega^{\frac{1}{D-3}}\,,\qquad \hat{\mathcal{F}}(\Omega)\rightarrow\sqrt{\Omega'(\bar{\omega})}\,\hat{\mathcal{F}}\left(\Omega(\bar{\omega})\right)\,.
\end{equation}

The net relationship between $\omega$ and $\bar{\omega}$ is
\begin{equation}
\bar{\omega}=\omega\times(1-\cos\theta)^{\frac{1}{D-3}}(\sin\theta)^{\frac{D-4}{D-3}}\,,
\end{equation}
which is equivalent, in real space, to a transformation of the time coordinate
\begin{equation}
\tau\rightarrow\bar{\tau}(\tau,\theta)=\tau\times(1-\cos\theta)^{-\frac{1}{D-3}}(\sin\theta)^{-\frac{D-4}{D-3}}\,.
\end{equation}
Apart from a $\theta$-dependent shift in $D=4$ (indeed an example of a \emph{supertranslation}~\cite{Coelho:2012sy}) $\bar{\tau}$ is exactly the same as in \eqref{tau_transform}.

In the next chapter we shall specialise this asymptotic solution to the surface case and compute all the terms that contribute to the inelasticity and depend linearly on the initial data. This will allows us to extract the isotropic coefficient $\epsilon_0$.

\chapter{Contribution from surface terms}
\label{ch:surface}

\epigraph{Shut up and calculate!}{David Mermin}

This famous quote is usually attributed to Richard Feynman, but is in fact due to David Mermin \cite{Mermin}. Indeed, a lot has been done and said in the previous chapters, but the physical quantities we wish to understand have not yet been computed. In this chapter we shall finally see some results by computing the surface terms, i.e. the linear contribution from the initial data to the news function and the inelasticity. This is the only term present at first order in perturbation theory, and is all we need to extract the isotropic coefficient $\epsilon_0$.

We begin in Section~\ref{numerical_review} with a review of the partial results of \cite{Herdeiro:2011ck}, where the first-order waveforms for even $D$ where computed numerically with a \verb!C++! code. These were later complemented by the results for odd $D$ in \cite{Coelho:2012sya}, where an empirical fit formula was found for $\epsilon_0$ as a function of $D$,
\begin{equation}
\epsilon_0=\frac{1}{2}-\frac{1}{D}\,.\label{fit_formula}
\end{equation}

In Section~\ref{analytical_surface} we obtain all surface waveforms analytically and, as a corollary, prove that Eq.~\eqref{fit_formula} is exact. Indeed, all contributions from surface terms to the inelasticity yield simple rational functions of $D$. However, for all but the isotropic term they are meaningless without the volume terms. Nevertheless, our results provide insight into an issue already encountered by D'Eath and Payne in $D=4$ \cite{D'Eath:1992hd,D'Eath:1992qu}: starting at second order, both surface and volume waveforms have non-integrable tails at late times. In Section~\ref{tails} we trace the origin of these tails to the Green's function and show that they are generic and may be present at all orders. D'Eath and Payne computed the news function at second order in $D=4$ and concluded that the tails cancel when the surface and volume terms are summed, hence we expect a similar cancellation in $D\geq4$. Again, many details are left to Appendix~\ref{app:surface_n_tails}.

\section{Review of numerical results}
\label{numerical_review}
In \cite{Herdeiro:2011ck}, a numerical method was set up, and coded in \verb!C++!, to evaluate the surface integral, i.e. the first term in Eq.~\eqref{formal_sol}. At first order the trace of the transverse metric perturbation vanishes so there is only one scalar function, $E$, to compute. The asymptotic limit was not taken analytically. Instead, the waveform was computed with $\theta=\pi-0.001$ (it vanishes if $\theta=\pi$ at finite $r$) for several (large) $r$, and the limit extracted numerically.

\begin{figure}
\hspace{-3mm}
\includegraphics[scale=0.66,clip=true,trim= 0 0 0 0]{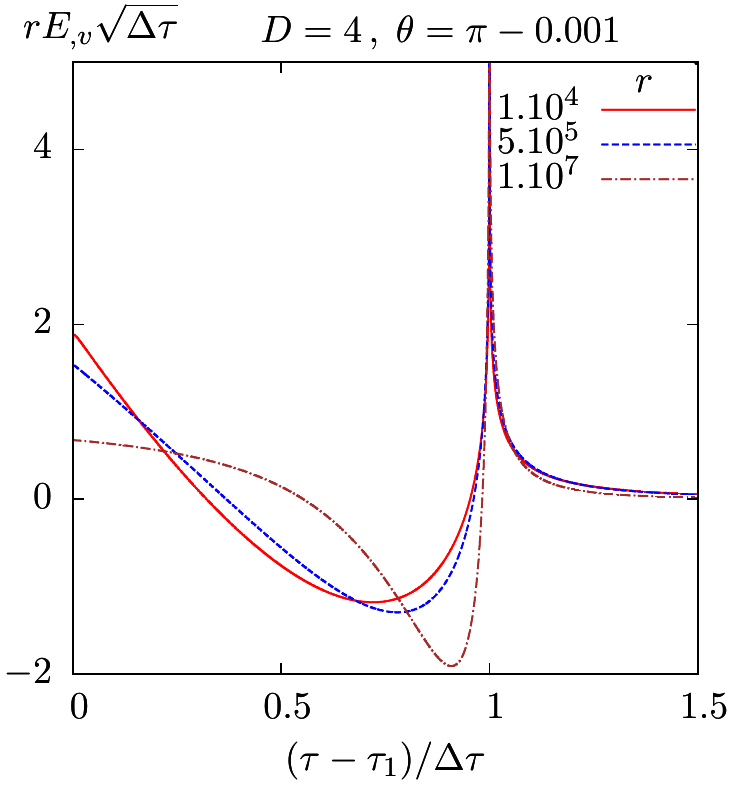} \hspace{1.5mm} 
\includegraphics[scale=0.66,clip=true,trim= 0 0 0 0]{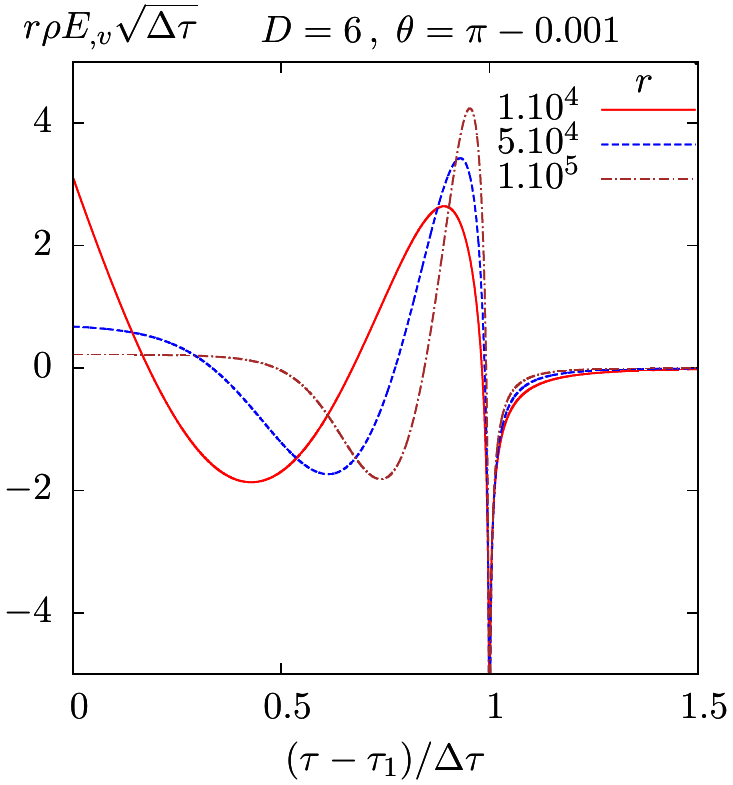} \hspace{1.5mm} 
\includegraphics[scale=0.66,clip=true,trim= 0 0 0 0]{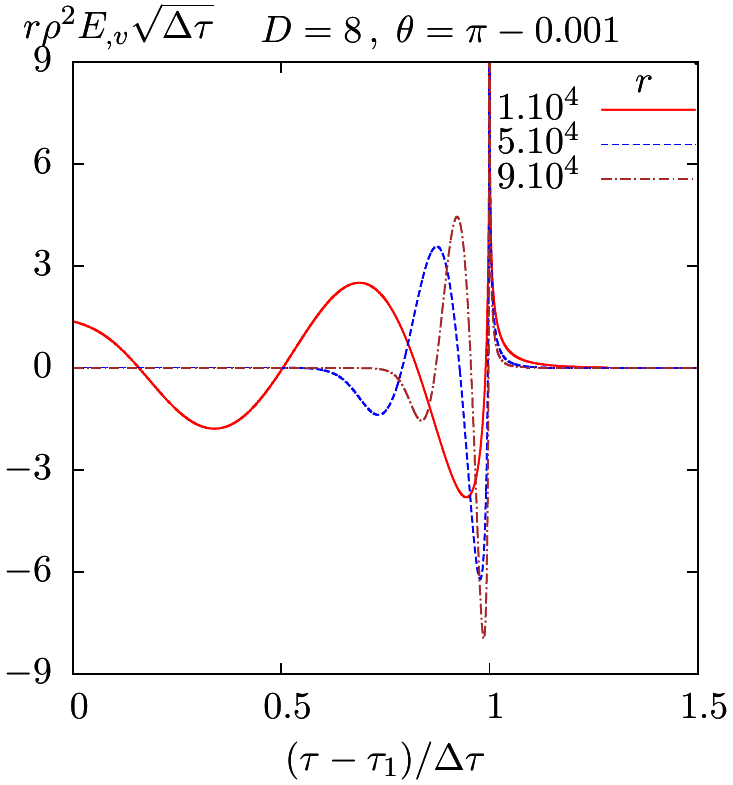}\hspace{1.5mm} 
\caption[First-order waveforms for even $D$.]{\label{evenD}Waveforms in a rescaled $\tau$ coordinate for even $D$: the signal starts at $\tau=\tau_1$ and peaks at $\tau=\tau_2$, which are the retarded times corresponding to rays 1 and 2 of Fig. \ref{rays}. \emph{From \cite{Herdeiro:2011ck}.}}
\end{figure}

The results are summarised in the first row of Fig.~\ref{evenD}, which shows plots of the (rescaled) waveform for $D=4,6,8$. As discussed in Section~\ref{time_domain}, the signal always begins upon reception of ray $1$ at $\tau_1$ and peaks at $\tau_2$ when the second optical ray arrives. The number of oscillations in the interval $\Delta\tau\equiv\tau_2-\tau_1$ increases with $D$.

\begin{figure}
\hspace{-3mm}
\includegraphics[scale=0.66,clip=true,trim= 0 0 0 0]{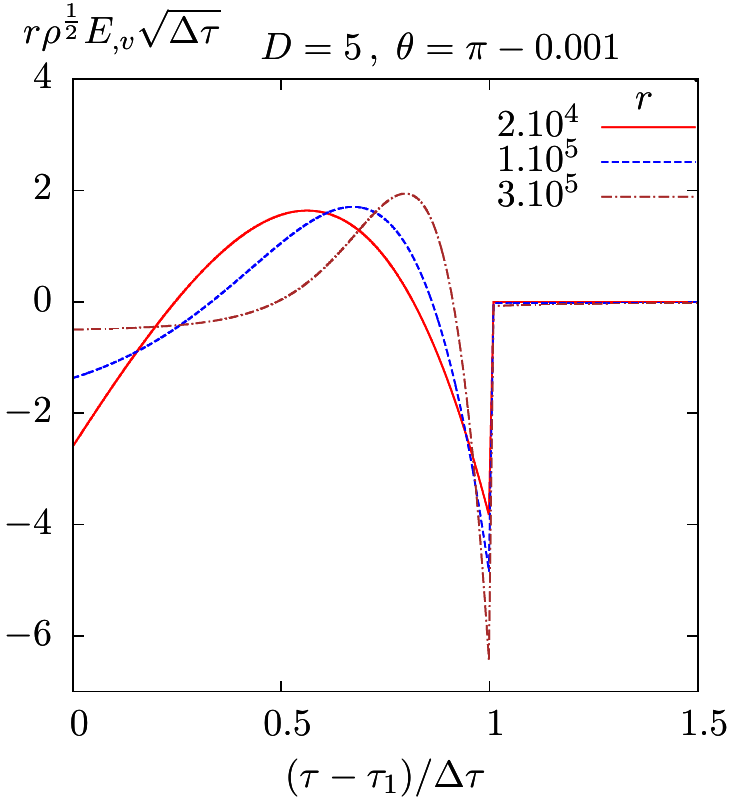}\hspace{1.5mm} 
\includegraphics[scale=0.66,clip=true,trim= 0 0 0 0]{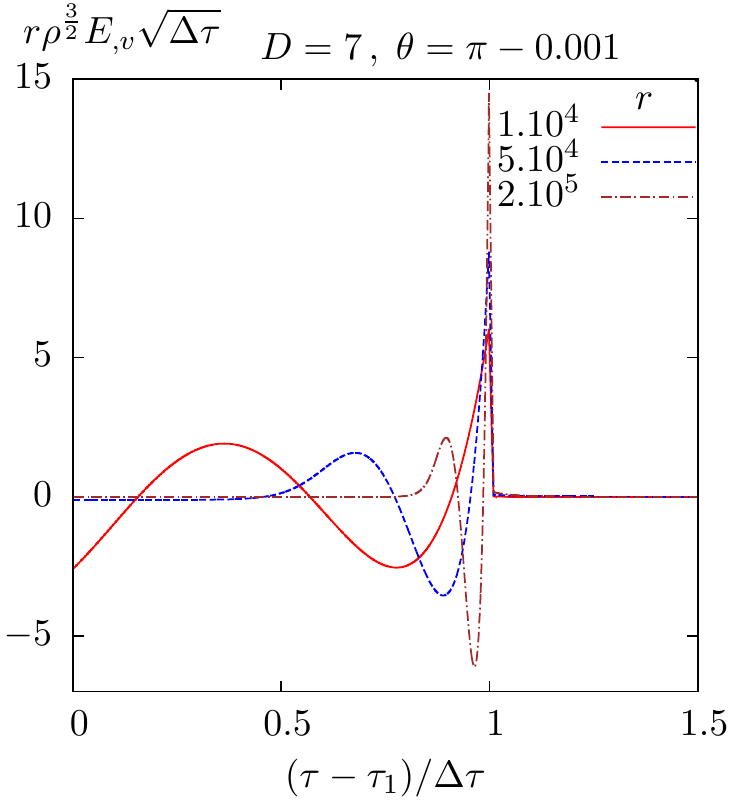}\hspace{1.5mm} 
\includegraphics[scale=0.66,clip=true,trim= 0 0 0 0]{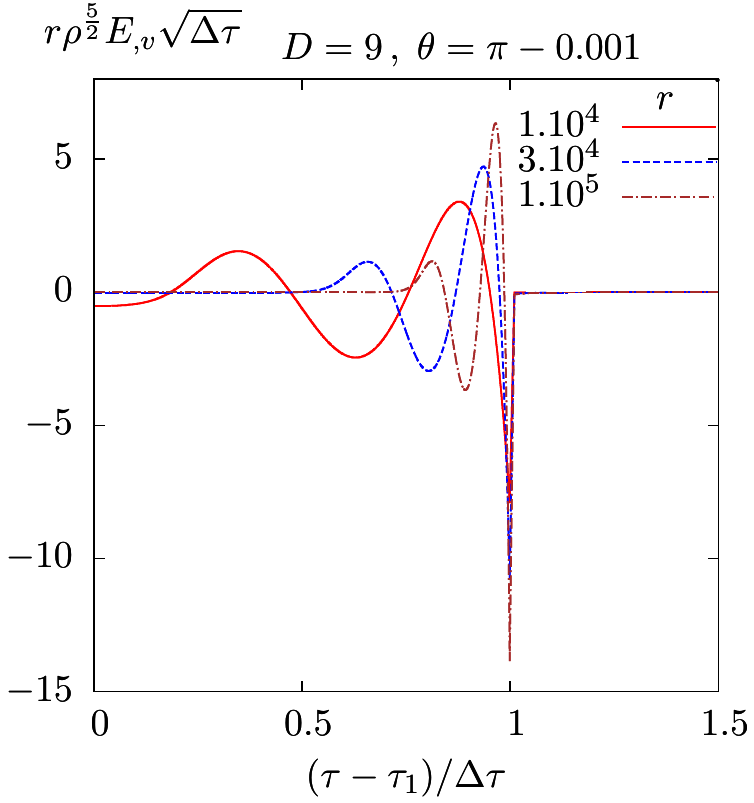}\hspace{1.5mm}
\caption[First-order waveforms for odd $D$.]{\label{oddD}Waveforms in a rescaled $\tau$ coordinate for odd $D$: the signal starts at $\tau=\tau_1$ and peaks at $\tau=\tau_2$, which are the retarded times corresponding to rays 1 and 2 of Fig. \ref{rays}. \emph{From \cite{Coelho:2012sya}.}}
\end{figure}

At the time, technical difficulties prevented the numerical integration for odd $D$. These were later overcome and the picture completed in \cite{Coelho:2012sya}, here shown in Fig.~\ref{oddD}.

\begin{figure}[h!]
\begin{center}
\includegraphics[scale=1.0,clip=true,trim= 0 0 0 0]{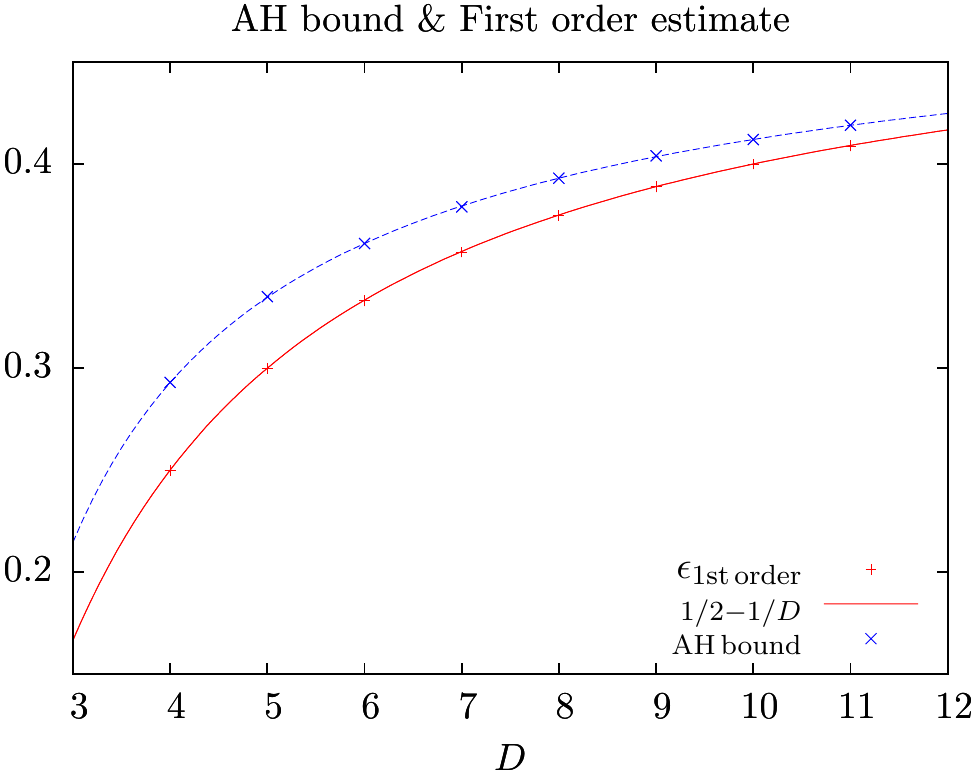}
\caption[Plot of the apparent horizon bound together with the first-order numerical estimate.]{\label{fig:ComparisonD}Plot of the apparent horizon bound, $\epsilon_{\text{AH}}$ (in blue), together with the first-order numerical estimate (red points). Observe the perfect fit with the formula for $\epsilon_0$ in Eq.~\ref{fit_formula} (red line). \emph{From \cite{Coelho:2012sya}.}}
\end{center}
\end{figure}

To get the inelasticity, another numerical integration was required. The results are summarised in Table~\ref{table1} and Fig.~\ref{fig:ComparisonD}, together with the apparent horizon bound, Eq.~\eqref{AHbound}, for comparison. Interestingly, not only is $\epsilon_0<\epsilon_{\text{AH}}$, but also both seem to converge when $D\rightarrow\infty$. Moreover, the numerical points fit nicely (within the numerical error $<0.1\%$) with a remarkably simple formula, Eq.~\eqref{fit_formula}.

\begin{table}[h]
\begin{center}
\begin{tabular}{||  c || c | c | c | c | c | c | c |c |}
\hline			
  $D$ & 4 & 5 & 6 & 7 & 8 & 9 & 10 & 11\\
  \hline
  $\epsilon_{\text{AH}}$ & 29.3 & 33.5 & 36.1 & 37.9 & 39.3 & 40.4 & 41.2& 41.9\\
  \hline
  $\epsilon_{0}$  & 25.0 & 30.0 & 33.3 & 35.7&37.5 & 38.9 & 40.0 & 40.9 \\
\hline
\end{tabular}
\caption[Comparison between the apparent horizon bound, $\epsilon_{\text{AH}}$, and the first-order estimate, $\epsilon_0$, obtained numerically.]{Comparison between the apparent horizon bound, $\epsilon_{\text{AH}}$, and the first-order estimate, $\epsilon_0$, obtained numerically. Note that $\epsilon_0\leq\epsilon_{\text{AH}}$.}
\label{table1}
\end{center}
\end{table}

We shall not give further details of the numerical methods and coding involved. A pedagogical presentation can be found in \cite{Sampaio:2013faa}. Instead, in the next section we will obtain the news function analytically and show that the fit formula for $\epsilon_0$ is exact.

\section{Analytical evaluation of surface terms}
\label{analytical_surface}

The initial data on $u=0$ were obtained in Chapter~\ref{ch:dynamics}. In particular, from Eqs.~\eqref{initial_data} we see that they all have the generic form
\begin{equation}
F^{(k)}(0,v,\rho)=f_0^{(k)}(\rho)(v-\Phi(\rho))^k\theta(v-\Phi(\rho))\,.\label{F_u0}
\end{equation}

However, the de Donder gauge condition, Eq.~\eqref{deDonderV2} is yet to be imposed. In Appendix~\ref{app:gauge} we show that the initial data for the transverse metric components, i.e. the scalars $E$ and $H$, are unaffected (actually, at first order, none of the components change), so we may move on. Details of the following calculation can be found in Appendix~\ref{app:surface}.

We start by feeding Eq.~\eqref{F_u0} to the simplified integral in Fourier space, Eq.~\eqref{Omega_form}, and compute
\begin{equation}
\hat{S}^{(k)}(\Omega^{-1},\Omega;R)=\frac{k!}{(i\Omega)^k}f_0^{(k)}(R)e^{-i\Omega\Phi(R)}\,.
\end{equation}

Next, one finds it convenient to Fourier transform again,
\begin{equation}
\mathcal{F}(t)\equiv\frac{1}{2\pi}\int d\Omega\, \hat{\mathcal{F}}(\Omega) e^{i\Omega t}\,.
\end{equation}

Given the non-linear relationship between $\omega$ and $\Omega$, this new $t$ is not proportional to $\tau$ (except in $D=4$) but, for all practical purposes, gives an equivalent representation of the waveform (not to be confused with the Minkowski time).

This was not necessary to compute the inelasticity. The integral could already be made in $\Omega$ as in Eq.~\eqref{oO_int}. The advantage is that $\mathcal{F}(t)$ can be written in closed form,
\begin{equation}
\mathcal{F}^{(k)}(t)=\sqrt{\frac{8}{D-3}}(-1)^kk!i^{\frac{D-4}{2}+m}\left[\frac{1}{\Phi'(R)}\frac{d}{dR}\right]^{k-1}\left(\Phi'(R)^{-1}R^{\frac{D-2}{2}}J_{\frac{D-4}{2}+m}(2R)f_0^{(k)}(R)\right)\,,\label{final_surface}
\end{equation}
where $R=\Phi^{-1}(t)$.

So, for each $E^{(k)}$, $H^{(k)}$ one must compute the corresponding $\mathcal{F}$ via \eqref{final_surface}, which we denote by
\begin{equation}
\mathcal{F}_{E^{(k)}}(t) \ ,  \qquad \mathcal{F}_{H^{(k)}}(t)\,.
\end{equation}

Then, from Eqs.~\eqref{news2} and \eqref{N_N-k}, the contribution to $\epsilon^{(N)}(\tfrac{\pi}{2})$ ($N=1,2,3$ for surface terms), is
\begin{equation}
\epsilon^{(N)}\left(\frac{\pi}{2}\right)=\frac{(D-2)(D-3)}{4} \sum_{k=1}^{N} \int dt\, \left[\mathcal{F}_{E^{(k)}}(t)+\mathcal{F}_{H^{(k)}}(t)\right]\left[\bar{\mathcal{F}}_{E^{(N+1-k)}}(t)+\bar{\mathcal{F}}_{H^{(N+1-k)}}(t)\right] \ ,
\label{epsn}
\end{equation}
where the bar denotes complex conjugation.

Finally, we perform the time integration as an integral in the auxiliary variable $R$,
\begin{equation}
\int dt\, \mathcal{F}_{H^{(k)}}(t)\bar{\mathcal{F}}_{E^{(N+1-k)}}(t)=\int_0^\infty dR\,|\Phi'(R)| \mathcal{F}_{H^{(k)}}(\Phi(R)) \bar{\mathcal{F}}_{E^{(N+1-k)}}(\Phi(R))\,,\label{F_inelasticity}
\end{equation}
and similarly for the other contributions.

Thus we arrive at the striking conclusion that all surface integral contributions to the inelasticity are given by integrals over Bessel functions. 

\begin{table}[h!]
\begin{center}
\begin{tabular}{||c|c|c||}
\hline
 $N$ & Term & contribution to $\epsilon^{(N)}(\tfrac{\pi}{2})$ \\
\hline
\hline
1 &  $\mathcal{F}_{E^{(1)}}\mathcal{F}_{E^{(1)}}$& $8\left(\frac{1}{2}-\frac{1}{D}\right)\phantom{\dfrac{D}{D}}$\\
\hline
2 &$2\mathcal{F}_{E^{(1)}}\mathcal{F}_{E^{(2)}}$ & $-32\left(\frac{1}{2}-\frac{1}{D}\right)\tfrac{D-4}{D+2}\phantom{\dfrac{D}{D}}$ \\
& $2\mathcal{F}_{E^{(1)}}\mathcal{F}_{H^{(2)}}$& $-32\left(\frac{1}{2}-\frac{1}{D}\right)\tfrac{D-3}{D-4}\phantom{\dfrac{D}{D}}$\\
\hline
3    & $\mathcal{F}_{E^{(2)}}\mathcal{F}_{E^{(2)}}$ & $64\left(\frac{1}{2}-\frac{1}{D}\right)\tfrac{(D-4)^2}{(D+2)(D+4)}\phantom{\dfrac{D}{D}}$\\
     & 2$\mathcal{F}_{E^{(2)}}\mathcal{F}_{H^{(2)}}$ & $64\left(\frac{1}{2}-\frac{1}{D}\right)\tfrac{D-3}{D+2}\phantom{\dfrac{D}{D}}$\\
    & $\mathcal{F}_{H^{(2)}}\mathcal{F}_{H^{(2)}}$ & $64\left(\frac{1}{2}-\frac{1}{D}\right)\tfrac{(D-3)^2}{(D-4)(D-8)}\phantom{\dfrac{D}{D}}$\\
\hline
\end{tabular}\vspace{2mm}\\
\end{center}\caption[Surface integral contributions to the news function series used to compute the inelasticity.]{ Surface integral contributions to the news function series used to compute the inelasticity, Eq.~\eqref{F_inelasticity}.}
\label{TabSurface}
\end{table}

Table~\ref{TabSurface} summarizes the contribution to $\epsilon^{(N)}(\tfrac{\pi}{2})$ from each term in $\dot{\mathcal{E}}^2\propto(\dot{E}+\dot{H})^2$. Details of the relevant integrals can be found in Appendix~\ref{app:R_integrals}. 

The first-order estimate for the inelasticity comes from $N=1$ which, from Eq.~\eqref{e0}, determines the isotropic term,
\begin{equation}
\epsilon_0=\frac{1}{8}\epsilon\left(\frac{\pi}{2}\right)^{(1)}=\frac{1}{2}-\frac{1}{D}\,.
\end{equation}

Thus we prove that the numerical fit was exact. Indeed, all the terms in Table~\ref{TabSurface} have been computed numerically with the same code used in \cite{Herdeiro:2011ck}, and the results agree with a relative error of less that $10^{-4}$.

For $N=2,3$, observe that they are all rational functions of $D$, and all contain $\epsilon_0$. Moreover, the minus sign at $N=2$ suggests that the series might be alternate, though this is of limited validity without the volume terms. The most striking observation, however, is that some terms have poles at $D=4$ or $D=8$. Indeed, in Appendix~\ref{app:R_integrals} we show that some of the integrals do not converge for all $D\geq4$. In the next section we will understand why.

\section{Late time tails}
\label{tails}
Let us examine the late time behaviour of $\mathcal{F}(t)$ in Eq.~\eqref{final_surface}. The far future $t\rightarrow\infty$ corresponds to $R\rightarrow0$ since
\begin{equation}
\Phi^{-1}(t)\propto\left\{
\begin{array}{ll}
 e^{-\frac{t}{2}}\ , &  D=4\  \vspace{2mm}\\
\displaystyle{ t^{-\frac{1}{D-4}}}\ , & D>4\ \label{Phi}
\end{array} \right. \ .
\end{equation} 
For the surface terms in $E$ and $H$ we get
\begin{equation}
\mathcal{F}_{E^{(1)}}(t)\simeq R^{D-2}\,,\qquad \mathcal{F}_{E^{(2)}}(t)\simeq R^{D-2}\,,\qquad \mathcal{F}_{H^{(2)}}(t)\simeq R^{D-6}\,,
\end{equation}

which means that $\mathcal{F}_{H^{(2)}}$ grows exponentially with $e^t$ in $D=4$. Indeed, only for $D>8$ does it decay faster than $t^{-1}$ which is the condition for integrability. This is in agreement with the findings of D'Eath and Payne~\cite{D'Eath:1992qu} in $D=4$: both the surface and volume terms have non-integrable, exponentially growing tails at late times but their sum is well behaved and integrable.

In Appendix \ref{app_tails} we trace the origin of these tails to the Green's function and show that they are generic. In particular, we confirm that the volume terms of $E^{(2)}$ and $H^{(2)}$ have the same behaviour as their surface counterparts above. Therefore we expect a similar cancellation in $D>4$.

\chapter{The two-dimensional problem}
\label{ch:volume}

\epigraph{Odd as it may seem, I am my remembering self, and the experiencing self, who does my living, is like a stranger to me.}{Daniel Kahneman\\ \emph{Thinking, Fast and Slow}}

In the previous chapter, we managed to compute, analytically, all the surface terms at null infinity and their contribution to the inelasticity. This allowed us to extract the isotropic coefficient $\epsilon_0$ and thus obtain a first-order estimate for the inelasticity. At higher orders, we need to take into account the non-linear volume terms as well. These higher-dimensional integrals do not appear to be calculable analytically (actually, the same applies to the surface integrals if the asymptotic limit is not taken). Thus, numerical methods must be used. 

In this chapter we shall begin preparing the problem for numerical integration in a computer. This requires, amongst others, careful study of the integration region and choice of appropriate coordinates.

We start in Section~\ref{red2D} by using the $CL$ symmetry to recast the general solution as a two-dimensional integral (one-dimensional for surface terms), which is an enormous computational advantage. In Section~\ref{int_domain} we study the integration domain and identify the regions where singularities might be present. Then, in Section~\ref{characteristics}, we perform an hyperbolicity analysis and find characteristic coordinates which, after compactification, produce a finite two-dimensional domain, presented in Carter-Penrose form in Section~\ref{CP_diagram}. Finally, in Section~\ref{asymptotics}, we consider an observer at null infinity and further simplify the Green's function. This is especially relevant for the second-order calculation of the next chapter.

\section{Reduction to two dimensions}
\label{red2D}
We start by recasting our dynamical problem in a two-dimensional form. From Eq.~\eqref{sol_3D}, we factor out the $\rho$ dependence using Eqs.~\eqref{f_pq} and \eqref{def_s} (details in Appendix~\eqref{app:Green2D}). We get that the solution to each reduced metric function $f^{(k)}(p,q)$ can be written as
\begin{equation}
f^{(k)}(p,q)=\int dq' \int dp'\, G^k_m(p,q;p',q')s^{(k)}(p',q')\,,\label{f_volume}
\end{equation}
where $G^{k}_m$ is the reduced Green's function,
\begin{equation}
G^{k}_m(p,q;p',q')=-\frac{1}{4}\int_0^\infty dy\, y^{\frac{D-4}{2}-(D-3)(2k+N_u-N_v)}I_m^{D,0}(x_\star)\,,
\end{equation}
and $x_\star$ now reads
\begin{eqnarray}
x_\star&=&\frac{1+y^2-(q-q'y^{D-2})(p-p'y^{-(D-4)}-\Psi(y))}{2y}\,,\label{x_star_pq}\\
\Psi(y)&\equiv&\left\{
\begin{array}{ll}
\Phi(y)\ , &  D=4\  \vspace{2mm}\\
\displaystyle{ \Phi(y)-\frac{2}{D-4}}\ , & D>4\ \label{psi}\,
\end{array} \right. \ .
\end{eqnarray}

Thus all the quantities that we might need, at any order in perturbation theory, can be computed numerically as a (at most) two-dimensional integral. Surface terms are a particular (and simpler) case. In Appendix~\ref{app:surface_pq} we show that they are given by a one-dimensional integral with a structure similar to the Green's function itself,
\begin{equation}
f^{(k)}_{S}(p,q)=(-1)^{D+1}k!f^{(k)}_0(1)\left(\frac{2}{q}\right)^k\int_0^\infty dy\,y^{\frac{D-4}{2}-(D-3)(k+N_u-N_v)}I_m^{D,k}(x_S)\,,\label{f_surface}
\end{equation}
where now
\begin{equation}
x_S\equiv x_\star(p'=0,q'=0)=\frac{1+y^2-q(p-\Psi(y))}{2y}\,.
\end{equation}

The causal structure is encoded in the functions $I_m^{D,n}(x_\star)$ and in $x_\star$ itself. In the next section we shall see how they determine the domain of integration.

\section{Integration domain}
\label{int_domain}
The integration domain is defined by two causality conditions. For an observation point $\mathcal{P}=(u,v,x^i)$ in the future of the collision, the integration point $\mathcal{P}'=(u',v',{x'}^i)$ must
\begin{enumerate}
\item be in the future light cone of the collision;
\item be in the past light cone of the observation point $\mathcal{P}$.
\end{enumerate}
We shall study these two conditions separately.

\subsection{The future light cone of the collision}
\label{future_cone}
Regarding the first one, recall from Chapter~\ref{ch:dynamics} that, in Rosen coordinates, it is defined as
\begin{equation}
\bar u=0\,\wedge\,\bar v\geq0\qquad \vee \qquad \bar u\geq0\,\wedge\,\bar v=0\,.
\end{equation}
In Brinkmann coordinates, from Eqs.~\eqref{Rosen_coords}, these become
\begin{equation}
u=0\,\wedge\,v\geq\Phi(\rho)\qquad \vee \qquad u\geq0\,\wedge\,\bar v=\Phi(\bar\rho)+\frac{u\Phi'(\bar\rho)^2}{4}\,.
\end{equation}

In the $(p,q)$ plane, these two conditions define two important curves. The first is simply where the initial data have support,
\begin{equation}
p\geq0 \qquad\wedge\qquad q=0\,,
\end{equation}
while the second separates the flat region from the curved region,
\begin{equation}
p(\zeta)=\Psi(\zeta)+\frac{\zeta-1}{\zeta^{D-3}}\qquad \wedge \qquad q(\zeta)=(\zeta-1)\zeta^{D-3}\,,
\end{equation}
where $\zeta\in[1,+\infty [$ is a parameter. This is actually ray 1 of Chapter~\ref{ch:dynamics} or the $+$ sign in Eq.~\eqref{eq_rays12}. The $-$ sign, corresponding to ray 2, is similarly given by
\begin{equation}
p(\zeta)=\Psi(\zeta)+\frac{\zeta+1}{\zeta^{D-3}}\qquad \wedge \qquad q(\zeta)=(\zeta+1)\zeta^{D-3}\,.\label{ray2_parametric}
\end{equation}
This curve is also important because the source is expected to be singular on it. Thus, extra care needs to be taken to integrate near this region.

Finally, we conclude that the lower bounds for the integration variables in Eq.~\eqref{f_volume} are
\begin{equation}
q'\geq0\,,\qquad p'\geq\left\{
\begin{array}{ll}
-\infty\ , &  D=4\  \vspace{2mm}\\
\displaystyle{-\frac{2}{D-4}}\ , & D>4\,
\end{array} \right. \ .\label{p_limits}
\end{equation}

\subsection{The past light cone of the observation point}
\label{past_cone}
For the second condition, we also begin the analysis in the more intuitive $(u,v,\rho)$ coordinates.

\subsubsection{Three-dimensional case}
\label{past_cone_3D}
As mentioned in Chapter~\ref{ch:dynamics} and explained in Appendix~\ref{app:Green}, the original Green's function $G(u,v,x^i)$ has support on the light cone ($\chi=0$) for even $D$ but also inside it ($\chi>0$) for odd $D$. From Eq.~\eqref{def_x_star}, these conditions are equivalent to $-1\leq x_\star \leq1$ for even $D$ and $x_\star\leq1$ for odd $D$ (indeed this is the domain where $I_m^{D,0}$ is non-vanishing). The region $x_\star>1$ is outside the light cone ($\chi<0$) and thus causally disconnected from the event $(u,v,x^i)$. 

If we define curves (indeed parabolas) $C_\pm(\rho')$ by
\begin{equation}
C_\pm(\rho')\equiv(\rho\pm\rho')^2-(u-u')(v-v')\,,
\end{equation}
we can write the above conditions as
\begin{equation}
C_-(\rho')\leq0 \Leftrightarrow x_\star-1\leq0\,,\qquad C_+(\rho')\geq0 \Leftrightarrow x_\star+1\geq0\,.
\end{equation}

In this way we can characterise the domain of the $\rho'$ integration in Eq.~\eqref{sol_3D}. Both curves start at
\begin{equation}
C_\pm(0)=\rho^2-(u-u')(v-v')\,,
\end{equation}
which can be either positive or negative (or zero). Then $C_+(\rho')$ is monotonically increasing while $C_-(\rho')$ has a minimum at $\rho'=\rho$,
\begin{equation}
C_-(\rho)=-(u-u')(v-v')\equiv-(\Delta\rho)^2\,.
\end{equation}
If this minimum is positive, the domain is empty. Thus $(u-u')(v-v')\geq0$ is an absolute condition for the event $(u',v',{x'}^i)$ to be causally connected with the event $(u,v,x^i)$. As this quantity increases, the domain is $\rho'\in\left[\rho-\Delta\rho,\rho+\Delta\rho\right]$. Note that both ends correspond to $x_\star=1$. The other condition, $C_+\geq0$, begins to matter when $(u-u')(v-v')=\rho^2$, i.e. when both curves start at the origin. After that, the domain is $\rho'\in\left[\Delta\rho-\rho,\Delta\rho+\rho\right]$, corresponding to $x_\star=-1$ to $x_\star=1$.

Even when the only condition is $C_-(\rho')\leq0$, the location of $C_+(0)=0$ is important as it corresponds to a coordinate singularity at $\rho'=0$: the angle $\phi'$ between $x^i$ and ${x'}^i$ is not defined if $\rho'=0$, so neither is $x_\star=\cos\phi'$. This is the first time that the $\rho'$ integration starts from the origin, and it corresponds to the first ray that came from the other side of the axis. The Green's function is expected to be singular there, and proper coordinates will need to be chosen to perform the integration near this region.

\subsubsection{Two-dimensional case}
\label{past_cone_2D}
In the two-dimensional reduced case, the picture is less intuitive but very analog. We define similar curves $C_\pm(y)$,
\begin{equation}
C_\pm(y)\equiv (y\pm1)^2 y^{D-4}-(q-q' y^{D-2})((p-\Psi(y))y^{D-4}-p')\,,
\end{equation}
such that, as before,
\begin{equation}
C_-(y)\leq0 \Leftrightarrow x_\star-1\leq0\,,\qquad C_+(y)\geq0 \Leftrightarrow x_\star+1\geq0\,.
\end{equation}

This time, both curves start at a non-negative value, $C_\pm(0)\geq0$, and grow to infinity for large $y$. Moreover, $C_+(y)\geq C_-(y)\, \forall\, y\in\mathbb{R}_0^+$. So, as before, the domain starts when $C_-(y)$ is tangent to the $y$ axis. After that, the condition $C_-(y)\leq0$ gives a finite domain for the $y$ integration. The other condition, $C_+(y)\geq0$, comes into play also when $C_+(y)$ is tangent to the $y$ axis. After that, the $y$ domain is broken in two.

The first condition defines the boundary of the light cone $(p',q')$ of the event $(p,q)$, whereas the second gives the location of the singularity of the reduced Green's function $G^k_m(p,q;p',q')$. They are the solutions to
\begin{equation}
C_\pm(y)=0 \qquad \wedge \qquad \frac{d}{dy}C_\pm(y)=0\,.
\end{equation}
For each case there are two solutions, parameterised by $y\in\mathbb{R}_0^+$ and labeled by $n=\pm1$, 
\begin{eqnarray}
p'&=&y^{D-4}\left[p-\Psi(y)-(1\pm y)\Delta_n(p,q)\right]\,,\\
q'&=&\frac{1}{y^{D-2}}\left[q-\frac{1\pm y}{\Delta_n(p,q)}\right]\,,
\end{eqnarray}
where
\begin{equation}
\Delta_\pm(p,q)=\frac{1\pm\sqrt{1+(2+(D-4)p)(D-2)q}}{(D-2)q}
\end{equation}

For the $-$ solution, $\Delta_\pm$ gives the two characteristics going through $(p,q)$ which delimit the light cone. For the $+$ solution (the curves where the Green's function has a singularity), one can check that the one with $\Delta_+$ is inside the past light cone whereas the one with $\Delta_-$ is inside the future light cone of $(p,q)$.

\begin{figure}
\hspace{-3mm}
\includegraphics[scale=0.38,clip=true,trim= 0 0 0 0]{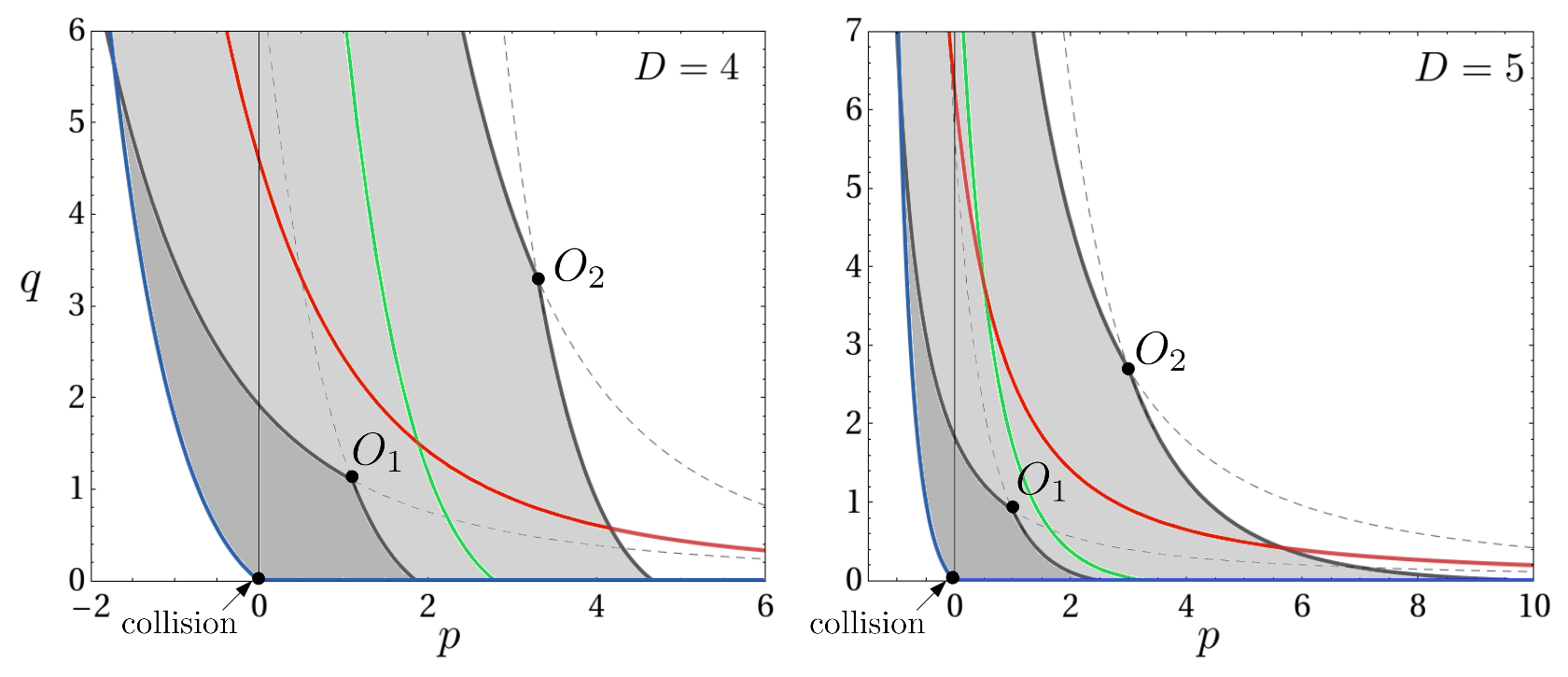}
\caption[Characteristic curves in the $(p,q)$ plane.]{\label{char_pq}Characteristic curves in the $(p,q)$ plane, showing the future light cone of the collision (in blue), the past/future light cones for two observers (in solid/dashed grey), and the singularities of the source (in red) and the Green's function (in green) for observer $O_2$. In $D=4$ (left diagram), the blue curve on the left goes to $p=-\infty$, whereas in $D>4$ (right diagram) is starts at $p=-2/(D-4)$.}
\end{figure}

In Fig.~\ref{char_pq} we illustrate all these curves in the $(p,q)$ plane for $D=4$ and $D=5$. The collision occurs at $(p,q)=(0,0)$ and the blue curves delimit its future light cone: the one to the right is where the initial data have support; the one to the left (ray 1) goes to $-\infty$ in $D=4$ and to a constant in $D>4$, as in Eq.~\eqref{p_limits}. Then we have two observers, $O_1$ and $O_2$, together with their past (solid grey) and future (dashed grey) light cones; their interiors are, respectively, to the left and to the right of those curves. Finally, the red curve corresponds to the second optical ray (ray 2), where we expect the sources to be singular, and the green curve is the location of the singularity of the Green's function inside the past light cone of observer $O_2$.

It is possible to write these curves in a non-parametric form. For that, it is useful to define yet a new set of coordinates $(P,Q)$ through
\begin{equation}
P\equiv
\left\{
\begin{array}{ll}
 p\ , &  D=4\  \vspace{2mm}\\
\displaystyle{p+\frac{2}{D-4}}\ , & D>4\ \label{def_PQ}\,
\end{array} \right. \ ,
\qquad Q\equiv (D-2)(2+(D-4)p)q\,.
\end{equation}
Then we can solve for the parameter $y$, 
\begin{equation}
y=\frac{\sqrt{1+Q}-n(D-3)}{\sqrt{1+Q'}\pm n(D-3)}\,.
\end{equation}
Inserting back in the parametric equations, we find that they can be written as
\begin{equation}
\mathcal{C}_{\pm n}(P',Q')=\mathcal{C}_{-n}(P,Q)\,,
\end{equation}
where 
\begin{equation}
\mathcal{C}_\pm(P,Q)\equiv\left\{
\begin{array}{ll}
 \dfrac{P-1}{2}+\ln\left(\dfrac{\sqrt{1+Q}\pm1}{2}\right)\pm\dfrac{1}{\sqrt{1+Q}\pm1}\ , &  D=4\ \vspace{5mm} \\
\displaystyle{\left(\dfrac{P}{\sqrt{1+Q}\pm1}\right)^{\frac{1}{D-3}}\frac{\sqrt{1+Q}\pm(D-3)}{D-2}-(D-4)^{-\frac{1}{D-3}}}\ , & D>4\ \label{L_PQ}\,
\end{array} \right. \ .
\end{equation}

To recapitulate, the light cone $(P',Q')$ of the event $(P,Q)$ is defined by
\begin{equation}
\mathcal{C}_\pm(P',Q')=\mathcal{C}_\pm(P,Q)\,,
\end{equation}
and the points $(P',Q')$ where the Green's function for the observation point $(P,Q)$ is singular are given by
\begin{equation}
\mathcal{C}_\pm(P',Q')=\mathcal{C}_\mp(P,Q)\,.
\end{equation}

In the next section we will find characteristic coordinates and confirm that all these curves are indeed characteristics of the reduced differential operator.

\section{Characteristic coordinates}
\label{characteristics}
For our initial problem, the coordinates $(u,v,x^i)$ were adapted to the initial data given on the characteristic surface $u=0$. In the two-dimensional reduced version, it would be useful to find characteristic coordinates $(\xi,\eta)$ such that the principal part of the differential operator contains only the mixed term $\partial_\xi\partial_\eta$, and the initial data is on a characteristic line given by, say, constant $\xi$.

In Appendix~\ref{app:Green2D} we obtain the differential operator acting on $f^{(k)}(p,q)$, Eq.~\eqref{app:op2D}. The terms with highest derivatives come from
\begin{equation}
-4\partial_p\partial_q+\left((2+(D-4)p)\partial_p-(D-2)q\partial_q\right)^2+\ldots\,,
\end{equation}
where we discard first derivatives. Transforming to $(P,Q)$,
\begin{equation}
\partial_p\rightarrow\partial_P+QP^{-1}\partial_Q\,,\qquad \partial_q\rightarrow(D-2)(D-4)P\partial_Q\,,
\end{equation}
this becomes
\begin{equation}
(D-4)^2P^2\partial_P^2+4Q\left(Q-(D-2)(D-4)\right)\partial_Q^2-4(D-4)P(Q+D-2)\partial_P\partial_Q+\ldots\,.
\end{equation}

Now we make a coordinate transformation,
\begin{equation}
(P,Q)\rightarrow \left(\xi(P,Q),\eta(P,Q)\right)\,,
\end{equation}
and the above operator takes the form
\begin{equation}
f(Z_\eta,Z_\xi)\dfrac{\partial \eta}{\partial P}\dfrac{\partial \xi}{\partial P}\dfrac{\partial^2}{\partial \eta \partial \xi}+\left(\dfrac{\partial \eta}{\partial P}\right)^2C(Z_\eta)\dfrac{\partial^2}{\partial \eta^2}+\left(\dfrac{\partial \xi}{\partial P}\right)^2C(Z_\xi)\dfrac{\partial^2}{\partial \xi^2}+\ldots\,,
\end{equation}
where $Z_X\equiv\partial_Q X/\partial_PX$ and the characteristic polynomial is
\begin{equation}
C(Z) \equiv (D-4)^2P^2+4Q(Q-(D-2)(D-4))Z^2-4(D-4)P(Q+D-2)Z\,.
\end{equation}

The coordinates $(\xi,\eta)$ are characteristics if $C(Z_\xi)=C(Z_\eta)=0$, i.e. if they obey
\begin{eqnarray}
\dfrac{2Q(Q-(D-4)(D-2))}{Q+(D-2)(1-\sqrt{1+Q})}\dfrac{\partial \eta}{\partial Q}&=&(D-4)P\dfrac{\partial \eta}{\partial P}\,, \\
\dfrac{2Q(Q-(D-4)(D-2))}{Q+(D-2)(1+\sqrt{1+Q})}\dfrac{\partial \xi}{\partial Q}&=&(D-4)P\dfrac{\partial \xi}{\partial P}\,.
\end{eqnarray}

These clearly allow for a solution by separation of variables. Moreover, they reproduce the results of D'Eath and Payne \cite{D'Eath:1992hd,Payne} in $D=4$\footnote{Note that $(D-4)P\rightarrow2$.}. With a convenient choice of normalisation, a possible solution is
\begin{equation}
\xi=\mathcal{C}_-(P,Q)\,,\qquad \eta=\mathcal{C}_+(P,Q)\,.
\end{equation}

Thus we conclude that all the curves found in the previous section are indeed characteristics: for an event $(\xi,\eta)$, its light cone $(\xi',\eta')$ is delimited by
\begin{equation}
\xi'=\xi\,,\qquad \eta'=\eta\,,
\end{equation}
and the Green's function is singular on
\begin{equation}
\xi'=\eta\,,\qquad \eta'=\xi\,,
\end{equation}
respectively in the past and future of $(\xi,\eta)$. The integration constants were chosen such that $\eta=0$ goes through $(p,q)=(0,0)$ (the left blue curve in Fig.~\ref{char_pq}) and $\xi=0$ corresponds to the second optical ray (ray 2), previously obtained in parametric form in Eq.~\eqref{ray2_parametric} (the red curve in Fig.~\ref{char_pq}).

\subsection{Compactified characteristic coordinates}
\label{compact}
From the above results it is straightforward to conclude that the ranges of the characteristic coordinates in the future of the collision are
\begin{equation}
\xi\in]-\infty,+\infty[\,,\qquad \eta\in[0,+\infty[\,,
\end{equation}
with $\xi\leq\eta$ inside the light cone. This suggests introducing compactified coordinates,
\begin{equation}
\hat\xi=\frac{\xi}{\sqrt{1+\xi^2}}\,,\qquad \hat\eta=\frac{\eta}{\sqrt{1+\eta^2}}\,,
\end{equation}
such that the integration domain becomes compact,
\begin{equation}
\hat\xi\in[-1,1]\,,\qquad \hat\eta\in[0,1]\,.
\end{equation}

Then the volume integrals, Eq.~\eqref{f_volume}, become
\begin{equation}
f^{(k)}(\hat\xi,\hat\eta)=\int_{-1}^{\hat\xi} d\hat\xi' \int^{\hat\eta}_{\max \{0,\hat\xi'\}} d\hat\eta'\, \left|\frac{\partial(p',q')}{\partial(\hat\xi',\hat\eta')}\right|G^k_m(\hat\xi,\hat\eta;\hat\xi',\hat\eta')s^{(k)}(\hat\xi',\hat\eta')\,,\label{f_etaxi}
\end{equation}
where the Jacobian determinant is
\begin{eqnarray}
\left|\frac{\partial(p,q)}{\partial(\xi,\eta)}\right|&=&\left|\frac{\partial(p,q)}{\partial(P,Q)}\right|\times\left|\frac{\partial(P,Q)}{\partial(\xi,\eta)}\right|\times\left|\frac{\partial(\xi,\eta)}{\partial(\hat\xi,\hat\eta)}\right|\,,\\
&=&\left(\frac{D-3}{D-4}\right)^2\frac{P^{-\frac{2}{D-3}}Q^{\frac{D-2}{D-3}}}{\sqrt{1+Q}}\frac{1}{(1-\hat\eta^2)^{\frac{3}{2}}}\frac{1}{(1-\hat\xi^2)^{\frac{3}{2}}}\,.\label{jacobian}
\end{eqnarray}

\section{Conformal Carter-Penrose diagram}
\label{CP_diagram}
\begin{figure}[h]
\hspace{-3mm}\centering
\includegraphics[scale=0.6,clip=true,trim= 0 0 0 0]{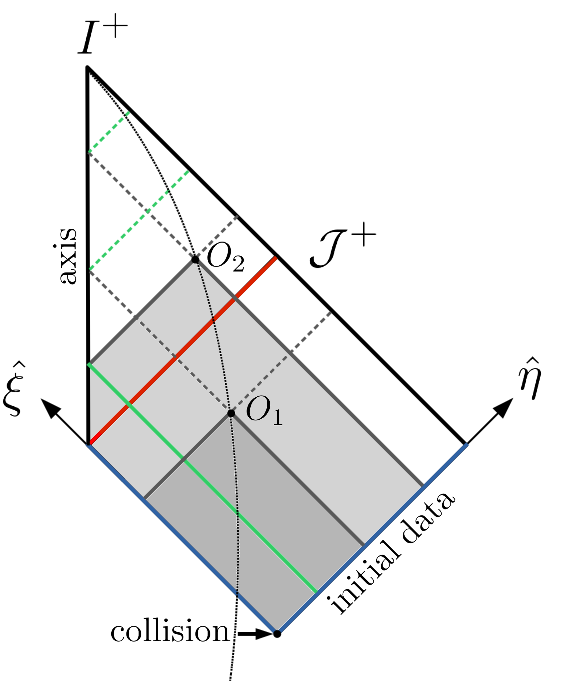}
\caption[Carter-Penrose diagram of the two-dimensional problem.]{\label{penrose}Carter-Penrose diagram of the effective two-dimensional space-time. The curves and colours match those of Fig.~\ref{char_pq}.}
\end{figure}

With the compactified characteristic coordinates $(\hat\xi,\hat\eta)$ we can produce a conformal Carter-Penrose diagram, Fig.~\ref{penrose}, of the effective two-dimensional space-time, where the causal structure and the important curves we have been discussing become very clear. 

The collision occurs at $(\hat\xi,\hat\eta)=(-1,0)$ and its future light cone is shown in blue: $\hat\xi=-1$ contains the initial data and $\hat\eta=0$ is the first optical ray. The left boundary is the axis, $\rho=0$, given here by $\hat\xi=\hat\eta$. Null infinity ($\mathcal{J}^+$) is at $\hat\eta=1$ and future time-like infinity ($I^+$) is at $(\hat\xi,\hat\eta)=(1,1)$.

The world line of a time-like observer is shown as a dotted black curve. When crossing the blue line it is hit by ray 1 and the radiation signal begins, peaking later when crossing the red line, $\hat\xi=0$, which corresponds to ray 2.

The light cones at two instants, $O_1$ and $O_2$, are also shown, with solid and dashed lines for the past and future branches, respectively. When these `reflect' (effectively cross) at the axis the Green's function becomes singular. The solid green line is the location of this singularity in the past light cone of $O_2$. Note that this does not exist for $O_1$ ($\hat\xi<0$) because ray 2 (the red line) is the first ray coming from the `other' side of the axis.

The radiation signal is extracted at $\mathcal{J}^+$, so the world line of a far-away observer is pushed to the right as $\hat\eta\rightarrow1$. That limit is the subject of the next section.

\section{An observer at null infinity}
\label{asymptotics}
Since we are ultimately interested in the extraction of radiation at null infinity, we shall see how that works in the two-dimensional picture. Indeed, the characteristic coordinates are well suited for such an asymptotic analysis. Taking the limit $r\rightarrow\infty$ with $\tau$ and $\theta$ fixed, we find that
\begin{equation}
\xi\rightarrow\left\{
\begin{array}{ll} \bar{\tau}(\tau,\theta)-1\ , &  D=4\ \vspace{5mm} \\
\displaystyle{\frac{1}{(D-4)^{\frac{1}{D-3}}}\left(\frac{D-4}{D-3}\bar\tau(\tau,\theta)-1\right)}\ , & D>4\,
\end{array} \right.\,, \qquad \eta\rightarrow\frac{2(D-3)}{(D-2)^{\frac{D-2}{D-3}}}\frac{1}{\hat q}\sim O(r)\,,\,
\end{equation}
where $\bar\tau(\tau,\theta)$ is the time coordinate in which the angular dependence of the news function factorises, Eq.~\eqref{tau_transform}. Thus we conclude that null infinity is when $\eta\rightarrow\infty$ for fixed $\xi$, and that the time integration at null infinity is equivalent to an integration in $\xi$.

Regarding the Green's function, observe that all the dependence on the observation point $(p,q)$ comes through $x_\star$. In $(P,Q)$ coordinates, from Eq.~\eqref{x_star_pq}, we get, in $D=4$,
\begin{eqnarray}
x_\star&=&\frac{4-Q(P-P'+2\ln y)}{8y}+\frac{4+Q'(P-P'+2\ln y)}{8}y\,,
\end{eqnarray}
and, in $D>4$,
\begin{eqnarray}
x_\star&=&
\frac{(D-2)(D-4)-Q}{2(D-2)(D-4)y}+\frac{(D-2)(D-4)-Q'}{2(D-2)(D-4)}y\nonumber\\
&&+\frac{QP'/P}{2(D-2)(D-4)y^{D-3}}+\frac{Q'P/P'}{2(D-2)(D-4)}y^{D-3}
\end{eqnarray}
In these coordinates it is manifest that $x_\star$ is actually a three-dimensional quantity, depending only on $Q$, $Q'$ and $P-P'$ or $P/P'$, respectively for $D=4$ and $D>4$.

In the limit $r\rightarrow\infty$, recall that $q\rightarrow0$ and $p\rightarrow\infty$ with $pq\rightarrow1$. This implies
\begin{equation}
P\rightarrow\infty\,,\qquad Q\rightarrow (D-2)(D-4)\,.
\end{equation}

Thus the term proportional to $P$ diverges unless we scale the integration variable $y$. A natural choice is
\begin{equation}
y=\left\{
\begin{array}{ll} \dfrac{Q}{Q'}\hat y			\ , &  D=4\ \vspace{5mm} \\
\displaystyle{	\left(\frac{P'}{P}\right)^{\frac{1}{D-3}}\hat y	}\ , & D>4\,\label{y_asympt}
\end{array} \right.\,,
\end{equation}

Then, asymptotically,
\begin{equation}
x_\star\rightarrow\left\{
\begin{array}{ll} \dfrac{1}{2\hat y}\left[\hat y^2-\dfrac{Q'}{2}\left(\ln\dfrac{4\hat y}{Q'}+1-\dfrac{\Delta'}{2}\right)\right]		\ , &  D=4\ \vspace{5mm} \\
\displaystyle{\frac{1}{2\hat y^{D-3}}\left(1+\frac{Q'\hat y^{2(D-3)}}{(D-2)(D-4)}-\frac{2(D-3)\hat y^{D-4}}{(D-4)\Delta'^{\frac{1}{D-3}}}\right)}\ , & D>4\,\label{x_asympt}
\end{array} \right.\,,
\end{equation}
where
\begin{equation}
{\Delta'}^{\frac{1}{D-3}}\equiv\left\{
\begin{array}{ll} P'-P+2+\dfrac{4}{Q}-2\ln Q \qquad\,\qquad  \rightarrow P'-2\xi\,, &  D=4\ \vspace{5mm} \\
\displaystyle{     \left(\frac{P'}{P}\right)^{\frac{1}{D-3}}\left(\frac{2(D-2)(D-3)}{Q-(D-2)(D-4)}\right) \rightarrow \frac{{P'}^{\frac{1}{D-3}}}{1+(D-4)^{\frac{1}{D-3}}\xi}    }\ , & D>4\,
\end{array} \right.\,.\label{Delta_p}
\end{equation}

Since $x_\star\rightarrow x_\star(\Delta',Q')$, if we factor out the trivial scaling of Eq.~\eqref{y_asympt}, the Green's function becomes effectively two-dimensional,
\begin{equation}
G_m^k(p,q;p',q')\rightarrow
\left\{
\begin{array}{ll}  \left(\dfrac{Q}{Q'}\right)^{1-(2k+N_u-N_v)}G_m^k(\Delta',Q')	, &  D=4\ \vspace{5mm} \\
\displaystyle{    \left(\dfrac{P'}{P}\right)^{\frac{1}{2}\frac{D-2}{D-3}-(2k+N_u-N_v)}G_m^k(\Delta',Q')    }\ , & D>4\,\label{GreenF_2D}
\end{array} \right.\,,
\end{equation}

Surface terms are a special case of the above, obtained by setting $q'=p'=0$ in $x_\star$, but the same procedure follows. We scale $y$ as in Eq.~\eqref{y_asympt} but without $Q'$ or $P'$. Thus, from Eq.~\eqref{f_surface},
\begin{equation}
q^k f_S^{(k)}(p,q)\rightarrow
\left\{
\begin{array}{ll}  Q^{1-(k+N_u-N_v)}f_S^{(k)}(\xi)	, &  D=4\ \vspace{5mm} \\
\displaystyle{    P^{-\frac{1}{2}\frac{D-2}{D-3}+(k+N_u-N_v)}f_S^{(k)}(\xi)    }\ , & D>4\,
\end{array} \right.\,.\label{surface_scri+}
\end{equation}

In particular, this can be used to compute the asymptotic behaviour of the second-order source $s^{(2)}(p',q')$ when $\eta'\rightarrow\infty$.

\subsection{The asymptotic metric functions}
\label{asymptotic_MF}
From Eq.~\eqref{GreenF_2D}, since $Q\sim r^{-1}$ in $D=4$ and $P\sim r^{D-3}$ in $D>4$, we conclude that the Green's function decays at null infinity with a power of $r$ (or, equivalently, of $\eta$) equal to
\begin{equation}
\frac{D-2}{2}-(D-3)(2k+N_u-N_v)\,.
\end{equation}

Therefore, we define the asymptotic metric function $\hat f^{(k)}(\hat\xi)$ as the finite limit
\begin{eqnarray}
\hat f^{(k)}(\hat\xi)&\equiv&\lim_{\eta\rightarrow\infty}f^{(k)}(\hat\eta,\hat\xi)\times\left\{
\begin{array}{ll}  \left(\dfrac{Q}{4}\right)^{1-(2k+N_u-N_v)}	, &  D=4\ \vspace{5mm} \\
\displaystyle{    \left(\frac{1}{P}\right)^{\frac{1}{2}\frac{D-2}{D-3}-(2k+N_u-N_v)}   }\ , & D>4\,
\end{array} \right.\,,\nonumber\\
&=&\int_{-1}^{\hat\xi} d\hat\xi' \int^{1}_{\max \{0,\hat\xi'\}} d\hat\eta'\, G^k_m(\Delta',Q')s^{(k)}(\hat\xi',\hat\eta')J^k(\hat\xi',\hat\eta')\,,\label{f_xi}
\end{eqnarray}
where
\begin{equation}
J^k(\hat\xi',\hat\eta')\equiv\left|\frac{\partial(p',q')}{\partial(\hat\xi',\hat\eta')}\right|\times\left\{
\begin{array}{ll}  \left(\dfrac{4}{Q'}\right)^{1-(2k+N_u-N_v)}	, &  D=4\ \vspace{5mm} \\
\displaystyle{    {P'}^{\frac{1}{2}\frac{D-2}{D-3}-(2k+N_u-N_v)}   }\ , & D>4\,
\end{array} \right.\,.\label{def_J_k}
\end{equation}

From the point of view of numerical integration, this means that both the Green's function and the source can be tabulated independently on a two-dimensional domain, prior to attempting the double integration. Indeed, there is a more natural choice of coordinates for the Green's function. Let us define
\begin{eqnarray}
\delta\xi&\equiv&\mathcal{C}_-(\Delta',Q')=\frac{\xi'-\xi}{1+(D-4)^{\frac{1}{D-3}}\xi}\,,\label{def_delta_xi}\\
\delta\eta&\equiv&\mathcal{C}_+(\Delta',Q')=\frac{\eta'-\xi}{1+(D-4)^{\frac{1}{D-3}}\xi}\,,\label{def_delta_eta}
\end{eqnarray}
together with their compactified versions,
\begin{equation}
\delta\hat\xi\equiv\frac{\delta\xi}{\sqrt{1+\delta\xi^2}}\,,\qquad \delta\hat\eta\equiv\frac{\delta\eta}{\sqrt{1+\delta\eta^2}}\,.
\end{equation}

Observe the analogy with the definition of $\eta$ and $\xi$ through $\mathcal{C}_\pm(P',Q')$ (but note that, whereas $P'\geq0$, $\Delta'$ can be negative). For a given observation time $\xi$, the required region of $(\delta\xi,\delta\eta)$ is $\xi$-dependent. What interests us, of course, is the full domain covered by all values of $\xi$. In $D=4$ this is simply
\begin{equation}
\delta\hat\xi\in [-1,0]\,,\qquad \delta\hat\eta\in[-1,1]\,,\qquad \text{with}\qquad \delta\hat\xi<\delta\hat\eta\,.
\end{equation}

For $D>4$ the picture is  complicated by the fact that the denominator of Eqs.~\eqref{def_delta_eta}-\eqref{def_delta_xi} changes sign when $\xi=\xi_0\equiv-(D-4)^{-\frac{1}{D-3}}$. 
One can show that it becomes
\begin{equation}
\delta\hat\xi\in[-1,0]\,,\qquad \delta\hat\eta\in [\hat\xi_0,1]\,,\qquad \text{with}\qquad \delta\hat\xi<\delta\hat\eta\,,
\end{equation}
plus an extra rectangle,
\begin{equation}
\delta\hat\xi\in[0,1]\,,\qquad \delta\hat\eta\in [-1,\hat\xi_0]\,.
\end{equation}

In the next Chapter we will apply these results to the second-order calculation. In particular, we shall obtain, numerically, the second-order Green's functions and sources for the metric functions $E^{(2)}$ and $H^{(2)}$.

\chapter{The second-order calculation}
\label{ch:second_order}

\epigraph{Nothing in the world can take the place of persistence. Talent will not; nothing is more common than unsuccessful men with talent. Genius will not; unrewarded genius is almost a proverb. Education will not; the world is full of educated derelicts. Persistence and determination alone are omnipotent. The slogan ``press on'' has solved and always will solve the problems of the human race.}{Calvin Coolidge}

In this chapter we conclude our journey with the first steps towards the computation of the second-order news function, $\dot{\mathcal{E}}^{(2)}$, and its contribution to the inelasticity's angular distribution, $\epsilon^{(2)}(\tfrac{\pi}{2})$. 

In terms of the asymptotic functions defined in Eq.~\eqref{f_xi}, the news function reads
\begin{equation}
\dot{\mathcal{E}}(\bar\tau,\tfrac{\pi}{2})=\sqrt{\frac{(D-2)(D-3)}{4}}\frac{d}{d\bar\tau}\left(\hat e(\hat\xi)+\hat{h}(\hat\xi)\right)\,,
\end{equation}
so our task is to compute the second-order metric functions $\hat e^{(2)}$ and $\hat h^{(2)}$.  

This would allows us to finally extract the second coefficient in the inelasticity's angular series, $\epsilon_2$, and thus obtain an improved estimate for the inelasticity of the collision. At the time of writing, we were not able to finish this demanding task. The results here presented are only partial but, nevertheless, constitute an important step towards the final goal.

The numerical implementation of the strategy described in this thesis was done in a \verb!C++! code by Marco Sampaio, building on the earlier code used in \cite{Herdeiro:2011ck,Coelho:2012sya} for the first order calculation. Therefore, we will not attempt to give a thorough description of the coding involved.\footnote{The interested reader may find a pedagogical discussion on the evaluation of one-dimensional integrals in \cite{Sampaio:2013faa}. Note that these make up not only the surface terms (and hence the second-order source) but also the second-order Green's function for the volume terms.}

We begin in Section~\ref{2nd_gf} by computing and tabulating the reduced Green's functions in their full domains. Then, in Section~\ref{2nd_sources}, we do the same for the sources. Finally, in Section~\ref{2nd_wf}, we comment on the outstanding two-dimensional integration, which we hope to perform in the near future.

\section{The second-order Green's functions}
\label{2nd_gf}
We begin by tabulating the two reduced Green's functions ($m=0$ and $m=2$)
\begin{eqnarray}
G^2_m(\Delta',Q')=-\frac{1}{4}\int_0^\infty d\hat y\, \hat y^{-4-\frac{7}{2}(D-4)}I^{D,0}_m(x_\star)\,,
\end{eqnarray}
with $x_\star$ as in Eq.~\eqref{x_asympt} and $\Delta'$ as in Eq.~\eqref{Delta_p}. We shall do this in the compact $(\delta\xi,\delta\eta)$ coordinates.

Before doing so, though,  we must ensure that they are finite within the whole domain, otherwise numerical errors will lead to loss of precision when interpolating the functions near the singular points. In Appendix~\ref{app:expansions} we show that the Green's function remains finite as $\eta'\rightarrow\infty$, but on the axis $\eta'=\xi'$ it vanishes for $D=4$ and diverges when $D>4$ (with a power law). Indeed, the strongest power occurs at the intersection of the axis with the light cone, $\xi'=\xi$. In the new coordinates this corresponds to $\delta\eta=\delta\xi$ and hence, to make the Green's function regular, we multiply it by an appropriate power of $(\delta\hat\eta-\delta\hat\xi)$, which can be read from Eq.~\eqref{power_axis}.

Regarding the green line ($\eta'=\xi$), which now occurs at $\delta\eta=0$, we have found empirically that the singularity there is at most logarithmic\footnote{This could probably be proved analytically in a manner similar to the asymptotic expansions of Appendix~\ref{app:expansions}, but we did not find it necessary to do so.}. To regularise, we choose to multiply by $\delta\hat\eta$ (indeed this makes the function go to zero at the green line).

Finally, to compensate the power of $Q'$ that is now part of $J^2$ (Eq.~\eqref{def_J_k}) in $D=4$, and which would make the Green's function grow close to the initial data $(\delta\hat\xi\rightarrow-1)$, we multiply by $(1+\delta\hat\xi)^\frac{3}{2}$.

So, putting everything together, we are going to tabulate numerically the functions
\begin{equation}
\mathcal{G}_m(\delta\hat\xi,\delta\hat\eta)\equiv G^2_m(\Delta',Q')\times\delta\hat\eta\,(\delta\hat\eta-\delta\hat\xi)^{\gamma_1}(1+\delta\hat\xi)^{\gamma_2}\,,\label{def_G_curl}
\end{equation}
where
\begin{equation}
\gamma_1=\begin{cases}
-\dfrac{7}{2}&,\; D=4 \\
3-\dfrac{1}{2(D-3)} &, \; D>4 
\end{cases}\;,\; \qquad
\gamma_2=\begin{cases}
\dfrac{3}{2}&,\; D=4 \\
0 &,\; D>4
\end{cases}\,.
\end{equation}

\begin{figure}
\includegraphics[scale=0.55,clip=true,trim= 0 0 0 0]{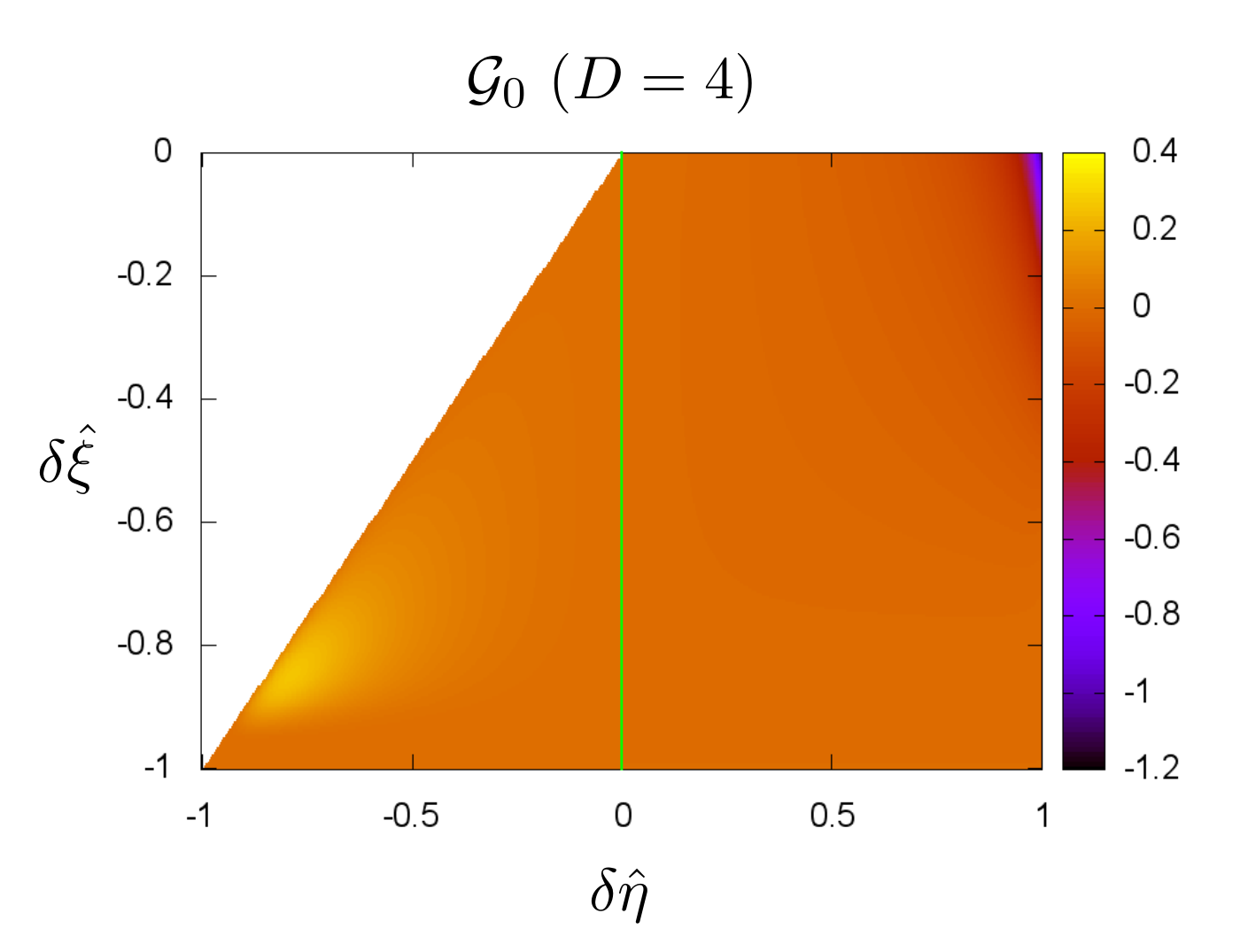}
\includegraphics[scale=0.55,clip=true,trim= 0 0 0 0]{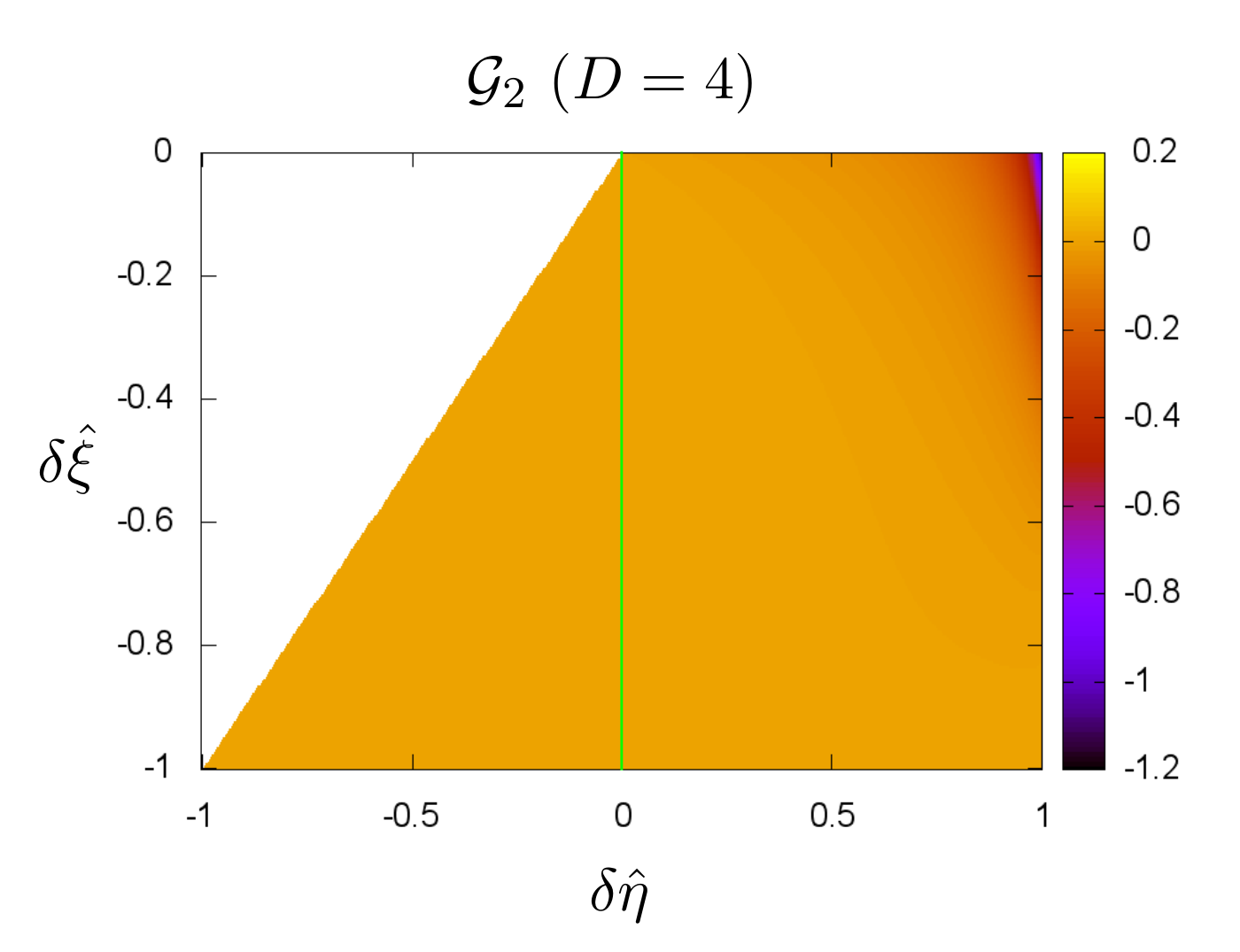} \newline
\includegraphics[scale=0.55,clip=true,trim= 0 0 0 0]{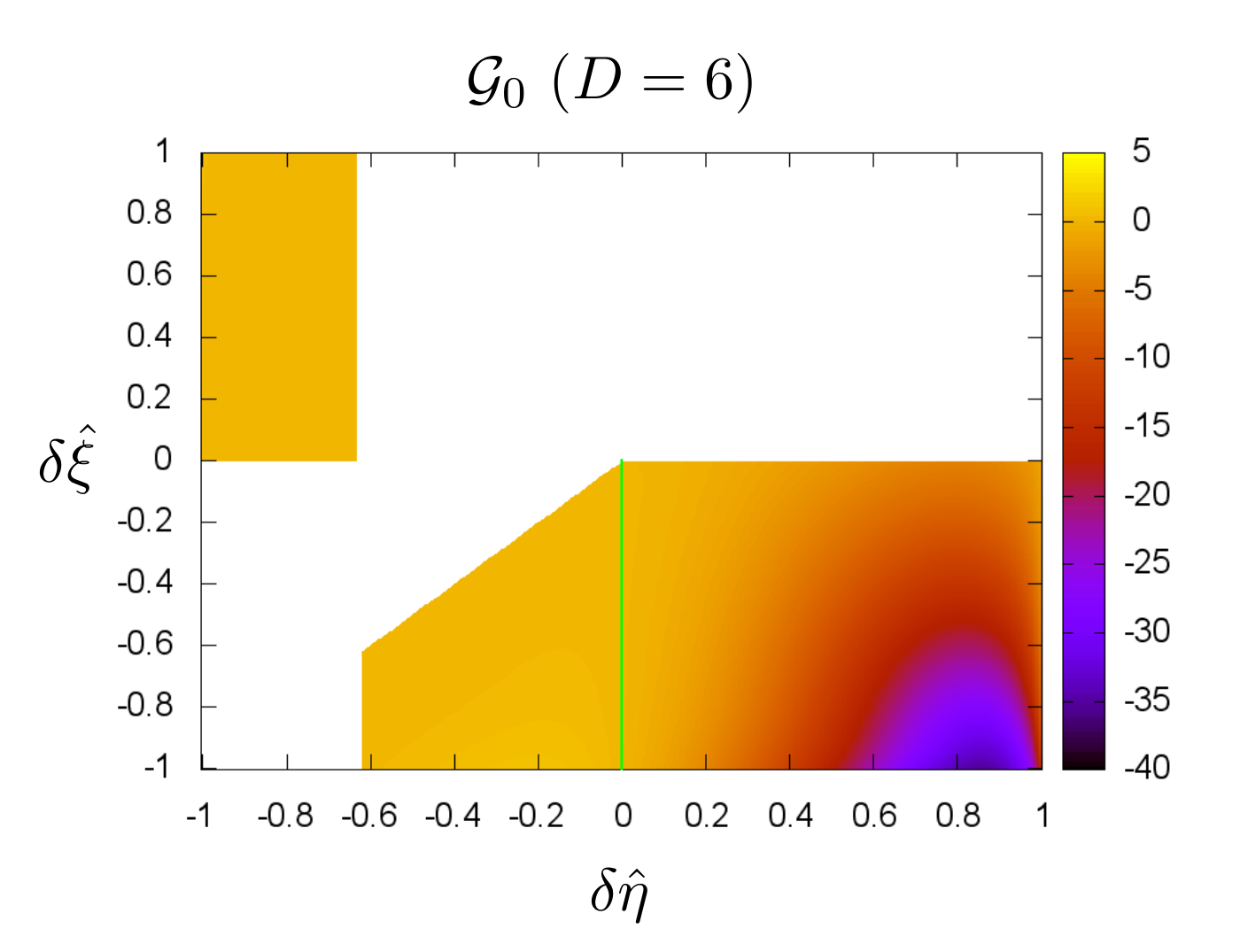} 
\includegraphics[scale=0.55,clip=true,trim= 0 0 0 0]{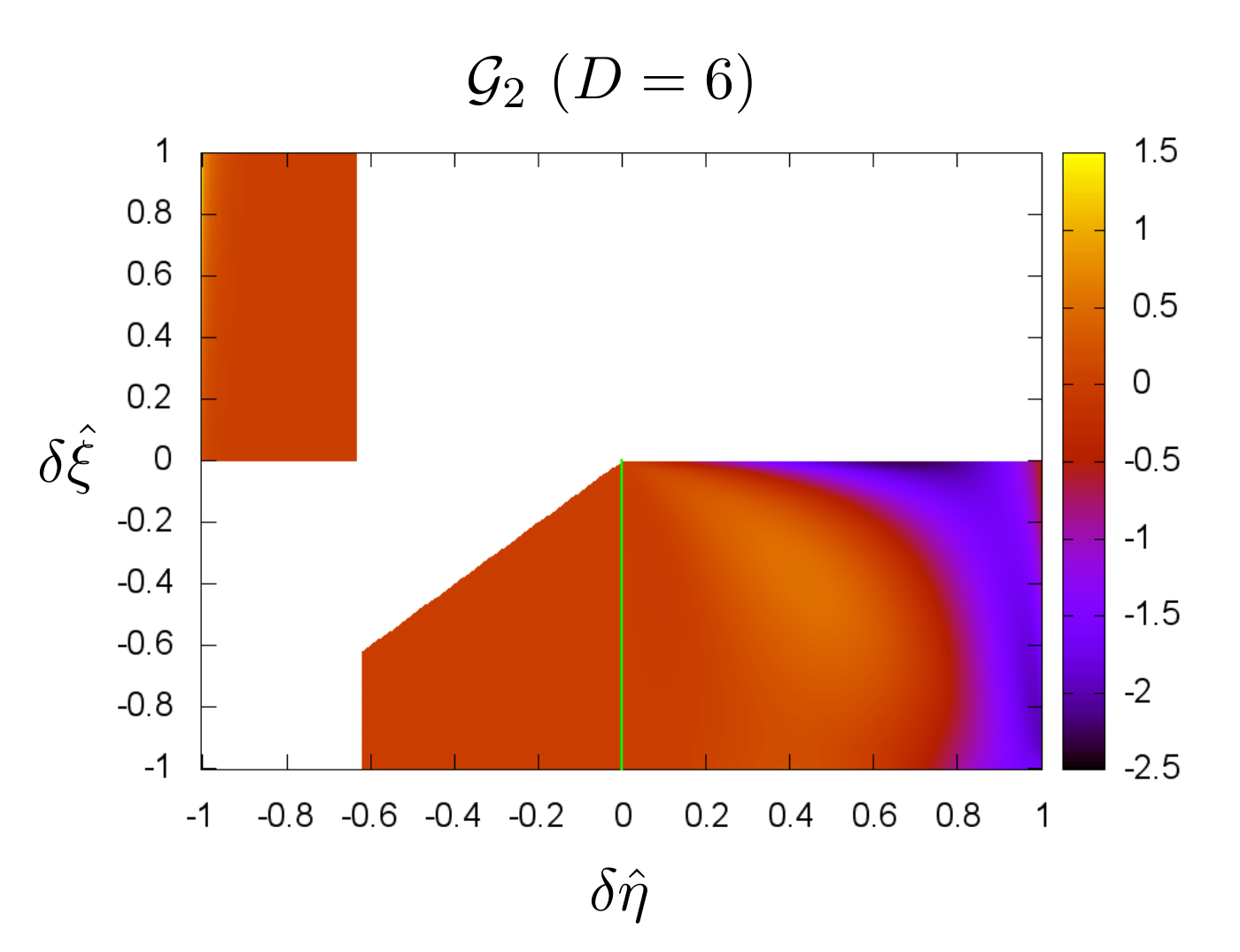} \newline
\includegraphics[scale=0.55,clip=true,trim= 0 0 0 0]{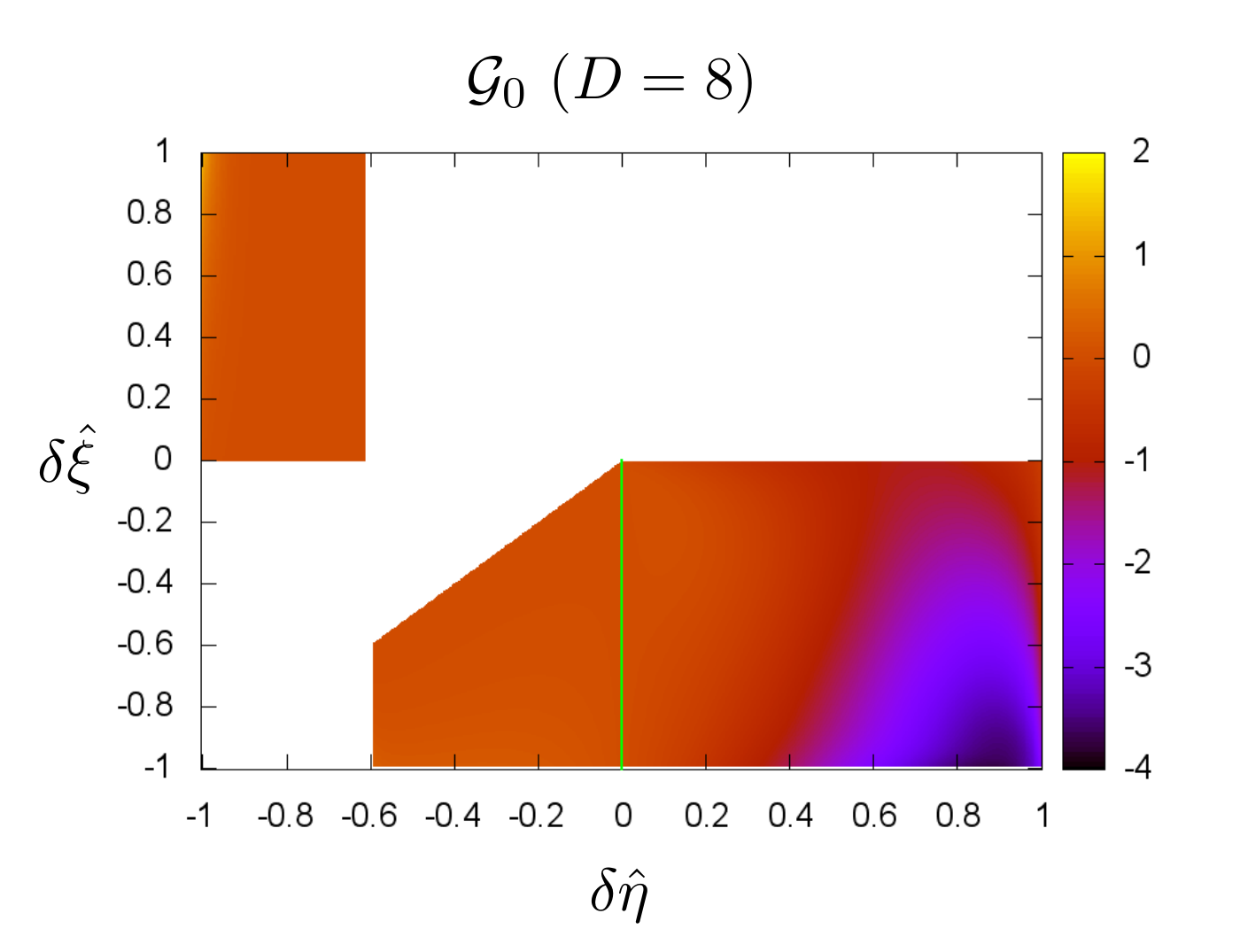} 
\includegraphics[scale=0.55,clip=true,trim= 0 0 0 0]{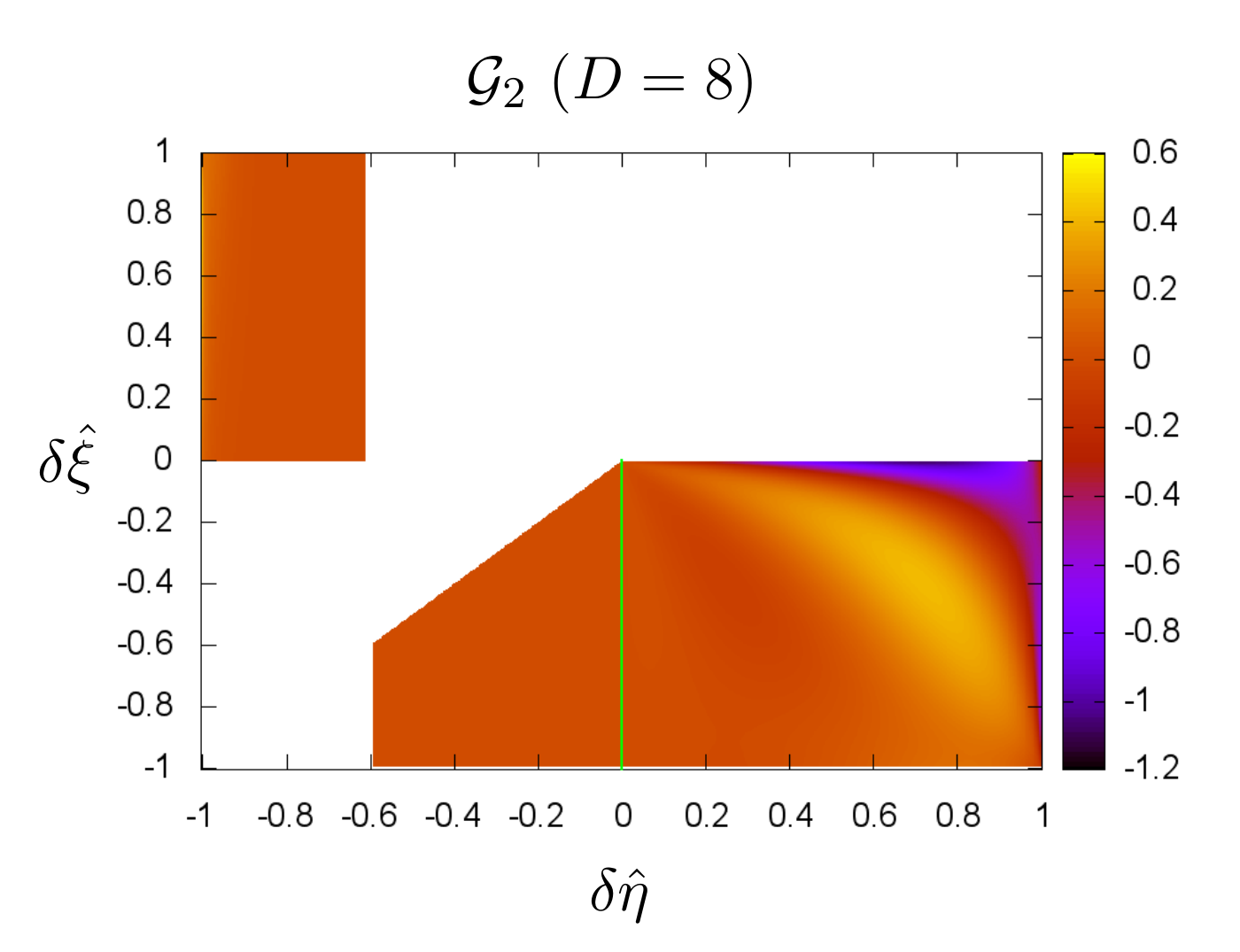} 
\caption[Second-order Green's functions $\mathcal{G}_0$ and $\mathcal{G}_2$ for even $D$.]{\label{G_even}Second-order Green's functions $\mathcal{G}_0$ (left column) and $\mathcal{G}_2$ (right column), for $D=4,6,8$, including the extra factors of Eqs.~\eqref{extra_1} and~\eqref{extra_2}. The domain is explained in Section~\ref{asymptotic_MF}. The green line shows the location of the singularity (which has been regularised).}
\end{figure}

\begin{figure}
\includegraphics[scale=0.55,clip=true,trim= 0 0 0 0]{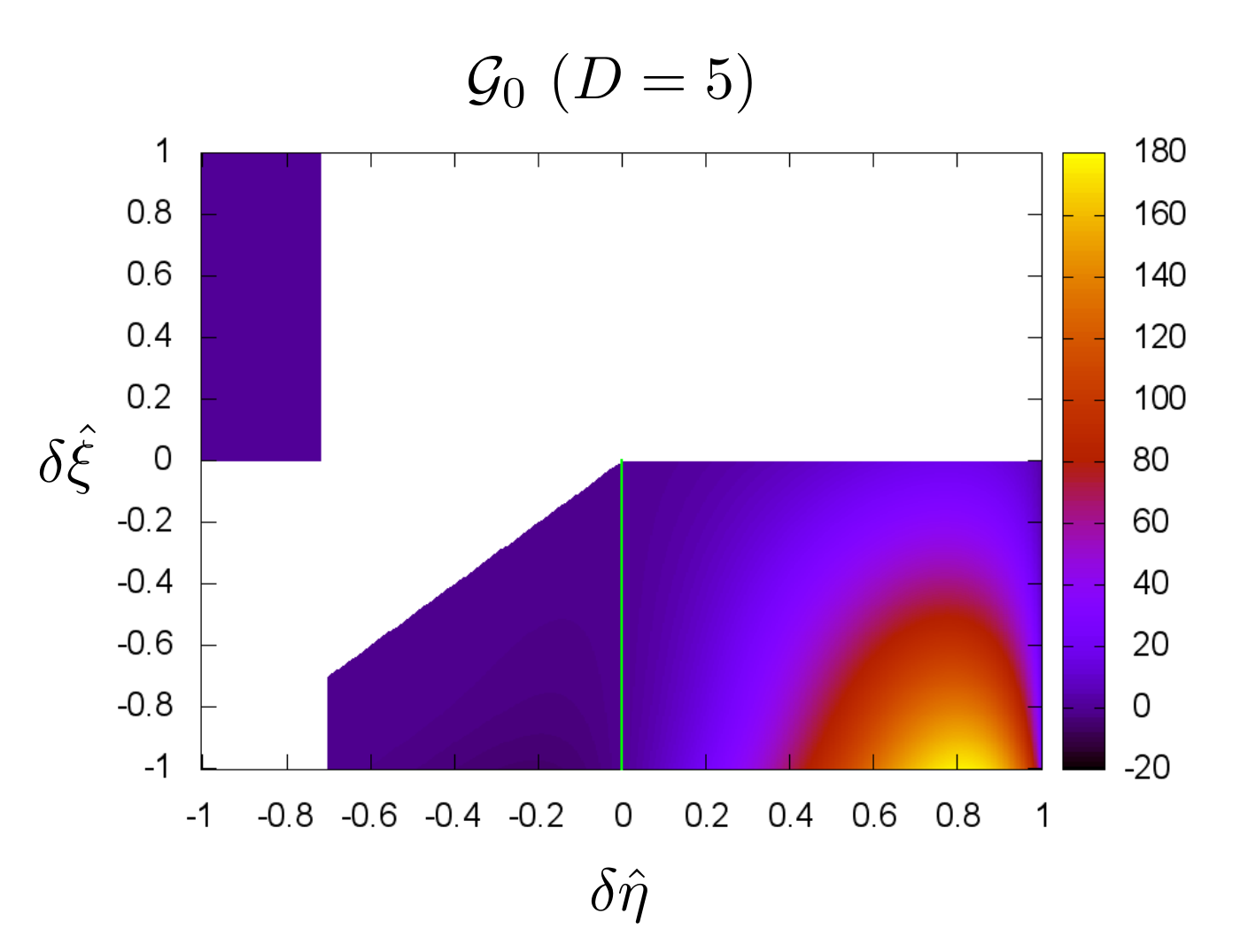}
\includegraphics[scale=0.55,clip=true,trim= 0 0 0 0]{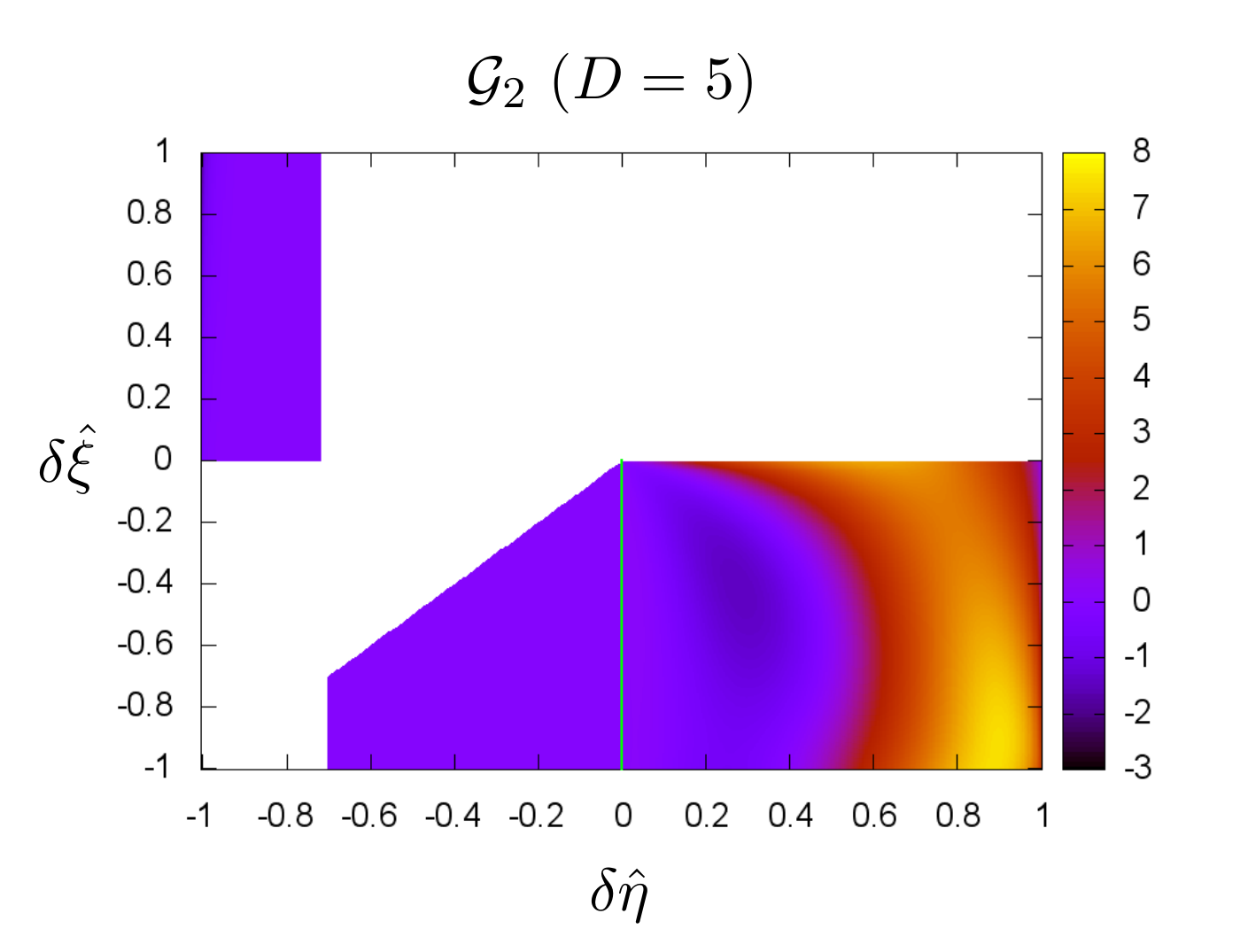} \newline
\includegraphics[scale=0.55,clip=true,trim= 0 0 0 0]{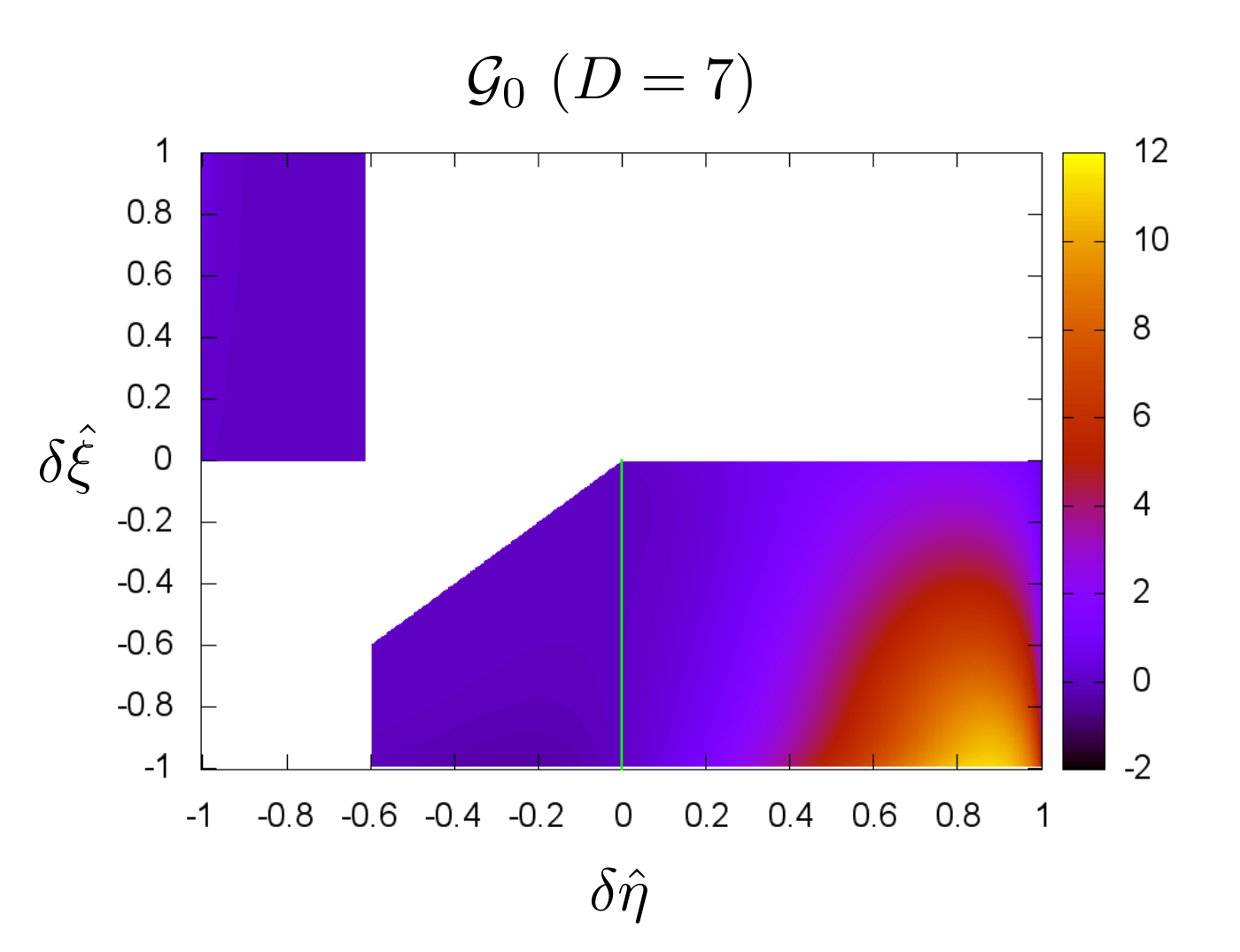} 
\includegraphics[scale=0.55,clip=true,trim= 0 0 0 0]{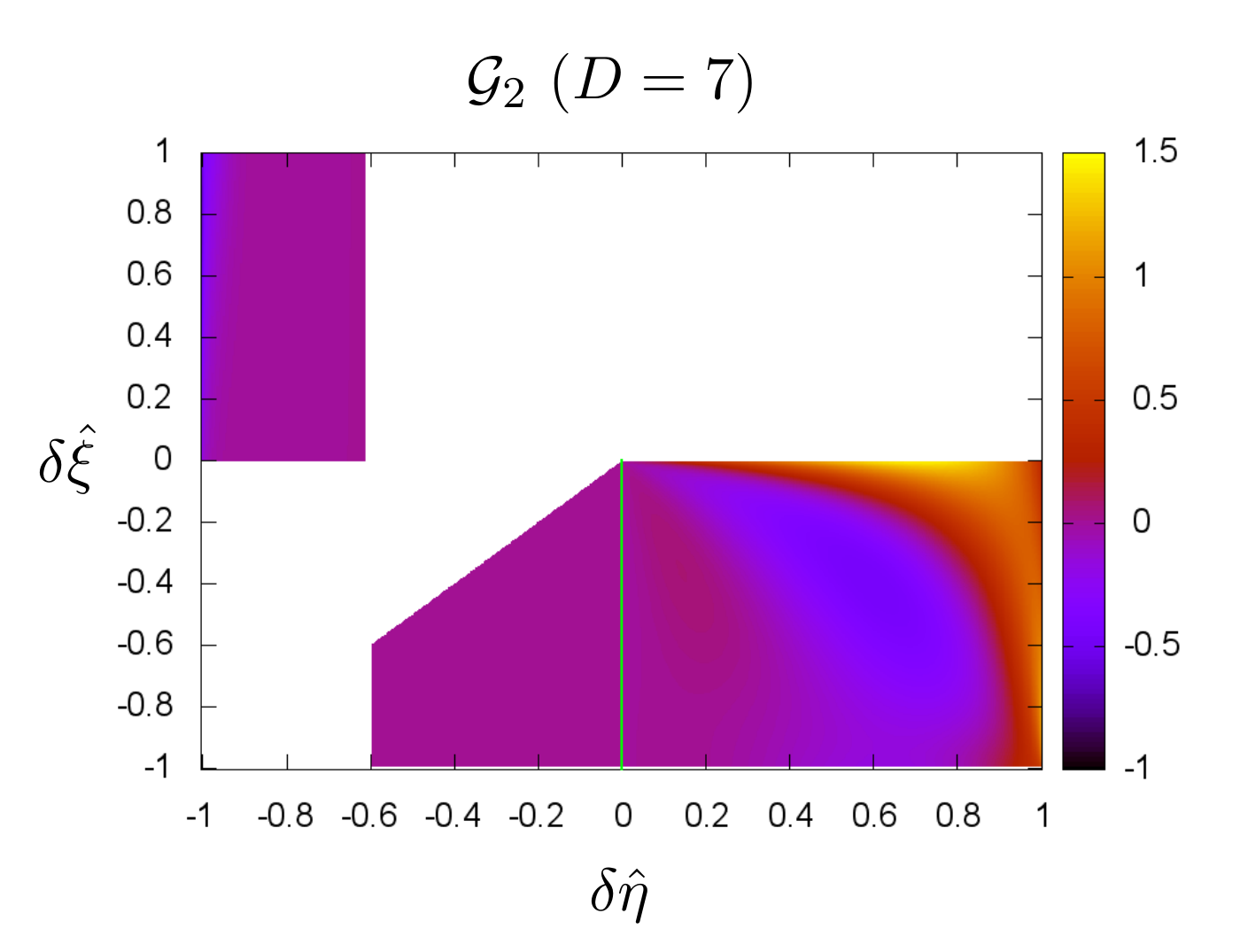} \newline
\includegraphics[scale=0.55,clip=true,trim= 0 0 0 0]{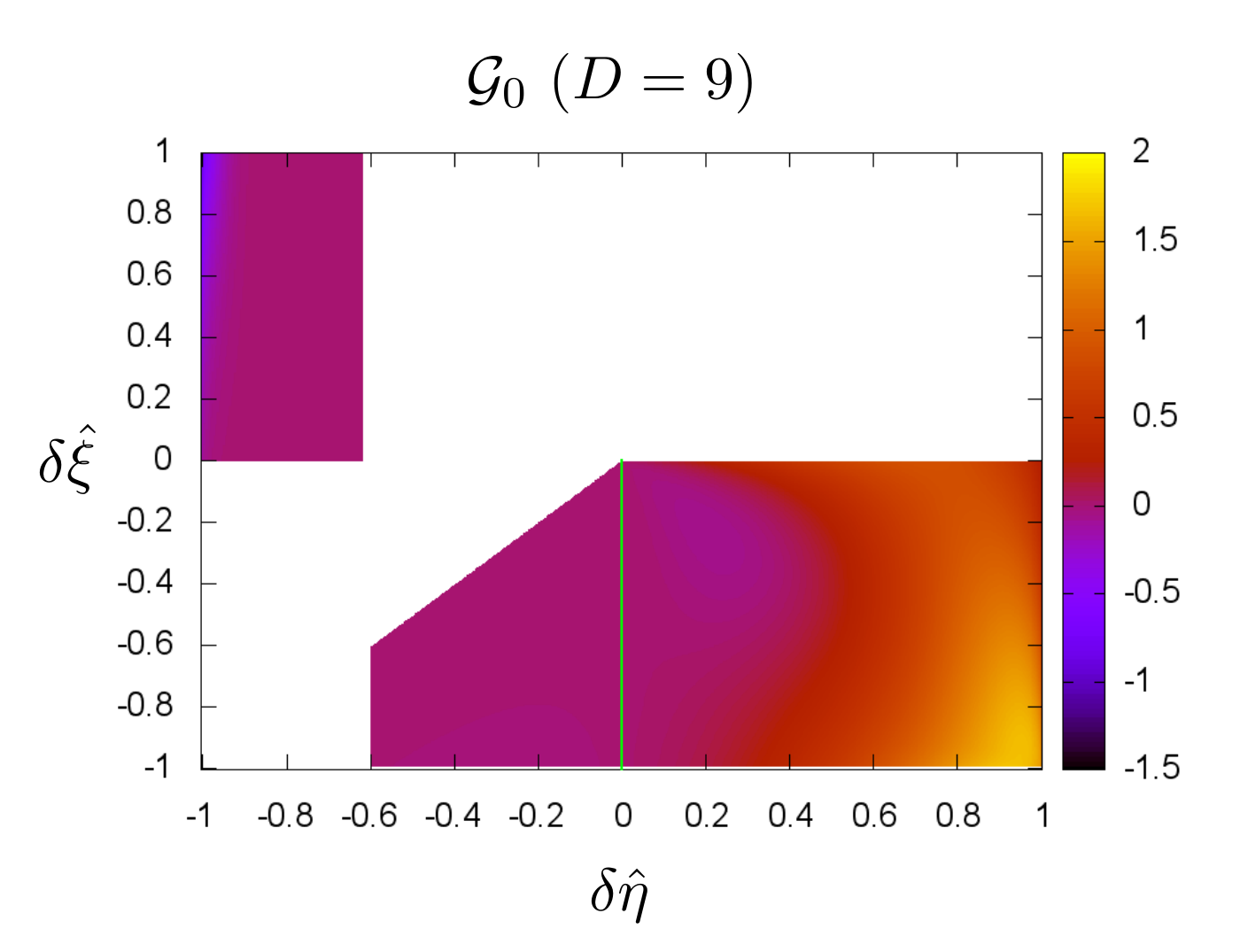} 
\includegraphics[scale=0.55,clip=true,trim= 0 0 0 0]{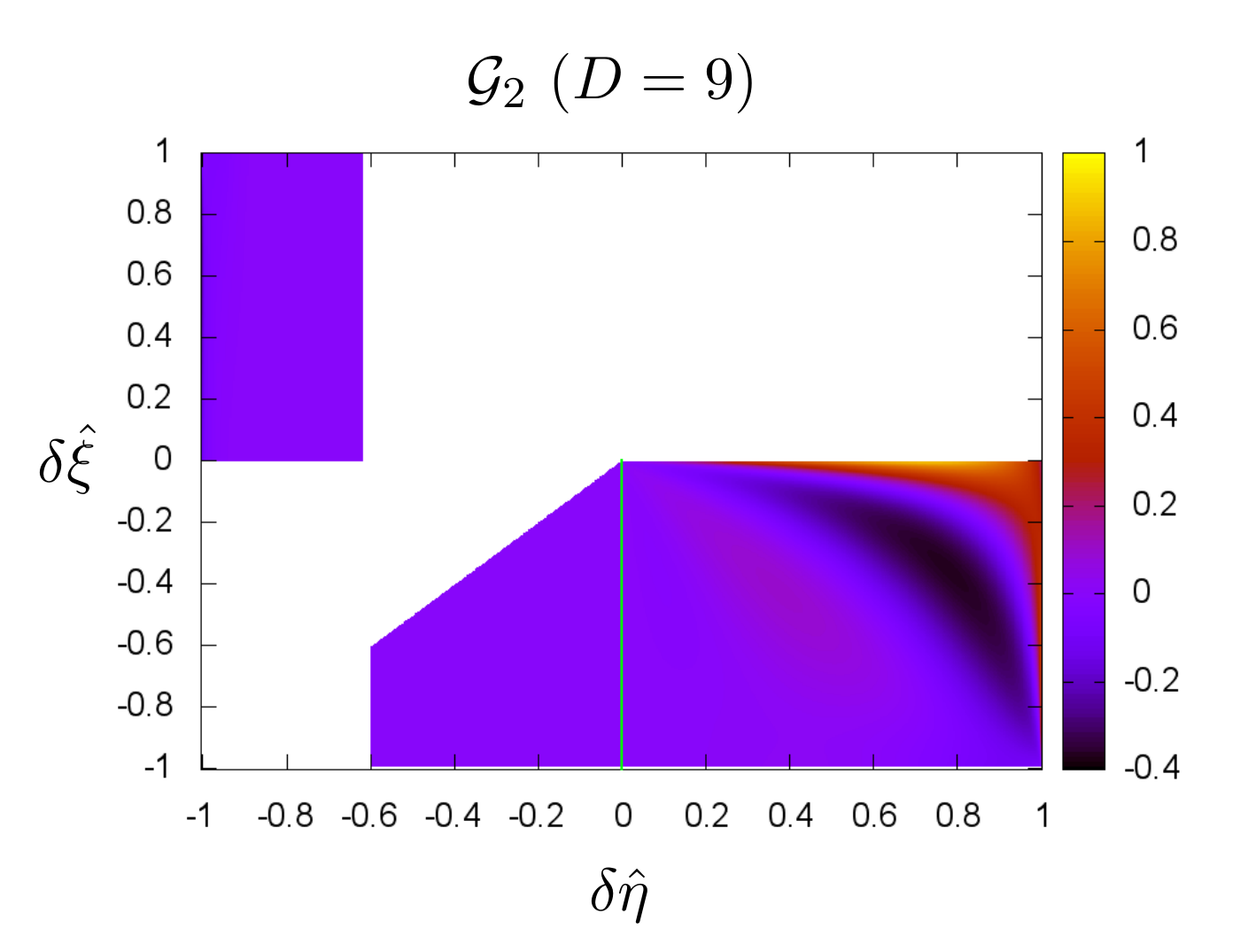}
\caption[Second-order Green's functions $\mathcal{G}_0$ and $\mathcal{G}_2$ for odd $D$.]{\label{G_odd}Second-order Green's functions $\mathcal{G}_0$ (left column) and $\mathcal{G}_2$ (right column), for $D=5,7,9$, including the extra factors of Eqs.~\eqref{extra_1} and~\eqref{extra_2}. The domain is explained in Section~\ref{asymptotic_MF}. The green line shows the location of the singularity (which has been regularised).}
\end{figure}

In Figs.~\ref{G_even} and~\ref{G_odd} we plot the numerical results for $\mathcal{G}_0$ and $\mathcal{G}_2$ for $D=4,6$ and $D=5,7$ respectively. In the horizontal axis we have $\delta\hat\eta$ and in the vertical axis $\delta\hat\xi$. The domain is as explained before at the end of Section~\ref{asymptotic_MF}. As usual, the green line depicts the location of the singularity, which has however been regularised.

In these plots we have also suppressed the late-time growth identified in Chapter~\ref{ch:surface}. In $D=4$ we chose a combination which goes as $e^{-2\xi}$ close to $(\delta\hat\xi,\delta\hat\eta)=(-1,-1)$,
\begin{equation}
\exp\left[\frac{1}{2}\left(\frac{\delta\hat\xi}{1+\delta\hat\xi}+\frac{\delta\hat\eta}{1+\delta\hat\eta}\right)\right]\,.\label{extra_1}
\end{equation}
Strictly speaking, this growth is only expected in $\mathcal{G}_0$, but we have used it for $\mathcal{G}_2$ as well to smooth out numerical errors very close to $(-1,-1)$.

For $4<D<8$ this growth manifests itself close to $\delta\eta=\xi_0$. Hence we multiply $\mathcal{G}_0$ (only) by
\begin{equation}
\left(\delta\hat\eta-\hat\xi_0\right)^{8-D}\,.\label{extra_2}
\end{equation}

We have not included these factors in the definition of $\mathcal{G}_m$ above because later when calling the function for the two-dimensional integration we shall simply divide the tabulated function by them, thus restoring the definition of Eq.~\eqref{def_G_curl}.

\section{The second-order sources}
\label{2nd_sources}
We have not yet written expressions for the second-order sources. Recall from Chapter~\ref{ch:dynamics} that the source for the wave equation obeyed by metric perturbations is an effective energy-momentum tensor built out of lower-order perturbations, Eq.~\eqref{wave_eq}. At second order, we have
\begin{equation}
\label{Tmunu1}
T^{(1)}_{\mu\nu}\equiv-h^{(1)\alpha\beta}\hspace{-1mm}\left[h^{(1)}_{\mu\alpha,\nu\beta}+h^{(1)}_{\nu\beta,\mu\alpha}-h^{(1)}_{\mu\nu,\alpha\beta}-h^{(1)}_{\alpha\beta,\mu\nu}\right]-h^{(1)\alpha}_{\phantom{(1)}\nu,\beta}\hspace{-1mm}\left[h^{(1)\beta}_{\phantom{(1)}\mu,\alpha}-h^{(1)\beta}_{\mu\alpha,}\right]+\tfrac{1}{2}h^{(1)\alpha\beta}_{\phantom{(1)\alpha},\mu}h^{(1)}_{\alpha\beta,\nu} \; .
\end{equation}

Using Eqs.~\eqref{gen_perts}, we can then read the scalar sources for the transverse metric functions $E^{(2)}$ and $H^{(2)}$, as defined in Eq.~\eqref{TE_TH},
\begin{eqnarray}
T_E^{(1)}(u,v,\rho)& =& \tfrac{1}{D-2}\left[(B_{,v})^2-\tfrac{2(D-3)}{\rho}B E_{,v}+2(D-3)B_{,\rho}E_{,v}-2BE_{,\rho v}\right]+\nonumber\\&& +\tfrac{4(D-3)}{\rho^2}E^2 -\tfrac{2(D-3)^2}{(D-2)\rho}E E_{,\rho}-\tfrac{2(D-3)}{D-2}E E_{,\rho\rho}+A E_{,vv}+\nonumber\\&& +(D-4)\left[2E_{,u}E_{,v}-\tfrac{D-1}{2(D-2)}(E_{,\rho})^2\right]\,,\\
T_H^{(1)}(u,v,\rho) &=&-\tfrac{2(D-3)}{D-2}\left[\tfrac{D-3}{\rho}BE_{,v}+B_{,\rho}E_{,v}+BE_{,\rho v}\right] +\nonumber\\&&+(D-3)\left[\tfrac{4(D-3)}{\rho^2}E^2+\tfrac{D }{2(D-2)}(E_{,\rho})^2-\tfrac{D^2-9D+16}{(D-2)\rho}E E_{,\rho}-\tfrac{D-4}{D-2}E E_{,\rho \rho}\right] \nonumber\\&& -\tfrac{1}{D-2}(B_{,v})^2-2(D-3)E_{,u}E_{,v}\,.
\end{eqnarray}

Finally, their two-dimensional form in $(p,q)$ coordinates, $t^{(1)}_E(p,q)$ and $t^{(1)}_H(p,q)$, is obtained by applying the definitions in Eqs.~\eqref{f_pq} and~\eqref{def_s}, and noting that the derivatives transform as in Eq.~\eqref{derivatives_transformation}.

The computation of these sources starts with the reduced functions $a^{(1)}(p,q)$, $b^{(1)}(p,q)$, $e^{(1)}(p,q)$ and their derivatives, all of which are given by one-dimensional integrals as in Eq.~\eqref{f_surface}. The initial data, i.e. the numbers $a_0^{(1)}$, $b_0^{(1)}$, $e_0^{(1)}$, are obtained from Eqs.~\eqref{initial_data} and from the transformation to de Donder gauge in Appendix~\ref{app:gauge} (which only affects $a_0^{(1)}$). Then, we build $t^{(1)}_E(p,q)$ and $t^{(1)}_H(p,q)$.

As for the Green's functions, we need to ensure that the tabulated functions are regular and smooth. We choose to include also the factor $J^2(\hat\xi',\hat\eta')$, so we must consider its behaviour as well.

Regarding the sources themselves, we found, in accordance with the asymptotical analysis of Section~\ref{asymptotics} (in particular Eq.~\eqref{surface_scri+}) that when $\eta\rightarrow\infty$ both grow with
\begin{equation}
\left(1-\hat\eta\right)^{-2-\frac{3}{2}(D-4)}\,.\label{source_div_scri+}
\end{equation}

At the axis, $\hat\eta=\hat\xi$, we found numerically that both sources go to zero. 

At the red line ($\xi=0$), we know from previous experience that the first order metric functions are square integrable, so the singularity (which is only present in even $D$) is regularised if we multiply by $\hat\xi$.

Regarding $J^2(\hat\xi',\hat\eta')$, from Eq.~\eqref{jacobian} we expect some singular behaviour when $\hat\eta'\rightarrow1$ or $\hat\xi'\rightarrow\pm1$. In the first case, we have
\begin{equation}
{P'}^{\frac{1}{D-3}}\sim\eta'\sim(1-\hat\eta')^{-\frac{1}{2}}\,,\qquad Q'-(D-2)(D-4)\sim{\eta'}^{-1}\sim(1-\hat\eta')^{\frac{1}{2}}\,,
\end{equation}
so we see that, at leading order, $J^2$ goes to zero as
\begin{equation}
(1-\hat\eta')^{1+\frac{7}{4}(D-4)}\,.
\end{equation}

Combining with the divergent behaviour of the source, Eq.~\eqref{source_div_scri+} above, and given that the Green's function is finite, we conclude that the full integrand goes with
\begin{equation}
(1-\hat\eta')^{-1+\frac{1}{4}(D-4)}\,.
\end{equation}
In $D=4$ this gives rise to the logarithmic terms discussed by Payne~\cite{Payne}, which we shall subtract and discuss in the next section. For the moment, we just multiply by $(1-\hat\eta')$ to make the tabulated function regular for all $D$.

In the second case, we found that multiplying by $\left(1-\hat{\xi'}^2\right)^\frac{3}{2}$ makes the result regular.

So, putting everything together, we shall tabulate numerically the functions
\begin{equation}
\mathcal{S}_{m}(\hat\xi',\hat\eta')\equiv J^2(\hat\xi',\hat\eta')t^{(1)}_{E,H}(p',q')\times\hat\xi'\left(1-\hat{\xi'}^2\right)^\frac{3}{2}(1-\hat\eta')(\hat\eta'-\hat\xi')^{-\gamma_1}(1+\hat\xi')^{-\gamma_2}\,.
\end{equation}

\begin{figure}
\hspace*{\fill}%
 \includegraphics[scale=0.45,clip=true,trim= 0 0 0 0]{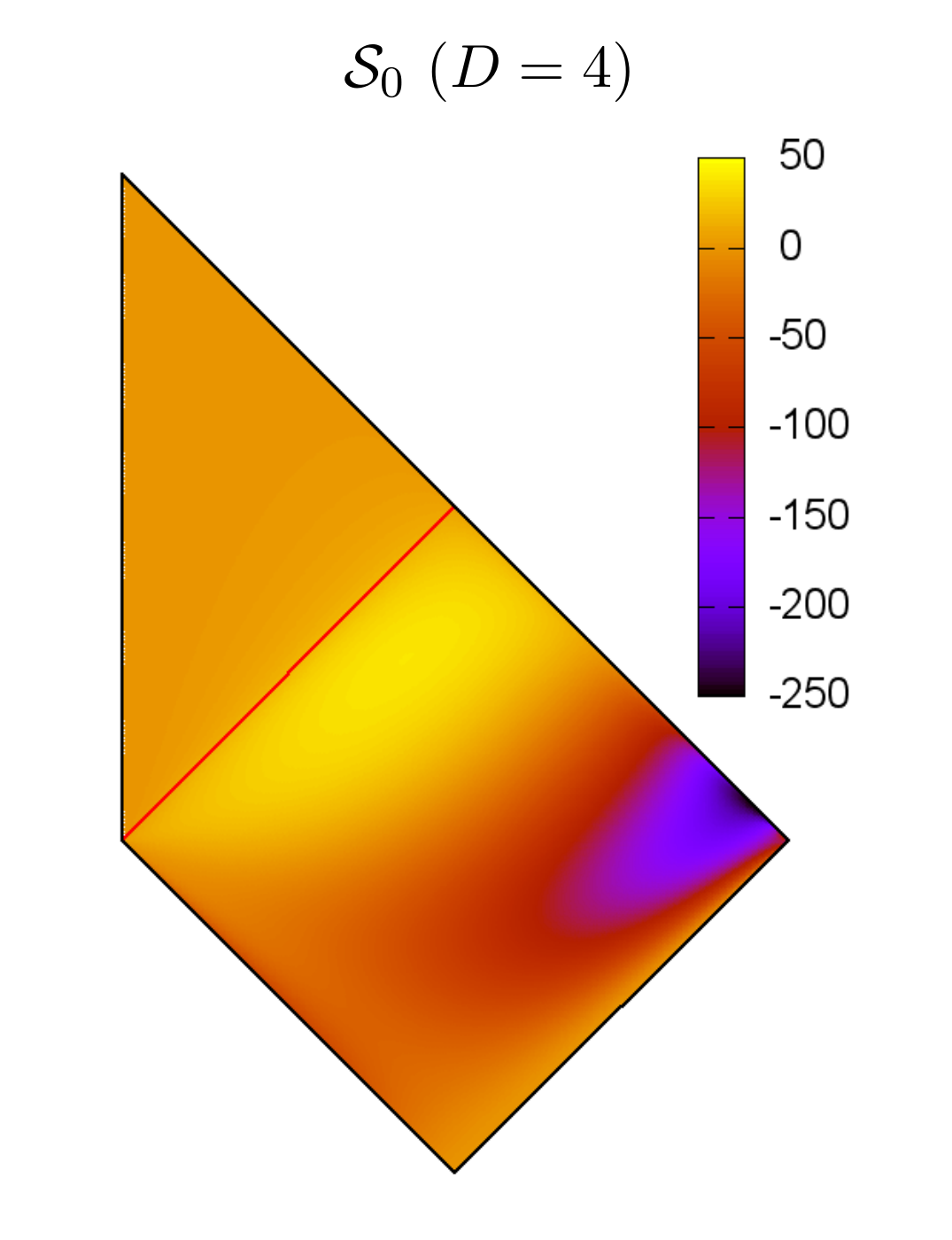}\hfill%
 \includegraphics[scale=0.45,clip=true,trim= 0 0 0 0]{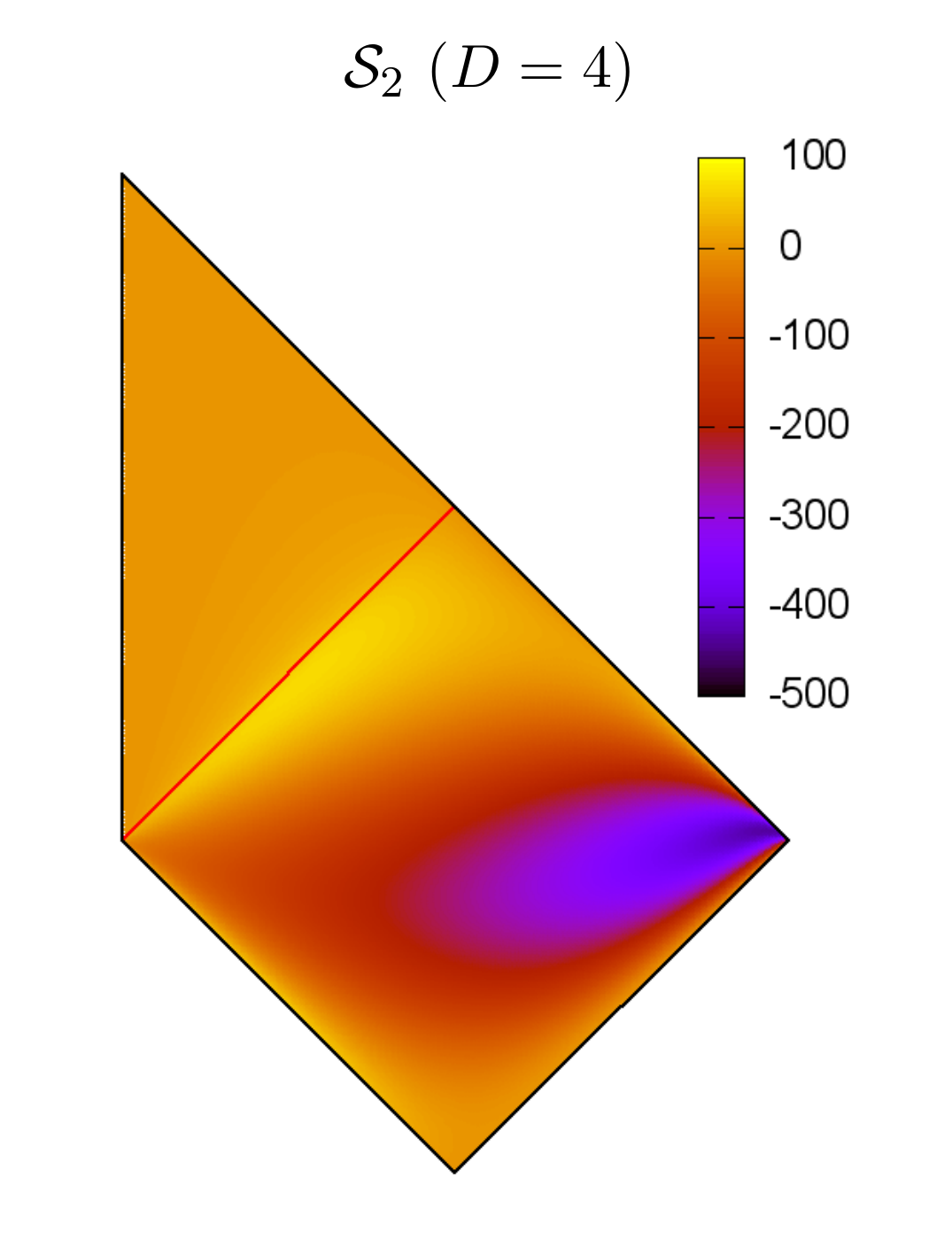}%
  \hspace*{\fill}\vspace*{\fill}\newline\vspace*{\fill}
  \hspace*{\fill}%
 \includegraphics[scale=0.45,clip=true,trim= 0 0 0 0]{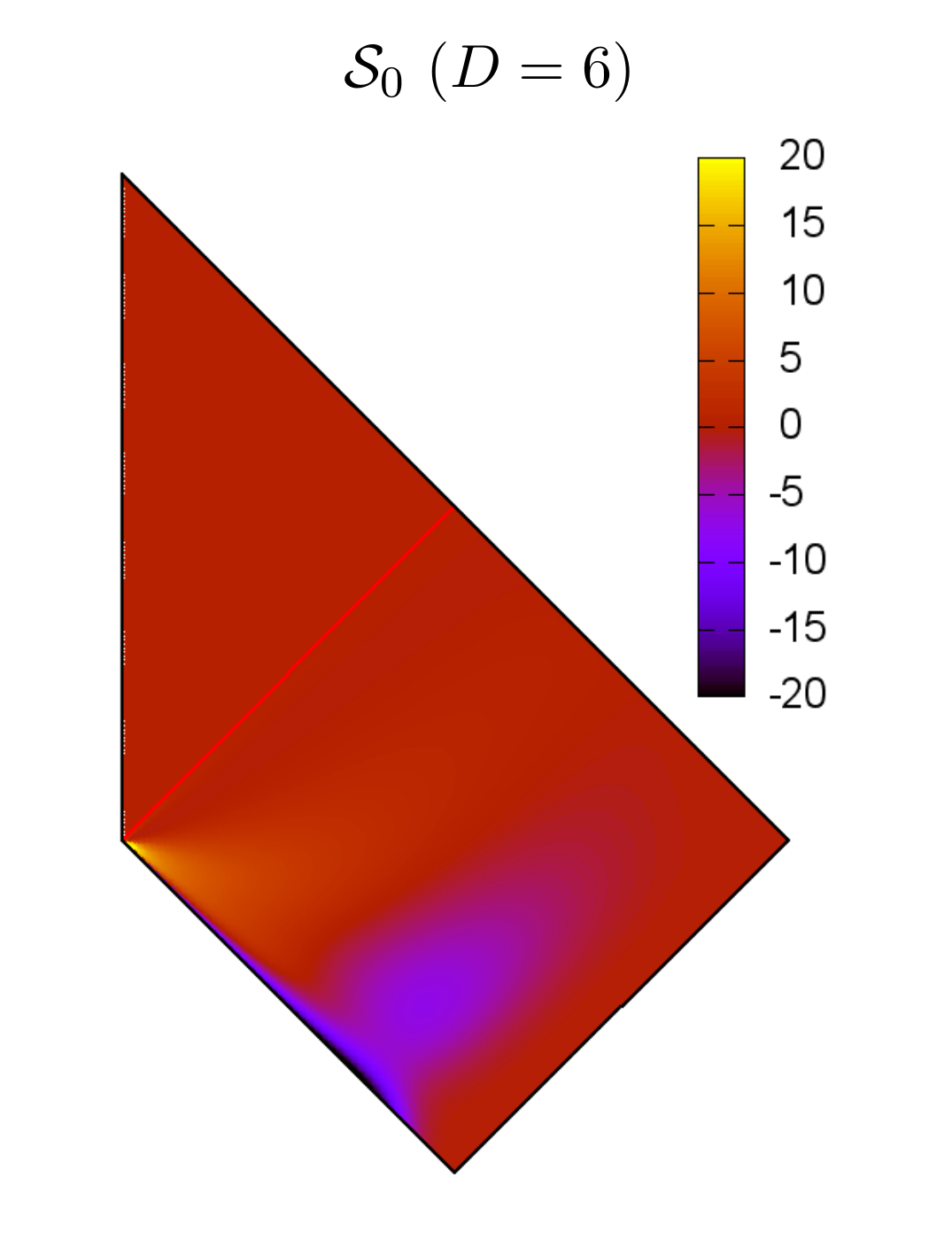}\hfill%
 \includegraphics[scale=0.45,clip=true,trim= 0 0 0 0]{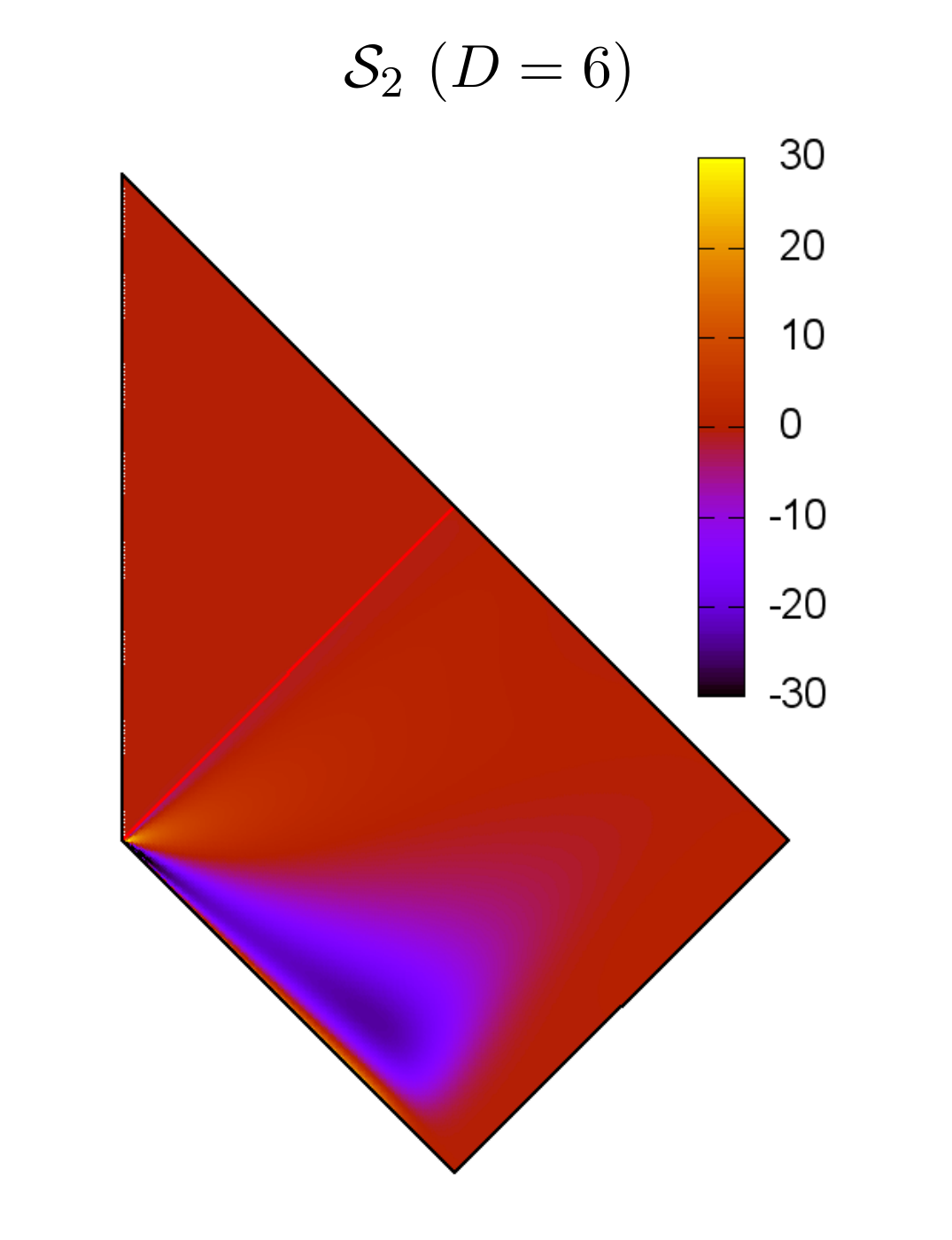}%
  \hspace*{\fill}\vspace*{\fill}\newline\vspace*{\fill}
    \hspace*{\fill}%
 \includegraphics[scale=0.45,clip=true,trim= 0 0 0 0]{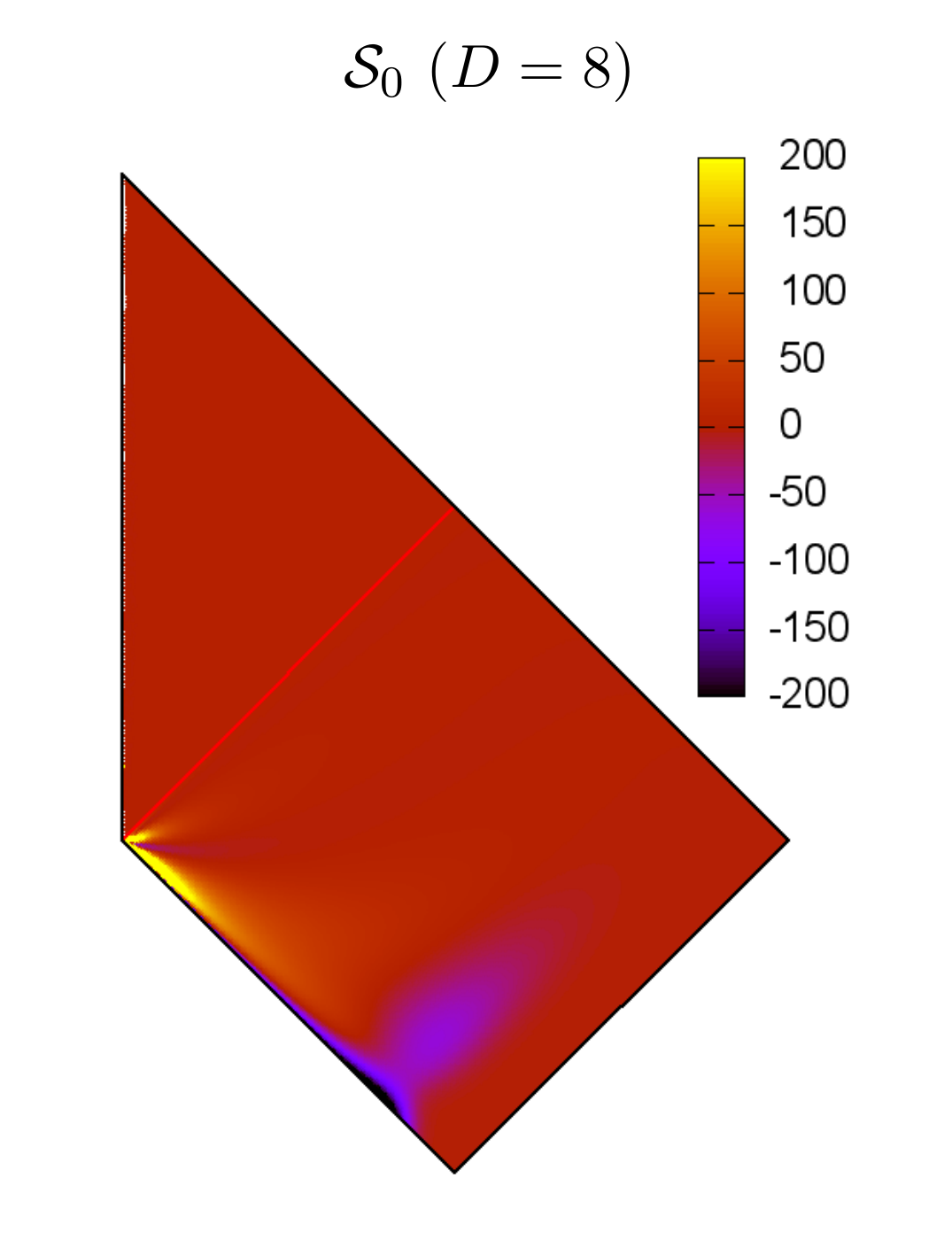}\hfill%
 \includegraphics[scale=0.45,clip=true,trim= 0 0 0 0]{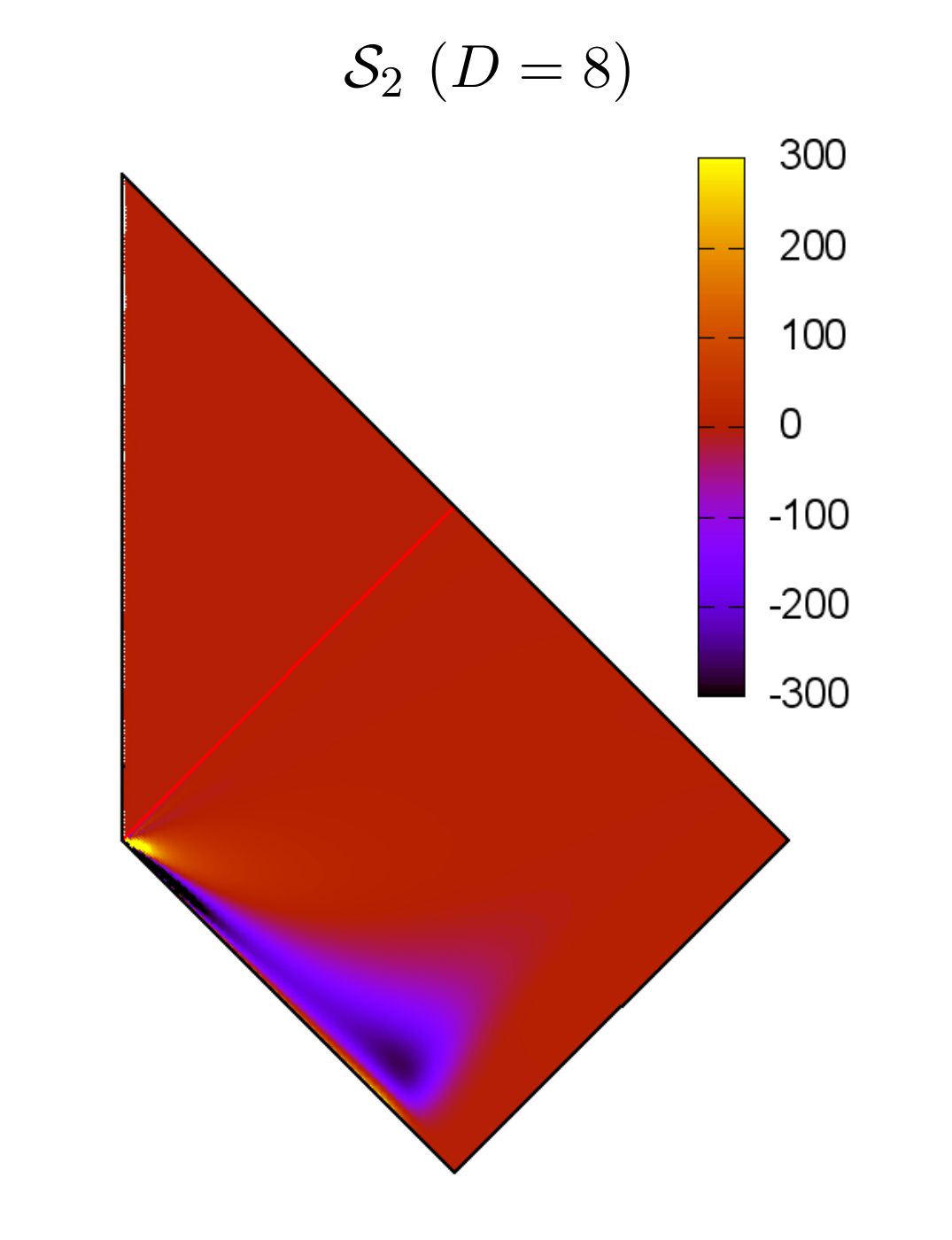}%
  \hspace*{\fill}%
\caption[Second-order sources $\mathcal{S}_0$ and $\mathcal{S}_2$ for even $D$.]{\label{S_even}Second-order sources $\mathcal{S}_0$ (left column) and $\mathcal{S}_2$ (right column), for $D=4,6,8$. The axes match the conformal diagram of Fig.~\ref{penrose}. The red line shows the location of the singularity (which has been regularised).}
\end{figure}

\begin{figure}
\hspace*{\fill}%
 \includegraphics[scale=0.45,clip=true,trim= 0 0 0 0]{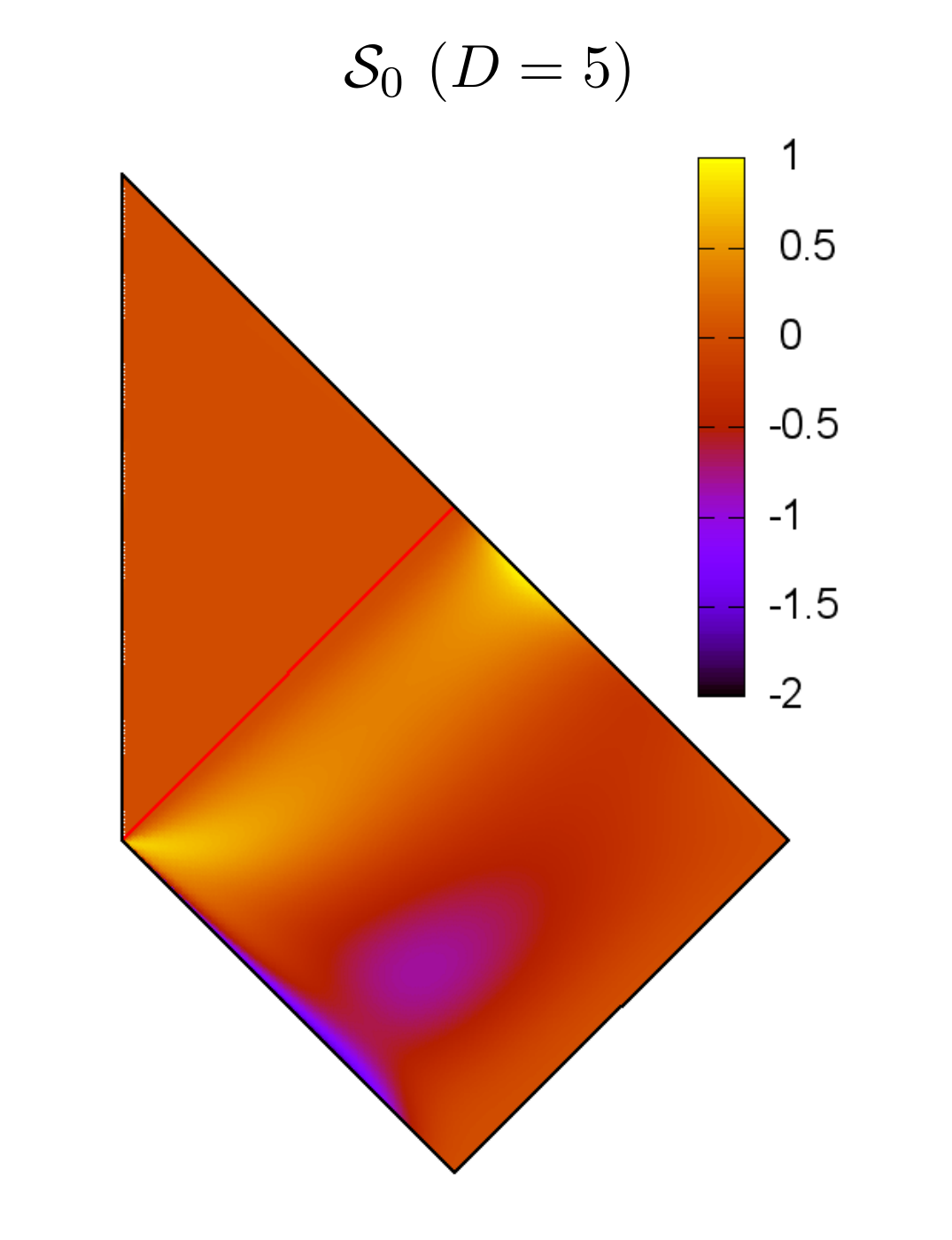}\hfill%
 \includegraphics[scale=0.45,clip=true,trim= 0 0 0 0]{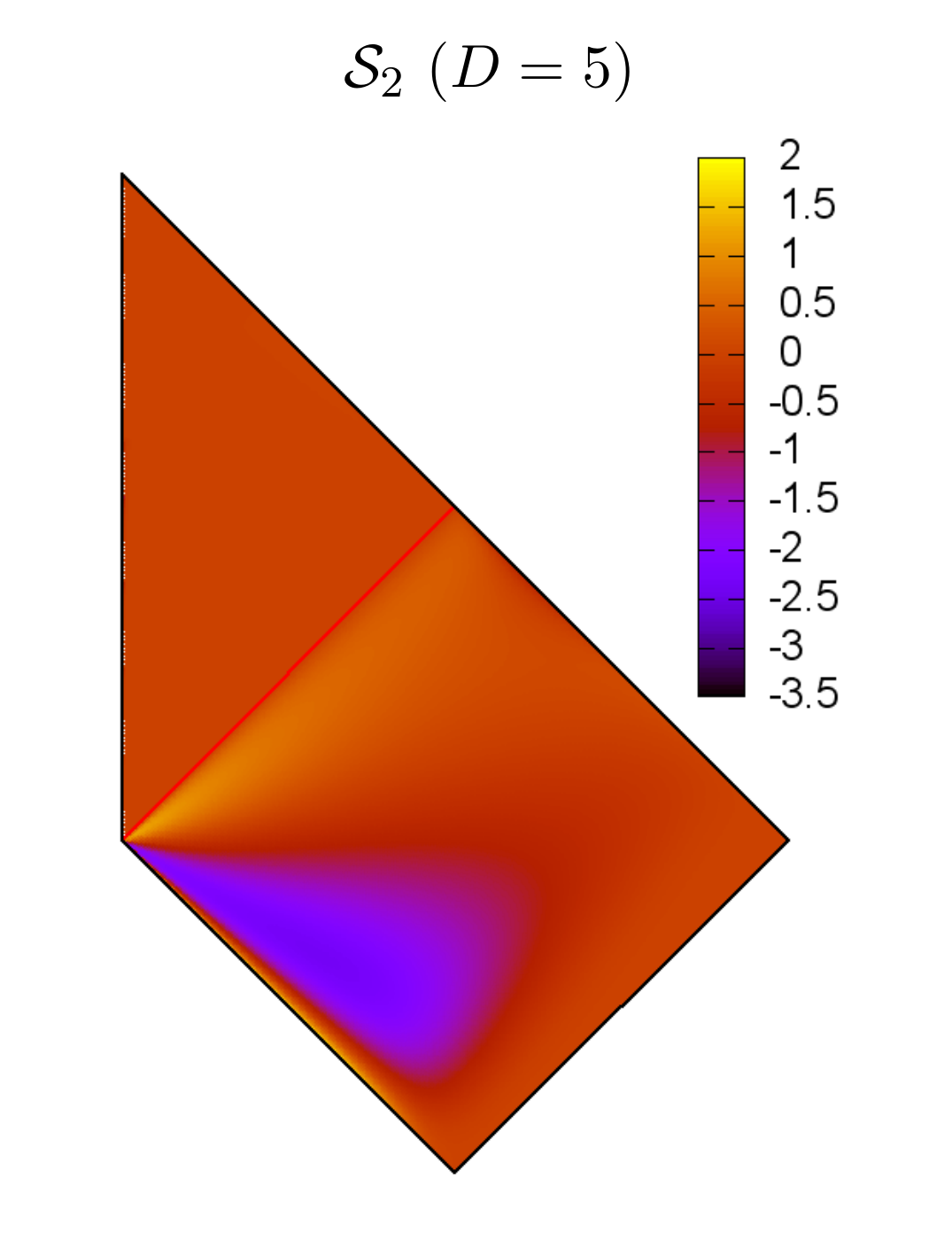}%
  \hspace*{\fill}\vspace*{\fill}\newline\vspace*{\fill}
  \hspace*{\fill}%
 \includegraphics[scale=0.45,clip=true,trim= 0 0 0 0]{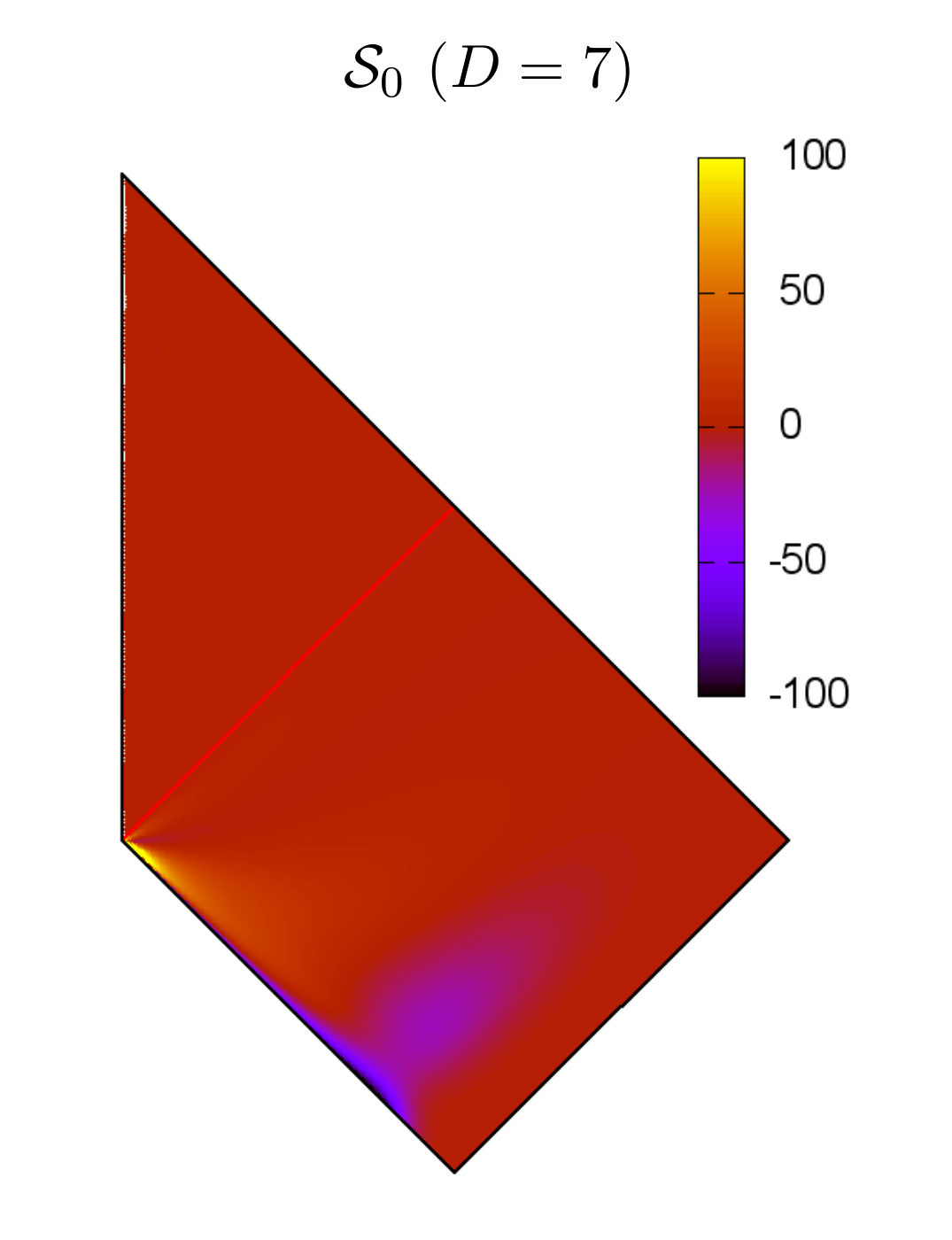}\hfill%
 \includegraphics[scale=0.45,clip=true,trim= 0 0 0 0]{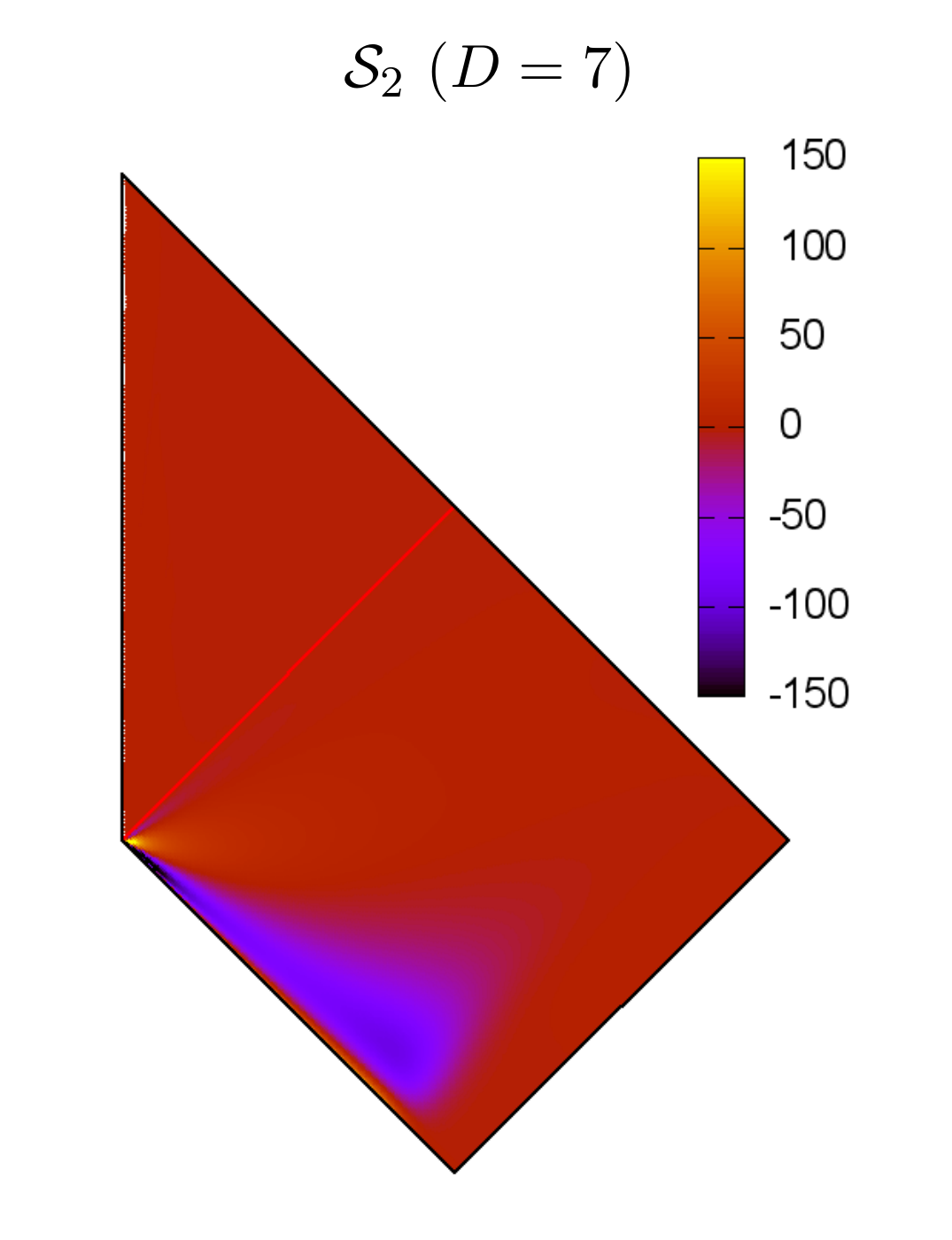}%
  \hspace*{\fill}\vspace*{\fill}\newline\vspace*{\fill}
    \hspace*{\fill}%
 \includegraphics[scale=0.45,clip=true,trim= 0 0 0 0]{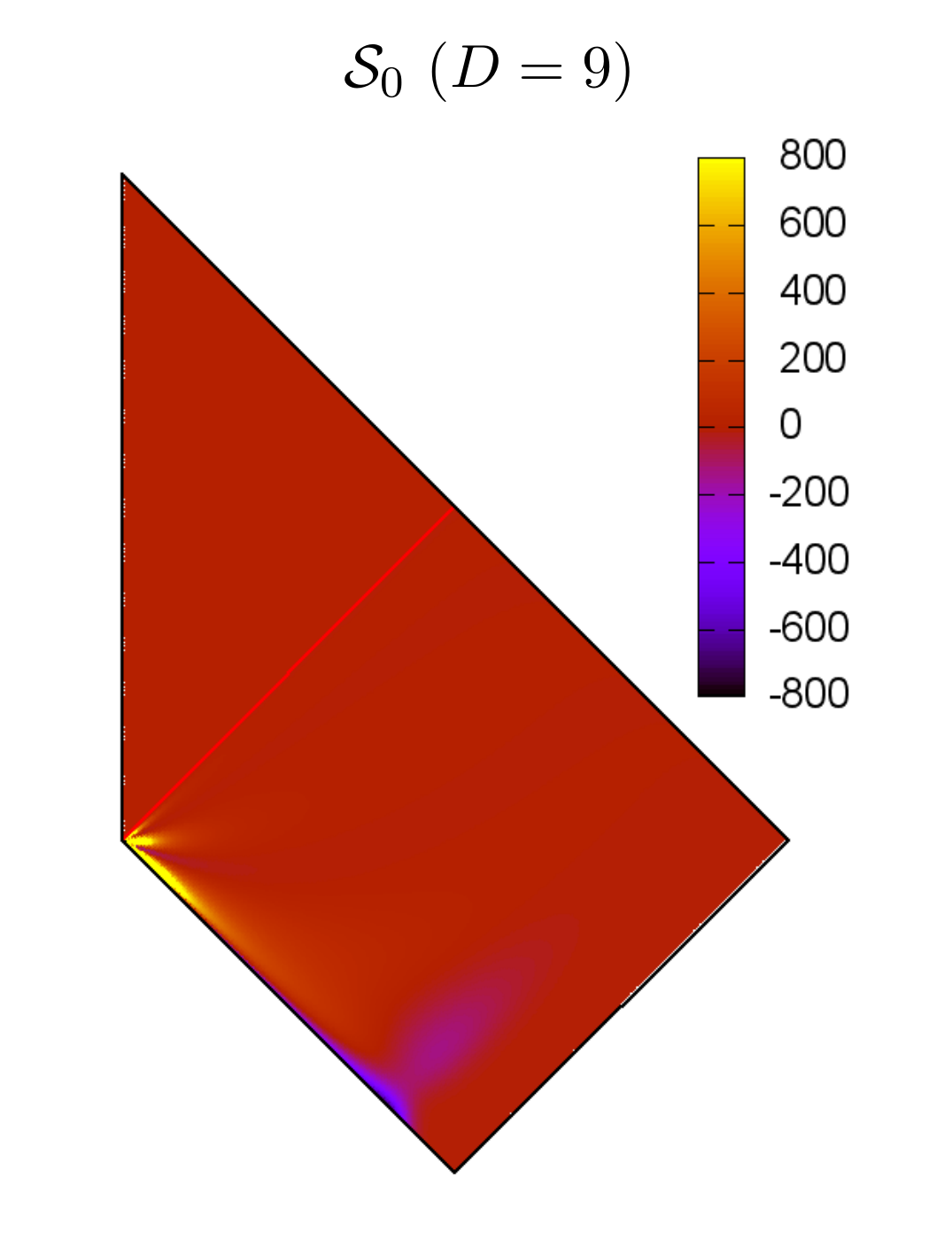}\hfill%
 \includegraphics[scale=0.45,clip=true,trim= 0 0 0 0]{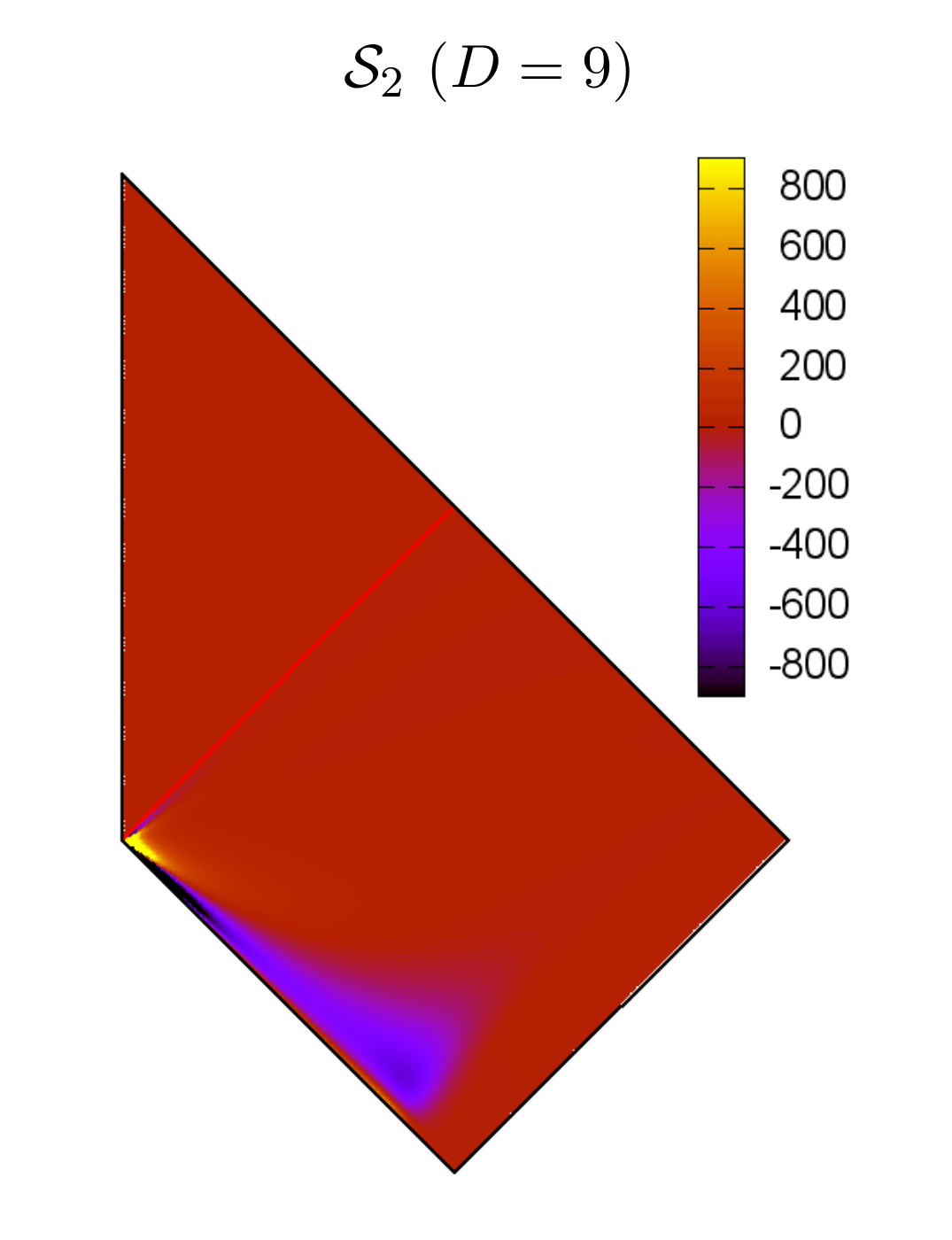}%
  \hspace*{\fill}%
\caption[Second-order sources $\mathcal{S}_0$ and $\mathcal{S}_2$ for odd $D$.]{\label{S_odd}Second-order sources $\mathcal{S}_0$ (left column) and $\mathcal{S}_2$ (right column), for $D=5,7,9$. The axes match the conformal diagram of Fig.~\ref{penrose}. The red line shows the location of the singularity (which has been regularised).}
\end{figure}

In Figs.~\ref{S_even} and~\ref{S_odd} we plot $\mathcal{S}_0$ and $\mathcal{S}_2$, for $D=4,6,8$ and $D=5,7,9$ respectively, on the conformal diagram of Fig.~\ref{penrose}. Note that the original singularity of the source at $\xi'=0$ has been regularised, so the red line is shown for reference only.

The apparent growth near the intersection of the axis with the red line is due to the extra factor that came from the Green's function. However, in the next section we will see that their product is smooth and finite.

\section{The two-dimensional integration}
\label{2nd_wf}
Comparing $\mathcal{G}_m\times\mathcal{S}_{m}$ with the original integral, we see that now we have
\begin{equation}
\hat f^{(2)}(\hat\xi)=\int_{-1}^{\hat\xi} d\hat\xi' \int^{1}_{\max \{0,\hat\xi'\}} d\hat\eta'\, \mathcal{J}\mathcal{G}_m(\delta\xi,\delta\eta)\mathcal{S}_{m}(\hat\xi',\hat\eta')\,,
\end{equation}
where
\begin{equation}
\mathcal{J}\equiv\frac{1}{1-\hat\eta'}\frac{1}{\left(1-\hat{\xi'}^2\right)^\frac{3}{2}}\frac{1}{\hat\xi'}\frac{1}{\delta\hat\eta'}\left(\frac{\eta'-\xi'}{\delta\hat\eta-\delta\hat\xi}\right)^{\gamma_1}\left(\frac{1+\hat\xi'}{1+\delta\hat\xi}\right)^{\gamma_2}\,,
\end{equation}
contains all the singularities of the original integrand: either they are integrable, or the principal value must be taken in accordance with the finite part prescription. The only major issue left is the logarithmic term in $D=4$. As seen in the previous section, the limit
\begin{equation}
\lim_{\hat\eta'\rightarrow1}(1-\hat\eta') \mathcal{J}\mathcal{G}_m(\delta\xi,\delta\eta)\mathcal{S}_{m}(\hat\xi',\hat\eta')\equiv\mathcal{L}(\hat\xi')
\end{equation}
is zero in $D>4$ but finite in $D=4$. This gives rise to logarithmically divergent terms when $\eta'\rightarrow1$. Payne~\cite{Payne} showed that these can be removed by an explicit transformation to Bondi gauge, so we can simply subtract them, i.e. in $D=4$ we write
\begin{eqnarray}
\hat f^{(2)}(\hat\xi)&=&\lim_{\hat\eta\rightarrow1}\left\{\int_{-1}^{\hat\xi} d\hat\xi' \int^{\hat\eta}_{\max \{0,\hat\xi'\}} \frac{d\hat\eta'}{1-\hat\eta'}\, \left[(1-\hat\eta')\mathcal{J}\mathcal{G}_m(\delta\xi,\delta\eta)\mathcal{S}_{m}(\hat\xi',\hat\eta')-\mathcal{L}(\hat\xi')\right]\right.\nonumber\\
&&\phantom{\lim}\left.+\int_0^{\hat\xi}d\hat\xi'\,\mathcal{L}(\hat\xi')\log(1-\hat\xi')-\log(1-\hat\eta)\int_{-1}^{\hat\xi}d\hat\xi'\,\mathcal{L}(\hat\xi')\right\}\,,
\end{eqnarray}
and we discard the last term.

\begin{figure}
\hspace*{\fill}%
 \includegraphics[scale=0.45,clip=true,trim= 0 0 0 0]{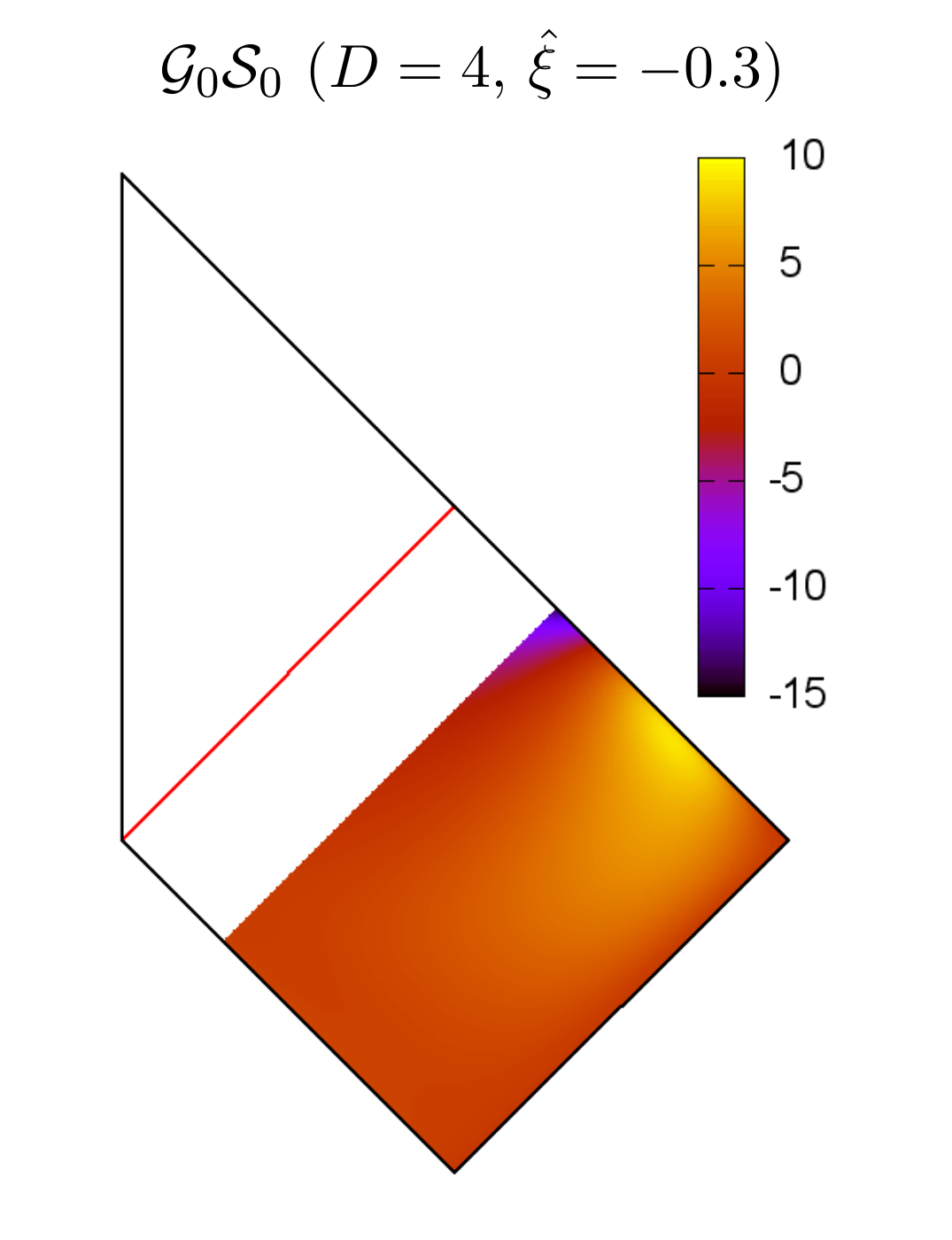}\hfill%
 \includegraphics[scale=0.45,clip=true,trim= 0 0 0 0]{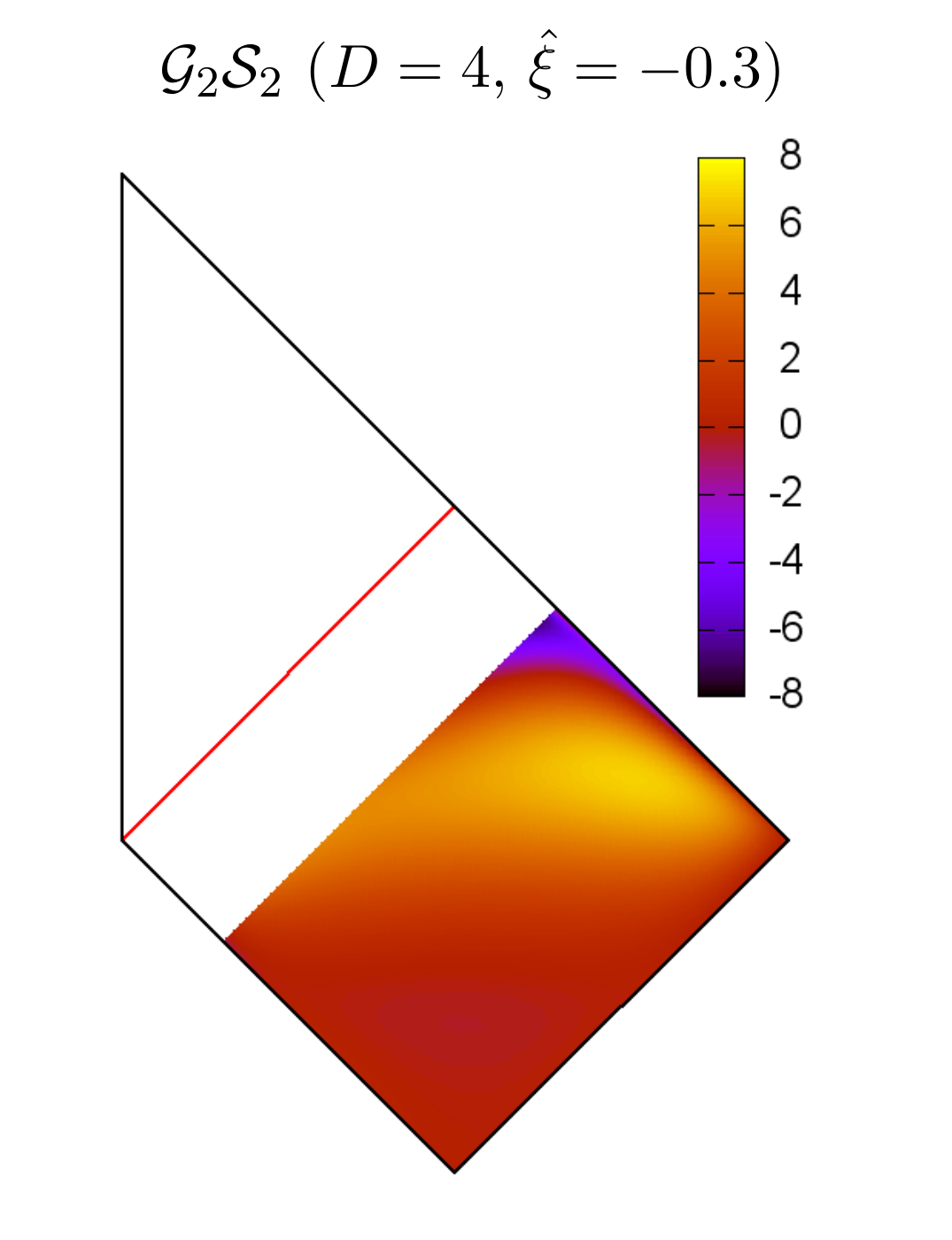}%
  \hspace*{\fill}\vspace*{\fill}\newline\vspace*{\fill}
  \hspace*{\fill}%
 \includegraphics[scale=0.45,clip=true,trim= 0 0 0 0]{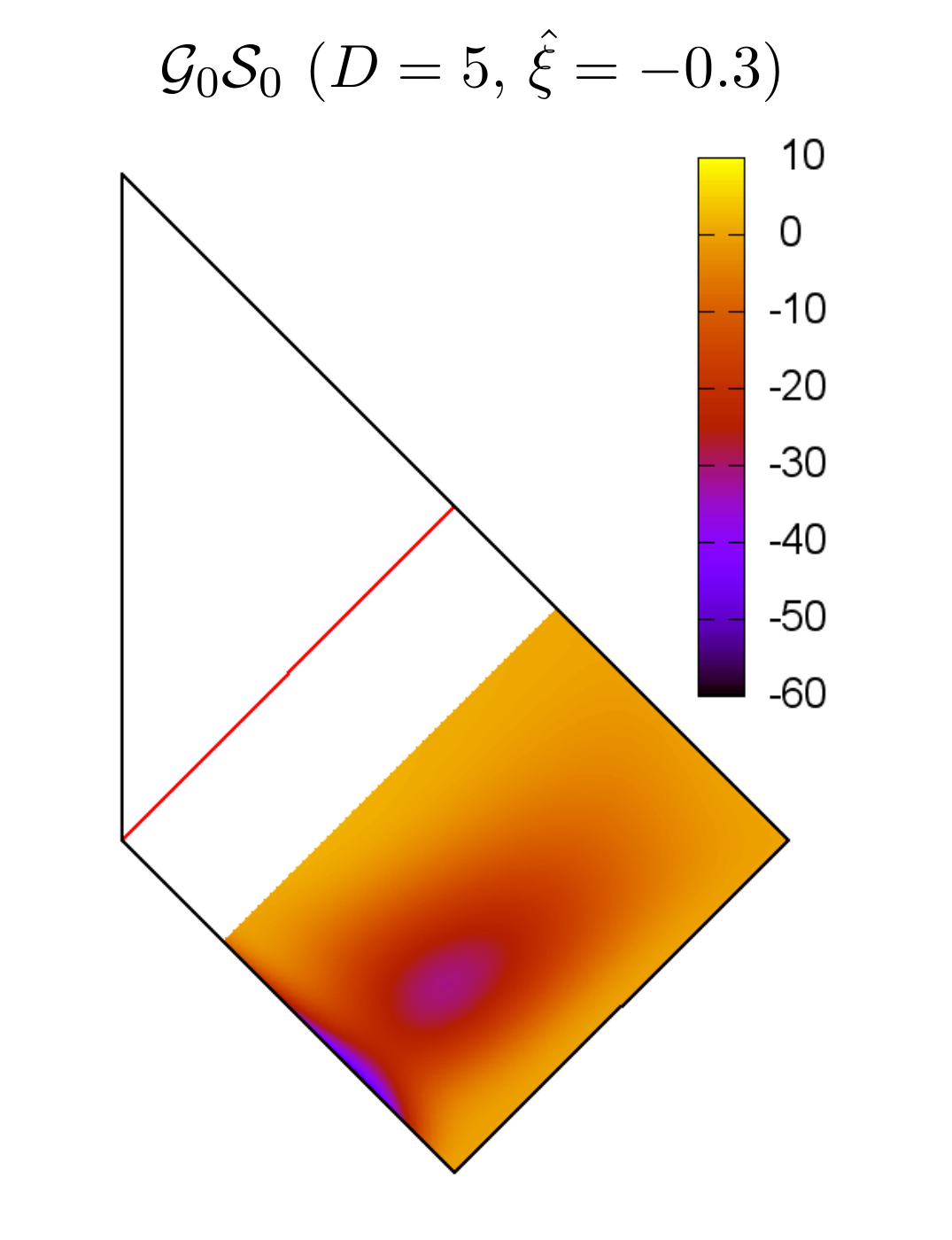}\hfill%
 \includegraphics[scale=0.45,clip=true,trim= 0 0 0 0]{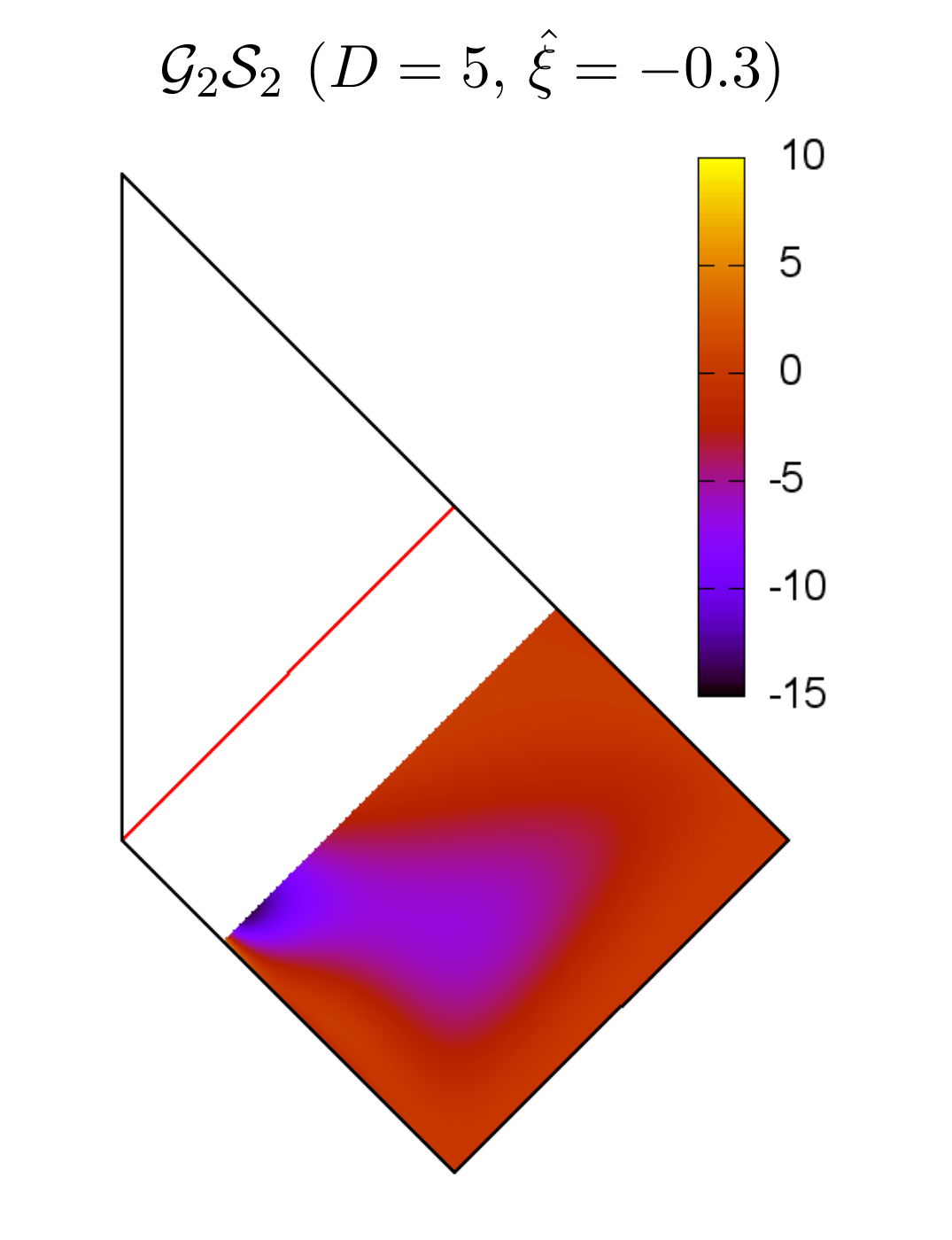}%
  \hspace*{\fill}\vspace*{\fill}\newline\vspace*{\fill}
    \hspace*{\fill}%
 \includegraphics[scale=0.45,clip=true,trim= 0 0 0 0]{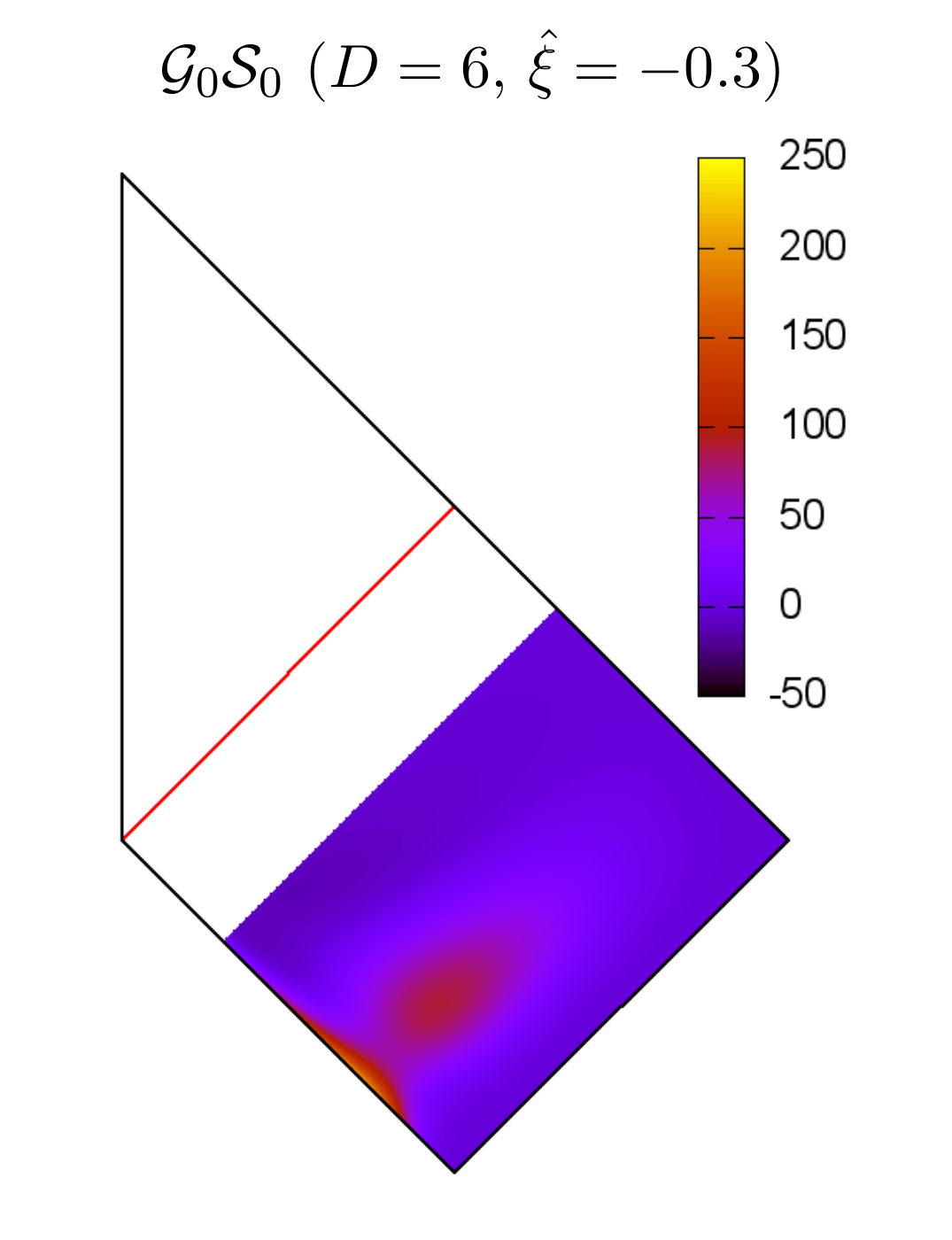}\hfill%
 \includegraphics[scale=0.45,clip=true,trim= 0 0 0 0]{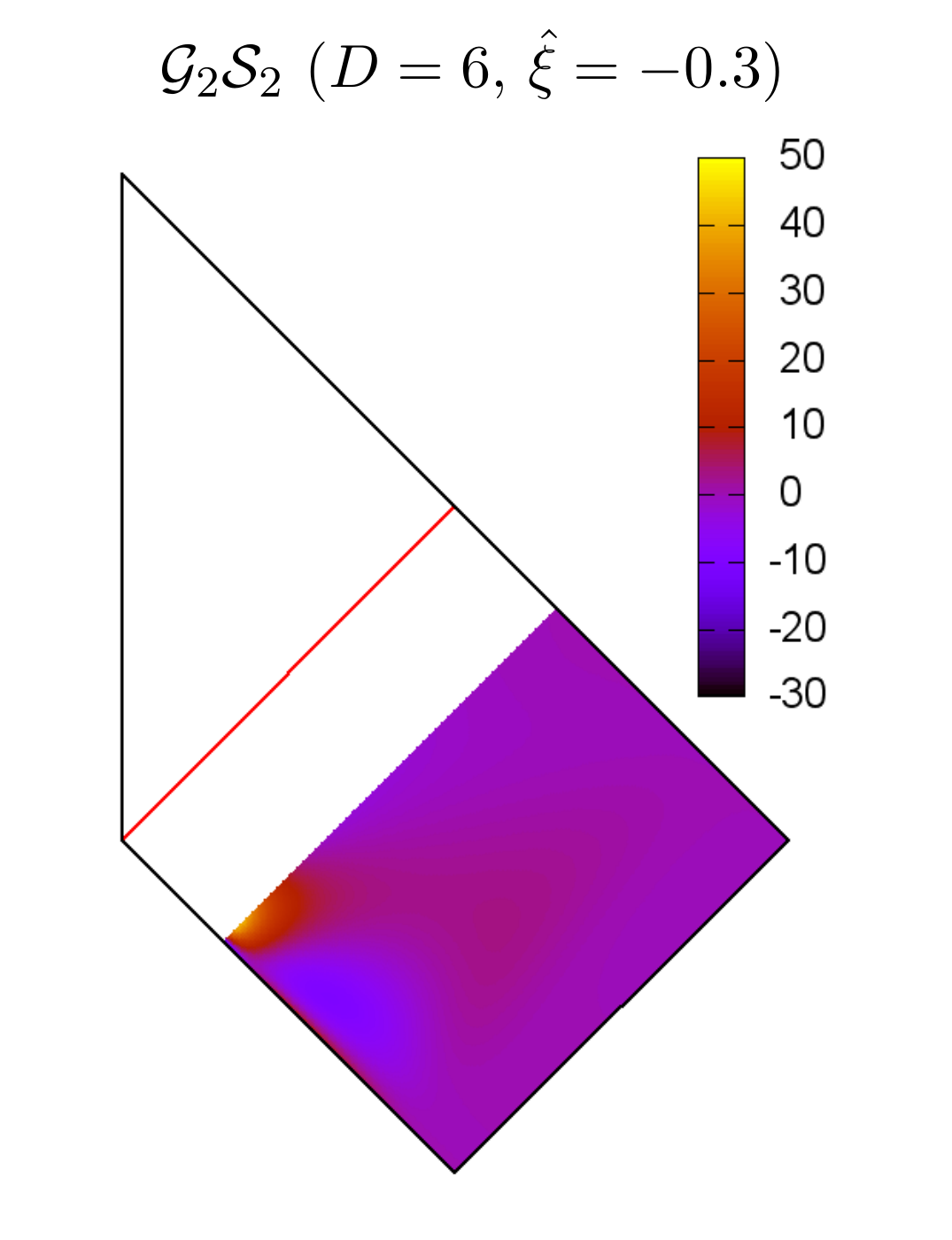}%
  \hspace*{\fill}%
  \caption[Second-order product $\mathcal{G}_m\times\mathcal{S}_m$ for an observation time $\tau<\tau_2$.]{\label{GS_before}Product of the second-order Green's functions with the second-order sources, $\mathcal{G}_0\times\mathcal{S}_0$ (left column) and $\mathcal{G}_2\times\mathcal{S}_2$ (right column), for $D=4,5,6$, for an observation time $\tau$ between $\tau_1$ and $\tau_2$, i.e. before the second optical ray. The axes match the conformal diagram of Fig.~\ref{penrose}. The green and red lines show the location of the singularities (which have been regularised).}
\end{figure}

\begin{figure}
\hspace*{\fill}%
 \includegraphics[scale=0.45,clip=true,trim= 0 0 0 0]{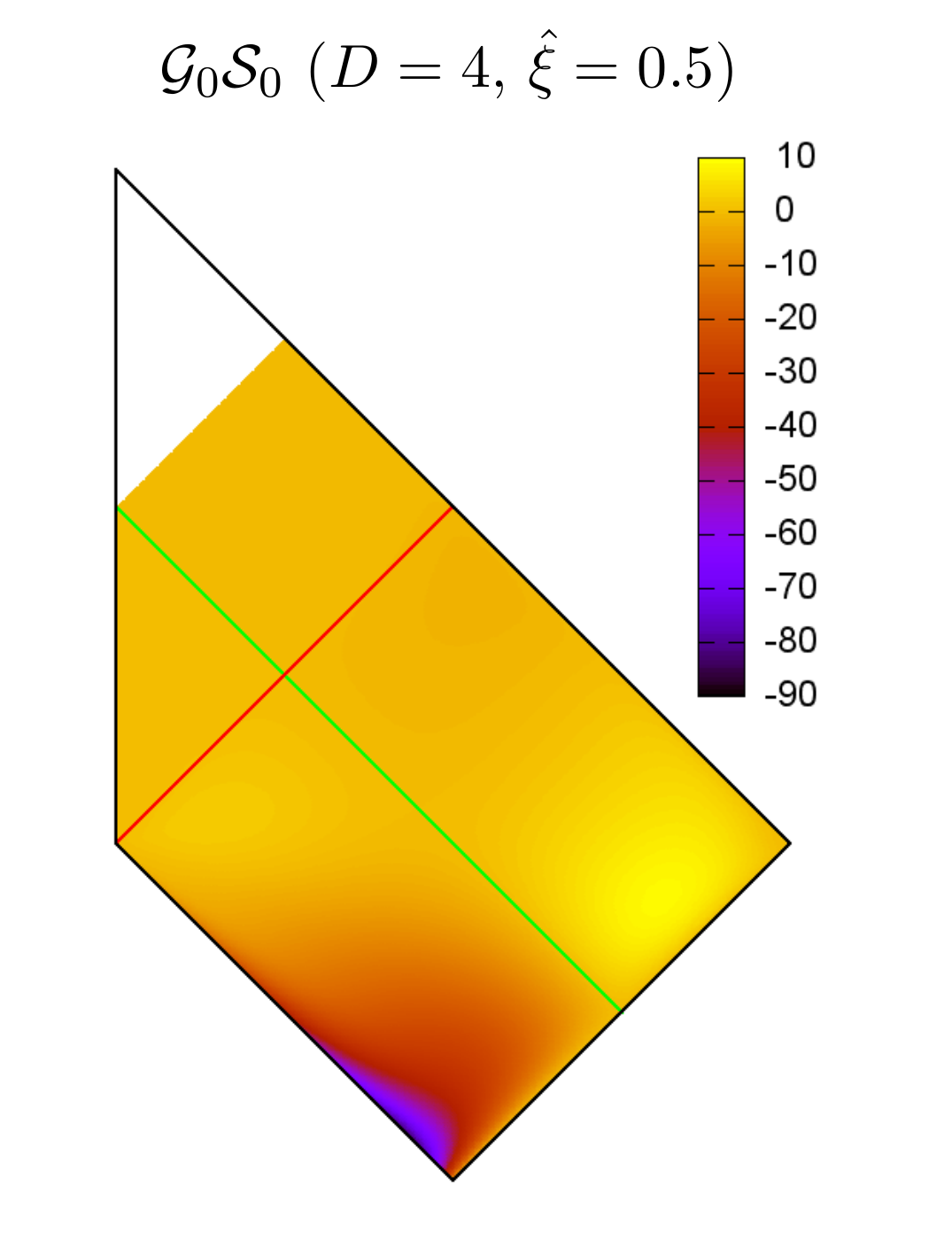}\hfill%
 \includegraphics[scale=0.45,clip=true,trim= 0 0 0 0]{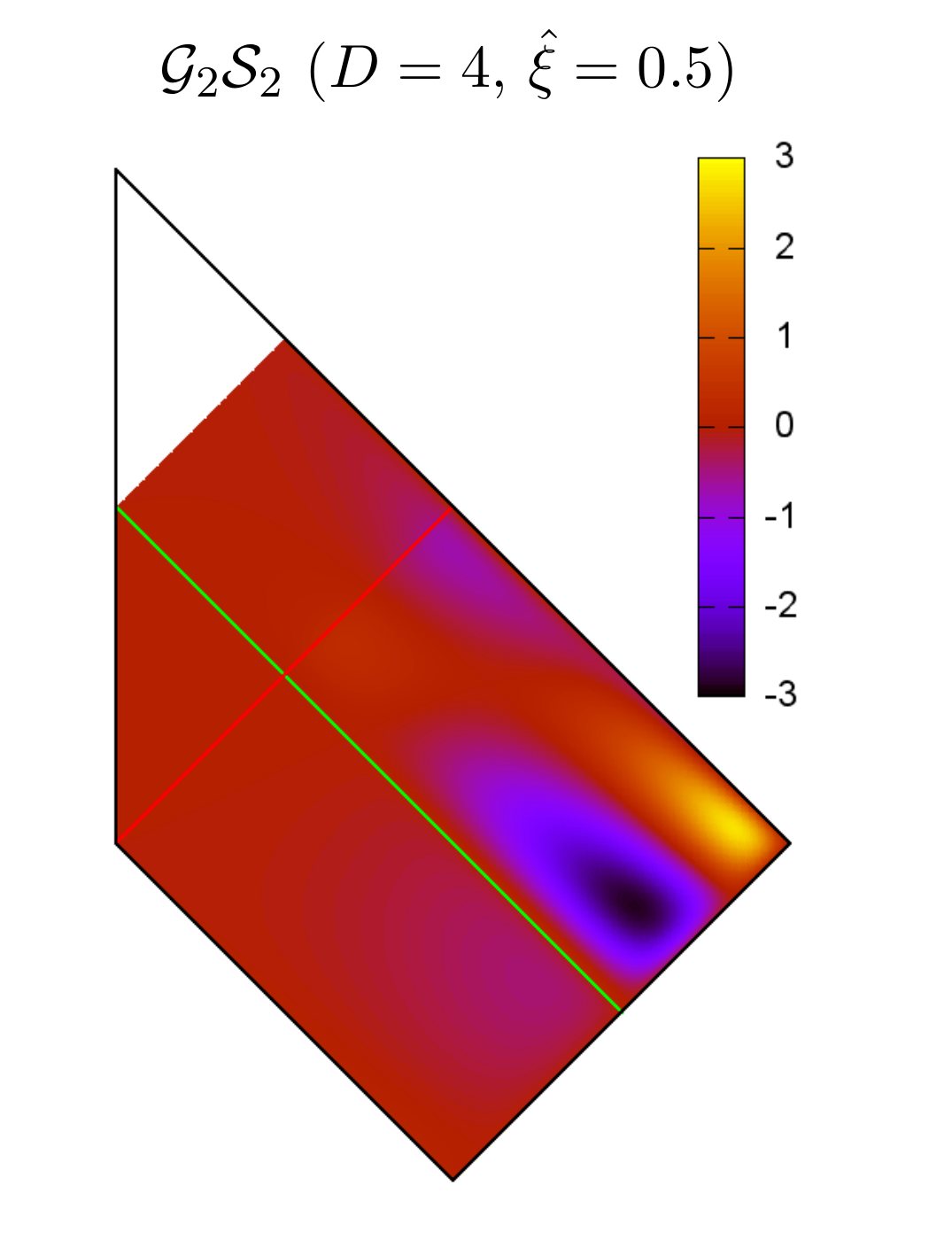}%
  \hspace*{\fill}\vspace*{\fill}\newline\vspace*{\fill}
  \hspace*{\fill}%
 \includegraphics[scale=0.45,clip=true,trim= 0 0 0 0]{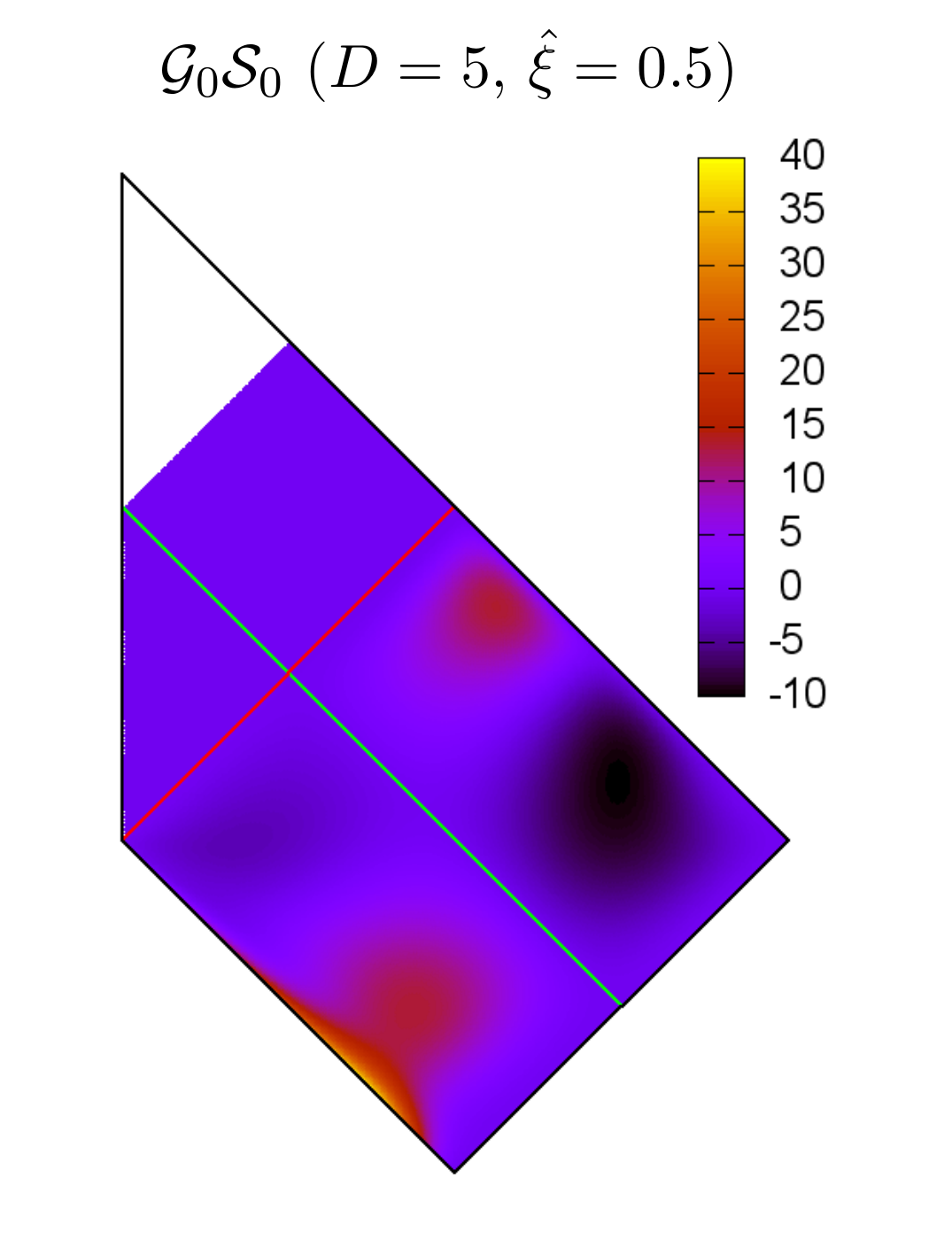}\hfill%
 \includegraphics[scale=0.45,clip=true,trim= 0 0 0 0]{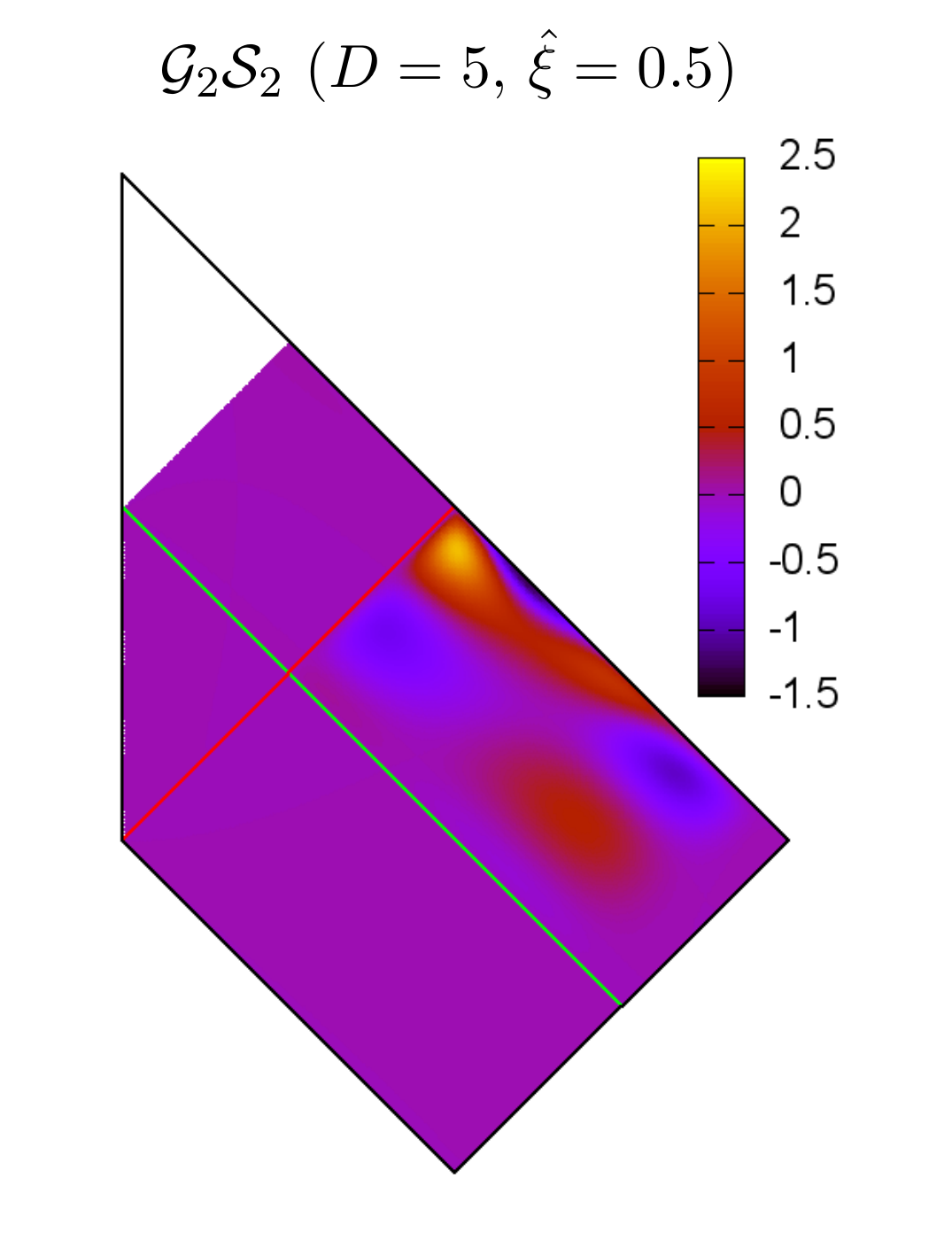}%
  \hspace*{\fill}\vspace*{\fill}\newline\vspace*{\fill}
    \hspace*{\fill}%
 \includegraphics[scale=0.45,clip=true,trim= 0 0 0 0]{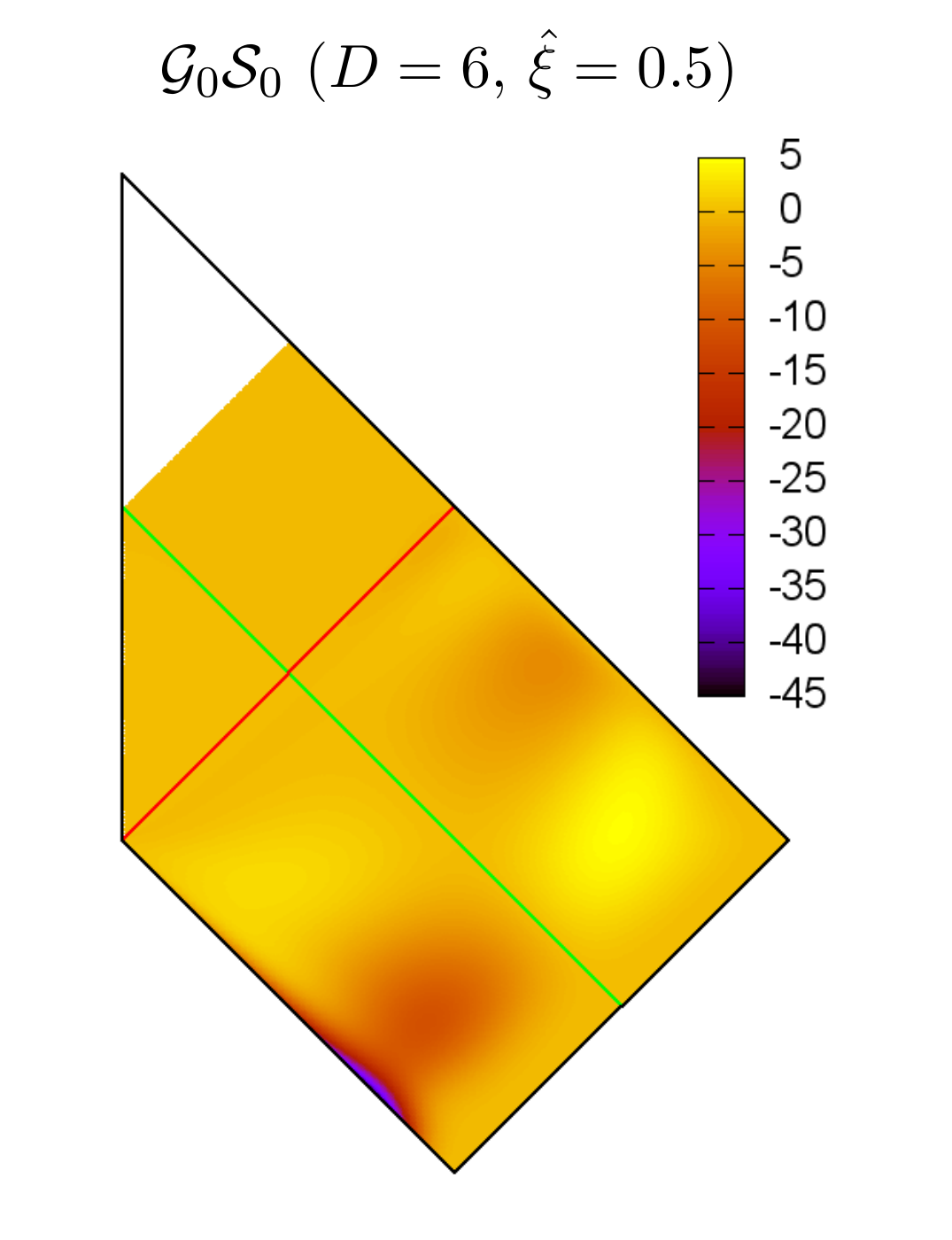}\hfill%
 \includegraphics[scale=0.45,clip=true,trim= 0 0 0 0]{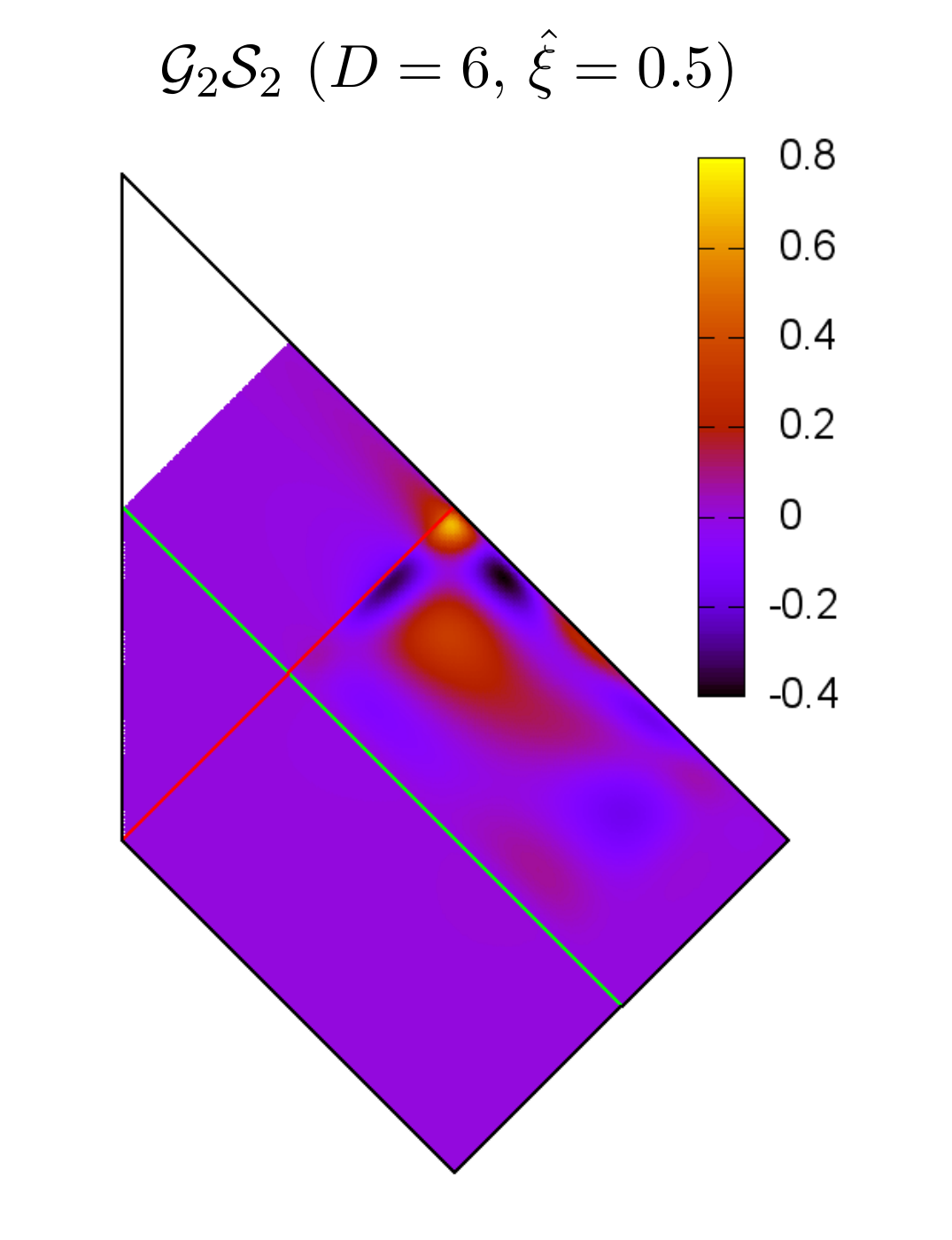}%
  \hspace*{\fill}%
  \caption[Second-order product $\mathcal{G}_m\times\mathcal{S}_m$ for an observation time $\tau>\tau_2$.]{\label{GS_after}Product of the second-order Green's functions with the second-order sources, $\mathcal{G}_0\times\mathcal{S}_0$ (left column) and $\mathcal{G}_2\times\mathcal{S}_2$ (right column), for $D=4,5,6$, for an observation time $\tau$ greater than $\tau_2$, i.e. after the second optical ray. The axes match the conformal diagram of Fig.~\ref{penrose}. The green and red lines show the location of the singularities (which have been regularised).}
\end{figure}

In Figs.~\ref{GS_before} and~\ref{GS_after} we plot the product $\mathcal{G}_{m}\times\mathcal{S}_m$, for $D=4,5,6$, on the conformal diagram of Fig.~\ref{penrose}. In Fig.~\ref{GS_before} the observation time $\tau$ is between $\tau_1$ and $\tau_2$, i.e. before the second optical ray, whereas in Fig.~\ref{GS_after} we have $\tau>\tau_2$. 

These functions are smooth in the whole domain, so we expect this two-dimensional integration to be performed in the near future. The major outstanding challenge is the achievement of the required numerical precision.

\chapter{Conclusion and outlook}
\label{ch:conclusion}

\epigraph{Perfection is achieved, not when there is nothing more to add, but when there is nothing left to take away.}{Antoine de Saint-Exup\'ery}

In this final chapter we conclude with some remarks and an outlook on open questions.

We believe this work constitutes a significant contribution to the understanding of gravitational shock wave collisions, which have been studied by many in the past few decades, with varied motivations.

It is quite remarkable that quantitative information from a non-perturbative process like black hole formation can be extracted using perturbation theory. The $CL$ symmetry, though not an \emph{exact} symmetry of the space-time, seems to play a major role to this end. The understanding of how it justifies perturbation theory, i.e. the correspondence between the perturbative series of the metric and an angular expansion of the news function is one of the major results of this work. 

Extra spatial dimensions may not exist, but the simple formula we obtained for the isotropic contribution, $\epsilon_0$, illustrates the advantage of considering the number of space-time dimensions, $D$, as a parameter in general relativity. The known four-dimensional result, $\epsilon_0=0.25$, would never look so understandable and appealing if the $D$-dimensional computation had not been performed.

The results of Table~\ref{TabSurface} for the second-order surface terms suggest that higher-order terms may exhibit similarly simple patterns, with increasing powers of $1/D$. If this is the case, it would be interesting to relate our angular series with the large $D$ limit of general relativity \cite{Emparan:2013moa}.

These remarkable analytical results were made possible by formulating the problem in Fourier space. It is tempting to think that an analytical solution might exist for the volume terms as well, so this is certainly an avenue worth pursuing. Indeed, given all the symmetries and simplifications that our work has revealed, perhaps there exists a more sensible approach to the problem which allows for an analytical solution.

The second-order calculation has proved to be very technical and challenging, both analytically and numerically. We expect the first angular correction, $\epsilon_1$, to be obtained very soon, shedding some light on the issues above.

Moreover, we hope our work may as well be the foundation stone for further studies. The most obvious candidate is the third (and higher) order contribution. From the numerical point of view, there is an increase in computer effort since three-dimensional Green's functions would have to be mapped, but the integrals from which they arise are essentially the same. 

As argued in Chapter~\ref{ch:intro}, the fact that the initial data are exact at second-order, and the good agreement of the four-dimensional result with numerical relativity, suggest that the second-order estimate for the inelasticity might be accurate enough. The third-order calculation could not only validate (or not) this statement, but also provide a better understanding of the angular distribution of the radiation.

Perhaps more interesting, though, would be to apply this method to different problems. Charged solutions have been considered in \cite{Coelho:2012sy}, but the results seem to indicate a breakdown of perturbation theory. The introduction of an impact parameter, however, is probably treatable. Scattering with large impact parameter has been recently addressed in \cite{Gruzinov:2014moa}, but otherwise the best estimates come from apparent horizon bounds \cite{Yoshino:2005hi}. It would be particularly interesting to see the threshold of black hole formation as the critical impact parameter is exceeded.

\appendix

\chapter{The infinite boost}
\label{app:boost}

This appendix contains details on the infinite boost of the Schwarzschild solution that yield the Aichelburg-Sexl shock wave. It supplements Chapter~\ref{ch:kinematics}. 

When taking the limit $v\rightarrow1$ in Eq.~\eqref{after_limit}, we expect a distribution similar to the delta function. Therefore, it is natural to take that limit on an integral of the distribution. Let us define
\begin{eqnarray}
F_v(t,z,\rho)&=&\int_{-\infty}^{z}dz'\,\gamma\left[\gamma^2\left(z'-v t\right)^2+\rho^2\right]^{-\frac{D-3}{2}}\,, \nonumber\\
&=&\frac{1}{\rho^{D-4}}\int_{-\infty}^{Z_v(z,t)}dZ'\,\left(Z'^2+1\right)^{-\frac{D-3}{2}}
\end{eqnarray}
with $Z_v(z,t)\equiv \gamma(z-v t)\rho^{-1}$. To take the limit, we first note that
\begin{equation}
\lim_{v\rightarrow 1}Z_v(z,t)=sign(z-t)\times \infty\,,
\end{equation}
so
\begin{equation}
\lim_{v\rightarrow 1}F_v(t,z,\rho)
=\dfrac{\theta(z-t)}{\rho^{D-4}}\int_{-\infty}^{+\infty}dZ'\dfrac{1}{\left(Z'^2+1\right)^{-\frac{D-3}{2}}}=\dfrac{\theta(z-t)}{\rho^{D-4}}\sqrt{\pi}\dfrac{\Gamma\left(\tfrac{D-4}{2}\right)}{\Gamma\left(\tfrac{D-3}{2}\right)}\,,
\end{equation}
where $\Gamma$ is the gamma function.

Finally, in Eq.~\eqref{after_limit},
\begin{equation}
\lim_{v\rightarrow 1}4\frac{D-2}{D-3}\gamma^2Z=\dfrac{16\pi G_D E}{(D-3)\Omega_{D-2}}\lim_{v\rightarrow 1}\dfrac{dF_v}{dz}=\dfrac{8\pi G_D E}{\Omega_{D-3}}\Phi(\rho)\delta(t-z)\,.
\end{equation}

\chapter{Green's functions}
\label{app:Green}

This appendix contains details on the Green's function for the d'Alembertian operator and on the general integral solution to the wave equation with a source with characteristic initial data. It mainly supplements Chapter~\ref{ch:dynamics} and Chapter~\ref{ch:volume}. In Section~\ref{app:GreenBox} we present the Green's function for the $D$-dimensional wave equation and show how to use it to solve a problem with characteristic initial data. In Section~\ref{app:Green3D} we perform the dimensional reduction to three dimensions and, in Section~\ref{app:Green2D}, to two dimensions. Finally, in Section~\ref{app:functions} we present expressions for the functions $I_m^{D,n}$ and their derivatives.

\section{Green's function for the d'Alembertian operator}
\label{app:GreenBox}
The d'Alembertian operator $\Box$ is defined, in cartesian coordinates $(t,z,x^i)$, as
\begin{equation}
\Box\equiv\eta^{\mu\nu}\partial_\mu\partial_\nu=-\partial_t^2+\partial_z^2+\partial^i\partial_i\,,\label{app_Green_cartesian}
\end{equation}
where $\eta^{\mu\nu}$ is the inverse Minkowski metric.

Its Green's function $G(t,z,x^i)$ is defined as the solution of
\begin{equation}
\Box G(t,z,x^i)=\delta(t)\delta(z)\delta(x^i)\,.
\end{equation}

Observe that this is not a coordinate-invariant definition since the delta function $\delta(x^\mu)$ transforms with the Jacobian determinant of a coordinate transformation,
\begin{equation}
\delta(x^\mu)=\left|\frac{\partial {x'}^\alpha}{\partial x^\beta}\right|\delta({x'}^\mu)\,.
\end{equation}

In particular, for $(t,z)\rightarrow(u,v)=(t-z,t+z)$,
\begin{equation}
\delta(t)\delta(z)=2\delta(u)\delta(v)\,.
\end{equation}

The Green's function in cartesian coordinates, i.e. the solution to Eq.~\eqref{app_Green_cartesian}, is \cite{Friedlander:112411}
\begin{equation}
G(t,z,x^i)=-\frac{1}{2\pi^{\frac{D-2}{2}}}\delta^{\left(\frac{D-4}{2}\right)}(t^2-z^2-x^ix_i)\,.
\end{equation}

Therefore, the solution to $\Box G(x^\mu)=\delta(x^\mu)$ in our coordinate system, $x^\mu=\{u,v,x^i\}$,
\begin{equation}
G(u,v,x^i)=-\frac{1}{4\pi^{\frac{D-2}{2}}}\delta^{\left(\frac{D-4}{2}\right)}(\chi)\,,\qquad \chi=-\eta_{\mu\nu}x^\mu x^\nu\,.
\end{equation}

For odd $D$, we have a fractional derivative of the delta function. This is to be understood in Fourier space, or formally as follows. For negative orders $(n>0)$,
\begin{equation}
\delta^{(-n)}(x)=\frac{1}{\Gamma(n)}x^{n-1}\theta(x)\,,\label{app_delta_theta}
\end{equation}
where $\Gamma$ is the gamma function. Then, acting with derivatives, we can extend this recursively to positive fractional orders. For instance,
\begin{equation}
\delta^{\left(\tfrac{1}{2}\right)}(x)=\frac{d}{dx}\delta^{\left(-\tfrac{1}{2}\right)}(x)=-\frac{1}{\sqrt{\pi|x|}}\left(\frac{\theta(x)}{2x}-\delta(x)\right)\,.
\end{equation}

This also shows that, while for even $D$ the Green's function has support on the light cone ($\chi=0$), for odd $D$ it has support also \emph{inside} it ($\chi\geq0$). An odd number of space-time dimensions behaves as a dispersive medium.

\subsection{Solution with characteristic initial data}
\label{app:char_data}
Given an equation $\Box F(u,v,x^i)=S(u,v,x^i)$ it is straightforward to show that a solution is
\begin{equation}
F(u,v,x^i)=\int d^Dx\, G(u-u',v-v',x^i-{x'}^i)S(u',v',{x'}^i)\,,\label{solFx}
\end{equation}
just by using the definition of $G(x^\mu)$. But what is the solution to $\Box F(u,v,x^i)=0$ \emph{given} $F(0,v,x^i)$, i.e. with characteristic initial data? Theorem 6.3.1 of \cite{Friedlander:112411} says that this is equivalent to a source
\begin{equation}
S(u,v,x^i)=4\delta(u)\partial_v F(0,v,x^i)\,.
\end{equation}
For a motivation of this result in Fourier space, see the appendix of \cite{Herdeiro:2011ck}.

\section{Reduction to three dimensions}
\label{app:Green3D}
In this section we specialise to the case of spherical symmetry in the $(D-2)$-dimensional transverse space spanned by the coordinates $x^i$. We choose spherical coordinates with radius $\rho=\sqrt{x^ix_ i}$ and $D-3$ angles $\phi^a$, and define a basis of tensors in that space,
\begin{equation}
\delta_{ij}\,,\qquad \Gamma_i\equiv\frac{x_i}{\rho}\,,\qquad \Delta_{ij}\equiv\delta_{ij}-(D-2)\Gamma_i\Gamma_j\,,
\end{equation}
with ranks $m=\{0,1,2\}$ respectively. Let $X_m(\Gamma_i)$ be one of such tensors. Our problem becomes to solve the equation
\begin{equation}
\Box \left(F(u,v,\rho)X_m(\Gamma_i)\right)=S(u,v,\rho)X_m(\Gamma_i)\,.\label{eqFX}
\end{equation}

First, observe that we can write
\begin{equation}
\Delta\equiv\partial^i\partial_i=\rho^{-(D-3)}\partial_\rho(\rho^{D-3}\partial_\rho)+\rho^{-2}\Delta_{S^{D-3}}\,,
\end{equation}
where $\Delta_{S^{D-3}}$ is the Laplace-Beltrami operator (or the spherical Laplacian) on the $(D-3)$-sphere. Indeed, our tensors $X_m$ are eigenfunctions of this operator,
\begin{equation}
\Delta_{S^{D-3}} X_m=-m(m+D-4)X_m\,,
\end{equation}
thus the equation for $F(u,v,\rho)$ implied by Eq.~\eqref{eqFX} is
\begin{equation}
\left(-4\partial_u\partial_v+\partial^2_\rho+(D-3)\rho^{-1}\partial_\rho-m(m+D-4)\rho^{-2}\right)F(u,v,\rho)=S(u,v,\rho)\,.\label{operator_3D}
\end{equation}

Instead of trying to find a Green's function for this operator directly, we can use the integral solution of Eq.~\eqref{solFx}. Contraction with the appropriate tensor on both sides yields
\begin{equation}
F(u,v,\rho)=\int d^{D}x'\, G(x-x')\Lambda_m\left(\frac{x^i x'_i}{\rho\rho'}\right)S(u',v',\rho')\,,\label{app:general_F}
\end{equation}
where $\Lambda_m(z)=\left\{1,z,(D-3)^{-1}\left((D-2)z^2-1\right)\right\}$, respectively for $m=\{0,1,2\}$.

Then again we choose spherical coordinates inside the integral, with the angles ${\phi'}^a$ aligned such that one of them is the angle between $x^i$ and ${x'}^i$. Then we integrate the remaining $D-4$ to get
\begin{equation}
F(u,v,\rho)=-\frac{\Omega_{D-4}}{2\pi^{\frac{D-2}{2}}}\int du' dv' d\rho'\,\rho'^{D-3}S(u',v',\rho')\int_{-1}^1dx\,(1-x^2)^{\frac{D-5}{2}+m}\Lambda_m(x)\delta^{\left(\frac{D-4}{2}\right)}(\chi)\,.
\end{equation}

Using the scaling properties of the delta function, and defining $x_\star$ through
\begin{eqnarray}
\chi&=&(u-u')(v-v')-(\rho^2+\rho'^2-2\rho\rho' x)\\
&\equiv&2\rho\rho'(x-x_\star)\,,
\end{eqnarray}
this becomes
\begin{equation}
F(u,v,\rho)=\int du'\int dv'\int d\rho'\,G_m(u-u',v-v';\rho,\rho')S(u',v',\rho')\,,\label{app:sol3D}
\end{equation}
where
\begin{equation}
G_m(u-u',v-v';\rho,\rho')=-\frac{1}{2\rho}\left(\frac{\rho'}{\rho}\right)^{\frac{D-4}{2}}I_m^{D,0}(x_\star)\,
\end{equation}
is the new Green's function and we have defined one-dimensional functions $I^{D,n}_m$ through
\begin{eqnarray}
I_m^{D,n}(x_\star)&\equiv&\frac{\Omega_{D-4}}{(2\pi)^{\frac{D-2}{2}}}\int_{-1}^1dx\,\Lambda_m(x)(1-x^2)^{\frac{D-4}{2}-\frac{1}{2}}\delta^{\left(\frac{D-4}{2}-n\right)}(x-x_\star)\,,\nonumber\\\label{app:def_I}
&=&\frac{\Omega_{D-4}}{(2\pi)^{\frac{D-2}{2}}}\lambda_m^{-1}\int_{-1}^1dx\,\partial^{m}(1-x^2)^{\frac{D-5}{2}+m}\delta^{\left(\frac{D-4}{2}-n\right)}(x-x_\star)\,,\label{app_functions}
\end{eqnarray}
where $\lambda_m=\{1,-(D-3),(D-1)(D-3)\}$.

\section{Reduction to two dimensions}
\label{app:Green2D}
In Chapter~\ref{ch:analytics} we show that the problem becomes effectively two-dimensional order by order. In this section we obtain the differential operator acting on $f^{(k)}(p,q)$ and its associated Green's function $G^k_m(p,q;p'q,q')$.

From the definition of $p,q$ in Eq.~\eqref{def_pq}, we see that
\begin{equation}
\partial_u\rightarrow\rho^{-(D-2)}\partial_q\,,\qquad \partial_v\rightarrow\rho^{D-4}\partial_p\,,\qquad \partial_\rho\rightarrow\partial_\rho-\frac{D-2}{\rho}q\partial_q+\frac{2+(D-4)p}{\rho}\partial_p\,.\label{derivatives_transformation}
\end{equation}

Inserting this in Eq.~\eqref{operator_3D}, together with Eqs.~\eqref{f_pq} and~\eqref{def_s}, we conclude that the operator acting on $f^{(k)}(p,q)$ is
\begin{eqnarray}
&&-4\partial_p\partial_q+\left((2+(D-4)p)\partial_p-(D-2)q\partial_q-(D-3)(2k+N_u+N_v)+D-4\right)\times\nonumber\\
&\times&\left((2+(D-4)p)\partial_p-(D-2)q\partial_q-(D-3)(2k+N_u+N_v)\right)-m(m+D-4)\,.\nonumber\\ \label{app:op2D}
\end{eqnarray}

As in the previous section, its Green's function can be found from the integral solution for $F(u,v,\rho)$, Eq.~\eqref{app:sol3D}. We factor out the $\rho$ dependence using Eqs.~\eqref{f_pq} and \eqref{def_s}, and change the integration variable $\rho'\rightarrow\rho y$. We get that the solution to each reduced metric function $f^{(k)}(p,q)$ can be written as
\begin{equation}
f^{(k)}(p,q)=\int dq' \int dp'\, G^k_m(p,q;p',q')s^{(k)}(p',q')\,,\label{app:f_volume}
\end{equation}
where $G^{k}_m$ is the reduced Green's function,
\begin{equation}
G^{k}_m(p,q;p',q')=-\frac{1}{4}\int_0^\infty dy\, y^{\frac{D-4}{2}-(D-3)(2k+N_u-N_v)}I_m^{D,0}(x_\star)\,,
\end{equation}
and now $x_\star$ reads
\begin{eqnarray}
x_\star&=&\frac{1+y^2-(q-q'y^{D-2})(p-p'y^{-(D-4)}-\Psi(y))}{2y}\,,\\
\Psi(y)&\equiv&\left\{
\begin{array}{ll}
 \Phi(y)\ , &  D=4\  \vspace{2mm}\\
\displaystyle{ \Phi(y)-\frac{2}{D-4}}\ , & D>4\ \label{psi}\,
\end{array} \right. \ .
\end{eqnarray}

\subsection{Surface terms}
\label{app:surface_pq}
Surface terms are a particular case of this general solution. For those, the source is as in Eq.~\eqref{S_surf}, where $f^{(k)}_0(\rho)\propto\rho^{-k(D-2)-(N_u-N_v)(D-3)}$. So,
\begin{equation}
s^{(k)}(p,q)=2k!f^{(k)}_0(1)\delta(q)\delta^{(-k)}(p)\,.
\end{equation}
Inserting this in Eq.~\eqref{app:f_volume}, the $q'$ integration is trivial, whereas the $p'$ integration, using Eq.~\eqref{app:def_I}, is essentially
\begin{equation}
\int dp'\,\delta^{(-k)}(p')\delta^{\left(\frac{D-4}{2}\right)}(x-x_\star)=(-1)^D\left(\frac{2y^{D-3}}{q}\right)^k \delta^{\left(\frac{D-4}{2}-k\right)}(x-x_S)\,.
\end{equation}

Therefore the solution for $f^{(k)}(p,q)$ in the surface case can be simplified to
\begin{equation}
f^{(k)}_{S}(p,q)=(-1)^{D+1}k!f^{(k)}_0(1)\left(\frac{2}{q}\right)^k\int_0^\infty dy\,y^{\frac{D-4}{2}-(D-3)(k+N_u-N_v)}I_m^{D,k}(x_S)\,,\label{app:f_surface}
\end{equation}
where now
\begin{equation}
x_S\equiv x_\star(p'=0,q'=0)=\frac{1+y^2-q(p-\Psi(y))}{2y}\,.
\end{equation}

\section{The functions $I_m^{D,n}$}
\label{app:functions}
In Eq.~\eqref{app_functions} we defined functions $I_m^{D,n}(x_star)$ given by one-dimensional integrals. These appear several times throughout this thesis. In this section we give explicit expressions for them. To simplify the expressions below, we shall commit the pre-factor
\begin{equation}
\frac{\Omega_{D-4}}{(2\pi)^{\frac{D-2}{2}}}\lambda_m^{-1}\,.
\end{equation}

\subsubsection{Even $D$}
If $D$ is even and $D-4\geq 2n$, we can integrate by parts $(D-4)/2-n$ times and get
\begin{eqnarray}
I_m^{D,n}(x_\star)&\propto&(-1)^{\frac{D-4}{2}-n}\partial^{\left(\frac{D-4}{2}-n+m\right)}\left[(1-x^2)^{\frac{D-5}{2}+m}\right]_{x=x_\star}\theta(|x_\star|\leq1)\\
&=&(-1)^{\frac{D-4}{2}-n}(1-x_\star^2)^{n-\frac{1}{2}}Q_m^{D,n}(x_\star)\theta(|x_\star|\leq1),
\end{eqnarray}
where
\begin{equation}
Q_m^{D,n}(x)\equiv\frac{\partial^{\left(\frac{D-4}{2}-n+m\right)}\left[(1-x^2)^{\frac{D-5}{2}+m}\right]}{(1-x_\star^2)^{n-\frac{1}{2}}}=\sum_{k=0}^{\frac{D-4}{2}+m-n}a_k x^k.
\end{equation}
If $D-4<2n$, which happens for $\{D=4,n=1,2\}$ and $\{D=6,n=2\}$, we use Eq.~\eqref{app_delta_theta} but must proceed case by case. We get
\begin{eqnarray}
I^{4,1}_0(x_\star)&\propto&\pi\theta(x_\star<-1)+\left[\frac{\pi}{2}-\arcsin(x_\star)\right]\theta(|x_\star|\leq1),\nonumber\\
I^{4,1}_1(x_\star)&\propto&-\sqrt{1-x_\star^2}\theta(|x_\star|\leq1),\nonumber\\
I^{4,1}_2(x_\star)&\propto& 3x_\star\sqrt{1-x_\star^2}\theta(|x_\star|\leq1),\nonumber\\
I^{4,2}_0(x_\star)&\propto&-x_\star\pi\theta(x_\star<-1)+\left[x_\star\left((\arcsin(x_\star)-\frac{\pi}{2}\right)+\sqrt{1-x_\star^2}\right]\theta(|x_\star|\leq1),\nonumber\\
I^{4,2}_1(x_\star)&\propto&-\frac{\pi}{2}\theta(x_\star<-1)-\left[\frac{\pi}{4}-\frac{1}{2}\left(\arcsin(x_\star)+x_\star\sqrt{1-x_\star^2}\right)\right]\theta(|x_\star|\leq1),\nonumber\\
I^{4,2}_2(x_\star)&\propto&(1-x_\star^2)^{\frac{3}{2}}\theta(|x_\star|\leq1),\nonumber\\
I^{6,2}_0(x_\star)&\propto&\left[\frac{\pi}{4}-\frac{1}{2}\left(\arcsin(x_\star)+x_\star\sqrt{1-x_\star^2}\right)\right]\theta(|x_\star|\leq1)+\frac{\pi}{2}\theta(x_\star<-1),\nonumber\\
I^{6,2}_1(x_\star)&\propto&-(1-x_\star^2)^{\frac{3}{2}}\theta(|x_\star|\leq1),\nonumber\\
I^{6,2}_2(x_\star)&\propto& 5x_\star(1-x_\star^2)^{\frac{3}{2}}\theta(|x_\star|\leq1).\nonumber
\end{eqnarray}

\subsubsection{Odd $D$}
If $D$ is odd and $D-5\geq 2n\geq0$, we can also integrate by parts $(D-5)/2-n$ times and get
\begin{eqnarray}
I_m^{D,n}(x_\star)&\propto&(-1)^{\frac{D-5}{2}-n}\int_{-1}^{1}\partial^{\left(\frac{D-5}{2}-n+m\right)}\left[(1-x^2)^{\frac{D-5}{2}+m}\right]\delta^{(\frac{1}{2})}(x-x_\star)\\
&=&(-1)^{\frac{D-5}{2}-n}\int_{-1}^{1}P_m^{D,n}(x)\delta^{(\frac{1}{2})}(x-x_\star),\label{app_odd_big}
\end{eqnarray}
where
\begin{equation}
P_m^{D,n}(x)\equiv\partial^{\left(\frac{D-5}{2}-n+m\right)}\left[(1-x^2)^{\frac{D-5}{2}+m}\right]=\sum_{k=0}^{\frac{D-5}{2}+m-n}a_k x^k.
\end{equation}
Then we use
\begin{equation}
\delta^{\left(\frac{1}{2}\right)}(x)=\frac{d}{dx}\delta^{\left(-\frac{1}{2}\right)}(x)=-\frac{1}{2\sqrt{\pi}}\left(\frac{\theta(x)}{x^{\frac{3}{2}}}-\frac{\delta(x)}{x}\right).
\end{equation}
The effect of the $\delta$-function is to remove infinities (poles), so we can ignore it provided we retain only the finite part of the integral with the $\theta$-function. So basically we have integrals of the form (finite part of)
\begin{equation}
\int_{-1}^{1}\frac{x^k}{(x-x_\star)^{\frac{3}{2}}}\theta(x-x_\star)dx.
\end{equation}
We change variables to $z=x-x_\star$ and expand the binomial, to get
\begin{eqnarray}
I_m^{D,n}(x_\star)&\propto&\frac{(-1)^{\frac{D-3}{2}+n}}{2\sqrt{\pi}}\sum_{k=0}^{\frac{D-5}{2}+m-n}a_k\sum_{j=0}^{k}\binom{k}{j}\frac{x_\star^{k-j}}{j-\frac{1}{2}}\times\\&\times&\left[(1-x_\star)^{j-\frac{1}{2}}\theta(|x_\star|\leq1)+\left((1-x_\star)^{j-\frac{1}{2}}-(-1-x_\star)^{j-\frac{1}{2}}\right)\theta(x_\star<-1)\right].\nonumber
\end{eqnarray}
If $0\leq D-5<2n$, which happens for $\{D=5,n=1,2\}$, $\{D=6,n=1,2\}$ and $\{D=7,n=2\}$, we define the negative-order derivative of $\delta$-function as before and get
\begin{eqnarray}
I_m^{D,n}(x_\star)&\propto&\frac{1}{\Gamma\left(n-\frac{D-4}{2}\right)}\sum_{k=0}^{D-5+m}a_k\sum_{j=0}^{k}\binom{k}{j}\frac{x_\star^{k-j}}{j+n-\frac{D-4}{2}}\times\nonumber\\&\nonumber\times&\left[(1-x_\star)^{j+n-\frac{D-4}{2}}\theta(|x_\star|\leq1)\right.\\&&\left.+\left((1-x_\star)^{j+n-\frac{D-4}{2}}-(-1-x_\star)^{j+n-\frac{D-4}{2}}\right)\theta(x_\star<-1)\right],
\end{eqnarray}
where now the coefficients $a_k$ are defined by
\begin{equation}
\partial^{(m)}(1-x^2)^{\frac{D-5}{2}+m}=\sum_{k=0}^{D-5+m}a_k x^k.
\end{equation}

For some derivatives of the metric functions, some $n=-1$ terms are actually required. For that case one obtains the same result as in Eq.~\eqref{app_odd_big}, plus an extra term
\begin{eqnarray}
I_{m}^{D,-1}(x)_{extra} & \propto & (-1)^{\frac{D-3}{2}}\left[P_{m}^{D,0}(1)\delta^{(\frac{1}{2})}(1-x)-P_{m}^{D,0}(-1)\delta^{(\frac{1}{2})}(-1-x)\right]\qquad\\
 & = & F.P.\dfrac{\kappa_{m}(-1)^{\frac{D-3}{2}}}{2\sqrt{\pi}}P_{m}^{D,0}(1)\left[\dfrac{\theta(1-x)}{(1-x)^{\frac{3}{2}}}+(-1)^{\frac{D-5}{2}+m}\dfrac{\theta(-1-x)}{(-1-x)^{\frac{3}{2}}}\right]\; .\; \nonumber
\end{eqnarray}

\subsubsection{Derivatives}
We will also need to compute derivatives of the metric functions which, in the two-dimensional form, can be $\partial_q$ or $\partial_p$ (any $\rho$-derivative is trivial since the $\rho$-dependence has been factorised). Acting on the general expression of Eq.~\eqref{app:f_volume} or Eq.~\eqref{app:f_surface}, there might be a term from the integration limits, and then there is the derivative of the integrand. Since all the dependence on $p$ and $q$ is through $x_\star$, and $\left(I^{D,n}_m\right)'=-I^{D,n-1}_m$,
\begin{equation}
\partial_\alpha I^{D,n}_m(x_\star)=-\partial_\alpha x_\star I^{D,n-1}_m(x_\star)\,.
\end{equation}
In this way, it is straightforward to build the sources, namely for second-order calculation of Chapter~\ref{ch:second_order}.

\chapter{The asymptotic waveform}
\label{app:asympWF}

This appendix contains details on the simplification of the integral formula for the metric functions in the limit $r\rightarrow\infty$, including the factorisation of the angular dependence. It mainly supplements Chapter~\ref{ch:analytics}. 

\section{The asymptotic limit in Fourier space}
\label{app:limit_Fourier}
Taking Eq.~\eqref{app:sol3D}, we define the asymptotic waveform $\dot{F}(\tau,\theta)$,
\begin{equation}
\dot{F}(\tau,\theta)\equiv\lim_{r\rightarrow\infty} r\rho^{\frac{D-4}{2}}\frac{d}{d\tau}F(u,v,\rho)\,,
\end{equation}
and $\bar{x}$ as the asymptotic form of $x_\star$,
\begin{equation}
\lim_{r\rightarrow\infty}x_\star=\frac{1}{\rho'\sin\theta}\left(-\tau+u'\frac{1+\cos\theta}{2}+v'\frac{1-\cos\theta}{2}\right)\equiv \bar{z}\,.\label{x_bar}
\end{equation}

Noting that $\left(I_m^{D,N}\right)'=-I_m^{D,N-1}$, we get
\begin{equation}
\dot{F}(\tau,\theta)=\frac{1}{2\sin\theta}\int du'\int dv'\int d\rho'\,\rho'^{\frac{D-4}{2}}S(u',v',\rho')I_m^{D,-1}(\bar{x})\frac{d \bar{x}}{d\tau}\,.\label{app:asymptotic_WF}
\end{equation}

Then we define the Fourier transform of $\dot{F}(\tau,\theta)$ with respect to the retarded time $\tau$,
\begin{equation}
\hat{\dot{F}}(\omega,\theta)=\int d\tau\,\dot{F}(\tau,\theta)e^{-i\omega\tau}\,.
\end{equation}

This integral is best done by inverting the function $\bar{x}(\tau)$ to $\tau(\bar{x})$ and switching the integration variable,
\begin{equation}
\int_{-\infty}^{+\infty}d\tau\,\frac{d \bar{x}}{d\tau}(\tau)I_m^{D,-1}(\bar{x}(\tau))e^{-i\omega\tau}=\int_{+\infty}^{-\infty}d\bar{x}\, I_m^{D,-1}(\bar{x})e^{-i\omega\tau(\bar{x})}=-\int dx\, I_m^{D,-1}(x)e^{-i\omega\tau(x)}\,.\label{x_int}
\end{equation}

The $x$ integration in Eq.~\eqref{x_int} is essentially
\begin{equation}
\int dx\, I_m^{D,-1}(x)e^{ixs}\,,\qquad s=\omega\rho'\sin\theta\,.
\end{equation}

From the definition in Eq.~\eqref{app:def_I}, this is
\begin{eqnarray}
&&\frac{\Omega_{D-4}}{(2\pi)^{\frac{D-2}{2}}}\int dx\,e^{ixs}\int_{-1}^{1} dy\,\Lambda_m(y)(1-y^2)^{\frac{D-4}{2}-\frac{1}{2}}\delta^{\left(\frac{D-2}{2}\right)}(y-x)\\
&=&\frac{\Omega_{D-4}}{(2\pi)^{\frac{D-2}{2}}}\int dx\,e^{ixs}\int_{-1}^{1} dy\,\lambda_m\partial^{(m)}(1-y^2)^{\frac{D-4}{2}+m-\frac{1}{2}}\delta^{\left(\frac{D-2}{2}\right)}(y-x)\\
&=&\frac{\Omega_{D-4}}{(2\pi)^{\frac{D-2}{2}}}(is)^{\frac{D-2}{2}}\lambda_m^{-1}\int_{-1}^{1} dy\,\partial^{(m)}(1-y^2)^{\frac{D-4}{2}+m-\frac{1}{2}}e^{iys}\\
&=&\frac{\Omega_{D-4}}{(2\pi)^{\frac{D-2}{2}}}(is)^{\frac{D-2}{2}+m}(-1)^m\lambda_m^{-1}\int_{-1}^{1} dy\,(1-y^2)^{\frac{D-4}{2}+m-\frac{1}{2}}e^{iys}\,,\label{poisson}
\end{eqnarray}
where
\begin{equation}
\lambda_0=1\,,\qquad \lambda_1=-(D-3)^{-1}\,,\qquad \lambda_2=(D-3)^{-1}(D-1)^{-1}\,.
\end{equation}

One recognizes in Eq.~\eqref{poisson} the Poisson integral representation of the Bessel function of the first kind,
\begin{equation}
J_\nu(x)=\frac{1}{\sqrt{\pi}\Gamma\left(\nu+\frac{1}{2}\right)}\left(\frac{x}{2}\right)^\nu\int_{-1}^1 dz\, (1-z^2)^{\nu-\frac{1}{2}}e^{izx}\,.
\end{equation}

Thus, inserting the volume of the $n$-sphere,
\begin{equation}
\Omega_n=\frac{2\pi^{\frac{n+1}{2}}}{\Gamma\left(\frac{n+1}{2}\right)}\,,
\end{equation}
we get
\begin{equation}
\int dx\, I_m^{D,-1}(x)e^{ixs}=i^{\frac{D-2}{2}+m}sJ_{\frac{D-4}{2}+m}(s)\,.
\end{equation}

Finally, 
\begin{equation}
\hat{\dot{F}}(\omega,\theta)=-i^{\frac{D-2}{2}+m}\omega\int_0^\infty d\rho\,\rho^{\frac{D-2}{2}}J_{\frac{D-4}{2}+m}(\omega\rho\sin\theta)\hat{S}\left(\omega\frac{1+\cos\theta}{2},\omega\frac{1-\cos\theta}{2};\rho\right)\,,\label{app:initial_volume}
\end{equation}
where
\begin{equation}
\hat{S}(x,y;\rho)=\frac{1}{2}\int du\int dv\, e^{-iux}e^{-ivy}\,S(u,v,\rho)\,.\label{app:S_hat}
\end{equation}

$\hat{S}$ is a double Fourier transform of the source with respect to $u$ and $v$. The $\rho$ integral can actually be cast as a $(D-2)$-dimensional Hankel transform, with argument $\omega\sin\theta$. This is not surprising since the Hankel transform arises whenever a Fourier transform is made of a function with spherical symmetry.

\section{Factorisation of the angular dependence}
\label{app:factorisation}
Now we shall see how the angular dependence factorises. Continuing from Eq.~\eqref{app:initial_volume}, we make the following transformations
\begin{equation}
\rho\rightarrow R\equiv\frac{1}{2}\omega\rho\sin\theta\,,\qquad \omega\rightarrow\Omega\equiv\omega^{D-3}\left(\frac{\sin\theta}{2}\right)^{D-4}\frac{1-\cos\theta}{2}\,,\label{transformation}
\end{equation}
and define a new transformed waveform
\begin{equation}
\hat{\mathcal{F}}(\Omega,\theta)\equiv\sqrt{\omega'(\Omega)}\hat{\dot{F}}(\omega(\Omega),\theta)\,,
\end{equation}
such that
\begin{equation}
\int d\omega\, |\hat{\dot{F}}(\omega,\theta)|^2=\int d\Omega\, |\hat{\mathcal{F}}(\Omega,\theta)|^2\,.
\end{equation}

We get
\begin{eqnarray}
\hat{\mathcal{F}}(\Omega,\theta)&=&-\frac{i^{\frac{D-2}{2}+m}}{\sqrt{D-3}}\left(\frac{2}{\sin\theta}\right)^2\left(\frac{1-\cos\theta}{2}\right)^{\frac{1}{2}}\times\Omega^{-1}\\
&\times&\int_0^\infty dR\,R^{\frac{D-2}{2}}J_{\frac{D-4}{2}+m}(2R)\hat{S}\left(\omega\frac{1+\cos\theta}{2},\omega\frac{1-\cos\theta}{2};\frac{2R}{\omega\sin\theta}\right)_{\omega=\omega(\Omega)}\,.
\end{eqnarray}

Now the $CL$ symmetry comes in handy. From Chapter~\ref{ch:analytics}, we know that if $F$ is $O(k)$,\footnote{Note that here $F$ is either $E$ or $H$ so $N_u=N_v=0$ in Eq.~\eqref{def_s}.}
\begin{equation}
S(u,v,\rho)=\rho^{-2k(D-3)-2}\,s^{(k)}(p,q)\,,
\end{equation}

Therefore, and using $dudv=\rho^2dpdq$,
\begin{equation}
\hat{S}(x,y;\rho)=\frac{1}{2}\rho^{-2k(D-3)}e^{-iy\Phi(\rho)}\int dp\int dq\,s^{(k)}(p,q)\,e^{-ixq\rho^{D-2}}\,e^{-iyp\rho^{-(D-4)}}\,.
\end{equation}

Now we have
\begin{eqnarray}
\rho^{-2k(D-3)}&=&(\omega\sin\theta)^{2k(D-3)}(2R)^{-2k(D-3)}\,,\\
y\Phi(\rho)&=&\Omega\Phi(R)\,,\\
x\rho^{D-2}&=&\Omega^{-1}R^{D-2}\,,\\
y\rho^{-(D-4)}&=&\Omega R^{-(D-4)}\,.
\end{eqnarray}

Thus,
\begin{eqnarray}
&&\hat{S}\left(\omega\frac{1+\cos\theta}{2},\omega\frac{1-\cos\theta}{2};\frac{2R}{\omega\sin\theta}\right)=\nonumber\\
&=&\frac{1}{2}\frac{(\omega\sin\theta)^{2k(D-3)}}{(2R)^{2k(D-3)}}\int dp\int dq\,s^{(k)}(p,q)\,e^{-iq\frac{R^{D-2}}{\Omega}}\,e^{-i\Omega\left(\Phi(R)+\frac{p}{R^{D-4}}\right)}\,,
\end{eqnarray}
and finally,
\begin{eqnarray}
\hat{\mathcal{F}}(\Omega,\theta)&=&-\frac{i^{\frac{D-2}{2}+m}}{\sqrt{D-3}}\times\left(\frac{1+\cos\theta}{1-\cos\theta}\right)^{k-1}\left(\frac{2}{1-\cos\theta}\right)^{\frac{3}{2}}\times\Omega^{2k-1}\times\nonumber\\
&\times&\frac{1}{2}\int_0^\infty dR\,R^{\frac{D-2}{2}-2k(D-3)}J_{\frac{D-4}{2}+m}(2R)\times\nonumber\\
&\times&\int dp\int dq\,s^{(k)}(p,q)\,e^{-iq\frac{R^{D-2}}{\Omega}-i\Omega\left(\Phi(R)+\frac{p}{R^{D-4}}\right)}\,.\label{app_transformed_F}
\end{eqnarray}

At this point, the $p,q$ coordinates are not necessarily advantageous since the $R$ integration cannot be done as it is. So we might as well go back to
\begin{equation}
u=qR^{D-2}\,,\qquad v=pR^{-(D-4)}+\Phi(R)\,,
\end{equation}
and rewrite Eq.~\eqref{app_transformed_F} as
\begin{eqnarray}
\hat{\mathcal{F}}(\Omega,\theta)&=&-\frac{i^{\frac{D-2}{2}+m}}{\sqrt{D-3}}\left(\frac{2}{1-\cos\theta}\right)^{\frac{3}{2}}\left(\frac{1+\cos\theta}{1-\cos\theta}\right)^{k-1}\times\\
&\times&\Omega^{2k-1}\int_0^\infty dR\,R^{\frac{D-2}{2}}J_{\frac{D-4}{2}+m}(2R)\hat{S}(\Omega^{-1},\Omega,R)\,.
\end{eqnarray}

This is Eq.~\eqref{Omega_form}.

\chapter{Gauge fixing of initial data}
\label{app:gauge}

This appendix contains details on the transformation of the initial data to de Donder gauge. It mainly supplements Chapter~\ref{ch:surface}. 

\section{General procedure}
\label{app:procedure}
Following the transformation in Eq.~\eqref{gauge_x}, we need to find $\xi^{(k)\mu}$ order by order, such that Eq.~\eqref{deDonderV2} is obeyed. The basic procedure is as follows:

\begin{enumerate}
\item
Find the new de Donder metric perturbations after the coordinate transformation in Eq.~\eqref{gauge_x} ($\mathcal{L}$ denotes a Lie derivative)
\begin{eqnarray}\label{app:gaugeTransfOrderbyO}
h^{(1)}_{\mu\nu}&\rightarrow&h^{(1)}_{\mu\nu}+\mathcal{L}_{\xi^{(1)}}\eta_{\mu\nu}\nonumber\\
h^{(2)}_{\mu\nu}&\rightarrow&h^{(2)}_{\mu\nu}+2\mathcal{L}_{\xi^{(1)}}h^{(1)}_{\mu\nu}+\left(\mathcal{L}_{\xi^{(1)}}\mathcal{L}_{\xi^{(1)}}+\mathcal{L}_{\xi^{(2)}}\right)\eta_{\mu\nu}\\
&\ldots&\nonumber\\
h^{(k)}_{\mu\nu}&\rightarrow&h^{(k)}_{\mu\nu}+{ k\choose {k-1}}\mathcal{L}_{\xi^{(1)}}h^{(k-1)}_{\mu\nu}+{ k\choose {k-2}}\left(\mathcal{L}_{\xi^{(1)}}\mathcal{L}_{\xi^{(1)}}+\mathcal{L}_{\xi^{(2)}}\right)h^{(k-2)}_{\mu\nu}+\ldots \nonumber 
\end{eqnarray}
\item Insert them in the gauge condition, Eq.~\eqref{deDonderV2}.
\item Use the wave equation to eliminate $u$-derivatives in favour of derivatives along the initial surface.
\item Now the gauge conditions are (schematically) in the form 
\begin{equation} \label{xiequations}
\mathcal{D}\xi^{(k)}_\mu=\zeta_\mu(v,x^i)\leftarrow \zeta_\mu\left[\xi^{(j<k)}_\alpha,\partial h^{(j\leq k)}_{\alpha\beta}\right]\; ,
\end{equation}
where $\mathcal{D}$ is some differential operator and $\zeta_\mu$ a known function of $(v,x^i)$, which is computed at $u=0^+$ from lower order gauge transformation vectors $\xi^{(m<n)}_\alpha$ and derivatives of the initial conditions along the initial surface (no $u$-derivatives).
\item Assume an ansatz
\begin{equation}
\xi_\mu^{(k)}(u,v,x^i)=\xi_\mu^{(k,0)}(v,x^i)+u\,\xi_\mu^{(k,1)}(v,x^i)+\dots \label{xi_ansatz}
\end{equation}
to solve~\eqref{xiequations} and find the functions $\xi_\mu^{(k,0)}(v,x^i)$ and $\xi_\mu^{(k,1)}(v,x^i)$.
\item Finally, insert those in Eqs.~\eqref{app:gaugeTransfOrderbyO}, together with Eqs.~\eqref{initial_data}, to obtain explicit expressions for the initial conditions in de Donder gauge.
\end{enumerate}

\section{First order}
\label{app:first_order}
The first-order metric perturbation transforms as
\begin{equation}
h_{\mu\nu}^{(1)}\rightarrow h_{\mu\nu}^{(1)}+\xi^{(1)}_{\mu,\nu}+\xi^{(1)}_{\nu,\mu}\,.
\end{equation}

Then the gauge condition in Eq.~\eqref{deDonderV2} becomes (dropping the superscript)
\begin{equation}
\Box \xi_{\alpha,v}=\frac{1}{2}h_{,\alpha v}-h^\beta_{\phantom{\beta}\alpha,\beta v}\,.\label{app_gauge_cond}
\end{equation}

To eliminate the $u$-derivative, we use the wave equation,
\begin{equation}
4h_{v\alpha,uv}=h_{v\alpha,ii}+\Box\xi_{v,\alpha}+\Box\xi_{\alpha,v}\,.\label{app_wave_eq}
\end{equation}

Finally, we get three equations for $\xi^{(1)}$,
\begin{equation}
\Box\xi_{[\alpha,v]}=\frac{1}{2}h_{,\alpha v}+2h_{u\alpha,vv}-h_{i\alpha,iv}+\frac{1}{2}h_{v\alpha,ii}\,.\label{app_3_eqs}
\end{equation}

$\alpha=v$ yields an identity. $\alpha=j$ reads
\begin{eqnarray}
\Box\xi_{[j,v]}&=&\frac{1}{2}h_{,j v}+2h_{uj,vv}-h_{ij,iv}+\frac{1}{2}h_{vj,ii}\nonumber\\
&=&0\,.
\end{eqnarray}

For $\alpha=u$, we need to eliminate the $u$-derivative from the trace. Again from the wave equation,
\begin{equation}
h_{,uv}=\frac{1}{4}h_{,ii}+\frac{1}{2}\Box\xi^{\beta}_{\phantom{\beta},\beta}\,,
\end{equation}
so we get
\begin{eqnarray}
\Box\left(\xi_{u,v}-\frac{1}{4}\xi_{i,i}\right)&=&\frac{1}{8}h_{,ii}+2h_{uu,vv}-h_{iu,iv}+\frac{1}{2}h_{vu,ii}\nonumber\\
&=&\frac{(D-2)(D-3)}{2}\left(\frac{\Phi'}{\rho}\right)^2\theta(v-\Phi)\,.
\end{eqnarray}

With the ansatz in Eq.~\eqref{xi_ansatz}, one possible solution is $\xi_\mu^{(1,0)}=0$ and
\begin{equation}
\xi_v^{(1,1)}=\xi_i^{(1,1)}=0\,,\qquad \xi_u^{(1,1)}=-\frac{(D-2)(D-3)}{16}\left(\frac{\Phi'}{\rho}\right)^2(v-\Phi)^2\theta(v-\Phi)\,.
\end{equation}

Therefore, the only component that changes is
\begin{equation}
h_{uu}^{(1)}\rightarrow h_{uu}^{(1)}-\frac{(D-2)(D-3)}{2}h(v,\rho)^2\,.
\end{equation}
In particular, the radiative components do not change.

The term $h(v,\rho)^2$ above makes it look like a second-order correction. However, when written in $(p,q,\rho)$ coordinates, the power of $\rho$ is the same as that of the original $h_{uu}^{(1)}$ and agrees with Eq.~\eqref{eq:separation_rho} for $k=1$ and $N_u=2, N_v=0$.

\section{Second order}
\label{app:second_order}
At second order, the first step is to compute the transformation induced by the previously computed $\xi^{(1)}$. From Eqs.~\eqref{app:gaugeTransfOrderbyO}, we must compute
\begin{equation}
2\mathcal{L}_{\xi^{(1)}}h^{(1)}_{\mu\nu}+\mathcal{L}_{\xi^{(1)}}\mathcal{L}_{\xi^{(1)}}\eta_{\mu\nu}\,,
\end{equation}
given the only non-vanishing components of $\xi^{(1)}$,
\begin{equation}
\xi^{(1)}_u=u\xi^{(1,1)}_u\,,\qquad \xi^{(1)v}=-2\xi^{(1)}_u=-2u\xi^{(1,1)}_u\,.
\end{equation}

The contribution from the first term is
\begin{equation}
-4u h_{\mu\nu,v}^{(1)}\xi_u^{(1,1)}\,,
\end{equation}
while that of the second is
\begin{equation}
-4u\xi^{(1,1)}_u\xi^{(1)}_{(\mu,\nu)v}-4\left(u\xi^{(1,1)}_u\right)_{,(\mu}\xi^{(1)}_{\nu),v}\,.
\end{equation}

All these vanish on $u=0$, so there is no contribution from $\xi^{(1)}$. For the second-order gauge transformation,
\begin{equation}
h_{\mu\nu}^{(2)}\rightarrow h_{\mu\nu}^{(2)}+\xi^{(2)}_{\mu,\nu}+\xi^{(2)}_{\nu,\mu}\,,
\end{equation}
we proceed in a manner similar to the first-order case. The gauge condition is the same, Eq.~\eqref{app_gauge_cond}, but Eq.~\eqref{app_wave_eq} acquires a new term from the source $T_{\mu\nu}$,
\begin{equation}
4h_{v\alpha,uv}=h_{v\alpha,ii}+\Box\xi_{v,\alpha}+\Box\xi_{\alpha,v}+T_{\alpha v}\,,
\end{equation}
which also modifies Eq.~\eqref{app_3_eqs} to
\begin{equation}
\Box\xi_{[\alpha,v]}=\frac{1}{2}h_{,\alpha v}+2h_{u\alpha,vv}-h_{i\alpha,iv}+\frac{1}{2}h_{v\alpha,ii}\,+\frac{1}{2}T_{\alpha v}.
\end{equation}

Again, $\alpha=v$ yields an identity, hence we choose $\xi^{(2,1)}_v=0$ (and $\xi^{(2,0)}_\mu=0$ as before). Then $\alpha=j$ and $\alpha=u$ yield equations of the form
\begin{eqnarray}
\xi^{(2,1)}_{j,vv}&=&f_j(v,x^i)\,,\\
\xi^{(2,1)}_{u,vv}&=&f_u(v,x^i)\,,
\end{eqnarray}
which can be integrated to find $\xi^{(2,1)}_{j}$ and $\xi^{(2,1)}_{u}$. There is no need for the explicit solution, since the radiative components, $h^{(2)}_{ij}$, are unaffected.

\chapter{Surface terms and late time tails}
\label{app:surface_n_tails}

This appendix contains details on the simplification of the computation of surface terms, i.e. the linear contribution from the initial data to the asymptotic metric functions, and on the late time behaviour of both surface and volume terms. It mainly supplements Chapter~\ref{ch:surface}. 

\section{Computation of surface terms}
\label{app:surface}
In this section we specialise the results of Appendix~\ref{app:asympWF} to the linear contribution from the initial data. The source for the surface terms is
\begin{equation}
S^{(k)}(u,v,\rho)=2\delta(u)\partial_vF^{(k)}(0,v,\rho)\,,\qquad F^{(k)}(0,v,\rho)=f_0^{(k)}(\rho)(v-\Phi(\rho))^k\theta(v-\Phi(\rho))\,,\label{S_surf}
\end{equation}
where $k=\{1,2\}$ and $f_0^{(k)}(\rho)\propto\rho^{-k(D-2)}$ for $E$ and $H$. Then, using the definition in Eq.~\eqref{app:S_hat}, we get
\begin{equation}
\hat{S}(\Omega^{-1},\Omega;R)=\frac{k!}{(i\Omega)^k}f_0^{(k)}(R)e^{-i\Omega\Phi(R)}\,.
\end{equation}

Inserting in Eq.~\eqref{app:initial_volume},
\begin{eqnarray}
\hat{\mathcal{F}}^{(k)}(\Omega)&=&\frac{k!i^{\frac{D+2}{2}+m-k}}{\sqrt{D-3}}\Omega^{k-1}\int_0^\infty dR\,R^{\frac{D-2}{2}}J_{\frac{D-4}{2}+m}(2R)f_0^{(k)}(R)e^{-i\Omega\Phi(R)}\,.\label{app_Omega}
\end{eqnarray}

As explained in Chapter~\ref{ch:surface}, we Fourier transform again to a new time coordinate $t$ (see discussion in Section~\ref{analytical_surface}),
\begin{equation}
\mathcal{F}(t)=\frac{1}{2\pi}\int d\Omega\, \hat{\mathcal{F}}(\Omega) e^{i\Omega t}\,.
\end{equation}

To perform this integral, we change the integration variable from $R$ to $T=\Phi^{-1}(R)$, i.e.
\begin{equation}
\int_0^\infty dR\sim \int_{\Phi(0)}^{\Phi(\infty)} dT\,\Phi'(R)^{-1}\,,\qquad R=\Phi^{-1}(T)\,.
\end{equation}

Then the $\Omega$ integral is
\begin{equation}
\frac{1}{2\pi}\int d\Omega\, \Omega^{k-1} e^{-i\Omega(T-t)}=i^{k-1}\delta^{(k-1)}(T-t)\,,
\end{equation}
yielding
\begin{eqnarray}
\mathcal{F}(t)&=&\frac{k!i^{\frac{D}{2}+m}}{\sqrt{D-3}}\int_{\Phi(0)}^{\Phi(\infty)} dT\,\Phi'(R)^{-1}R^{\frac{D-2}{2}}J_{\frac{D-4}{2}+m}(2R)f_0^{(k)}(R)\delta^{(k-1)}(T-t)\,,\nonumber\\
&=&\frac{(-1)^kk!i^{\frac{D-4}{2}+m}}{\sqrt{D-3}}\left[\frac{1}{\Phi'(R)}\frac{d}{dR}\right]^{k-1}\left(\frac{R^{\frac{D-2}{2}}f_0^{(k)}(R)}{\Phi'(R)}J_{\frac{D-4}{2}+m}(2R)\right)\,,
\end{eqnarray}
where $R=\Phi^{-1}(t)$.

From the initial data in Eq.~\eqref{initial_data}, we read
\begin{eqnarray}
&E^{(1)}:&k=1\,,\qquad m=2\,,\qquad f_0(\rho)=-\frac{\Phi'(\rho)}{\rho}=-\frac{2}{\rho^{D-2}}\,,\\
&E^{(2)}:&k=2\,,\qquad m=2\,,\qquad f_0(\rho)=-(D-4)\frac{\Phi'(\rho)^2}{4\rho^2}=-\frac{(D-4)}{\rho^{2(D-2)}}\\
&H^{(2)}:&k=2\,,\qquad m=0\,,\qquad f_0(\rho)=(D-3)\frac{\Phi'(\rho)^2}{4\rho^2}=\frac{(D-3)}{\rho^{2(D-2)}}\,,
\end{eqnarray}
and obtain
\begin{eqnarray}
\mathcal{F}_{E^{(1)}}(\Phi(R))&=&\frac{-i^{\frac{D}{2}}}{\sqrt{D-3}}\,R^{\frac{D-4}{2}}J_{\frac{D}{2}}(2R)\,,\\
\mathcal{F}_{E^{(2)}}(\Phi(R))&=&i^{\frac{D}{2}}\frac{D-4}{\sqrt{D-3}}\,R^{\frac{D-6}{2}}J_{\frac{D}{2}+1}(2R)\,,\\
\mathcal{F}_{H^{(2)}}(\Phi(R))&=&i^{\frac{D-4}{2}}\sqrt{D-3}\,R^{\frac{D-8}{2}}\left(J_{\frac{D-4}{2}}(2R)+RJ_{\frac{D-2}{2}}(2R)\right)\,.
\end{eqnarray}

\section{The contribution to the inelasticity}
\label{app:R_integrals}
The contribution of $\mathcal{F}(t)$ to the inelasticity is in Eq.~\eqref{epsn}. Recall that $\dot{\mathcal{E}}^2\propto(\dot{E}+\dot{H})^2$.

At $N=1$ there is only one term, $\dot{E}^{(1)}\dot{E}^{(1)}$. The relevant integral is
\begin{equation}
\int_0^\infty dR\, \frac{1}{R}J_{\frac{D}{2}}(2R)^2=\frac{1}{D}\,,
\end{equation}
which converges for all $D$.

At $N=2$ there are two terms. In $\dot{E}^{(1)}\dot{E}^{(2)}$ the relevant integral is
\begin{equation}
\int_0^\infty dR\, \frac{1}{R^2}J_{\frac{D}{2}}(2R)J_{\frac{D}{2}+1}(2R)=\frac{2}{D}\frac{1}{D+2}\,.
\end{equation}
This integral converges for $D>0$.

In $\dot{E}^{(1)}\dot{H}^{(2)}$ the relevant integral is
\begin{equation}
\int_0^\infty dR\, \frac{1}{R^3}J_{\frac{D}{2}}(2R)\left(J_{\frac{D-4}{2}}(2R)+R\, J_{\frac{D-2}{2}}(2R)\right)=\frac{2}{D}\frac{1}{D-4}\,.
\end{equation}
This integral converges for $D>4$.

At $N=3$ there are three terms. In $\dot{E}^{(2)}\dot{E}^{(2)}$ the relevant integral is
\begin{equation}
\int_0^\infty dR\, \frac{1}{R^3}J_{\frac{D}{2}+1}(2R)^2=\frac{8}{D}\frac{1}{(D+2)(D+4)}\,.
\end{equation}
This integral converges for $D>0$.

In $\dot{E}^{(2)}\dot{H}^{(2)}$ the relevant integral is
\begin{equation}
\int_0^\infty dR\,\frac{1}{R^4}J_{\frac{D}{2}+1}(2R)\left(J_{\frac{D-4}{2}}(2R)+R\,J_{\frac{D-2}{2}}(2R)\right)=\frac{4}{D}\frac{1}{(D-4)(D+2)}\,.
\end{equation}
This integral converges for $D>4$.

Finally, in $\dot{H}^{(2)}\dot{H}^{(2)}$ the relevant integral is
\begin{equation}
\int_0^\infty dR\, \frac{1}{R^5}\left(J_{\frac{D-4}{2}}(2R)+R\,J_{\frac{D-2}{2}}(2R)\right)^2=\frac{8}{D}\frac{1}{(D-4)(D-8)}\,.
\end{equation}
This integral converges for $D>8$.

These results are summarised in Table~\ref{TabSurface}.

\section{Late time tails}
\label{app_tails}
In this section we compute the late time behaviour of the waveforms, both for surface and volume terms. Following the strategy of Section~\ref{app:limit_Fourier} for the surface terms, we do the inverse Fourier transform from $\Omega$ to the new time coordinate $t$. The relevant integral is
\begin{equation}
\frac{1}{2\pi}\int d\Omega\, \Omega^{2k-1}e^{-i\Omega P-iQ\Omega^{-1}}=(-i\partial_t)^{2k-1}\frac{1}{2\pi}\int d\Omega\, e^{-i\Omega P-iQ\Omega^{-1}}\,.
\end{equation}
where
\begin{equation}
P=\Phi(R)+pR^{-(D-4)}-t\,,\qquad Q=qR^{D-2}\,.
\end{equation}

This integral can be made and equals
\begin{equation}
i^{-2k-1}\partial_t^{(2k)}\,\theta(X)J_0\left(2\sqrt{X}\right)\,,\qquad X\equiv PQ\,,
\end{equation}

Therefore, we get
\begin{equation}
\mathcal{F}(t,\theta)\propto \int dp\int dq\,s^{(k)}(p,q)\, G^k_m(p,q;t)\,,
\end{equation}
where $G^k_m(p,q;t)$ is essentially the asymptotic (i.e. for an observer at null infinity) Green's function in $p,q$ coordinates and for the new time $t$,
\begin{equation}
G^k_m(p,q;t)=\left(\frac{d}{dt}\right)^{2k}\int_0^\infty dR\,R^{\frac{D-2}{2}-2k(D-3)}J_{\frac{D-4}{2}+m}(2R)\theta(X)J_0\left(2\sqrt{X}\right)\,,\label{G_pqt}
\end{equation}
where
\begin{eqnarray}
X\equiv PQ=qR^{D-2}\left(\left(p+\frac{2}{D-4}\right)R^{-(D-4)}-t\right)\,.
\end{eqnarray}

The action of the $2k$ derivatives distributes as
\begin{eqnarray}
&&\left(qR^{D-2}\right)^{2k}\sum_{j=0}^{2k}{2k\choose j}\frac{(-1)^j}{j!X^{\frac{j}{2}}}\delta^{(2k-j-1)}(X)J_j\left(2\sqrt{X}\right)\\
&=&\left[\sum_{j=0}^{2k-1}{2k\choose j}\frac{(-qR^{D-2})^j}{j!}\delta^{(2k-j-1)}(P)\right]+\frac{\left(qR^{D-2}\right)^{2k}}{(2k)!X^k}\theta(X)J_{2k}\left(2\sqrt{X}\right)\,.
\end{eqnarray}

The $R$ integral in (\ref{G_pqt}) can be readily done for all but the very last term. So the contribution of the series inside the square brackets for $G^k_m(p,q;t)$ is
\begin{equation}
\sum_{j=0}^{2k-1}{2k\choose j}\frac{q^j}{\left(\frac{D-4}{2}p+1\right)^{2k-j}}\left(\frac{1}{\Phi'(T)}\frac{d}{dR}\right)^{2k-j-1}\left(\frac{R^{\frac{D-2}{2}-2k(D-3)+j(D-2)}}{\Phi'(R)}J_{\frac{D-4}{2}+m}(2R)\right)\,,\label{series_G_pqt}
\end{equation}
where
\begin{equation}
R=\left\{
\begin{array}{ll}
\Phi^{-1}(t-p)\ , &  D=4\  \vspace{2mm}\\
\displaystyle{\Phi^{-1}\left(t\left(\frac{D-4}{2}p+1\right)^{-1}\right)}\ , & D>4\,
\end{array} \right. \ .
\end{equation} 

Late times $t>>1$ correspond to $R<<1$. Thus we can study the late time tails by making a series expansion of Eq.~\eqref{series_G_pqt} for $R\simeq0$.

The most relevant term is $j=0$, since it has the highest powers of $R$ (indeed, for $k=1$ this is the only surviving term because of the $\delta(q)$ in $s^{(k)}(p,q)$). The waveform will only be square-integrable if its tail decays at least with $t^{-1}$. Thus we conclude that $E^{(1)}$ and $E^{(2)}$ are square integrable for all $D\geq4$, but $H^{(2)}$ is not: it grows exponentially in $D=4$ and with $t^{-\alpha}$ for higher $D$. The condition $\alpha\geq1$ is only true for $D>8$.

\chapter{Asymptotic expansions for the Green's function}
\label{app:expansions}

In this appendix we study the behaviour of the Green's function for an observer at null infinity, when the source point is also near null infinity (Section~\ref{app:behaviour_scri+}) and when it is close to the axis (Section~\ref{app:behaviour_axis}). This is important because a rapid growth or decay can induce numerical instabilities and loss of precision. It mainly supplements Chapter~\ref{ch:second_order} (the second-order calculation), but we present results valid to all orders.

\section{Source point near null infinity}
\label{app:behaviour_scri+}
When the source point is near null infinity, $\eta'\rightarrow\infty$ or, equivalently, $Q'\rightarrow(D-2)(D-4)$ (for fixed $\xi'$). Thus a convenient small parameter is
\begin{equation}
\varepsilon'\equiv\frac{Q'-(D-2)(D-4)}{2(D-2)}\,.
\end{equation}

Then we have
\begin{equation}
{\Delta'}^{\frac{1}{D-3}}\varepsilon'=\left\{
\begin{array}{ll} 1+2\varepsilon'-2\varepsilon'\ln4\varepsilon'+2(\xi'-\xi)\varepsilon'+O(\varepsilon'^2)\ , &  D=4\ \vspace{5mm} \\
\displaystyle{	\frac{1+(D-4)^{\frac{1}{D-3}}\xi'}{1+(D-4)^{\frac{1}{D-3}}\xi}+O(\varepsilon')		}\ , & D>4\,
\end{array} \right.\,,
\end{equation}
and $x_\star$ becomes
\begin{equation}
x_\star=f_0(\hat y)-\varepsilon' f_1(\hat y)+O(\varepsilon'^2)\,,\label{eq_x_f}
\end{equation}
where
\begin{eqnarray}
f_0(\hat y)&=&
\left\{
\begin{array}{ll} 	\dfrac{1+\hat y^2}{2\hat y}\,	,	\, &  D=4\ \vspace{5mm} \\
\displaystyle{		\dfrac{1+\hat y^{2(D-3)}}{2\hat y^{D-3}}		}\ , & D>4\,
\end{array} \right.\,.\\
f_1(\hat y)&=&
\left\{
\begin{array}{ll} 	\dfrac{\xi-\xi'+\ln 4y}{\hat y}	\,,	\, &  D=4\ \vspace{5mm} \\
\displaystyle{		\frac{-\Xi\hat y^{2(D-3)}+(D-3)\Xi^{-1}\hat y^{D-4}}{(D-4)\hat y^{D-3}}	}\ , & D>4\,
\end{array} \right.\,,
\end{eqnarray}
and we have defined a short-hand
\begin{equation}
\Xi\equiv\frac{1+(D-4)^{\frac{1}{D-3}}\xi'}{1+(D-4)^{\frac{1}{D-3}}\xi}\,.
\end{equation}

In the limit $\varepsilon'\rightarrow0$ the integration domain collapses to the point $\hat y=1\Leftrightarrow x_\star=1$. To show that the integral remains finite, let us define new variables $w$ and $w_\star$,
\begin{equation}
\hat y=1+\sqrt{\varepsilon'}w\,,\qquad x_\star=1-\varepsilon' w_\star\,.\label{w_x_scri+}
\end{equation}

Then, expanding Eq.~\eqref{eq_x_f} around $\hat y=1$,
\begin{equation}
w^2=\frac{2}{f_0''(1)}\left(f_1(1)-w_\star\right)\,,\qquad 0\leq w_\star\leq f_1(1)\,.
\end{equation}

This gives a finite domain for the integration (either in $w$ or $w_\star$), and the shrinking measure $d\hat y$ cancels the divergent integrand,
\begin{equation}
\int\,d\hat y\,\hat y^\alpha I_m^{D,0}(x_\star)\sim\int \frac{d\hat y}{\sqrt{1-x^2_\star}}\sim\int \frac{\sqrt{\varepsilon'}dw_\star}{\sqrt{2w_\star\varepsilon'}}\sim O(1)\,,
\end{equation}

Thus we conclude that there is no extra suppression factor in $G_m^k(\Delta',Q')$ as $\eta'\rightarrow\infty$.

\section{Source point near the axis}
\label{app:behaviour_axis}
The axis $\rho'=0$ corresponds to $\eta'=\xi'$, so we take the limit $\xi'\rightarrow\eta'$ for fixed $\eta'\leq\xi$. A convenient small parameter for this expansion is $\varepsilon\equiv Q'^{-\frac{1}{2}}$. Indeed, in $D=4$,
\begin{eqnarray}
\xi'-\eta'&=&-2\frac{1+Q'}{Q'}+\ln\frac{\sqrt{1+Q'}-1}{\sqrt{1+Q'}+1}\,,\\
&\simeq&-4\varepsilon+O(\varepsilon^2)\,,
\end{eqnarray}
and, in $D>4$,
\begin{eqnarray}
\frac{1+(D-4)^{\frac{1}{D-3}}\xi'}{1+(D-4)^{\frac{1}{D-3}}\eta'}&=&\frac{\sqrt{1+Q'}-D+3}{\sqrt{1+Q'}+D-3}\left(\frac{Q'}{Q'+2(1-\sqrt{1+Q'})}\right)^{\frac{1}{D-3}}\,,\\
&\simeq&1-\frac{2(D-2)(D-4)}{(D-3)}\varepsilon+O(\varepsilon^2)\,.
\end{eqnarray}

Then we solve $\xi'(P',Q')$ for $P'(\xi',Q')$ and substitute in $\Delta'$,
\begin{equation}
\Delta'=
\left\{
\begin{array}{ll} 	\dfrac{\sqrt{1+Q'}+1}{\sqrt{1+Q'}-1}-2\ln\dfrac{\sqrt{1+Q'}-1}{2}	-2(\xi-\xi')\,,	\, &  D=4\ \vspace{5mm} \\
\displaystyle{		\left(\frac{1+(D-4)^{\frac{1}{D-3}}\xi'}{1+(D-4)^{\frac{1}{D-3}}\xi}\right)^{D-3}\left(\frac{D-2}{\sqrt{1+Q'}-(D-3)}\right)^{D-3}\frac{\sqrt{1+Q'}-1}{D-4}		}\ , & D>4\,
\end{array} \right.\,.
\end{equation}

Expanding, we find, in $D=4$,
\begin{equation}
\Delta'=1-2(\xi-\eta')+\ln(4\varepsilon^2)-4\varepsilon+O(\varepsilon^2)\,,
\end{equation}
and, in $D\geq4$,
\begin{equation}
\Delta'Q'^{\frac{D-4}{2}}\simeq\frac{(D-2)^{D-3}}{D-4}\zeta^{D-3}\Lambda^{D-3}\,.
\end{equation}
where
\begin{equation}
\zeta\equiv\frac{1+(D-4)^{\frac{1}{D-3}}\eta'}{1+(D-4)^{\frac{1}{D-3}}\xi}\leq1\,,\qquad \Lambda\equiv1-\frac{(D-2)(D-4)}{(D-3)}\varepsilon+O(\varepsilon^2)\,.
\end{equation}

Then we scale the integration variable $\hat y$,
\begin{equation}
\hat y=
\left\{
\begin{array}{ll} 	\dfrac{\sqrt{Q'}}{2} z\,,	\, &  D=4\ \vspace{5mm} \\
\displaystyle{		\left(\frac{D-4}{\sqrt{Q'}} \right)^{\frac{1}{D-3}}z	}\ , & D>4\,
\end{array} \right.\,.\label{y_scaling}
\end{equation}

such that Eq.~\eqref{x_asympt} becomes
\begin{eqnarray}
\varepsilon x_\star
&\equiv&f_0(z)+\varepsilon f_1(z)+O(\varepsilon^2)\,,
\end{eqnarray}
where
\begin{eqnarray}
f_0(z)&=&
\left\{
\begin{array}{ll} 	\dfrac{-1-2(\xi-\eta')+z^2-2\ln z}{4z}\,	,	\, &  D=4\ \vspace{5mm} \\
\displaystyle{		\frac{(D-4)z^{2(D-3)}-2(D-3)\zeta^{-1}z^{D-4}+(D-2)}{2(D-2)(D-4)z^{D-3}}		}\ , & D>4\,
\end{array} \right.\,.\\
f_1(z)&=&
\left\{
\begin{array}{ll} 	-\dfrac{1}{z}	\,,	\, &  D=4\ \vspace{5mm} \\
\displaystyle{		-\dfrac{1}{\zeta z}	}\ , & D>4\,
\end{array} \right.\,.
\end{eqnarray}

\subsection{Inside the light cone}
Let us start with the case $\xi_0<\xi$. For a consistent, well-defined solution in the limit $\varepsilon\rightarrow0$, we define a new integration variable $w$ through
\begin{equation}
z=z_0+\varepsilon w\,,\label{def_w_axis_in}
\end{equation}
where $z_0$ solves $f_0(z_0)=0$. Then the equation for $x_\star$ becomes
\begin{equation}
x_\star=f'_0(z_0)w+f_1(z_0)+O(\varepsilon)\,,\label{w_x}
\end{equation}
which we can invert for $w(x_\star)$.

Then, in the integral which defines the asymptotic Green's function,
\begin{equation}
G_m^k(\Delta',Q')\propto\int_0^\infty d\hat y\, \hat y^\alpha I_m^{D,0}(x_\star)\,,\qquad \alpha\equiv\frac{D-4}{2}-(D-3)(2k+N_u-N_v)\,,\label{eqGF_I_axis}
\end{equation}
we substitute Eqs.~\eqref{y_scaling},~\eqref{def_w_axis_in} and~\eqref{w_x} and transform to an integral in $x_\star$,
\begin{equation}
G_m^k(\Delta',Q')\propto\varepsilon^\beta\int dx_\star\,I_m^{D,0}(x_\star)\left(1+\varepsilon w(x_\star)\right)^\alpha\,+\ldots\,,
\end{equation}
where
\begin{equation}
\beta=\left\{
\begin{array}{ll} 	-(\alpha+1)	\,,	\, &  D=4\ \vspace{5mm} \\
\displaystyle{		\frac{\alpha+1}{D-3}	}\ , & D>4\,
\end{array} \right.\,.
\end{equation}

Now observe that (see Eq.~\eqref{app:def_I}),
\begin{equation}
\int\,dx_\star\,x_\star^n I_m^{D,0}(x_\star)\propto\int dx_\star\,x_\star^n\delta^{\left(\frac{D-4}{2}+m\right)}(x-x_\star)\,,
\end{equation}
vanishes for $n<(D-4)/2+m$\footnote{This is obvious in even $D$ but by analytic continuation is also true for odd $D$.}. If each extra power of $x_\star$ comes with another power of $\varepsilon$, then the first non-zero term will have a power of
\begin{equation}
\beta+\frac{D-4}{2}+m\,.
\end{equation}

\subsection{On the light cone}
When $\eta'=\xi$, we have
\begin{equation}
z_0=1\,,\qquad f_0(1)=f'_0(1)=0\,,
\end{equation}
so we require a different definition for $w$, similar to Eq.~\eqref{w_x_scri+},
\begin{equation}
z=z_0+\sqrt{\varepsilon}w\,.
\end{equation}
We get
\begin{equation}
w^2=\frac{2}{D-3}(x_\star+1)\,.
\end{equation}
Thus, in Eq.~\eqref{eqGF_I_axis}, the transformation from $\hat{y}$ to $w$ yields a power of $\varepsilon$ equal to
\begin{equation}
\beta+\frac{1}{2}\,.\label{power_axis}
\end{equation}

For the second-order calculation of Chapter~\ref{ch:second_order}, this is stronger than outside the light cone. 

\chapter{List of publications}

This thesis is based on the following published works by the author:
\begin{enumerate}

\item
{\bf ``Radiation from a $D$-dimensional collision of shock waves: a remarkably simple fit formula''}
  \\{}F.~S.~Coelho, C.~Herdeiro and M.~O.~P.~Sampaio.
  \\{}\href{http://arxiv.org/abs/1203.5355}{\tt arXiv:1203.5355 [hep-th]}
  \\{}\href{http://link.aps.org/doi/10.1103/PhysRevLett.108.181102}{Phys.\ Rev.\ Lett.\  {\bf 108} (2012) 181102}

\item
{\bf ``Radiation from a $D$-dimensional collision of shock waves: Higher-order setup and perturbation theory validity''}
  \\{}F.~S.~Coelho, C.~Herdeiro, C.~Rebelo and M.~Sampaio.
  \\{}\href{http://arxiv.org/abs/1206.5839}{\tt arXiv:1206.5839 [hep-th]}
  \\{}\href{http://link.aps.org/10.1103/PhysRevD.87.084034}{Phys.\ Rev.\ D {\bf 87} (2013) 8,  084034}  

\item
{\bf ``Radiation from a $D$-dimensional collision of shock waves: proof of first order formula and angular factorisation at all orders''}
  \\{}F.~S.~Coelho, C.~Herdeiro and M.~O.~P.~Sampaio.
  \\{}\href{http://arxiv.org/abs/1410.0964}{\tt arXiv:1410.0964 [hep-th]}
  \\{}\href{http://doi.org/10.1007/JHEP12(2014)119}{JHEP\  {\bf 12} (2014) 119}

\end{enumerate}

\newpage
Other publications in international refereed journals (not including conference proceedings):

\begin{enumerate}

\item
{\bf ``Relativistic Euler's 3-body problem, optical geometry and the golden ratio''}
  \\{}F.~S.~Coelho and C.~A.~R.~Herdeiro.
  \\{}\href{http://arxiv.org/abs/0909.4413}{\tt arXiv:0909.4413 [gr-qc]}
  \\{}\href{http://dx.doi.org/10.1103/PhysRevD.80.104036}{Phys.\ Rev.\ D {\bf 80} (2009), 104036}

\item
{\bf ``Scalar graviton in $n$-DBI gravity''}
  \\{}F.~S.~Coelho, C.~Herdeiro, S.~Hirano and Y.~Sato.
  \\{}\href{http://arxiv.org/abs/1205.6850}{\tt arXiv:1205.6850 [hep-th]}
  \\{}\href{http://dx.doi.org/10.1103/PhysRevD.86.064009}{Phys.\ Rev.\ D {\bf 86} (2012), 064009}  

\item
{\bf ``$n$-DBI gravity, maximal slicing and the Kerr geometry''}
  \\{}F.~S.~Coelho, C.~Herdeiro and M.~Wang.
  \\{}\href{http://arxiv.org/abs/1301.1070}{\tt arXiv:1301.1070 [gr-qc]}
  \\{}\href{http://dx.doi.org/10.1103/PhysRevD.87.047502}{Phys.\ Rev.\ D {\bf 87} (2013),  047502}
  
  \item
{\bf ``Parametrized post-Newtonian expansion and FRW scalar perturbations in $n$-DBI gravity''}
  \\{}F.~S.~Coelho, C.~Herdeiro, S.~Hirano and Y.~Sato.
  \\{}\href{http://arxiv.org/abs/1307.4598}{\tt arXiv:1307.4598 [gr-qc]}
  \\{}\href{http://dx.doi.org/10.1103/PhysRevD.90.064040}{Phys.\ Rev.\ D {\bf 90} (2014),  064040}
  
\end{enumerate}



\cleardoublepage
\phantomsection
\addcontentsline{toc}{chapter}{Bibliography}
\bibliographystyle{myutphys}
\bibliography{tese}

\end{document}